\documentclass[prd,superscriptaddress,preprintnumbers,tightenlines,nofootinbib, eqsecnum]{revtex4-2}

\usepackage{amsmath}
\usepackage{amsfonts}
\usepackage{amssymb}
\usepackage{physics}
\usepackage{bm}
\usepackage{mathrsfs}
\usepackage{graphicx}
\usepackage[colorlinks]{hyperref}
\usepackage[usenames]{color}
\usepackage{tikz}
\usepackage{mathtools}

\definecolor{darkgreen}{rgb}{0,0.5,0}
\hypersetup{urlcolor=darkgreen}
\usepackage{empheq}
\usepackage{ulem}
\usepackage{orcidlink}
\hypersetup{
 unicode=false,          
 pdftoolbar=true,        
 pdfmenubar=true,        
 pdffitwindow=false,     
 pdfstartview={FitH},    
 pdftitle={My title},    
 pdfauthor={Author},     
 pdfsubject={Subject},   
 pdfcreator={Creator},   
 pdfproducer={Producer}, 
 pdfkeywords={keyword1} {key2} {key3}, 
 pdfnewwindow=true,      
 colorlinks=true,       
 linkcolor=red,          
 citecolor=cyan,        
 filecolor=magenta,      
 urlcolor=darkgreen,           
 linktocpage=true
}

\allowdisplaybreaks

\DeclareSymbolFontAlphabet{\mathrsfs}{rsfs}
\DeclareMathAlphabet{\mathcal}{OMS}{cmsy}{m}{n}

\newcommand{\be}{\begin{equation}}
\newcommand{\ee}{\end{equation}}
\newcommand{\bse}{\begin{subequations}}
\newcommand{\ese}{\end{subequations}}
\newcommand{\ba}{\begin{align}}
\newcommand{\ea}{\end{align}}
\newcommand{\nn}{\nonumber}
\newcommand{\p}{\partial}
\newcommand{\calO}{\mathcal{O}}
\newcommand{\di}{\mathrm{i}} 
\newcommand{\de}{\mathrm{e}} 
\newcommand{\dM}{\mathrm{M}}

\newcommand{\dI}{\mathrm{I}}
\newcommand{\dJ}{\mathrm{J}}
\newcommand{\dW}{\mathrm{W}}
\newcommand{\dX}{\mathrm{X}}
\newcommand{\dY}{\mathrm{Y}}
\newcommand{\dZ}{\mathrm{Z}}

\newcommand{\gam}{\bar{\gamma}}
\newcommand{\gamInv}{\bar{\gamma}^{-1}}

\newcommand{\Sm}{\mathcal{S}_{-}}
\newcommand{\Sp}{\mathcal{S}_{+}}
\newcommand{\bep}{\bar{\beta}_{+}}
\newcommand{\bem}{\bar{\beta}_{-}}
\newcommand{\dep}{\bar{\delta}_{+}}
\newcommand{\dem}{\bar{\delta}_{-}}

\newcommand{\chip}{\bar{\chi}_{+}}
\newcommand{\chim}{\bar{\chi}_{-}}

\makeatletter
\g@addto@macro\bfseries{\boldmath}
\makeatother

\defcitealias{BBT22}{paper~I}
\defcitealias{T24_QK}{paper~II}
\begin{document}
\interfootnotelinepenalty=10000
\title{Gravitational waves from quasielliptic compact binaries in scalar-tensor theory to one-and-a-half post-Newtonian order}

\author{David \textsc{Trestini}\,\orcidlink{0000-0002-4140-0591}}\email{david.trestini@southampton.ac.uk}
\affiliation{CEICO, Institute of Physics of the Czech Academy of Sciences, Na Slovance 2, 182 21 Praha 8, Czechia}
\affiliation{School of Mathematical Sciences and STAG Research Centre, University of Southampton, Southampton SO17 1BJ, United Kingdom}

\date{\today}

\begin{abstract}

The orbit-averaged fluxes of energy and angular momentum generated by a compact binary system of nonspinning particles are obtained in a popular class of massless scalar-tensor theories with first-and-a-half post-Newtonian (1.5PN) accuracy, i.e., 2.5PN accuracy beyond the leading $-1$PN dipolar radiation. The  secular evolution of the orbital elements (frequency and eccentricity) are then also obtained with 1.5PN accuracy.  Three technical advances were necessary to obtain these results: (i) the decomposition of the scalar dipole moment as a Fourier series at 1PN order, along with the other moments at Newtonian order, (ii) the derivation of the formula for the passage to the center-of-mass frame at 2.5PN, including a novel hereditary term arising from the contribution of the radiation, and (iii) the complete treatment of the memory-like term appearing in the angular momentum flux.
 Moreover, as a byproduct of this work, I revisit in the~Appendix the treatment \emph{in general relativity} of the memory integral appearing  at 2.5PN order in the angular momentum flux \mbox{\href{https://doi.org/10.1103/PhysRevD.80.124018}{[Phys. Rev. D \textbf{80}, 124018 (2009)]}}, and find that the spurious divergence in the initial eccentricity~$e_1$ found there is cured by a careful split between AC and DC terms.

\end{abstract}


\maketitle
\section{Introduction}

General relativity (GR) still passes with flying colors all precision astrophysical tests so far, such as solar system tests~\cite{Bertotti:2003rm,Fomalont:2003pd, Kopeikin:2005vf, Fomalont:2009vn,Williams:2004qba}, binary pulsar tests~\cite{Dgef92,Fujii:2005,DeFelice:2010aj,DoublePulsar,Kramer:2021jcw} and gravitational wave constraints~\cite{ligoscientific:2021sio}. However, it is expected that GR is only the low-energy limit of some UV complete theory of quantum gravity, and deviations to GR would first manifest themselves as some extra degrees of freedom, such as a scalar field. This could be useful in explaining phenomena such as the accelerating universe, inflation, the Hubble tension~\cite{Smith:2019ihp}, etc.

Our understanding of GR itself has also dramatically improved. Focusing on gravitational waveforms, we now have post-Newtonian waveform templates at very high 4.5PN order~\cite{bfhlt_letter,bfhlt_ii}, numerical relativity templates are now full blown~\cite{Ajith:2012az, Ferguson:2023vta, Boyle:2019kee}, gravitational self-force techniques have yielded generic waveform template at adiabatic order~\cite{Hughes:2021exa, Isoyama:2021jjd} and waveforms for circular orbits at second post-adiabatic order~\cite{Wardell:2021fyy}, and methods such as effective-one-body (EOB) \cite{Buonanno:1998gg,Ramos-Buades:2023ehm, Gamba:2023mww} or phenomenological models \cite{Pratten:2020fqn,Williams:2024twp} have been very successful at unifying these different waveforms.

In parallel, there has been a recent effort towards developing a complete template bank in alternative theories of gravity. To date, the most studied alternative theory is a class of scalar-tensor theories introduced by Damour and Esposito-Farèse~\cite{Damour:1996ke} (see Ref. \cite{MW13} for an equivalent formulation, which I will be using), which generalizes Brans-Dicke gravity~\cite{BransDicke, Fierz:1956zz}. This theory has first  been  studied within the post-Newtonian framework, focusing on the equations of motion~\cite{MW13,B18_i, B18_ii} and waveform templates for nonspinning circular orbits~\cite{Lang:2014osa, Lang:2013fna, Sennett:2016rwa, BBT22}. There has also been recent interest in studying the post-Newtonian dynamics within the EFT framework \cite{Almeida:2024uph, Diedrichs:2023foj}, in which the spinning case \cite{Almeida:2024cqz} and tidal effects \cite{Bernard:2023eul} were recently studied. These PN results have also been used to inform effective-one-body models~\cite{Julie:2017pkb, Julie:2017ucp, Julie:2022qux, Jain:2022nxs, Jain:2023fvt}.
Finallly, this class of theories has also recently been studied in the context of numerical relativity~\cite{Ma:2023sok,Corman:2024vlk}, which found good agreement with post-Newtonian templates in the region where they overlap.

The present work aims at generalizing some of the results obtained for nonspinning \emph{quasicircular} orbits in Ref.~\cite{BBT22}, henceforth referred to as \citetalias{BBT22}, to the case of nonspinning \emph{quasielliptic} orbits. It crucially relies on the recent quasi-Keplerian parametrization at 2PN order for ST theories obtained in Ref.~\cite{T24_QK}, henceforth referred to as  \citetalias{T24_QK}. The main result I will be deriving is the orbit-averaged fluxes of energy and angular momentum at 1.5PN (i.e. 2.5PN beyond the leading $-1$PN dipolar radiation), along with the secular evolution of the orbital parameters (e.g. the frequency and eccentricity) at 2.5PN. The latter result generalizes the ``chirp'' to the case of quasielliptic orbits, and is thus the main GW observable. To this end, three technical advances were needed. First, the expansion of scalar dipole moment as a Fourier series in time is computed for the first time with 1PN accuracy, along with the other required moments at Newtonian order. Secondly, the formula for the passage to the center-of-mass (CM) frame is obtained for the first time at 2.5PN order, including a novel hereditary term that arises from the contribution of the radiation to the definition of the CM frame, through the fluxes of linear momentum and CM position.  This extra contribution was discovered only very recently  in the context of  GR~\cite{BFT24}. Thirdly, the tail sector of the flux is computed for the first time in terms of enhancement functions, whose small-eccentricity expansions are also presented. Finally, the memory piece of the angular momentum flux is treated consistently, where the AC part is computed using the Fourier expansion of the scalar dipole moment, while the DC part is treated thanks to the secular evolution equations of the orbital parameters derived in Ref.~\cite{T24_QK}.

Despite the stringent constraints on this class of theories, stemming mostly from the constraints on the existence of dipolar radiation~\cite{Barausse:2016eii}, the motivation for this work is two-fold. First, future ground-based detectors such as Einstein Telescope \cite{Maggiore:2019uih} or Cosmic Explorer \cite{LIGOScientific:2016wof} will have drastically higher sensitivities than current ones, and will be able to place much more stringent constraints on deviations from GR. In this context, it is interesting to have alternative template banks for popular theories that cover the entire parameter space (masses, spins, eccentricity, tidal effects,~...), e.g., in the event that we detect an alternative signal with very small coupling constants. Moreover, it is not excluded that a subpopulation of signals exhibit GW signals that differ greatly from GR, and have thus been missed by matched filtering \cite{Magee:2023muf} --- alternative templates would be very useful in this regard. The second motivation is that this class of theories is simple enough to explore the phenomenology of extra degrees of freedom at high post-Newtonian order, and serve as a testbed for more complicated theories. For example, there has been recent interest for scalar-Gauss-Bonnet theories, which build on results in ST theories to obtain their waveforms~\cite{Shiralilou:2021mfl, Julie:2019sab, Julie:2022qux, Julie:2024fwy, Corman:2024vlk}. Moreover, more general theories, e.g. in the Horndeski class, can exhibit screening mechanisms which cannot yet be correctly treating within the post-Newtonian framework \cite{Boskovic:2023dqk, deRham:2024xxb}. Thus, I view this program as an incremental step towards the ultimate goal of obtaining waveforms for some robust future extension of general relativity.

The plan of the paper is as follows. After notational reminders, I will introduce in Sec.~\ref{sec:STtheory} the ST theory under scrutiny, and review the quasi-Keplerian parametrization at 2PN for ST theories in Sec.~\ref{sec:PK}. In Sec.~\ref{sec:Fourier}, I will go over the crucial step of decomposing the multipole moments into Fourier series. In Sec.~\ref{sec:CM}, I will discuss the passage to the CM frame, with the appearance of non-local terms first discovered in Ref.~\cite{BFT24}. The fluxes of energy and angular momentum for eccentric orbits will be derived  in Sec.~\ref{sec:flux} at 1.5PN order beyond quadrupolar radiation (i.e. 2.5PN beyond the leading $-1$PN dipolar radiation), and will be used in Sec~\ref{sec:orbital_elements} to obtain the secular evolution of the orbital elements through 2.5PN order. After a short conclusion in Sec.~\ref{sec:conclusion}, many lengthy coefficients are relegated in the Appendices. Most results are also given in machine-readable form in the Supplemental Material~\cite{SuppMaterial}. Finally, I complete \emph{in general relativity}  the memory piece of the angular momentum flux in App.~\ref{app:memory_in_GR}, curing the spurious divergence in the initial eccentricity appearing in the previous treatment of  Ref.~\cite{ABIS09}.

\subsection*{Main notations and summary of parameters}
\label{sec:not}

The convention in this work is that all stated PN orders are, by default, relative to the the Newtonian dynamics and to the standard quadrupole radiation in GR. Thus, the dominant dipole radiation enters at~$-0.5$PN order in the waveform and at~$-1$PN order in the energy flux. The QK parametrization at second post-Newtonian order computed in \citetalias{T24_QK} is therefore next-to-next-to-leading-order, while the Newtonian fluxes and waveforms are next-to-leading order.

	After a 3+1 decomposition of spacetime, and using the convention that boldface letters represent three-dimensional Euclidean vectors, the field-point spatial vector in radiative coordinates is denoted by $\bm{X} = R \,\bm{N}$, where $\bm{N}$ has unit norm. In spherical coordinates, the coordinates of the field point are denoted by $(R,\Theta,\Phi)$. Time is denoted by $t$, and retarded time is defined as  \mbox{$t_{\text{ret}}=t-R/c$}, so as to not confuse it with the eccentric anomaly of the QK motion, $u$.

The positions of particles 1 and 2 are denoted by $\bm{y}_1$ and $\bm{y}_2$, and \mbox{$\bm{x} = \bm{y}_1-\bm{y}_2$} is the separation vector. The orbital radius is given by \mbox{$r = |\bm{x}|$}, and the unit vector \mbox{$\bm{n} = \bm{x}/r$} is introduced.  The relative velocity is given by \mbox{$\bm{v}=\dd\bm{x}/\dd t$}. One can then introduce the rotating orthonormal triad \mbox{$(\bm{n}, \bm{\lambda},\bm{\ell})$} and the nonrotating reference triad \mbox{$(\bm{n}_0, \bm{\lambda}_0,\bm{\ell}_0)$}, which are defined rigorously in Sec.~\ref{sec:Fourier}. For nonspinning binaries, there is no precession, so $\ell = \ell_0$ and one can describe the orbital motion with polar coordinates $(r,\phi)$ in the orbital plane\footnote{The notation $\phi$ is used both for the scalar field and the polar angle, but its meaning will always be clear in context. Moreover, $r$ will be used at one occasion to denote the radial coordinate of the field point in harmonic coordinates, at this will be stated clearly.} such that $\bm{n} = \cos \phi \,\bm{n}_0+\sin \phi \,\bm{\lambda}_0$ and $\bm{\lambda} = -\sin(\phi)\bm{n}_0 + \cos(\phi)\bm{\lambda}_0$. Notice that the relative velocity is given by \mbox{$\bm{v} = \dot{r} \bm{n} + r\dot{\phi} \bm{\lambda}$}. The following relations then hold:
\mbox{$\bm{n}\cdot\bm{v}=\dot{r}$}, \mbox{$v^2=\dot{r}^2 + r^2\dot{\phi}^2$}, and \mbox{$\bm{n}\times\bm{v}=r\,\dot{\phi}\,\bm{\ell}$}. 

The notation $L=i_1\cdots i_\ell$ stands for multi-index with $\ell$ spatial indices (and $K$ would stand for a multi-index with $k$ spatial indices). One can then write $\partial_L = \partial_{i_1}\cdots\partial_{i_\ell}$, $\partial_{aL-1} = \partial_a\partial_{i_1}\cdots\partial_{i_{\ell-1}}$ and so on;  similarly,  $n_{L} = n_{i_1}\cdots n_{i_\ell}$, $n_{aL-1} = n_a n_{i_1}\cdots n_{i_{\ell-1}}$. The symmetric trace-free (STF) part is indicated using a hat or angled brackets: for instance, $\mathrm{STF}_L [\partial_L] = \hat{\partial}_L = \partial_{\langle i_1} \partial_{i_2}\cdots \partial_{ i_\ell \rangle}$, $\mathrm{STF}_L [n_L] = \hat{n}_L = n_{\langle i_1} n_{i_2}\cdots n_{ i_\ell \rangle}$, and $\mathrm{STF}_L [x_L] = \hat{x}_L = r^\ell n_{\langle i_1} n_{i_2}\cdots n_{ i_\ell \rangle}$.
The $n$th time derivative of a function $F(t)$ is denoted $F^{(n)}(t) = \dd^n F / \dd t^n$.

The constant asymptotic value of the scalar field at spatial infinity is denoted $\phi_0$, and the normalized scalar field is defined by $\varphi\equiv\phi/\phi_{0}$.  The Brans-Dicke-like scalar function $\omega(\phi)$ is expanded around the asymptotic value $\omega_0 \equiv \omega(\phi_0)$ and  the mass functions $m_A(\phi)$ (see Section~\ref{subsec:fieldEquations} for a definition) are expanded around the asymptotic values $m_A \equiv m_A(\phi_0)$, where $A\in\{1,2\}$.  In the CM frame, one then defines the asymptotic total mass $m=m_1+m_2$, reduced mass $\mu = m_1 m_2 / m$, symmetric mass ratio $\nu= \mu/ m \in \ ]0,1/4]$, and relative mass difference $\delta = (m_1-m_2)/m \in \ ]1,1[ $. Note that the asymptotic symmetric mass ratio and the relative mass difference are linked by the relation $\delta^2 = 1-4\nu$.

Following~\cite{B18_i, BBT22}, a number of parameters describing these expansions are introduced in~Table~\ref{table}. The ST parameters are defined directly from the expansions of the Brans-Dicke-like scalar function $\omega(\phi)$ and of the mass functions $m_A(\phi)$. The PN parameters are combinations of the ST parameters that naturally extend and generalize the usual PPN parameters to the case of a general ST theory~\cite{will1972conservation,Will:2018bme}. Moreover, these parameters can be mapped back to those of the formulation of Damour and Esposito-Farèse~\cite{Damour:1996ke}, see e.g. Table I of Ref.~\cite{JR24} for the complete correspondence.

\begin{small}
\begin{center}\begin{table}[h]
\begin{tabular}{|c||cc|}
	\hline
	& \multicolumn{2}{|c|}{\textbf{ST parameters}} \\[2pt]
	\hline &&\\[-10pt]
	general & \multicolumn{2}{c|}{$\omega_0=\omega(\phi_0),\qquad\omega_0'=\eval{\frac{\dd\omega}{\dd\phi}}_{\phi=\phi_0}, \qquad\omega_0''=\eval{\frac{\dd^2\omega}{\dd\phi^2}}_{\phi=\phi_0},\qquad\varphi = \frac{\phi}{\phi_{0}},\qquad\tilde{g}_{\mu\nu}=\varphi\,g_{\mu\nu},$} \\[12pt]
	& \multicolumn{2}{|c|}{$\tilde{G} = \frac{G(4+2\omega_{0})}{\phi_{0}(3+2\omega_{0})},\qquad \zeta = \frac{1}{4+2\omega_{0}},$} \\[8pt]
	& \multicolumn{2}{|c|}{$\lambda_{1} = \frac{\zeta^{2}}{(1-\zeta)}\left.\frac{\dd\omega}{\dd\varphi}\right\vert_{\varphi=1},\qquad \lambda_{2} = \frac{\zeta^{3}}{(1-\zeta)}\left.\frac{\dd^{2}\omega}{\dd\varphi^{2}}\right\vert_{\varphi=1}, \qquad \lambda_{3} = \frac{\zeta^{4}}{(1-\zeta)}\left.\frac{\dd^{3}\omega}{\dd\varphi^{3}}\right\vert_{\varphi=1}.$} \\[7pt]
	\hline &&\\[-7pt]
	~sensitivities~ & \multicolumn{2}{|c|}{$s_A = \eval{\frac{\dd \ln{m_A(\phi)}}{\dd\ln{\phi}}}_{\phi=\phi_0},\qquad s_A^{(k)} = \eval{\frac{\dd^{k+1}\ln{m_A(\phi)}}{\dd(\ln{\phi})^{k+1}}}_{\phi=\phi_0},\qquad(A=1,2)$} \\[9pt]
    & \multicolumn{2}{|c|}{$s'_A = s_A^{(1)},\qquad s''_A = s_A^{(2)},\qquad s'''_A = s_A^{(3)},$} \\[5pt]
	& \multicolumn{2}{|c|}{$\Sp = \frac{1-s_1 - s_2}{\sqrt{\alpha}}\,,\qquad \Sm= \frac{s_2 - s_1}{\sqrt{\alpha}}.$} \\[7pt]	\hline\hline 
	Order & \multicolumn{2}{|c|}{\textbf{PN parameters}} \\[2pt]
	\hline &&\\[-10pt]
	N & \multicolumn{2}{|c|}{$\alpha= 1-\zeta+\zeta\left(1-2s_{1}\right)\left(1-2s_{2}\right)$}   \\[5pt]
	\hline &&\\[-10pt]
	1PN & $\overline{\gamma} = -\frac{2\zeta}{\alpha}\left(1-2s_{1}\right)\left(1-2s_{2}\right),$ & Degeneracy \\[5pt]
	&~~$\overline{\beta}_{1} = \frac{\zeta}{\alpha^{2}}\left(1-2s_{2}\right)^{2}\left(\lambda_{1}\left(1-2s_{1}\right)+2\zeta s'_{1}\right),$~~~~&  $\alpha(2+\overline{\gamma})=2(1-\zeta)$ \\[5pt]
	& $\overline{\beta}_{2} = \frac{\zeta}{\alpha^{2}}\left(1-2s_{1}\right)^{2}\left(\lambda_{1}\left(1-2s_{2}\right)+2\zeta s'_{2}\right),$~~~~& \\[5pt]
	&  $\overline{\beta}_+ = \frac{\overline{\beta}_1+\overline{\beta}_2}{2}, \qquad \overline{\beta}_- = \frac{\overline{\beta}_1-\overline{\beta}_2}{2}.$ &  \\[5pt]
	\hline &\\[-10pt]
	2PN & $\overline{\delta}_{1} = \frac{\zeta\left(1-\zeta\right)}{\alpha^{2}}\left(1-2s_{1}\right)^{2}\,,\qquad \overline{\delta}_{2} = \frac{\zeta\left(1-\zeta\right)}{\alpha^{2}}\left(1-2s_{2}\right)^{2},$ & Degeneracy \\[5pt]
	&  $\overline{\delta}_+ = \frac{\overline{\delta}_1+\overline{\delta}_2}{2}, \qquad \overline{\delta}_- = \frac{\overline{\delta}_1-\overline{\delta}_2}{2},$ &  $16\overline{\delta}_{1}\overline{\delta}_{2} = \overline{\gamma}^{2}(2+\overline{\gamma})^{2}$\\[5pt]
	& $~~\overline{\chi}_{1} = \frac{\zeta}{\alpha^{3}}\left(1-2s_{2}\right)^{3}\left[\left(\lambda_{2}-4\lambda_{1}^{2}+\zeta\lambda_{1}\right)\left(1-2s_{1}\right)-6\zeta\lambda_{1}s'_{1}+2\zeta^{2}s''_{1}\right],~~$  &  \\[5pt]
	& $\overline{\chi}_{2} = \frac{\zeta}{\alpha^{3}}\left(1-2s_{1}\right)^{3}\left[\left(\lambda_{2}-4\lambda_{1}^{2}+\zeta\lambda_{1}\right)\left(1-2s_{2}\right)-6\zeta\lambda_{1}s'_{2}+2\zeta^{2}s''_{2}\right],$ &  \\[5pt]
	&  $\overline{\chi}_+ = \frac{\overline{\chi}_1+\overline{\chi}_2}{2}, \qquad \overline{\chi}_- = \frac{\overline{\chi}_1-\overline{\chi}_2}{2}.$ &  \\[5pt]
\hline
\end{tabular}
\caption{Summary of parameters for the general ST theory and notations for PN parameters. \label{table}}\end{table}
\end{center}
\end{small}\newpage


\section{Massless scalar-tensor theories}\label{sec:STtheory}


\label{subsec:fieldEquations}
As in \citetalias{BBT22}, I will consider a generic class of ST theories in which a single massless scalar field~$\phi$ minimally couples to the metric~$g_{\mu\nu}$, described by the action:
\begin{align}\label{eq:STactionJF}
S_{\mathrm{ST}} &= \frac{c^{3}}{16\pi G} \int\dd^{4}x\,\sqrt{-g}\left[\phi R - \frac{\omega(\phi)}{\phi}g^{\alpha\beta}\p_{\alpha}\phi\p_{\beta}\phi\right]  +S_{\mathrm{m}}\left(\mathfrak{m},g_{\alpha\beta}\right)\,,
\end{align}
where $R$ and $g$ are, respectively, the Ricci scalar and the determinant of the metric, $\omega$ is a function of the scalar field and $\mathfrak{m}$ stands generically for the matter fields. Since these theories break the strong equivalence principle, I will follow the standard prescription by Eardley~\cite{Eardley1975} (see also~\cite{Nordtvedt:1990zz}), which consists in modeling the inertial masses of each particle as a function of the scalar field, namely $m_A(\phi)$, where $A\in\{1,2\}$ for a binary system. The effective matter action for point particles thus reads
\be\label{eq:matteract}
S_{\mathrm{m}} = - c \sum_{A} \int\,m_{A}(\phi) \sqrt{-\left(g_{\alpha\beta}\right)_{A}\dd y_{A}^{\alpha}\,\dd y_{A}^{\beta}}\,,
\ee
where $y_A^\alpha$ denotes the space-time positions of the particles, and $\left(g_{\alpha\beta}\right)_{A}$ is the metric evaluated\footnote{Divergences are treated with Hadamard regularization~\cite{BFreg}, which is equivalent, at this order, to dimensional regularization~\cite{BDE04,BDEI05dr}.} at the position of particle~$A$. Despite minimal coupling, the point particle effective model introduces a dependence in the scalar field.
The action of Eq.~\eqref{eq:STactionJF} is usually called the Jordan-frame action, as the matter only couples to the Jordan or ``physical'' metric $g_{\alpha\beta}$. It is then very useful to introduce a rescaled scalar field and a conformally related metric,
\be\label{eq:def_gt}
\varphi\equiv \frac{\phi}{\phi_{0}}\qquad\mathrm{and}\qquad\tilde{g}_{\alpha\beta}\equiv \varphi\,g_{\alpha\beta}\,,
\ee
where $\phi_0$ is the constant cosmological background value of the scalar field, such that the physical and conformal metrics have the same asymptotic behavior at spatial infinity. Quantities expressed in terms of the pair $(\varphi, \tilde{g}_{\alpha\beta})$ are said to be in the ``Einstein frame.''
%
%
From these, one defines the scalar and metric perturbation variables  $\psi\equiv\varphi-1$ and $h^{\mu\nu}\equiv \sqrt{-\tilde{g}}\tilde{g}^{\mu\nu}-\eta^{\mu\nu}$, where $\eta^{\mu\nu}~=~\text{diag}(-1,1,1,1)$ is the Minkowski metric. Within this perturbative setup, the arbitrary Brans-Dicke coupling function will be Taylor-expanded around the cosmological background value $\phi_0$, namely \mbox{$\omega(\phi) = \omega_0 + (\phi-\phi_0) \omega_0' + \calO\left((\phi-\phi_0)^2\right)$}\,. Moreover, the mass functions $m_A(\phi)$ will be described perturbatively in terms of their sensitivites, defined as
\be\label{sAk}
s_A^{(k)} \equiv \eval{\frac{\dd^{k+1}\ln{m_A(\phi)}}{\dd(\ln{\phi})^{k+1}}}_{\phi=\phi_0}\,, 
\ee
where $s_A \equiv  s_A^{(1)}$. 
In these theories, no hair theorems~\cite{Hawking:1972qk, Sotiriou:2011dz} impose that isolated black holes carry no scalar charge. This strongly suggests that black holes in a binary system also have vanishing scalar charge, and this is partially confirmed by numerical relativity computations~\cite{Healy:2011ef}. Since the scalar charge $\alpha_A$ of the particle $A$ is related to its sensitivity by the relation~\cite{Sennett:2016rwa}
\be \alpha_A = \frac{1-2s_A}{\sqrt{3+2\omega_0}} \,,\ee
this means that black holes in a binary system have sensitivity $s_A = 1/2$. In all post-Newtonian computations, it was indeed found that the results of GR are recovered by setting $s_1 = s_2 = 1/2$.


With all these tools in hand, the field equations in the Landau-Lifschitz formulation~\cite{LL} can now be rewritten~\cite{MW13}
\begin{subequations}\label{eq:rEFE}
\begin{align}
& \Box_{\eta}\,h^{\mu\nu} = \frac{16\pi G}{c^{4}}\tau^{\mu\nu}\,,\\
& \Box_{\eta}\,\psi = -\frac{8\pi G}{c^{4}}\tau_{s}\,,
\end{align}
\end{subequations}
where $\Box_{\eta}$ denotes the ordinary flat space-time d'Alembertian operator, and where the source terms read
\begin{subequations}
\begin{align}\label{eq:taumunu}
& \tau^{\mu\nu} = \frac{\varphi}{\phi_{0}} (- g) T^{\mu\nu} +\frac{c^{4}}{16\pi G}\Lambda^{\mu\nu}[h,\psi]\,,\\*
\label{taus} & \tau_{s} = -\frac{\varphi}{\phi_{0}(3+2\omega(\phi))}\sqrt{-g}\left(T-2\varphi\frac{\p T}{\p \varphi}\right) -\frac{c^{4}}{8\pi G}\Lambda_s[h,\psi]\,.
\end{align}
\end{subequations}
%
Here $T^{\mu\nu}= 2 (-g)^{-1/2}\delta S_{\mathrm{m}}/\delta g_{\mu\nu}$ is the matter stress-energy tensor, $T\equiv g_{\mu\nu}T^{\mu\nu}$ is its trace and $\p T/\p \varphi$ is defined as the partial derivative of $T(g_{\mu\nu}, \varphi)$ with respect to $\varphi$ holding $g_{\mu\nu}$ constant. Moreover, $\Lambda^{\mu\nu}$ and $\Lambda^s$ are given explicitly in (2.8) and (2.9) of \citetalias{BBT22} as functionals of the $h^{\mu\nu}$ and $\psi$ that are at least quadratic in the fields.  

The field equations of \eqref{eq:rEFE} are solved using the PN-MPM construction \cite{BD86,Blanchet:2013haa}. The reader is invited to refer to \citetalias{BBT22} for a detailed description of this construction in the case of ST theories.  The main point is that the fields in vacuum zone exterior to the source the are written as a formal expansion in the bookkeeping parameter $G$, namely $h^{\mu\nu} = G h_1^{\mu\nu} + G^2 h_2^{\mu\nu} + \calO(G^3)$ and $\psi = G \psi_1 + G^2 \psi_2+ \calO(G^3)$. The linearized piece is parametrized by the so-called source moments $\dI_L$, $\dJ_L$  and $\dI_L^s$, along with some gauge moments $\dW_L$, $\dX_L$, $\dY_L$, and $\dZ_L$, which are pure functions of retarded time and are symmetric trace-free (STF) in their indices. These moments uniquely parametrize the linearized metric  $h_1^{\mu\nu}$ and the linearized scalar field $\psi_1$. For example, the linearized scalar field reads explicitly
\be 
\psi_1 = - \frac{2}{c^2}\sum_{\ell=0}^{+\infty}\frac{(-)^\ell}{\ell!}\,\partial_L\!\left[\frac{\dI_L^s(t-r/c)}{r}\right]\,,\label{psi1}
\ee
where in this particular case, $r$ denotes the radial coordinate of the field point in harmonic coordinates (rather than the radial separation of the binary). 
For a given source, these moments can be computed explicitly as functions of the orbital variables, see e.g. App.~(B.1) of \citetalias{BBT22}. A feature of ST theory is that the scalar monopole $\dI^s$ is not constant but its time-variation will be a small PN effect, i.e., $\dd \dI^s/\dd t = \mathcal{O}(c^{-2})$. This is why the leading-order radiation is dipolar, and not monopolar. It is thus practical to define 
\bse\be
\dI^s  = \frac{1}{\phi_0}\left[ m^s+\frac{E^s}{c^2} \right]\,,
\ee
where 
\be 
m^s = - \frac{1}{3+2\omega_0}\sum_A m_A (1- 2s_A)
\ee\ese
is constant and $E^s$ is the time-varying, PN correction.

The parametrized linearized piece is reinjected iteratively into the field equations \eqref{eq:rEFE}, where $T^{\mu\nu} = 0$, since only the vacuum zone is considered here. This yields field equations for $h_n^{\mu\nu}$ and $\psi_{n}$ (for $n\ge 2$) which can be iteratively solved at quadratic \cite{BD92, B98quad}, cubic \cite{B98tail, TLB22, TB23_ToM} and quartic order \cite{MBF16}. Once this MPM metric is constructed at desired order, one can extract its leading order expression in the large radii expansion (this is the only thing an asymptotic observer will see). This asymptotic metric, written in terms of some radiative coordinates $(t,R,\bm{N})$ [or alternatively  $(t,R,\Theta,\Phi)$] which describe the position of the field point, admits a multipolar decomposition as well, which is uniquely parametrized, after TT projection,  by three families of \emph{radiative} moments~\cite{Blanchet:2013haa, BBT22}, denoted $\mathcal{U}_L$, $\mathcal{V}_L$ and $\mathcal{U}^s_L$. Introducing $t_{\text{ret}} = t - R/c$, namely retarded time in the radiative coordinate system \cite{B87, TLB22}, one find that that the asymptotic metric reads in a transverse-traceless gauge 
\begin{subequations}\label{eq:radiative_expansion}
\begin{align}
 	h_{ij}^\text{TT} &= - \frac{4G}{c^2 R} \perp^\text{TT}_{ijab} \sum_{\ell=2}^{+\infty} \frac{1}{c^\ell \ell!}\Bigg[ N_{L-2}\,\mathcal{U}_{ab L-2}(t_{\text{ret}}) - \frac{2\ell}{c(\ell+1)} N_{c L-2} \epsilon_{cd(a}\mathcal{V}_{b)dL-2}(t_{\text{ret}})\Bigg] + \mathcal{O}\Big(\frac{1}{R^2}\Big)\,,\label{seq:tensor_radiative_expansion}\\
 	\psi &= - \frac{2G}{c^2 R}\sum_{\ell=0}^{+\infty} \frac{1}{c^\ell \ell!} N_L \, \mathcal{U}_L^s(t_{\text{ret}}) + \mathcal{O}\Big(\frac{1}{R^2}\Big)\,, \label{seq:scalar_radiative_expansion}
\end{align}
\end{subequations}
where \mbox{$\perp_{ijab}^\mathrm{TT} = \perp_{a(i}\perp_{j)b}-\frac{1}{2}\!\perp_{ij}\perp_{ab}$} with \mbox{$\perp_{ij} =\delta_{ij}-N_i N_j$}.
The radiative moments are generically related to the source moments by the relations 
\begin{align}
 \mathcal{U}_L &=  \overset{(\ell)}{\dI}_{\!\!L}+\mathcal{O}\left(\frac{1}{c^3}\right)\,,&\quad
 \mathcal{V}_L &=  \overset{(\ell)}{\dJ}_{\!\!L} +\mathcal{O}\left(\frac{1}{c^3}\right)\,,&\quad
\mathcal{U}^s_L &= \overset{(\ell)}{\dI}{\!\!}^{s}_{L}+\mathcal{O}\left(\frac{1}{c^3}\right)\,,
 \end{align}
where the small $\mathcal{O}(1/c^3)$ corrections correspond to the nonlinear corrections of the MPM construction. The full expressions needed for the 1.5PN fluxes were obtained in \citetalias{BBT22}, and read
\bse\begin{align}
\mathcal{U}_{ij} &= \overset{(2)}{\dI_{ij}} + \frac{2G \dM}{\phi_0 c^3} \int_{0}^\infty \!\dd \tau \, \overset{(4)}{\dI}{\!}_{ij}(t_{\text{ret}}-\tau) \left[ \ln{\left( \frac{c\tau}{2b_0} \right)} + \frac{11}{12} \right] \nn \\
&\qquad+ \frac{G(3+2\omega_0)}{3c^3}\Bigg( -\dI^s_{\langle i}\overset{(3)}{\dI}{\!}^s_{j\rangle}- \overset{(1)}{\dI}{\!}^s_{\langle i}\overset{(2)}{\dI^s_{j\rangle}} -\frac{1}{2}\dI^s \overset{(3)}{\dI}{\!}^s_{ij} +  \int_{0}^\infty \!\dd \tau\, \Bigl[\overset{(2)}{\dI}{\!}^s_{\langle i} \overset{(2)}{\dI}{\!}^s_{j\rangle}\Bigr](t_{\text{ret}}-\tau)\Bigg)+ \calO\left(\frac{1}{c^5}\right) \,. \\
\mathcal{U}^s &= \dI^s + \frac{2 G \dM}{\phi_0 ^2 c^5} \int_{0}^\infty \dd \tau\, \overset{(2)}{E}{\!}^s(t_{\text{ret}}-\tau)\ln\left(\frac{c\tau}{2b_0}\right)  + \frac{G}{c^5}\left(1-\frac{\phi_0 \omega_0'}{3+2\omega_0}\right) \Bigg[\frac{2}{9} \dI_k^s\overset{(3)}{\dI}{\!}_k^s -2\, \dI^s \frac{\overset{(1)}{E}{\!}^s}{\phi_0}\Bigg]+\calO\left(\frac{1}{c^7}\right)\,,\\
\mathcal{U}_i^s &= \overset{(1)}{\dI}{\!}_i^s + \frac{2G\dM}{\phi_0 c^3} \int_{0}^\infty \dd \tau\, \overset{(3)}{\dI}{\!}_i^s(\tau) \left[ \ln{\left(\frac{c\tau}{2b_0} \right)} + 1 \right] \nn\\*
&\qquad\ +\frac{G}{c^5} \Bigg[ -\frac{1}{5}\overset{(1)}{\dI}{\!}^s_{k}\overset{(3)}{\dI}{\!}_{ik} -\frac{1}{5}\overset{(2)}{\dI}{\!}^s_{k}\overset{(2)}{\dI}{\!}_{ik} +\frac{3}{5}\overset{(3)}{\dI}{\!}^s_{k}\overset{(1)}{\dI}{\!}_{ik}  +\frac{3}{5}\overset{(4)}{\dI}{\!}^s_{k} \dI_{ik} 
- \epsilon_{iab}\dJ_a \overset{(3)}{\dI}{\!}^s_b
- 4\overset{(1)}{\dW}\overset{(2)}{\dI}{\!}^s_i
- 4 \overset{(2)}{\dW}\overset{(1)}{\dI}{\!}^s_i
+ 4 \dI^s \overset{(2)}{\dY_i} \nn\\*[-0.1cm]
&\qquad\qquad\quad\ + \Big( 1-\frac{\phi_0\omega_0'}{3+2\omega_0}\Big)\Bigg( 
-2 \frac{\overset{(1)}{E}{\!}_s }{\phi_0}\overset{(1)}{\dI}{\!}^s_i 
-2 \frac{\overset{(2)}{E}{\!}_s }{\phi_0}\dI^s_i 
+ \frac{2}{5} \overset{(3)}{\dI}{\!}^s_k \overset{(1)}{\dI}{\!}^s_{ik} 
+ \frac{2}{5} \overset{(4)}{\dI}{\!}^s_k \dI^s_{ik} \Bigg)\Bigg] + \calO\left(\frac{1}{c^7}\right)\,, \\
\mathcal{U}_{ij}^s &= \overset{(2)}{\dI}{\!}_{ij}^s  + \frac{2 G \dM}{\phi_0 c^3} \int_{0}^\infty \dd \tau \,\overset{(4)}{\dI}{\!}^s_{ij}(t_{\text{ret}}-\tau)\left[ \ln{\left(\frac{c\tau}{2b_0} \right)} + \frac{3}{2} \right]  - \frac{G}{c^3} \dI^s \overset{(3)}{\dI}{\!}_{ij}  +\calO\left(\frac{1}{c^5}\right) \,,
\end{align}
\ese
where the instantaneous moments are implicitly evaluated at $t_{\text{ret}}$.
Note the presence of hereditary tails (associated as usual to the arbitrary length scale $b_0$) as well and memory integrals in the latter expressions.

 \section{The quasi-Keplerian parametrization}	
 \label{sec:PK}

In previous works \cite{Lang:2013fna, Lang:2014osa, BBT22, JR24}, the fluxes of energy and angular momentum were computed in terms of the orbital variables of the binary in the CM frame, namely the relative position $\bm{x} = r \bm{n}$ and the relative velocity $\bm{v}$. However, in order to provide a ready-to-use post-Newtonian waveform template, one still has to solve equations of motions such as to obtain a parametrization of the orbital motion, which is then explicitly injected into these fluxes. This parametrization is very easy if one assumes that the orbit is bound and quasicircular \cite{Sennett:2016klh, BBT22}, which is relevant astrophysically, since compact binaries are expected to have circularized by the time they enter the frequency band of ground based GW detectors. However, other mechanisms, such as von Zeipel–Lidov–Kozai resonances \cite{Blaes:2002cs} or dynamical captures, can lead to non-negligeable eccentricities even within the detector's frequency band, and there are hints that eccentric binaries have in fact already been detected \cite{Romero-Shaw:2022xko}. For an quasielliptic orbit, one of the standard prescriptions in general relativity is the quasi-Keplerian (QK) parametrization \cite{DD85}, which have been fully extended up to 3PN order \cite{DS88,SW93,MGS04}, and partially up to 4PN order \cite{ChoTanay2022} (see also~\cite{Tucker:2018rgy} for an equivalent osculating-orbit prescription). In \citetalias{T24_QK}, I have extended the QK parametrization to the case of scalar-tensor theories at 2PN order, which is exactly what will be needed for this work.
At 2PN order, the parametrization reads
\bse \label{eq:PK_equations}
\begin{align}
r &= a_r(1- e_r \cos(u)) \,, \label{seq:radial_equation}\\
\ell=n(t-t_0) &= u - e_t \sin(u) + f_t \sin(v) + g_t(v-u) \,,  \label{seq:kepler_equation}\\
\phi-\phi_0 &= K\Big[ v + f_\phi \sin(2v) + g_\phi \sin(3v) \Big] \,, \label{seq:angular_equation}
\end{align}
\ese
where $t$ is time, $(r,\phi)$ are the polar coordinates in the orbital plane, $u$ is an affine parameter called the \emph{eccentric anomaly}, and the \emph{true anomaly} is defined as 
\begin{align}\label{eq:v}
v &= 2 \arctan\Bigg[ \sqrt{\frac{1+e_\phi}{1-e_\phi}}\tan\Big(\frac{u}{2}\Big) \Bigg] + 2\pi\bigg\lfloor \frac{u+\pi}{2\pi}\bigg\rfloor \nonumber\\
&= u + 2 \arctan\Bigg[\frac{\beta(e_\phi) \sin(u)}{1-\beta(e_\phi) \cos(u)}\Bigg] \,,
\end{align}
where
\be \label{eq:beta}\beta(e) \equiv \frac{e}{1+\sqrt{1-e^2}} \,.\ee
The quasi-Keplerian parameters are the mean motion $n$, the rate of pericenter advance $K$, the semi major axis $a_r$, the time eccentricity $e_t$, the radial eccentricity $e_r$, the angular eccentricity  $e_\phi$, the rate of pericenter advance $K$, and the higher-order parameters $f_\phi$, $f_t$, $g_\phi$, and $g_t$. The motion is characterized by two generically incommensurable periods, namely the radial period $P$ [i.e. the time between two pericenter passages], which is related to the mean motion by $n = 2\pi/  P$ ; and the angular period [i.e. the time needed for $\phi$ to go from $0$ to $2\pi$], related to the angular frequency $\omega = K n $. The mean anomaly is then defined as $\ell = n(t-t_0)$, and the advance in the pericenter is typically decomposed as $K = 1+k$, where $k$ is a small PN quantity. Moreover,  the three eccentricities $e_r$, $e_\phi$ and $e_t$ differ by small PN corrections, and are thus equal in the Newtonian limit. It will also be extremely useful to introduce the small frequency-related  PN parameter
\be \label{eq:x} x = \left(\frac{G m \omega}{c^3}\right)^{2/3} = \left(\frac{G m n K}{c^3}\right)^{2/3} \,.\ee

Crucially, these various parameters are explicitly expressed in terms of  a set of two parameters that entirely characterize the orbits, e.g. the dimensionless energy and angular moment $(\varepsilon, j)$, which are defined from the mechanical energy and angular momentum $(E, J)$ as
\be \varepsilon =   -\frac{2 E}{\mu c^2} \qquad\text{and}\qquad j = - \frac{2 J^2 E}{\mu^3 (\tilde{G}\alpha m)^2} \,. \ee
This is the main result of \cite{T24_QK}, and the interested reader is encouraged to refer to it for more details. Another useful pair of  gauge invariant variables is $(x,\iota)$ where $\iota \equiv x  \mathcal{B} /k$ and 
\be \label{eq:calB}  \mathcal{B} = 3 + 2 \gam -\bep +  \bem \delta \,.\ee 
The variable $\iota$ is defined
such that $\iota \rightarrow j$ when $x \rightarrow 0$. However, the limiting case of  circular orbit limit is not immediate to obtain with this choice. Thus, from now on, I will  work with the pair of variables $(x,e_t)$, which has the good property of yielding the results for circular orbits by taking the limit $e_t \rightarrow 0$, but has the unfortunate property of not being gauge invariant.

Moreover, for the purpose of the following derivations, it will be very practical to define
\begin{align} \label{eq:w} w = v(u) - u &= 2 \arctan\left[\frac{\beta(e_\phi) \sin(u)}{1-\beta(e_\phi) \cos(u)}\right] \,, \\
\label{eq:wTilde} \tilde{w} &= 2 \arctan\left[\frac{\beta(e_t) \sin(u)}{1-\beta(e_t) \cos(u)}\right] \,,
\end{align}
where $\beta$ is defined in Eq.~\eqref{eq:beta}.
We then denote \mbox{$\tilde{v} = u+\tilde{w}$}, such that $w$ and $\tilde{w}$ represent the periodically oscillating pieces of $v$ and $\tilde{v}$.
Finally, I will often need to apply orbit averaging. For any time-depending quantity $f$, I define its orbit average over a radial period as
\be \left\langle f \right\rangle = \frac{1}{P} \int_{0}^{P} \dd t \, f(t) = \frac{1}{2\pi} \int_0^{2\pi} \dd \ell\, f(\ell) = \frac{1}{2\pi} \int_0^{2\pi} \dd u \,\frac{\dd \ell}{\dd u} \, f(u)  \,, \ee
where the 2PN-accurate expression for $\dd \ell/\dd u$ can be derived from the results presented in \citetalias{T24_QK}, and reads 
\begin{align}\label{eq:dldu}
\frac{\dd \ell}{\dd u} &=  1 - e_t \cos u + \frac{x^2}{8} \Bigg\{\frac{(1-e_t^2)(15+8\gam-\nu) \nu}{\big[1-e_t \cos u\big]^2} \nn\\*
&\quad\qquad\qquad\qquad\qquad + \frac{1}{1-e_t\cos u} \bigg[60 + 60\gam +15 \gam^2 - 4 \dep - 4 \dem \delta + \nu\,\big(\!-39 - 24 \gam + 8 \bep - 8 \bem \delta\big) + \nu^2  \bigg]\nn\\*
&\quad\qquad\qquad\qquad\qquad - \frac{1}{\sqrt{1-e_t^2}} \bigg[60 + 60\gam +15 \gam^2 - 4 \dep - 4 \dem \delta  + \nu\,\big(\!-24 - 16 \gam + 8 \bep - 8 \bem \delta\big)  \bigg]\Bigg\}+\mathcal{O}(x^3) \,.
\end{align}

\section{The Fourier decomposition of the multipolar moments}
\label{sec:Fourier}

One of the challenges when computing high-order PN waveforms is the treatment of hereditary integrals, such as the tails or the memory. These integrals span over the whole history of the binary, and require a model for its past evolution. Moreover, these integrals can be easily computed analytically in the case of quasicircular orbits, but the treatment of elliptic orbits is more complicated. The standard strategy is to use the periodicity (or double-periodicity) of the multipolar moments in order to express their time-evolution as a Fourier series.  In general relativity, the Fourier coefficients required for the 3PN flux were computed in \cite{ABIQ08tail}. These results were then improved by Loutrel and Yunes \cite{LY16}, and I will follow the latter reference, adapting it to the case of ST theories. 

\subsection{Moments at Newtonian order}
\label{subsec:FourierN}

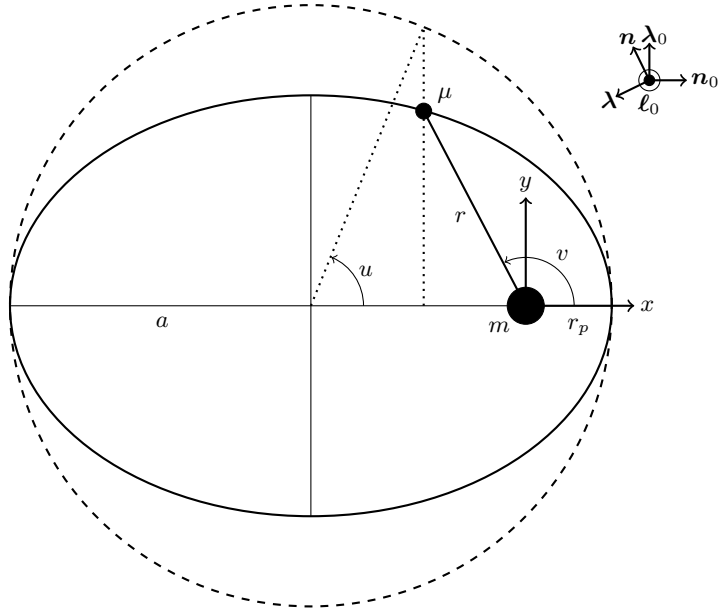
\begin{figure}[h]
\begin{tikzpicture}
\draw [dashed,thick] (0,0) circle (4);
\draw [thick] (0,0) ellipse (4 and 2.8);
\draw (-4,0) -- (4,0) node[xshift=-170,yshift=-6]{$a$};
\draw (0,-2.8) -- (0,2.8);
\filldraw [black] (1.5,2.59) circle (3pt) node[xshift=8,yshift=6]{$\mu$};
\draw [dotted,thick] (1.5,0) -- (1.5,3.71);
\draw [dotted,thick]  (0,0) -- (1.5,3.71) ;
\filldraw [black] (2.856571371, 0) circle (7pt) node[xshift=-10,yshift=-8]{$m$} node[xshift=20,yshift=-8]{$r_p$}; 
\draw [thick] (2.856571371, 0) -- (1.5,2.59) node[xshift=14,yshift=-40]{$r$};
\draw[->] (0.7,0) arc (0:67:0.7) node[xshift=13pt,yshift=-5]{$u$};
\draw[->] (3.5,0) arc (0:116:0.643428629) node[xshift=22,yshift=3]{$v$};
\draw [ thick,->] (2.856571371, 0) -- (4.3,0) node[xshift=5,yshift=0pt]{$x$};
\draw [ thick,->] (2.856571371, 0) -- (2.856571371, 1.443428629) node[xshift=0pt,yshift=5pt]{$y$};
\filldraw [black] (0.5+4,3) circle (2pt);
\draw (0.5+4,3) circle (4pt) node[xshift=0,yshift=-9]{$\bm{\ell}_0$};
\draw [->,thick](0.5+4,3)--(0.5+4,3.5) node[xshift=2,yshift=3]{$\bm{\lambda}_0$} ;
\draw [->,thick](0.5+4,3)--(0.5+4.5,3) node[xshift=7,yshift=0]{$\bm{n}_0$};
\draw [->,thick](0.5+4,3)--(0.5+3.780814427,3.449397023) node[xshift=-2,yshift=3]{$\bm{n}$} ;
\draw [->,thick](0.5+4,3)--(0.5+3.550602977,2.780814427) node[xshift=-2.5,yshift=-1]{$\bm{\lambda}$} ;
\end{tikzpicture}
\caption{Elliptical motion}\label{fig:ellipse}
\end{figure}

At Newtonian order, the three eccentricities $(e_t, e_r, e_\phi)$ are interchangeable and denoted $e$.
Moments at Newtonian order are periodic in $\ell = n(t-t_0)$, so they can be decomposed as a Fourier series as
\be \label{eq:genericNewtonianFourierSeries} \mathrm{K}_L(t) = \sum_{p\in\mathbb{Z}}{}_p\widetilde{\mathrm{K}}_L\,\de^{\di p\ell} \,,\ee 
where $\mathrm{K}_L$ stands either for  $\dI_L$, $\dJ_L$,  $\dI^s_L$, or $E^s$. The Fourier coefficients are given by 
\be \label{eq:genericNewtonianFourierCoefficient} {}_p \widetilde{\mathrm{K}}_L = \frac{1}{2\pi}\int_0^{2\pi} \dd \ell \,\mathrm{K}_L (t) \de^{-\di p \ell}\,.\ee
In practice, the multipolar moments, which are known in terms of $r$, $\dot{r}$, $\phi$, and $\dot\phi$ [see (4.6) and (4.7) of \citetalias{BBT22}], are first  expressed in terms of the eccentric anomaly $u$ using the Keplerian representation, and projected onto the time-independent orthonormal triad $(\bm{n}_0, \bm{\lambda}_0, \bm{\ell}_0)$, which corresponds, respectively, to the components $(x,y,z)$. As illustrated in Fig.~\ref{fig:ellipse}: $\bm{n}_0$ is parallel to the major axis and points from the center of the ellipse towards the periastron; $\bm{\lambda}_0$ is parallel to the minor axis, and oriented such that $v^i  \lambda_0^i > 0$ at each periastron passage; $\bm{\ell}_0$ is orthogonal to the orbital plane, and oriented such that such that $(\bm{n}_0, \bm{\lambda}_0, \bm{\ell}_0)$ is a positively-oriented orthonormal basis. I will also use the corotating orthonormal basis $(\bm{n}, \bm{\lambda}, \bm{\ell})$ defined by $\bm{n} = \cos(\phi)\bm{n}_0 + \sin(\phi) \bm{\lambda}_0$, $\bm{\lambda} = -\sin(\phi)\bm{n}_0 + \cos(\phi)\bm{\lambda}_0$ and $\bm{\ell} = \bm{\ell}_0$. Since the Kepler equation reads at Newtonian order $\ell= 1 - e \cos u$, after a change of variables, the only integrals we will have to evaluate can in fact be reduced to Bessel functions, namely \cite{LY16}
%
\be \label{eq:BesselJ} J_p(x) = \frac{1}{2\pi}\int_0^{2\pi}\dd u\,\de^{\di(pu-x \sin u)} \,.\ee

Moreover, following Loutrel and Yunes~\cite{LY16}, the expressions of the Fourier coefficients are simplified by exploiting identities between the Bessel functions and their derivatives [see their Eqs. (46-48)]. Thus, the following rules were recursively applied, valid for $x \neq 0$:
\bse \label{eq:identitiesBessel}\begin{align}
    &J_{p+n}(x) =\frac{2 (p+n-1)}{x} J_{p+n-1}(x) - J_{p+n-2}(x) \qquad\qquad \text{if}\quad n\ge 2 \,,\\
    &J_{p - n}(x) =\frac{2 (p-n+1)}{x} J_{p-n+1}(x) - J_{p-n+2}(x) \qquad\qquad \text{if}\quad n \le 1 \,,\\
    &J_{p+1}(x) = \frac{p}{x} J_{p}(x) - J'_{p}(x)\,.
\end{align}\ese


To simplify the presentation, the Fourier coefficients are normalized as
\be \label{eq:genericNewtonianFourierCoefficientNormalization} {}_p \widetilde{K}_L = \mathcal{K}_{\ell}  \times {}_p\widehat{\mathrm{K}}_L \,, \ee
where the generic normalization $\mathcal{K}_{\ell}$ is explicitly given for each moment by 
\bse \label{eq:FourierNormalizationFactor} \begin{align}
\label{seq:FourierNormalizationFactorEs0} \mathcal{E}^s_0 &= \frac{8 \sqrt{\alpha} \zeta m \nu c^2 x}{3(1-\zeta)} \Big[\Sp- \frac{1}{4}  \Sm \delta - 3 \gam^{-1} (\Sp \bep + \Sm \bem)\Big] \,, \\
\label{seq:FourierNormalizationFactorIs1} \mathcal{I}^s_1 &= - \frac{2 \Sm \zeta \tilde{G}\alpha^{3/2}m^2 \nu  }{c^2 (1- \zeta)\phi_0 x} \,,\\
\label{seq:FourierNormalizationFactorIs2} \mathcal{I}^s_2 &= - \frac{(\Sp-\Sm \delta)\zeta \tilde{G}^2\alpha^{5/2}m^3 \nu  }{c^4 (1- \zeta)\phi_0 x^2} \,, \\
\label{seq:FourierNormalizationFactorI2} \mathcal{I}_2 &= \frac{\tilde{G}^2 \alpha^2 m^3 \nu}{c^4 \phi_0 x^2} \,.
\end{align}\ese

Omitting the terms that vanish or that can immediately be deduced from the fact that the moments are symmetric in their indices, I find, for $p\neq 0$ (the $p=0$ terms will never be needed):
\bse  \label{eq:normalizedFourierCoefficients} \begin{align}
{}_p \widehat{E}^s &= J_p(ep) \,,\\*
{}_p \widehat{\dI}^s_{x} &= \frac{1}{p}J'_p(ep) \,,\\*
{}_p \widehat{\dI}^s_{y} &= -\frac{\di \sqrt{1- e^2}}{ep}J_p(ep) \,,\\*
{}_p \widehat{\dI}^s_{xx} &= {}_p \widehat{\dI}_{xx} =  \frac{2}{ep}\bigg[(1-e^2)J'_p(e p)+ \frac{(e^2-3)}{3 e p}J_p(ep)  \bigg] \,,\\*
{}_p \widehat{\dI}^s_{xy} &= {}_p \widehat{\dI}_{xy} =  \frac{2 \di \sqrt{1-e^2}}{e p}\bigg[\frac{1}{p}J'_p(e p) -\frac{1-e^2}{e} J_p(ep)\bigg] \,,\\*
{}_p \widehat{\dI}^s_{yy} &={}_p \widehat{\dI}_{yy} =   - \frac{2}{ep}\bigg[(1-e^2)J'_p(e p) + \frac{(2 e^2-3)}{3 e p}J_p(ep)  \bigg] \,, \\*
{}_p \widehat{\dI}^s_{zz} &= {}_p \widehat{\dI}_{zz} =  \frac{2}{3 p^2}J_p(ep)  \,,
\end{align}\ese
where it is immediate to verify that \mbox{${}_p \widehat{\dI}_{xx} + {}_p \widehat{\dI}_{yy} + {}_p \widehat{\dI}_{zz} = 0$} and  \mbox{${}_p \widehat{\dI}^s_{xx} + {}_p \widehat{\dI}^s_{yy} + {}_p \widehat{\dI}^s_{zz} = 0$}.
%
%

\subsection{The scalar dipole moment at 1PN}
\label{subsec:Fourier1PN}
The Fourier decomposition of $\dI^s_i$ will also be required at 1PN order. For this, the decomposition will be slightly different from the Newtonian case, because the relativistic precession of the periapsis requires that the doubly periodic structure of the motion be  correctly account for. Since the ellipse is precessing, the $(\bm{n}_0, \bm{\lambda}_0,\bm{\ell}_0)$ triad is now defined based on the osculating ellipse at current time $t$, and will remain constant throughout the past motion. Thus, at times $t' < t$, the major and minor axes are not necessarily aligned with $(\bm{n}_0, \bm{\lambda}_0)$.

Following Arun \textit{et al.} \cite{ABIQ08tail}, I first express $\dI_i^s$ in terms of $r$ and $\phi$, and use the decomposition
\be \phi = K \ell + W(\ell) = k \ell + \ell + W(\ell) \,,\ee
where, at this order, $W(\ell) = K(v - \ell)$ is a $2\pi$-periodic function of $\ell$. The explicit expression of  $\dI_i^s$ is solely composed of terms proportional to $\de^{\pm \di \phi}$, so it is immediate to factor out the dependence in $k \ell$, namely
\be \dI_i^s = \sum_{m \in\{-1,1\}} {}^m \widetilde{\dI}_i \,\de^{\di m k \ell}  = \mathcal{I}_1\!\!\!\!\! \sum_{m \in\{-1,1\}} {}^m \widehat{\dI}_i \,\de^{i m k \ell} \,, \ee
where the normalization $\mathcal{I}_1$ is the same as for the Newtonian case. The residual ${}^m \widehat{\dI}_i $ terms can then be written as a sum of terms proportional to $\de^{\pm \di [\ell + W(\ell) ]}$, which are all $2\pi$-periodic in $\ell$ because $W(\ell)$ is. Thus, it can be decomposed into a Fourier series. The final expansion reflects the double periodicity of the motion, and reads
\be \dI_i^s = \mathcal{I}_1 \!\!\!\!\!  \sum_{m \in\{-1,1\}} \sum_{p\in\mathbb{Z}} \,{}^{m}_{\,\,p}\widehat{\dI}_i^s \de^{i (m k+p) \ell} \,, \ee
where 
\begin{align}
\,{}^{m}_{\,\,p} \widehat{\dI}_i^s &= \frac{1}{2\pi} \int_0^{2\pi}\!\!\dd\ell  \ {}^{m} \widehat{\dI}_i^s \, \de^{- \di p \ell}= \frac{1}{2\pi} \int_0^{2\pi}\!\!\dd u \, \frac{\dd \ell}{\dd u}  \, {}^{m} \widehat{\dI}_i^s \,  \de^{- \di p (u-e_t \sin u)} \,,
\end{align}
and where  we recall from \eqref{eq:dldu} that $\dd \ell/\dd u = 1 -e_t \cos u + \mathcal{O}(c^{-4})$.
In order to explicitly compute ${}^{m}_{\,\,p} \widehat{\dI}_i^s$, recall that~${}^m \widehat{\dI}^s_i(t)$ only contains terms proportional to $\de^{\pm i [\ell + W(\ell)]}$,  and use  the 1PN-accurate relation 
\be \ell+W(\ell) = v + k(w + e_t \sin u) + \mathcal{O}(c^{-4}) \,.\ee
The exponential is PN-expanded to find
\be\de^{\pm i [\ell + W(\ell)]} = \de^{\pm i v}\Big(1 \pm  \di k [\tilde{w}+e_t \sin u] + \mathcal{O}(c^{-4})\Big)\,, \ee
where I have replaced at this order $w$ by  $\tilde{w}$, and where $k$  can be replaced by its leading order expression in terms of $x$ and $e_t$. 
The exponential term is easily computed using trigonometric functions, namely
\be\de^{\pm i  v}  = \frac{\cos u - e_\phi \pm \di   \sqrt{1-e_\phi^2} \sin u}{1-e_\phi \cos u} \,, \ee
where $e_\phi$ should then be replaced in terms of its 1PN expression in terms of $x$ and $e_t$.
After all replacements are made, I obtain a 1PN accurate expression in terms only of $x$, $e_t$ and $u$. From the structure of this expression, it appears that the Fourier coefficients can only be computed by introducing, on top of the Bessel functions, the real-valued\footnote{By a $u' = -u$ variable change, it is clear that \mbox{$ \Theta_{-p}^{-q}(e_t) =  -\Theta_{p}^{q}(e_t)$}, by oddness of $\arctan$. This also means that \mbox{$ \Theta_{p}^{q}(e_t) =  \Theta_{p}^{q}(e_t)^*$}, so~$\Theta_{p}^{q}(e_t) \in \mathbb{R}$.}  special function
\begin{align}\label{eq:Theta_pq}
     \Theta_{p}^q(e_t) &\equiv \frac{i}{2\pi}  \int_0^{2\pi} \dd u \, \de^{- \di[(p+q)u - p e_t \sin u] } \tilde{w}(u,e_t)\,,
\end{align}
which is not entirely novel since it appears in a different form in (68) of \cite{LY16}.
%
%
Finally, following \cite{ABIQ08tail, LY16}, it is useful for presentation purposes to expand the Fourier coefficients in the variables $x$ and $\nu$, so as to present coefficients that depend only on the eccentricities. The decomposition is denoted:
\be\,{}^{m}_{\,\,p} \widehat{\dI}_i^s = \,{}^{m}_{\,\,p} \widehat{\dI}_i^{s,00} + x\Big[ \,{}^{m}_{\,\,p} \widehat{\dI}_i^{s,01} + \nu  \,{}^{m}_{\,\,p} \widehat{\dI}_i^{s,11}\Big] + \mathcal{O}(x^2) \,.\ee
First, I found that $\forall p \in \mathbb{Z}, \,{}^{\pm1}_{\,\,p} \widehat{\dI}_z^s = 0$, and that the following relation holds
\be \label{eq:relationFourierComponentsXY} {}^{\pm 1}_{\,\,\,\,p} \widehat{\dI}_y^s = (\mp \di){}^{\pm 1}_{\,\,\,\,p} \widehat{\dI}_x^s \,.\ee
Thus, recalling the definition \eqref{eq:calB} for the recurrent parameter $\mathcal{B}$, all the information is contained in the following expressions for the $x$-component: 
\bse\begin{align}
\,{}^{\pm 1}_{\,\,\,p} \widehat{\dI}^{s,{00}}_{x}  &=  \pm \frac{\sqrt{1-e_t^2}}{2 p e_t}J_p(p e_t) + \frac{1}{2 p} J'_p(p e_t) \,,\\*
\,{}^{\pm 1}_{\,\,\,p} \widehat{\dI}^{s,{11}}_{x}&= \pm \frac{14+e_t^2}{30 p e_t \sqrt{1-e_t^2} }J_p(p e_t) + \frac{7}{15 p} J'_p(p e_t) \,,\\*
\,{}^{\pm 1}_{\,\,\,p} \widehat{\dI}^{s,{01}}_{x}&= J_p(p e_t)\Bigg[ \mp \frac{\di \mathcal{B}}{p e_t } + (1-e_t^2) \,\frac{\bep \Sm+\bem \Sp -\delta(\bep \Sp+\bem \Sm )- \frac{1}{4}\gam(\Sm-2 \Sp \delta) }{e_t \gam \Sm}\nn \\*
&\qquad\qquad +\frac{\di \mathcal{B}}{p^2 e_t \sqrt{1-\smash{e_t}^2} } \mp \frac{ \sqrt{1-\smash{e_t}^2} \Sp\delta }{10 p \Sm e_t}\mp\frac{39+10 \gam + 20 \bar(\beta_{+}- \bem\delta) + e_t^2(21 + 10 \gam) }{60 p  e_t\sqrt{1-\smash{e_t}^2}}   \Bigg]
\nn\\*
& + J'_p(p e_t)  \Bigg[ \pm \frac{\di\mathcal{B}}{p^2 \left(1-\smash{e_t}^2\right) }  - \frac{ \Sp\delta }{10 p \Sm}   -\frac{39+10 \gam + 20 \bar(\beta_{+}- \bem\delta) - e_t^2(99 + 50 \gam) }{60 p  (1-\smash{e_t}^2)}  \nn\\*
&\qquad\qquad  - \frac{\di \mathcal{B}}{p (1-e_t)^2 } \pm \sqrt{1-\smash{e_t}^2} \,\frac{\bep \Sm+\bem \Sp -\delta(\bep \Sp+\bem \Sm )- \frac{1}{4}\gam(\Sm-2 \Sp\delta) }{\gam \Sm}  \Bigg] \nn \\*
&  \mp \frac{\mathcal{B}}{8(1-e_t^2)}\Bigg[  e_t \left(1 \mp \sqrt{1-\smash{e_t}^2}\right) \Theta_p^{+2}(e_t)+ e_t \left(1 \pm\sqrt{1-\smash{e_t}^2}\right) \Theta_p^{-2}(e_t)\nn  \\*
&\qquad\qquad  - 2\left(1+e_t^2 \mp \sqrt{1-\smash{e_t}^2} \right)\Theta_p^{+1}(e_t) - 2\left(1+e_t^2 \pm \sqrt{1-\smash{e_t}^2}\right) \Theta_p^{-1}(e_t)  + 6 e_t \, \Theta_{p}^0(e_t)  \Bigg] \,.
\end{align}
 \ese
It is easy to verify that, 
\begin{align*}
 \,{}_{p} &\widehat{\dI}_x^s \equiv \de^{+\di k \ell}\left({}^{+1}_{\,\,\,\,p} \widehat{\dI}_x^s\right) +  \de^{-\di k \ell}\left({}^{-1}_{\,\,\,\,p} \widehat{\dI}_x^s\right)  \underset{c \rightarrow + \infty}{\longrightarrow} \,{}^{+1}_{\,\,\,\,p} \widehat{\dI}_x^{s,00} + \,{}^{-1}_{\,\,\,\,p}\widehat{\dI}_x^{s,00} = \frac{1}{p}J'_p(p e_t)  \,,
 \end{align*}
which is indeed the expression I found in the Newtonian case --- the same holds for the $y$-component.
Finally, the following identities, valid for any $(p,q) \in \mathbb {Z}^2$, will prove very useful\footnote{
The proof goes as follows. Performing an explicit summation over the contracted indices, I find that $\forall (m,s) \in \{-1,1\}^2$,
\begin{align*}
(m+s) \left({}^{m}_{\,\,p} \widehat{\dI}_i^{s}\right)\left(\!{}^{m}_{\,\,q} \widehat{\dI}_i^{s}\right) &= (m+s)(1-ms) \left({}^{m}_{\,\,p} \widehat{\dI}_x^{s}\right)\left(\!{}^{m}_{\,\,q} \widehat{\dI}_x^{s}\right)  \,, \\
 (m+s)\,\epsilon_{ijk}\! \left({}^{m}_{\,\,p} \widehat{\dI}_j^{s}\right)\left(\!{}^{m}_{\,\,q} \widehat{\dI}_k^{s}\right) &= \di   (m+s)(m-s)\,\epsilon_{ixy} \left({}^{m}_{\,\,p} \widehat{\dI}_x^{s}\right)\left(\!{}^{m}_{\,\,q} \widehat{\dI}_x^{s}\right)  \,,
\end{align*}
where I have used \eqref{eq:relationFourierComponentsXY} and the fact that the $z$ component of the dipole vanishes. 
But the right hand sides of the previous equations vanish identically for any allowed values of $m$ and $s$, which concludes the proof.
} in Sec.~\ref{sec:flux}:
\bse\begin{align}
&\sum_{\substack{m\in\{-1,1\}\\s\in\{-1,1\}}}(m+s)\left({}^{m}_{\,\,p} \widehat{\dI}_i^{s}\right)\left(\!{}^{m}_{\,\,q} \widehat{\dI}_i^{s}\right) = 0  \,, \\
&\sum_{\substack{m\in\{-1,1\}\\s\in\{-1,1\}}}(m+s)\,\epsilon_{ijk}\!\left({}^{m}_{\,\,p} \widehat{\dI}_j^{s}\right)\left(\!{}^{m}_{\,\,q} \widehat{\dI}_k^{s}\right) = 0  \,.
 \end{align}\ese


\subsection{Special functions}

\begin{figure}
\includegraphics[trim={1.9cm 1.3cm 0 0},clip,width=0.49\columnwidth]{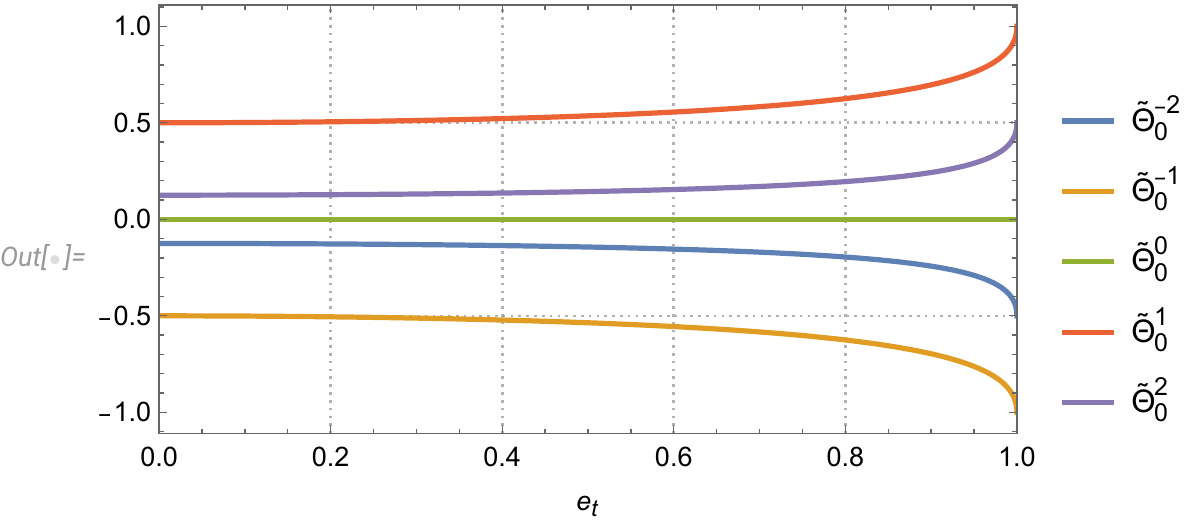}
\includegraphics[trim={1.9cm 1.3cm 0 0},clip,width=0.49\columnwidth]{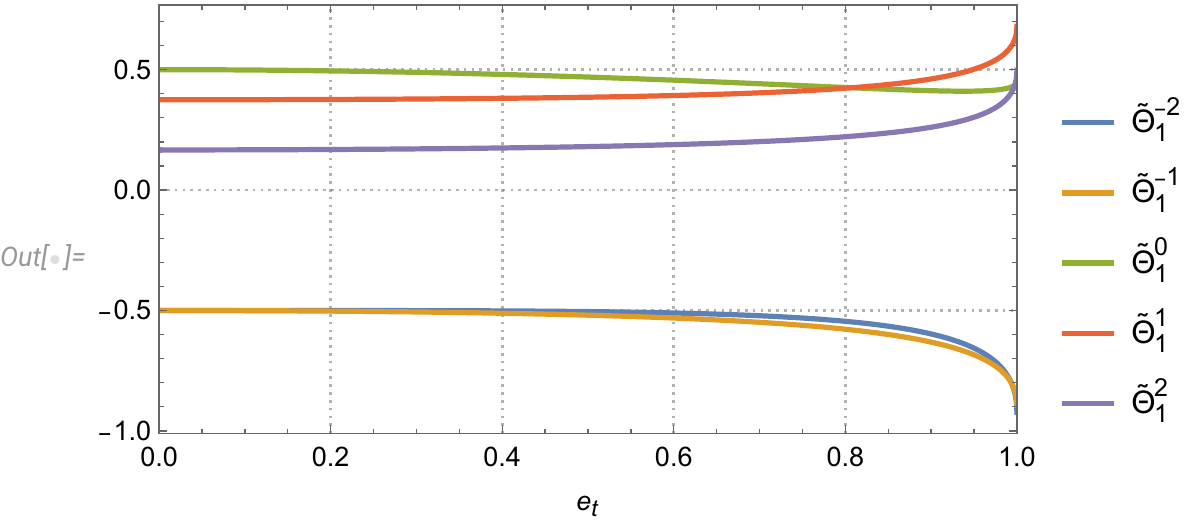}
\includegraphics[trim={1.9cm 1.3cm 0 0},clip,width=0.49\columnwidth]{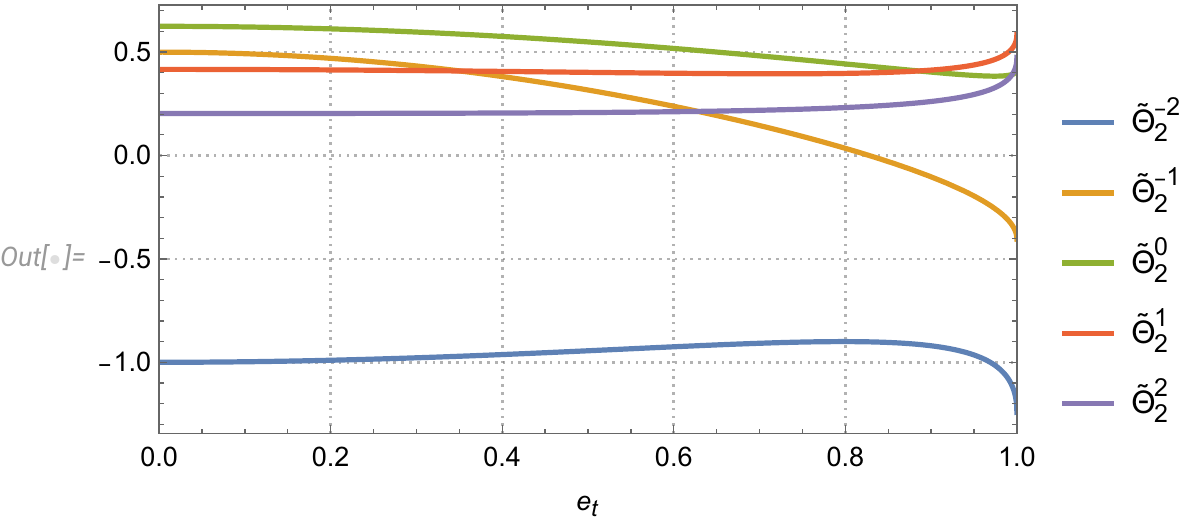}
\includegraphics[trim={1.9cm 1.3cm 0 0},clip,width=0.49\columnwidth]{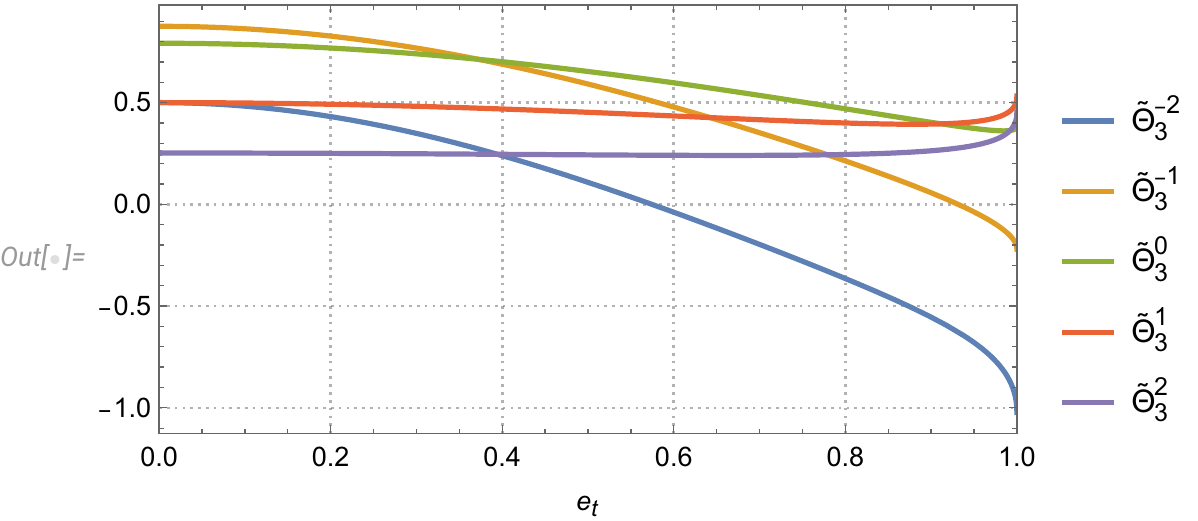}
\includegraphics[trim={1.9cm 1.3cm 0 0},clip,width=0.49\columnwidth]{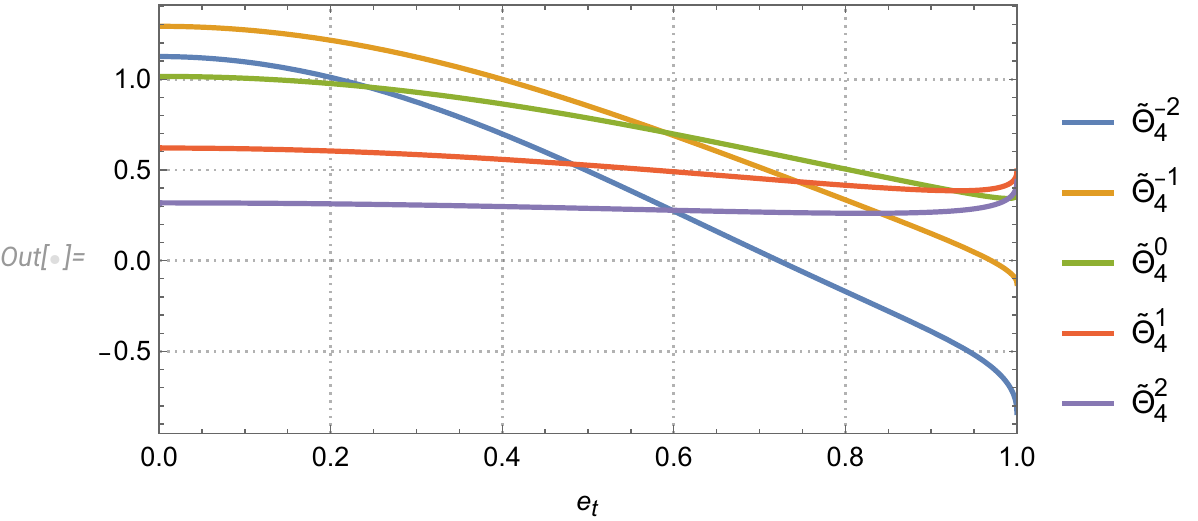}
\includegraphics[trim={1.9cm 1.3cm 0 0},clip,width=0.49\columnwidth]{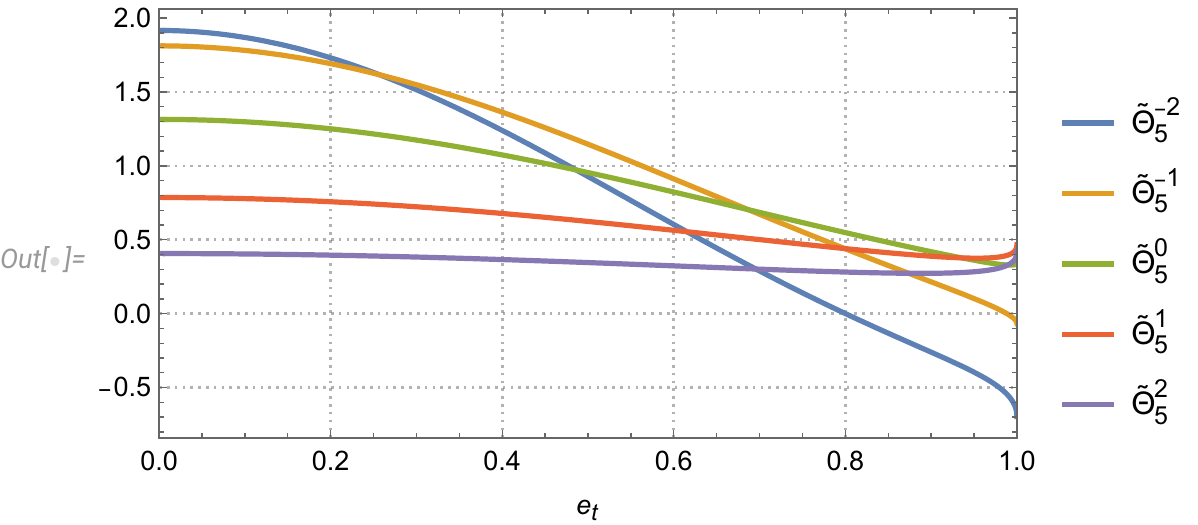}
\includegraphics[trim={1.9cm 1.3cm 0 0},clip,width=0.49\columnwidth]{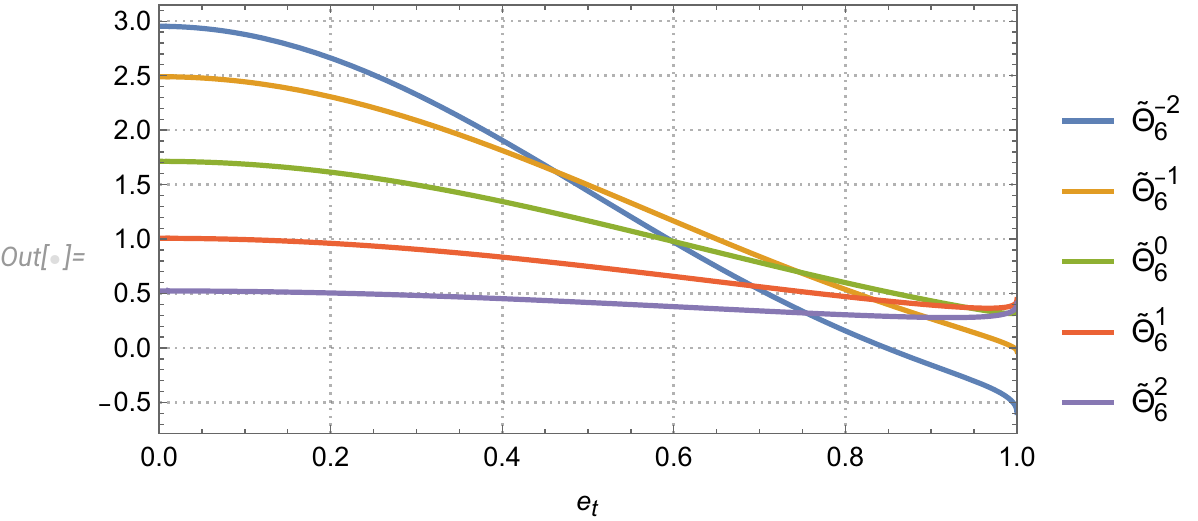}
\includegraphics[trim={1.9cm 1.3cm 0 0},clip,width=0.49\columnwidth]{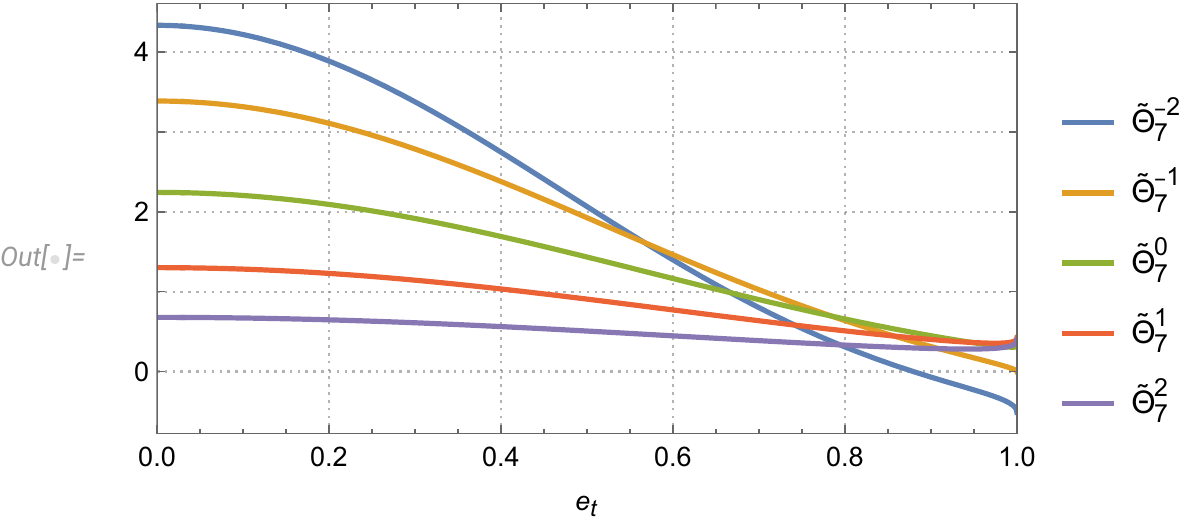}
\includegraphics[trim={1.9cm 1.3cm 0 0},clip,width=0.49\columnwidth]{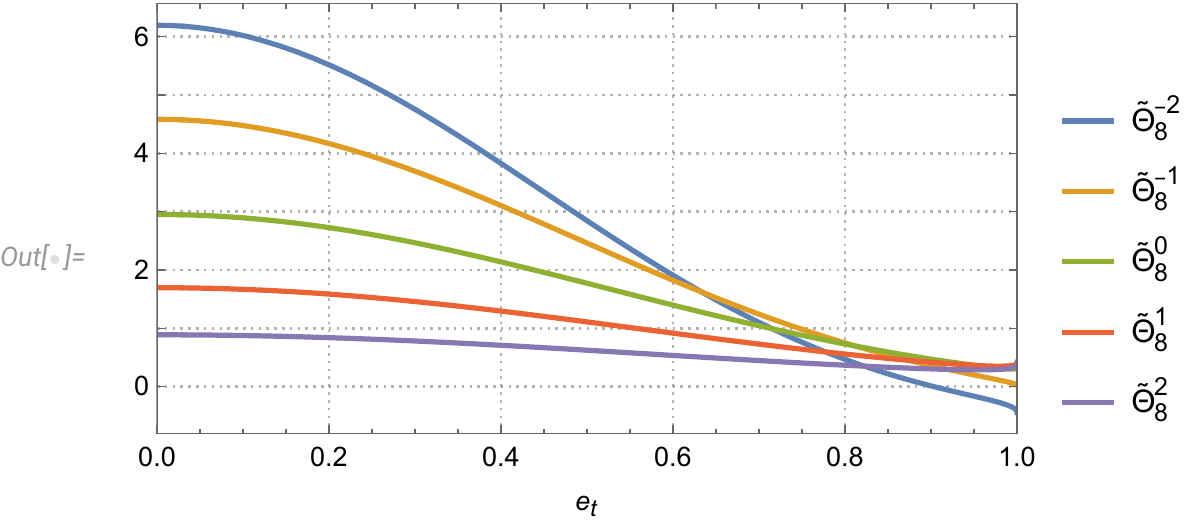}
\includegraphics[trim={1.9cm 1.3cm 0 0},clip,width=0.49\columnwidth]{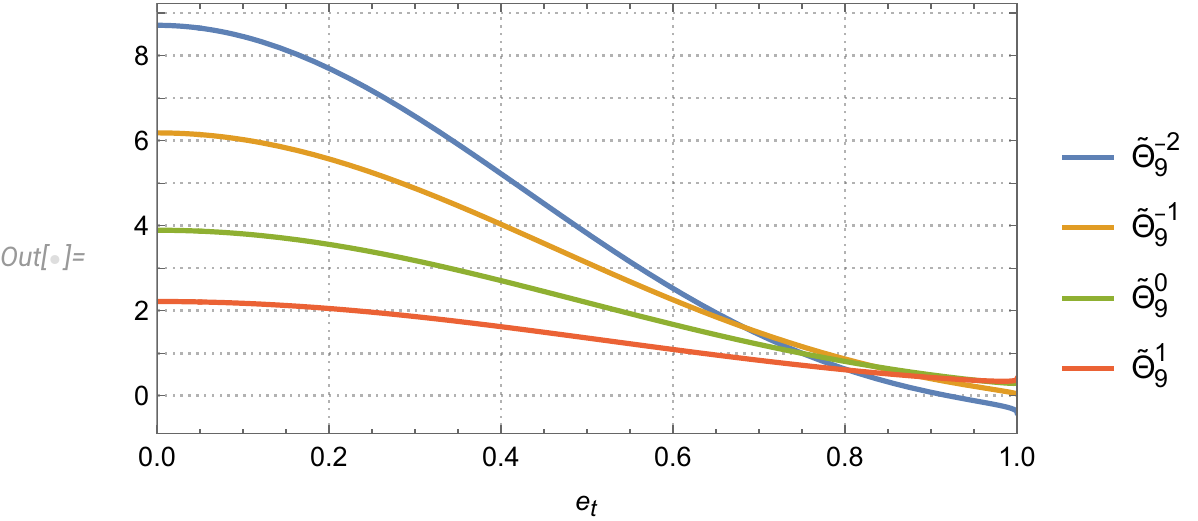}
\includegraphics[trim={1.9cm 0 0 0},clip,width=0.49\columnwidth]{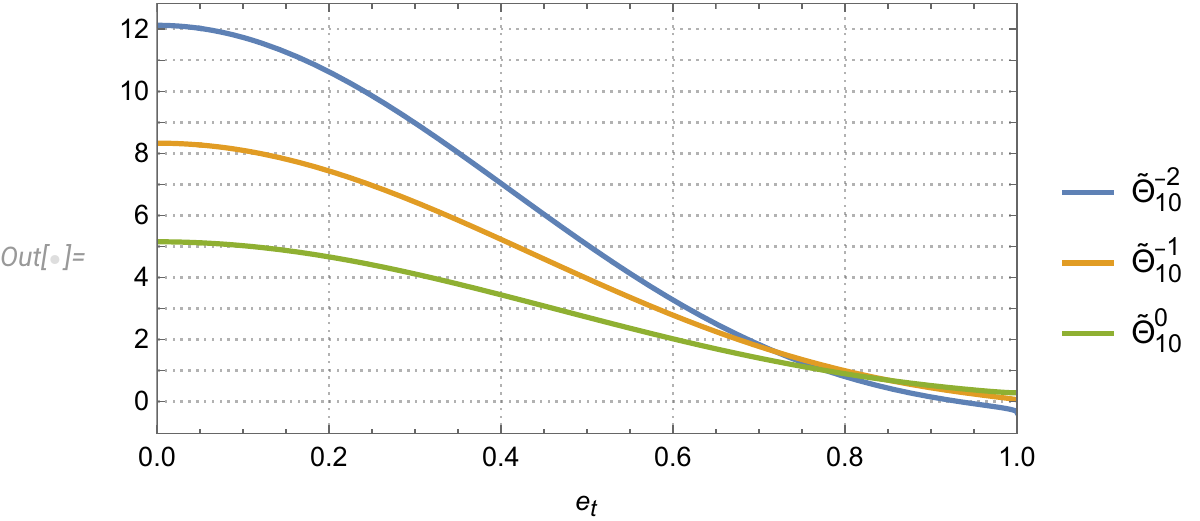}
\includegraphics[trim={1.9cm 0 0 0},clip,width=0.49\columnwidth]{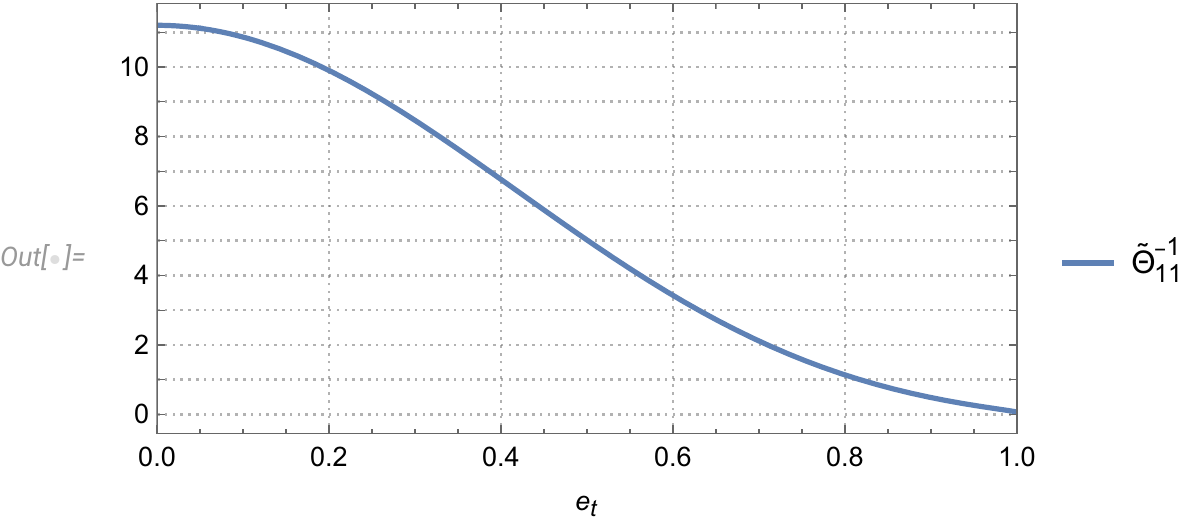}
\caption{Numerical plots of the $\widetilde{\Theta}_{p}^{q}(e_t)$ required for this computation, as well as the special case $p=0$. \label{fig:Theta_tilde}}
\end{figure}

The Fourier coefficients computed previously feature some special functions. The most common ones are the Bessel functions, which are well known, but the new special function  $\Theta_{p}^q(e_t)$ deserves some attention. For $p+q \neq 0$, it behaves as \mbox{$\Theta_{p}^{q}(e_t) = \mathcal{O}(e_t^{|p+q|})$} when as $e_t \rightarrow 0$, whereas \mbox{$\Theta_{p}^{-p}(e_t) = \mathcal{O}(e_t^{2})$}. It is therefore useful to define the renormalized function
\be\label{eq:Theta_tilde_pq}
\widetilde{\Theta}_{p}^{q}(e_t) =  \begin{cases}
	e_t^{-2}  \Theta_{p}^{q}(e_t) &\quad \text{if}\ p + q  = 0 \\
      e_t^{-|p+q|} \Theta_{p}^{q}(e_t)  &\quad \text{otherwise}
    \end{cases}\,,
\ee
such that $\tilde{\Theta}^{p}_{q}(0)$ is always finite and nonzero (except for $p=q=0$, since $\Theta^0_0(e_t)$ is identically zero).
These special functions only appear in the unnormalized enhancement functions $\rho_1^{s,01}(e_t)$ and $\tilde{\rho}_1^{s,01}(e_t)$  [see Eqs. \eqref{eq:decomposition_rho1s_tilderho1s}], which means that they only  appear in the normalized enhancement functions $\alpha_1^{s}(e_t)$ and $\tilde{\alpha}_1^{s}(e_t)$ [see Eqs. \eqref{eq:def_normalized_enhancement_functions_scalar}]. In Fig.~\ref{fig:Theta_tilde}, I have plotted the exact values of all the special functions which are needed to evaluate these enhancement functions, along with the case $p=0$, which was never needed but is nonetheless instructive.

In view of obtaining fully analytical results, it is also useful to control the small eccentricity expansion of these special functions. First, recall that the Bessel function and its derivative can be expanded for $p\neq 0$ as follows: 
\begin{align}
    \mathcal{J}_{p}(p e_t) &= \sum_{n=0}^\infty \frac{(-1)^{n}}{n!(n+|p|)!} \Big(\frac{|p| e_t}{2}\Big)^{2n+|p|} \,,\\*
    \mathcal{J}'_{p}(p e_t) &=  \mathrm{sgn}(p) \sum_{n=0}^\infty \frac{(-1)^{n}(2n+|p|)}{2 (n!)(n+|p|)!} \Big(\frac{|p| e_t}{2}\Big)^{2n+|p|-1}\,,
\end{align}
where I have introduced the sign function $\mathrm{sgn}(p) = 1$ if $p > 0$,  $\mathrm{sgn}(p) = -1$ if $p < 0$.

The Taylor expansion of $\Theta_{p}^{q}(e_t) $ as $e_t \rightarrow 0$ can be easily obtained  at any order for given values of $p$ and $q$ (e.g. with \emph{Mathematica}), but its general term is very cumbersome. Here, I will only present the results which are required to obtain the expansions of the normalized enhancement functions $\alpha_1^{s}(e_t)$ and $\tilde{\alpha}_1^{s}(e_t)$ at tenth order in the eccentricity [see Eqs. \eqref{eq:expansion_normalized_enhancement_functions} and App.~\ref{app:smallEccentricityExpansion}]. The corresponding expansions are:
\bse\begin{align}
\Theta_{1}^{-2}(e_t) &= - \frac{1}{2}e_t - \frac{23}{384}e_t^5 - \frac{181}{4608}e_t^7 - \frac{43093}{1474560}e_t^9 + \calO(e_t^{11}) \\
\Theta_{1}^{-1}(e_t) &= - \frac{1}{2}e_t^2 - \frac{1}{16} e_t^4 - \frac{59}{1152}e_t^6 - \frac{203}{6144} e_t^8 - \frac{58183}{2457600} e_t^{10}+ \calO(e_t^{12})  \\
\Theta_{1}^{0}(e_t) &= \frac{1}{2} e_t  - \frac{1}{8} e_t^3 + \frac{1}{128} e_t^5 + \frac{41}{9216} e_t^7 + \frac{2143}{491520} e_t^9 + \calO(e_t^{11})\\
\Theta_{1}^{1}(e_t) &=  \frac{3}{8} e_t^2 + \frac{1}{32} e_t^4 + \frac{113}{3072} e_t^6 + \frac{4589}{184320} e_t^8  +\calO(e_t^{10})\\
\Theta_{1}^{2}(e_t) &= \frac{1}{6} e_t^3 + \frac{35}{768} e_t^5  + \frac{25}{278} e_t^7 + \calO(e_t^9) \\
\Theta_{2}^{-2}(e_t) &= - e_t^2 + \frac{1}{4} e_t^4 - \frac{7}{72} e_t^6 - \frac{7}{192} e_t^8 + \calO(e_t^{10}) \\
\Theta_{2}^{-1}(e_t) &=  \frac{1}{2} e_t - \frac{3}{4} e_t^3 + \frac{1}{12} e_t^5 - \frac{53}{1152} e_t^7 - \frac{25}{1152} e_t^9 + \calO(e_t^{11})\\
\Theta_{2}^{0}(e_t) &= \frac{5}{8}e_t^2 - \frac{5}{16}e_t^4 + \frac{1}{24} e_t^6 - \frac{13}{11520}e_t^8 + \calO(e_t^{10})\\
\Theta_{2}^{1}(e_t) &= \frac{5}{12}e_t^3 - \frac{13}{192} e_t^5 + \frac{67}{1920} e_t^7 + \calO(e_t^9)\\
\Theta_{2}^{2}(e_t) &= \frac{13}{64}e_t^4 + \frac{1}{96} e_t^6 + \calO(e_t^8) \\
\Theta_{3}^{-2}(e_t) &= \frac{1}{2} e_t - \frac{7}{4} e_t^3 + \frac{101}{128} e_t^5 - \frac{59}{256} e_t^7 + \calO(e_t^9)\\
\Theta_{3}^{-1}(e_t) &= \frac{7}{8} e_t^2 - \frac{39}{32} e_t^4+ \frac{367}{1024} e_t^6  - \frac{1867}{20480} e_t^8 + \calO(e_t^{10})\\
\Theta_{3}^{0}(e_t) &= \frac{19}{24} e_t^3  - \frac{151}{256} e_t^5 + \frac{741}{5120} e_t^7 + \calO(e_t^9)\\
\Theta_{3}^{1}(e_t) &= \frac{1}{2} e_t^4 - \frac{53}{256} e_t^6 + \calO(e_t^8)\\
\Theta_{3}^{2}(e_t) &= \frac{323}{1280} e_t^5 + \calO(e_t^7)\\
\Theta_{4}^{-2}(e_t) &= \frac{9}{8}e_t^2 - \frac{47}{16} e_t^4 + \frac{691}{384} e_t^6 + \calO(e_t^8)\\
\Theta_{4}^{-1}(e_t) &= \frac{31}{24} e_t^3 - \frac{47}{24} e_t^5 + \frac{1687}{1920}e_t^7 + \calO(e_t^9)\\
\Theta_{4}^{0}(e_t) &= \frac{65}{64}e_t^4 - \frac{193}{192} e_t^6 + \calO(e_t^8)\\
\Theta_{4}^{1}(e_t) &= \frac{149}{240} e_t^5 + \calO(e_t^7)\\
\Theta_{5}^{-2}(e_t) &= \frac{23}{12}e_t^3 - \frac{3661}{768}e_t^5 + \calO(e_t^7)\\
\Theta_{5}^{-1}(e_t) &= \frac{29}{16} e_t^4 - \frac{2365}{768} e_t^6 + \calO(e_t^8) \\
\Theta_{5}^{0}(e_t) &= \frac{1683}{1280}e_t^5 + \calO(e_t^7)\\
\Theta_{6}^{-2}(e_t) &= \frac{189}{64} e_t^4 + \calO(e_t^6) \\
\Theta_{6}^{-1}(e_t) &= \frac{797}{320} e_t^5 + \calO(e_t^5) 
\end{align}\ese

\section{Passage to the center-of-mass frame at 2.5PN order}
\label{sec:CM}
In order to use the quasi-Keplerian parametrization, but also to yield simple and unambiguous results, the fluxes will be presented in the CM frame. However, the transformations from a general frame to the CM frame (including both conservative and dissipative contributions) are currently known only up to 2PN in scalar tensor theory \cite{BBT22}, but the computation of the angular momentum flux requires these laws at 2.5PN. This is the object the present section. Importantly, it was recently rediscovered (in general relativity) that it is vital to define the CM frame as the barycenter of both the matter and radiation \cite{BFT24}, the latter entering at high PN order as nonlocal integrals over the radiation escaping at infinity. Previous results \cite{GII97,Leibovich:2023xpg} that ignored the radiation contribution were therefore erroneous. In this section, I will therefore be careful to include these contributions.

\subsection{Flux balance laws for the linear momentum and position of the center of mass}
\label{subsec:flux_balance}
Due the the symmetries of Minkowski spacetime, the ten Poincaré invariants of the binary system (its energy $E$, angular momentum $J^i$, linear momentum $P^i$ and CM position $G^i$) are conserved in the absence of gravitational radiation. When including the radiation, these quantities are only approximately conserved, and are described by flux-balance laws. For the energy and angular momentum, these are described in \citetalias{T24_QK} for the case of ST theory, whereas the linear momentum, $P^i$, and the position of the CM, $G^i$, satisfy 
\bse \label{eq:balanceLawsPiGi} \begin{align}
\label{seq:balanceLawsPi} \frac{\dd P^i}{\dd t} &= - \mathcal{F}_{\bm{P}}^i  - \mathcal{F}_{s, \bm{P}}^i  \,,\\*
\label{seq:balanceLawsGi} \frac{\dd G^i}{\dd t} &=P^i  - \mathcal{F}_{\bm{G}}^i - \mathcal{F}_{s,\bm{G}}^i \,.
\end{align}\ese
The tensorial fluxes, $\mathcal{F}_{\bm{P}}^i $ and $ \mathcal{F}_{\bm{G}}^i$, are given (up to global factor $\phi_0$) in terms of the radiative moments by (4.12b-4.15b) of \cite{COS19}, and the scalar fluxes are defined~by
\bse \label{eq:fluxesIntegralsSpherePseudotensor}\begin{align}
\mathcal{F}^i_{s,\bm{P}} &= \frac{c^4 \phi_0 }{16 \pi G} \int \dd^2\! S_j \,\Lambda_\phi^{ij} \,, \\*
\mathcal{F}^i_{s,\bm{G}} &= \frac{c^3\phi_0 }{16 \pi G} \int \dd^2\! S_j \,\Big[ x^i \Lambda_\phi^{0j} - r \Lambda_\phi^{ij} \Big] \,,
\end{align}\ese
where $\int \dd^2 \! S_j $ denotes integration over the sphere. When taking the $r \rightarrow 0$ limit, only the asymptotic expression of $\Lambda_\phi$ are needed, namely
\begin{equation}\Lambda^{\mu\nu}_\phi = (3+\omega_0)\left(\eta^{\mu\alpha} \eta^{\nu \beta} - \frac{1}{2} \eta^{\mu\nu} \eta^{\alpha \beta}\right)\partial_\alpha \partial_\beta \psi + \mathcal{O}\left(\frac{1}{r^3}\right) \,.\end{equation}
Discarding the $\mathcal{O}(1/r^3)$ term which is contains at least cubic terms in the multipolar moments,${}^\text{\ref{footnote:uncontrolled_subleading_terms}}$
 and plugging this expression into \eqref{eq:fluxesIntegralsSpherePseudotensor}, I find
\bse\begin{align}
\mathcal{F}^i_{s,\bm{P}} &= \frac{c^2(3+2\omega_0)\phi_0r^2}{16\pi G}\int\dd \Omega \,n_i \dot{\psi}^2 \,, \\
\mathcal{F}^i_{s,\bm{G}} &= -\frac{c^3(3+2\omega_0)\phi_0r^3}{16\pi G}\int\dd \Omega \, \Bigg[n_j  \p_i \psi \p_j \psi + \frac{n_i n_j}{c} \dot{\psi}\p_j \psi  - \frac{1}{2} n^i \left(\p_k \psi \p_k \psi - \frac{1}{c^2}\dot{\psi}^2\right)\Bigg]\,,
\end{align}\ese
where $\int \dd \Omega$ denotes integration over angles. Finally, replacing $\psi$ by its multipolar expansion \eqref{seq:scalar_radiative_expansion} yields the final expressions\footnote{\label{footnote:uncontrolled_subleading_terms}Just as in \cite{Thorne:1980ru}, the uncontrolled $O(R^{-2})$ subleading term of the asymptotic multipolar expansion is assumed not to contribute. This assumption was rigorously proven in \cite{COS19} within the Bondi-Sachs framework in the case of GR, and I will assume without proof that the same arguments hold in the scalar sector.}
\bse\label{eq:FsP_FsG_radiative_moments}\begin{align}
\label{seq:FsP_radiative_moments} \mathcal{F}^i_{s,\bm{P}} &= G \phi_0 (3+2\omega_0) \sum_{\ell=0}^{\infty} \frac{2}{c^{2\ell+3}\ell ! (2\ell+3)!!}\overset{(1)}{\mathcal{U}}{\!}{}_L^s \overset{(1)}{\mathcal{U}}{}_{iL}^s \,,\\*
\label{seq:FsG_radiative_moments}   \mathcal{F}^i_{s,\bm{G}} &= G \phi_0 (3+2\omega_0) 
 \sum_{\ell=0}^{\infty}  \frac{\ell+2}{c^{2\ell+3}\ell!(2\ell+3)!!}\Big(\overset{(1)}{\mathcal{U}}{\!}{}_L^s{\mathcal{U}}_{iL}^s - \frac{\ell}{\ell+2}\overset{(1)}{\mathcal{U}}{\!}{}_{iL}^s{\mathcal{U}}_{L}^s\Big) \,.
\end{align}\ese
Note that, since the monopole is constant at Newtonian order and only acquires time-dependent contributions at subleading 1PN order, $\dd \mathcal{U}^s / \dd t$ is actually a 1PN quantity. Thus, the leading contributions to the scalar fluxes for $P^i$ and $G^i$ are of order $\mathcal{O}(1/c^5)$, and read 

\bse\label{eq:FsG_FsP_source_moments}\begin{align}
\label{seq:FsP_source_moments}\mathcal{F}^i_{s,\bm{P}} &= \frac{G \phi_0 (3+2\omega_0)}{c^5}\bigg[\frac{2}{3} \frac{\overset{(1)}{E}{}^s}{\phi_0} \overset{(2)}{\dI}{\!}_i^s + \frac{2}{15}\overset{(2)}{\dI}{\!}^s_j \overset{(3)}{\dI}{\!}^s_{ij}\bigg] + \calO\left(\frac{1}{c^7}\right)\,, 
\\*
\label{seq:FsG_source_moments}\mathcal{F}^i_{s,\bm{G}} &=  \frac{G \phi_0 (3+2\omega_0)}{c^5}\bigg[\frac{2}{3} \frac{\overset{(1)}{E}{}^s}{\phi_0} \overset{(1)}{\dI}{\!}_i^s +  \frac{1}{5}\overset{(2)}{\dI}{\!}^s_j \overset{(2)}{\dI}{\!}^s_{ij}- \frac{1}{15} \overset{(1)}{\dI}{\!}^s_{j} \overset{(3)}{\dI}{\!}^s_{ij} \bigg] + \calO\left(\frac{1}{c^7}\right) \,.
\end{align}\ese

The expression \eqref{seq:FsG_radiative_moments} of the scalar flux of CM position is analogous to the tensorial flux of CM position found by \cite{KNQ18, COS19}, rather than the one obtained by \cite{Blanchet:2018yqa}. For our purposes, both expressions are equivalent, because they differ by a total time derivative, see the discussion in Sec. V of \cite{BFT24}.

\subsection{Passage to the center of mass at 2.5PN order}
\label{subsec:passage_CM}
For the clarity of the discussion, the linear momentum and position of the CM are first decomposed into different PN contributions, namely
\bse \label{eq:PNdecompositionPiGi} \begin{align}  
\label{seq:PNdecompositionPi}  \bm{P} &= \bm{P}^\text{N} +\bm{P}^\text{1PN} +\bm{P}^\text{1.5PN} +\bm{P}^\text{2PN} + \bm{P}^\text{2.5PN} + \mathcal{O}(1/c^6)\,,\\*
\label{seq:PNdecompositionGi}  \bm{G} &= \bm{G}^\text{N} +\bm{G}^\text{1PN} +\bm{G}^\text{1.5PN} +\bm{G}^\text{2PN} + \bm{G}^\text{2.5PN} + \mathcal{O}(1/c^6)\,.
\end{align}\ese
Before this work, the even (conservative) contributions to $\bm{G}$ and $\bm{P}$ were derived in Ref. \cite{B18_ii}. The 1.5PN dissipative contributions have been previously computed in order to obtain (A.2) of \cite{BBT22}, but were not published. They read:
\bse\begin{align}
G_i^\text{1.5PN} &= \frac{\tilde{G} \alpha m^2 \zeta \Sm \Sp (2\nu-\delta-1) v_{12}^i }{3c^3} \,, \\
P_i^\text{1.5PN} &= \frac{2 \tilde{G}^2\alpha^2 m^3 \zeta \Sm \Sp(1+\delta-3\nu - \nu \delta)n_{12}^i}{3 c^3 r_{12}^2} \,,
\end{align}\ese
and are defined such that $\dd G^i / \dd t = P_i + \mathcal{O}(1/c^5)$ and $\dd P^i / \dd t =  \mathcal{O}(1/c^5)$  when using the full 2PN equations of motion, including both conservative and dissipative contributions. The latter were first computed in~\cite{MW13}, but the 1.5PN term is given in a more explicit form in (A.1a) of \citetalias{BBT22}. At 2PN order, the equation for the passage to the CM is then obtained by solving $G^i = \mathcal{O}(1/c^5)$, and is given by (A.2) of \citetalias{BBT22}.

	I am now going to obtain the 2.5PN-accurate equation for the passage to the CM frame, but without needing to determine $\bm{G}^\text{2.5PN}$ in a general frame~\cite{BlanchetCommunication}. At 2.5PN order enter the associated fluxes for $\bm{P}$ and $\bm{G}$, so the CM frame (associated to both the matter content and gravitational waves) will now be generically defined by the condition \cite{BFT24}
\begin{align}
G_i + \Gamma_i + \Gamma_i^s = 0 \,,
\end{align}
where the radiation contribution is described by
\bse \label{eq:defGammaGammas} \begin{align}
\Gamma_i &= \int_{-\infty}^t \!\!\!\! \dd t' \mathcal{F}_{\bm{G}}^i(t')+ \int_{-\infty}^t\!\!  \dd t' \!\! \int_{-\infty}^{t'}\!\! \dd t''   \mathcal{F}_{\bm{P}}^i(t'') \,,  \\*
\Gamma_i^s &= \int_{-\infty}^t \!\!\!\! \dd t' \mathcal{F}_{s,\bm{G}}^i(t')+ \int_{-\infty}^t \!\! \dd t' \!\! \int_{-\infty}^{t'} \!\! \dd t''   \mathcal{F}_{s,\bm{P}}^i(t'') \, .
\end{align}\ese
It will also prove useful to introduce
\bse \label{eq:defPiPis}\begin{align}
\Pi_i &= \int_{-\infty}^t \!\! \dd t' \mathcal{F}_{\bm{P}}^i(t') \,, \\*
\Pi_i^s &= \int_{-\infty}^t \!\! \dd t' \mathcal{F}_{s,\bm{P}}^i(t') \,,
\end{align}\ese
such that $P_i + \Pi_i + \Pi_i^s = 0$ is satisfied in the CM frame (in fact, this equation is the \emph{definion} of a center-of-inertia frame).
Note that at 2.5PN, the tensor nonlocal quantities $\Gamma_i$ and $\Pi_i$ do not contribute. Now, using the decomposition \eqref{eq:PNdecompositionPiGi}, one can rewrite the balance equations for the linear momentum \eqref{seq:balanceLawsPi} as
\begin{align} \label{eq:balanceLawPiReshuffled} 
&\frac{\dd P_i^\text{2.5PN}}{\dd t} = - \bigg(\frac{\dd  P_i^\text{N}}{\dd t} + \frac{\dd P_i^\text{1PN}}{\dd t} +\frac{\dd P_i^\text{1.5N}}{\dd t}+ \frac{\dd P_i^\text{2PN}}{\dd t} + \mathcal{F}^i_{\bm{P}} + \mathcal{F}^i_{s,\bm{P}}\bigg) \,.
\end{align}
Since ${\dd  P_i^\text{N}}/{\dd t} = \mathcal{O}(1/c^2)$, all quantities on the right hand side of \eqref{eq:balanceLawPiReshuffled} can be rewritten with 2.5PN accuracy in terms of the CM variables with the sole knowledge of the $1.5$PN-accurate formula for the passage to the CM frame, which is given by (3.6-8) of \cite{B18_ii} for the conservative piece and (A.2a) of \cite{BBT22} for the dissipative piece. By construction, the right-hand side will be of order $\mathcal{O}(1/c^5)$ in the CM frame. 

For any quantity $\mathcal{Z}$ expressed in terms of $(\bm{y_1}, \bm{y_2}, \bm{v_1}, \bm{v}_2)$, I will now define the related quantity $\left(\mathcal{Z}\right)_\text{CM}$, which depends only on $(x^i, v^i)$ and is obtained by applying to $\mathcal{Z}$ the formulas for the passage to the CM (e.g. $y_1^i = ...$) at the relevant order. Note that in the following, I will often be able to control $\left(\mathcal{Z}\right)_\text{CM}$ without controlling the full expression of $\mathcal{Z}$.

The quantity I am now interested in obtaining is $\left(P_i^\text{2.5PN}\right)_\text{CM}$, expressed at leading 2.5PN order. To this end, the equation which should be solved reads
\be \left. \frac{\dd \left(P_i^\text{2.5PN}\right)_\text{CM}}{\dd t} \right|_{a_i^{\text{CM},\text{N}}} = \Big(\text{RHS of \eqref{eq:balanceLawPiReshuffled}}\Big)_\text{CM} \,, \ee
where $\text{RHS}$ denotes the ``right hand side'' of the equation, and where I have indicated that only the Newtonian relative acceleration in the CM frame $a_i^\text{CM,N}$ is needed when taking the time derivative. This is easily solved in the CM frame, and the solution reads:
\begin{align}
 \left(P_i^\text{2.5PN}\right)_\text{CM} &= \left[\frac{p_1}{r^3} + p_2 \frac{(nv)^2}{r^2} + p_3 \frac{v^2}{r^2}\right]n^i + p_ 4 \frac{(nv) v^i}{r^2}\,,
\end{align}
where the coefficients $p_n$ are given in App.~\ref{app:passageCM}. Repeating this process for the CM balance equation~\eqref{seq:balanceLawsGi} leads to
\begin{align}\label{eq:balanceLawGiReshuffled}
 \frac{\dd G_i^\text{2.5PN}}{\dd t} &=   P_i^\text{N}+ P_i^\text{1PN}+ P_i^\text{1.5N}+ P_i^\text{2PN} + P_i^\text{2.5PN}   - \bigg(\frac{\dd  G_i^\text{N}}{\dd t} + \frac{\dd G_i^\text{1PN}}{\dd t} +\frac{\dd G_i^\text{1.5N}}{\dd t}+ \frac{\dd G_i^\text{2PN}}{\dd t}  + \mathcal{F}^i_{\bm{G}} + \mathcal{F}^i_{s,\bm{G}}\bigg) \,.
\end{align}

Since $P_i^\text{N} - \dd G_i^\text{N}/\dd t = \mathcal{O}(1/c^2)$, and since $\left(P_i^\text{2.5PN}\right)_\text{CM}$ has been computed in the previous step, one can again easily express the right hand side of \eqref{eq:balanceLawGiReshuffled} in terms of the CM variables using solely the $1.5$PN-accurate formula for the passage to the CM. Again, by construction, the right-hand side is thus of order $\mathcal{O}(1/c^5)$ in the CM frame, and one is left to solve 
\be\left. \frac{\dd \left(G_i^\text{2.5PN}\right)_\text{CM}}{\dd t} \right|_{a_i^{\text{CM},\text{N}}} =  \Big(\text{RHS of \eqref{eq:balanceLawGiReshuffled}}\Big)_\text{CM}   \,,\ee
Again, this is easily solved for in the CM frame, and I find:
\begin{align}
 \left(G_i^\text{2.5PN}\right)_\text{CM} &=   q_1\frac{ (nv) n^i}{r}+\left[\frac{q_2}{r} + q_3 v^2\right]v^i \,,
\end{align}
where the coefficients $q_n$ are given in App.~\ref{app:passageCM}. With this expression, I am now in a position to solve for $y_1^i$ with 2.5PN accuracy in the equation
\begin{align} &G_i^\text{N} + \Big(G_i^\text{1PN} + G_i^\text{1.5PN}+ G_i^\text{2PN}+ G_i^\text{2.5PN}+\Gamma_i +\Gamma_i^s \Big)_\text{CM} =  \mathcal{O}\left(\frac{1}{c^6}\right) \,, \end{align}
where $G_i^\text{N} = m_1 y_1^i + m_2 y_2^i = m y_1^i - m_2 r n^i$.
This yields the final expression
\bse \label{eq:CMformulay1y2}\begin{align}
\label{seq:CMformulay1} y_1^i &= \left(X_2 + \nu \mathcal{P} \right)x^i + \nu \mathcal{Q}  v^i + \mathcal{R}_i \,\\*
\label{seq:CMformulay2} y_2^i &= \left(- X_1 + \nu \mathcal{P} \right)x^i + \nu \mathcal{Q}  v^i + \mathcal{R}_i \,,
\end{align}\ese
where $X_1 = m_1/m = (1+\delta)/2$ and $X_2 = m_2/m = (1-\delta)/2$. The $\mathcal{R}_i$ corresponds to the contribution arising from the radiation (in line with the conventions of Ref.~\cite{BFT24}), and simply reads
\be \mathcal{R}^i = - \frac{\Gamma_i^s}{m} + \calO\left(\frac{1}{c^7}\right) \,,\ee
where it should be noted that $\Gamma_i^s$ is intrinsically a 2.5PN quantity, i.e. $\Gamma_i^s=  \calO\left(c^{-5}\right)$.
Conversely, the $\mathcal{P}$ and $\mathcal{Q}$ contributions are the contributions arising from matter [i.e. what one would find if one solved $G^i = 0$]. They are decomposed as 
\bse\begin{align}
\mathcal{P} &=  \mathcal{P}_\text{1PN}+ \mathcal{P}_\text{1.5PN}+ \mathcal{P}_\text{2PN}+ \mathcal{P}_\text{2.5PN} + \calO\left(\frac{1}{c^6}\right) \,,  \\
\mathcal{Q} &= \mathcal{Q}_\text{1PN}+ \mathcal{Q}_\text{1.5PN}+ \mathcal{Q}_\text{2PN}+ \mathcal{Q}_\text{2.5PN} + \calO\left(\frac{1}{c^6}\right) \,,
\end{align}\ese
where the conservative contributions $\mathcal{P}_\text{1PN}$, $\mathcal{Q}_\text{1PN}$, $\mathcal{P}_\text{2PN}$ and  $\mathcal{Q}_\text{2PN}$, are given by (3.7) and (3.8) of \cite{B18_ii}, and where the dissipative contributions read\footnote{\label{footnote:1.5PN_CoM_passage} See also (A.2) of  \cite{BBT22} for the 1.5PN piece, where the 1.5PN contributions of $y_1^i$ and $v_1^i$ are expressed with a less compact choice of ST variables.} 
\bse \label{eq:CMdissipativePQ}\begin{align}
\mathcal{P}_\text{1.5PN} &= 0 \,,\\*
\mathcal{Q}_\text{1.5PN} &= \frac{2 \tilde{G} \alpha m \zeta \Sm (\Sp + \Sm \delta )}{3 c^3}\,, \\
\mathcal{P}_\text{2.5PN} &= \frac{(\tilde{G}\alpha m)^2\, (nv)}{15 c^5 r^2}\Big[10 \bem - 5 \zeta \Sm\Sp  + 24  \zeta \Sm \Sp \nu   +  \delta (5 \gam - 10 \bep - 5 \zeta \Sm^2)  + 26  \zeta \Sm^2  \nu \delta   \Big] \,,\\
\mathcal{Q}_\text{2.5PN} &= \frac{ \tilde{G}\alpha m  v^2}{15 c^5}\Big[ 5 \zeta \Sm \Sp + \delta(12 + 5\gam +5 \zeta \Sm^2)  -23 \zeta \Sm \Sp \nu - 17 \zeta \Sm^2 \nu \delta \Big]\nn\\* 
& + \frac{(\tilde{G}\alpha m)^2}{15 c^5 r}\bigg[ - 10 \bem - 5 \zeta \Sm\Sp   + \delta(-24 - 15 \gam +10 \bep - 5 \zeta \Sm^2 ) \nn\\*
&\qquad\qquad\quad\ \,  + \nu \Big(160 \gamInv \zeta  \Sm ( \bep \Sp +\bem \Sm)   - 18 \zeta \Sm \Sp \Big) + 28 \zeta \Sm^2 \nu \delta  \bigg]\,.
\end{align}\ese

Taking the time derivative of \eqref{eq:CMformulay1y2}, using the 2.5PN-accurate acceleration \cite{MW13,B18_ii, BBT22}, as well as  the identity $\dd \Gamma^s_i  / \dd t = \Pi^s_i + \mathcal{F}^i_{s,\bm{G}}$, one obtains the expressions for the passage to the CM for $v_1^i$ and $v_2^i$.
For the reader's convenience, the expressions of  $\mathcal{F}^i_{s,\bm{P}}$ (which enter the definition of $\bm{\Pi}^s$)  and of $\mathcal{F}^i_{s,\bm{G}}$ (which enters the definitions of $\bm{\Pi}^s$ and $\bm{\Gamma}^s$) are given at leading order and in the CM frame as follows:  
\bse \label{eq:expressionFPiFGi}\begin{align}
\label{seq:expressionFPi} &\mathcal{F}^i_{s,\bm{P}} = \frac{8 (\tilde{G}\alpha m)^3 m  \nu^2 \zeta \Sm  }{15 c^5 r^4}\Bigg[\Big(  20 \gamInv (\bep  \Sp + \bem  \Sm)   - 8  \Sp  + 3 \Sm \delta \Big) (nv) n^i + 2 \big(\Sp - \Sm \delta  \big) v^i \Bigg] \,,\\
\label{seq:expressionGi} &\mathcal{F}^i_{s,\bm{G}} = \frac{4 (\tilde{G}\alpha m)^2 m  \nu^2 \zeta \Sm }{15 c^5 r^2}  \Bigg[(\Sp - \Sm \delta)\bigg(\frac{2 \tilde{G}\alpha m}{r} - 3 (nv)^2 + 3 v^2 \bigg)n^i \nn\\
& \qquad\qquad\qquad\qquad\qquad\quad - \Big(  40\gamInv ( \bep \Sm + \bem \Sm)    - 12  \Sp + 2 \Sm \delta  \Big)  (nv) v^i\Bigg] \,.
\end{align}\ese
Note that only the 1.5PN-accurate formulas for the passage to the CM frame were required when deriving the 2.5PN equations of motion in the CM frame, so previous results \cite{MW13, BBT22} are unaffected by the inclusion of the radiation contributions.
\subsection{Expression of $\Gamma_i^s$ and $\Pi_i^s$ for an elliptic orbit}
\label{subsec:nonlocal_term_elliptic}
The appearance of nonlocal terms in the formula for the passage to the CM frame introduces a practical difficulty, because it depends on the whole past history of the binary. In this section, I will show that assuming the two-body system has eternally been isolated makes it possible to ``localize'' these terms, in a similar way to other nonlocal effects like tails or memory. I will consider the case of a  binary on an elliptic orbit (only the leading Newtonian order is required here), and discuss its reduction it to the special case of circular orbits at the end.

Since these quantities are to be computed only at leading order, I assume that the motion has been eternally elliptic, where the orbital parameters adiabatically evolve due to radiation reaction. Thus, at this level of accuracy, the semimajor axis $a$ and the eccentricity $e$ asymptote, respectively, to $+\infty$ and $1$ in the infinite past. The fluxes of linear momentum \eqref{seq:FsP_source_moments} and CM position \eqref{seq:FsG_source_moments} are controlled at any point in the past, which I subdivide into an orbit-averaged contribution and a zero-average oscillating contribution,  namely \mbox{$\mathcal{F}_{s,\bm{P}}^i = \langle \mathcal{F}_{s,\bm{P}}^i \rangle +\widetilde{\mathcal{F}}_{s,\bm{P}}^i$} and  \mbox{$\mathcal{F}_{s,\bm{G}}^i = \langle \mathcal{F}_{s,\bm{G}}^i \rangle +\widetilde{\mathcal{F}}_{s,\bm{G}}^i$}. Since only the leading-order contribution in  $\Pi_i^s$ and $\Gamma_i^s$ are required, I will consider that the motion is Keplerian, up to leading-order radiation reaction terms. This leads to the decomposition \mbox{$\Pi^s_i = \Pi^{s,\text{DC}}_i+ \Pi^{s,\text{AC}}_i $} and  \mbox{$\Gamma^s_i = \Gamma^{s,\text{DC}}_i+ \Gamma^{s,\text{AC}}_i $}, where 
\begin{align}
 \Pi_i^{s,\text{DC}} &= \int_{-\infty}^{t} \dd t' \langle \mathcal{F}_{s,\bm{P}}^i \rangle (t')\,,\qquad\qquad\qquad&
 \Gamma_i^{s,\text{DC}} &= \int_{-\infty}^{t} \dd t' \langle \mathcal{F}_{s,\bm{G}}^i \rangle (t') +  \int_{-\infty}^t \!\! \dd t' \!\! \int_{-\infty}^{t'} \!\! \dd t''   \left\langle \mathcal{F}_{s,\bm{P}}^i \right\rangle(t'')  \,,\nn\\*
  \Pi_i^{s,\text{AC}} &= \int_{-\infty}^{t} \dd t' \widetilde{ \mathcal{F}}_{s,\bm{P}}^i  (t')  \,,\qquad\qquad\qquad&
 \Gamma_i^{s,\text{AC}} &= \int_{-\infty}^{t} \dd t' \widetilde{ \mathcal{F}}_{s,\bm{G}}^i  (t') + \int_{-\infty}^t \!\! \dd t' \!\! \int_{-\infty}^{t'} \!\! \dd t''   \widetilde{\mathcal{F}}_{s,\bm{P}}^i(t'') \,.
 \end{align}

\subsubsection{The AC part}

I will first consider the AC contribution. The oscillatory part of the flux can be written in terms of the Fourier decomposition of the source moments (see Sec.~\ref{sec:Fourier}), and reads
\bse\begin{align}
\widetilde{\mathcal{F}}^i_{s,\bm{P}} &= \frac{ G \phi_0 (3+2\omega_0)}{c^5} \sum_{p + q \neq 0}\de^{\di (p+q)\ell}  \bigg[\frac{ 2(\di p n)(\di q n)^2}{3 \phi_0}  \Big({}_p \widetilde{E}^s\Big)  \Big( {}_q\widetilde{\dI}^s_{i} \Big)   + \frac{2(\di p n)^2 (\di q n)^3}{15}  \Big( {}_p \widetilde{\dI}^s_{j}\Big) \Big( {}_q \widetilde{\dI}^s_{ij}\Big) \bigg] \,, \\
\widetilde{\mathcal{F}}^i_{s,\bm{G}} &= \frac{G \phi_0  (3+2\omega_0)}{c^5}   \sum_{p + q \neq 0}\de^{\di (p+q)\ell}  \bigg[\frac{2}{3}\frac{ (\di p n)(\di q n)}{ \phi_0}  \Big({}_p \widetilde{E}^s\Big)  \Big( {}_q\widetilde{\dI}^s_{i} \Big)   + \Big(\frac{(\di p n)^2 (\di q n)^2}{5} -  \frac{(\di p n) (\di q n)^3}{15}  \Big) \Big( {}_p \widetilde{\dI}^s_{j}\Big) \Big( {}_q \widetilde{\dI}^s_{ij}\Big) \bigg] \,, 
\end{align}\ese
where the $p=q$ case has been excluded from the sum, as it corresponds exactly to the DC piece, which will be treated separately.
It is then immediate to compute its time integral, namely 

\bse\begin{align}
\Pi_i^{s,\text{AC}} &= \frac{ G \phi_0 (3+2\omega_0)}{c^5}  \sum_{p + q  \neq 0}\de^{\di (p+q)\ell}  \bigg[\frac{2}{3 \phi_0}\frac{ (\di p n )(\di q n)^2}{\di (p+q) n}  \Big({}_p \widetilde{E}^s\Big)  \Big( {}_q\widetilde{\dI}^s_{i} \Big)    + \frac{2}{15} \frac{(\di p n)^2 (\di q n)^3}{\di (p+q) n}  \Big( {}_p \widetilde{\dI}^s_{j}\Big) \Big( {}_q \widetilde{\dI}^s_{ij}\Big) \bigg] \, ,\\
\Gamma_i^{s,\text{AC}} &= \frac{G \phi_0  (3+2\omega_0)}{c^5}  \sum_{p + q \neq 0}\de^{\di (p+q)\ell}  \bigg[ \frac{2}{3 \phi_0} \bigg( \frac{ (\di p n)(\di q n)}{\di (p+q) n} + \frac{(\di p n)(\di q n)^2}{\left[\di (p+q)n\right]^2}\bigg)  \Big({}_p \widetilde{E}^s\Big)  \Big( {}_q\widetilde{\dI}^s_{i} \Big) \nn\\
&\qquad\qquad\qquad\qquad\qquad\qquad   +\frac{1}{5} \bigg(\frac{(\di p n)^2 (\di q n)^2}{\di (p+q) n} -  \frac{1}{3}\frac{(\di p n) (\di q n)^3}{\di (p+q) n} + \frac{2}{3} \frac{(\di p n)^2 (\di q n)^3}{\left[\di(p+q) n\right]^2} \bigg) \Big( {}_p \widetilde{\dI}^s_{j}\Big) \Big( {}_q \widetilde{\dI}^s_{ij}\Big) \bigg] \,, 
\end{align}\ese
where the $-\infty$ bound does not contribute and has been discarded (see \cite{BS93,ABIQ04} for a justification). Projecting onto the $(x,y,z)$ components, leveraging the fact that $\Pi_i^{s,\text{AC}}  \in \mathbb{R}$, and using the real or purely imaginary character of the Fourier coefficients in Eq.~\eqref{eq:normalizedFourierCoefficients}, I find that \mbox{$\Gamma_i^{s,\text{AC}}  = \Gamma_x^{s,\text{AC}} n_0^i + \Gamma_y^{s,\text{AC}} \lambda_0^i$} and  \mbox{$\Pi_i^{s,\text{AC}}  = \Pi_x^{s,\text{AC}} n_0^i + \Pi_y^{s,\text{AC}} \lambda_0^i$}, where 
\bse\begin{align}
\Pi_x^{s,\text{AC}} &=  - \frac{2  G  \phi_0(3+2\omega_0)}{3 c^5}  \sum_{p + q \neq 0} \cos\Big( (p+q)\ell \Big)\Bigg\{\frac{n^2}{\phi_0}\,\frac{p q^2}{p+q} \, {}_p \widetilde{E}^s \  {}_q \widetilde{\dI}_x^s   - \frac{n^4}{5} \,\frac{p^2 q^3}{p+q} \,\bigg({}_p \widetilde{\dI}_x^s  \   {}_q\widetilde{\dI}_{xx}^s + {}_p \widetilde{\dI}_y^s \ {}_q  \widetilde{\dI}_{xy}^s \bigg) \Bigg\} \,, \\
\Pi_y^{s,\text{AC}} &= -\frac{2 \di G  \phi_0(3+2\omega_0)}{3 c^5}  \sum_{p + q \neq 0} \sin\Big( (p+q)\ell \Big)\Bigg\{\frac{n^2}{\phi_0}\,\frac{p q^2}{p+q} \, {}_p  \widetilde{E}^s \  {}_q \widetilde{\dI}_y^s   - \frac{n^4}{5} \,\frac{p^2 q^3}{p+q} \,\bigg({}_p \widetilde{\dI}_x^s \  {}_q  \widetilde{\dI}_{xy}^s + {}_p \widetilde{\dI}_y^s  \ {}_q\widetilde{\dI}_{yy}^s \bigg)\! \Bigg\} \,, \\
\Gamma_x^{s,\text{AC}} &=   \frac{ G  \phi_0(3+2\omega_0)}{c^5}   \sum_{p + q \neq 0} \sin\Big( (p+q)\ell \Big)\Bigg\{- \frac{2n}{3\phi_0}\,\frac{p q(p+2q)}{(p+q)^2} \, {}_p \widetilde{E}^s \  {}_q \widetilde{\dI}_x^s   \nn\\
& \qquad\qquad\qquad\qquad\qquad\qquad\qquad\qquad+ \frac{n^3}{15}\, \frac{pq^2(3p^2 + 4 p q - q^2)}{(p+q)^2}\bigg({}_p \widetilde{\dI}_x^s  \  {}_q\widetilde{\dI}_{xx}^s + {}_p \widetilde{\dI}_y^s \ {}_q  \widetilde{\dI}_{xy}^s  \bigg) \Bigg\} , \\
\Gamma_y^{s,\text{AC}} &=  - \frac{\di G  \phi_0(3+2\omega_0)}{c^5}  \sum_{p + q \neq 0} \cos\Big( (p+q)\ell \Big)\Bigg\{- \frac{2n}{3\phi_0}\,\frac{p q(p+2q)}{(p+q)^2} {}_p\widetilde{E}^s \  {}_q \widetilde{\dI}_y^s    \nn\\
& \qquad\qquad\qquad\qquad\qquad\qquad\qquad\qquad\quad + \frac{n^3}{15}\, \frac{pq^2(3p^2 + 4 p q - q^2)}{(p+q)^2} \bigg({}_p \widetilde{\dI}_x^s  \  {}_q\widetilde{\dI}_{xy}^s + {}_p \widetilde{\dI}_y^s \ {}_q  \widetilde{\dI}_{yy}^s \bigg) \Bigg\} \,.
\end{align}\ese
Note that the latter contributions vanish upon orbit averaging. Using the QK parametrization \eqref{eq:PK_equations}, it is possible to expand the previous expressions in the small eccentricity. These expressions are quite cumbersome, so I will only include the leading correction in the eccentricity. They read:
\bse\begin{align}
\Pi_i^{s,\text{AC}} &= \frac{16}{15} c \,\zeta m \nu^2 \Sm\, x^3 \Bigg\{\left(\Sp - \Sm \delta \right) \Big[ \cos(\ell) \,n_0^i + \sin(\ell)\, \lambda_0^i \Big] \nn\\
& \qquad+ e  \Bigg[- \frac{5}{2} \gam^{-1}(\bep \Sp + \bem \Sm) + \frac{5}{2}  \Sp - \frac{15}{8}  \Sm \delta\Bigg] \Bigl(\cos(2\ell)\,n_0^i + \sin(2\ell) \, \lambda_0^i\Bigr) + \calO(e^2)
\Bigg\}\, , \\
\Gamma_i^{s,\text{AC}} &= \frac{12}{5 c^2}\tilde{G}\alpha m^2 \nu^2 \zeta \Sm x^{3/2} \Bigg\{\left(\Sp - \Sm \delta \right) \Big[ \sin(\ell) \,n_0^i - \cos(\ell)\, \lambda_0^i \Big] \nn\\
&  \qquad+ e  \Bigg[- \frac{5}{3} \gam^{-1}(\bep \Sp + \bem \Sm) + \frac{5}{3}  \Sp - \frac{5}{4}  \Sm \delta\Bigg] \Bigl(\sin(2\ell)\,n_0^i - \cos(2\ell) \, \lambda_0^i\Bigr) + \calO(e^2) \Bigg\} \,. 
\end{align}\ese

I can then take the limit of circular orbits, re-express everything in terms of the corotating basis $(\bm{n},\bm{\lambda},\bm{\ell})$ and then substitute $\bm{\lambda}$ and $x$ by their expressions in terms of $\bm{v}$ and $r$. I then find
\bse\begin{align}
 \Big[\Pi_i^{s,\text{AC}}\Big]_\text{circ} &= \frac{16 (\tilde{G} \alpha m)^3 m \nu^2 \zeta \Sm(\Sp - \Sm \delta)}{15 c^5 r^3}n^i \,,\\*
  \Big[\Gamma_i^{s,\text{AC}}\Big]_\text{circ} &= -\frac{12 (\tilde{G} \alpha m) m \nu^2 \zeta \Sm(\Sp - \Sm \delta)}{5 c^5 r}v^i \,.
 \end{align}\ese
%
I have checked that the simpler calculation that assumes circular orbits from the start (and does not use the Fourier decomposition of the multipolar moments) yields the same result, which is an important consistency test of the result and of the Fourier decomposition.
\subsubsection{The DC part}

Injecting the Keplerian parametrization into Eqs. \eqref{eq:expressionFPiFGi}, and performing the orbit averaging as usual, I find that the orbit-averaged fluxes of linear momentum and CM position read
\bse\label{eq:orbit_averaged_FsG_FsP}\begin{align}
\langle \mathcal{F}_{s,\bm{G}}^i \rangle  &=\frac{4(\tilde{G} \alpha m)^{3}m \nu^2 \zeta \Sm }{15 \gam c^5} (20 \bep \Sp+20 \bem \Sm   - 2 \gam \Sp   - 3 \gam \Sm \delta) \, \frac{e}{(1-e^2)^{3/2} a^{3}} n_0^i  \,,\\*
\langle \mathcal{F}_{s,\bm{P}}^i \rangle  &= \frac{(\tilde{G} \alpha m)^{7/2}m \nu^2 \zeta \Sm }{3 \gam c^5} (4 \bep \Sp+4 \bem \Sm  - \gam \Sm \delta)\,\frac{e(4+e^2)}{(1-e^2)^3 a^{9/2}} \lambda_0^i \,.
\end{align}\ese
%
%
Thanks to the leading-order relations \cite{T24_QK}
\bse\label{eq:PetersMathews_ST}\begin{align}
\label{seq:dedt} \left\langle \frac{\dd e}{\dd t} \right\rangle &=  - \frac{(\tilde{G}\alpha m)^2 \zeta\mathcal{S}_-^2 \nu}{c^3}\times \frac{2e}{a^3 (1-e^2)^{3/2}} \,,  \\
\label{seq:relation_a_e} a(e) &=  \frac{c_0 e^{4/3}}{1-e^2} \,,
\end{align}\ese
one  can perform a change of variables in the integrals of \eqref{eq:orbit_averaged_FsG_FsP}, namely transform an integral over time into an integral over eccentricity. In the far past, the eccentricity asymptotes towards $1$ (at leading order in the post-Newtonian expansion), and I found that the associated bound converges in this limit. These integrals can in fact be computed analytically, and I find that $ \Pi_i^{s,\text{DC}} =  \Pi_y^{s,\text{DC}} \lambda_0^i$ and  $\Gamma_i^{s,\text{DC}} = \Gamma_x^{s,\text{DC}} n_0^i + \Gamma_y^{s,\text{DC}} \lambda_0^i$, where 
\bse\begin{align}
 \Pi_y^{s,\text{DC}} &= \frac{(\tilde{G} \alpha m)^{3/2}m \nu  }{6  \gam \Sm c^2} (4 \bep \Sp+4 \bem \Sm  - \gam \Sm \delta) \frac{e(4+e)}{a^{3/2}(1+e)\sqrt{1-e^2} }\,, \\
\Gamma_x^{s,\text{DC}} &= \frac{2(\tilde{G}\alpha m) m\nu }{15 \gam \Sm c^2} \Big(20 \bep \Sm + 20 \bem \Sm   - 2 \gam \Sp - 3 \gam \Sm \delta\Big)(1-e)  \,,\\
  \Gamma_y^{s,\text{DC}} &= \frac{m c}{24(\tilde{G} \alpha m)^{1/2}\gam \zeta\Sm^3}\left(4 \bep\Sp + 4 \bem \Sm - \gam \Sm \delta\right) \,  \frac{a^{3/2}(1-e^2)}{e^2}\Biggl[\!12 - \! 5e - 6e^2 - \! e^3   - \! 10 \sqrt{1-e^2}\arctan\sqrt{\frac{1-e}{1+e}}\Biggr].
\end{align}\ese
%
%
In the $e\rightarrow 0$ limit (with $a$ kept constant), these functions behave as
\begin{align}
\Pi_y^{s,\text{DC}} &= \calO(e)\,, & \Gamma_x^{s,\text{DC}} &= \calO(1)\,, &  \Gamma_y^{s,\text{DC}} &= \calO \left(\frac{1}{e^2}\right) \,,
\end{align}
whereas in the $e\rightarrow 1$ limit, they behave as
\begin{align}
\Pi_y^{s,\text{DC}} &= \calO\left(\frac{1}{\sqrt{1-e}}\right)\,, & \Gamma_x^{s,\text{DC}} &= \calO(1-e)\,,  &  \Gamma_y^{s,\text{DC}} &= \calO \left((1-e)^2\right) \,.
\end{align}

One can interpret these quantities as the boost and shift associated to the secular piece of the gravitational recoil. The boost term $\bm{\Pi}^{s,\text{DC}}$ vanishes in the limit of vanishing \emph{current} eccentricity, which smoothly connects with the predictions that, for an eternally quasicircular orbit, there is no secular boost, at least in the post-Newtonian, adiabatic regime. However, the $y$-component of the shift term $\bm{\Gamma}^{s,\text{DC}}$  blows up as the current eccentricity goes to zero, whilst the $x$-component goes to a nonzero finite limit. This behavior appears to be paradoxical, since in the quasicircular limit, the shift is predicted to be strictly zero. This paradox is also observed in the case of general relativity \cite{TrestiniUnpublished24}, although with more complicated analytical expressions. Since only $\bm{\Pi}^{s,\text{DC}}$ will be required in this work, the complete resolution of this paradox is left to future work. However, I will state a few mitigating arguments. First, assuming quasicircular orbits artificially imposes a symmetry which does not exist in the general case, because an ellipse exhibits a preferred direction, but not a circle. Moreover, the quasicircular model implies that the eccentricity was zero in the infinite past --- whereas assuming that the current eccentricity is nonzero but arbitrarily small implies that the eccentricity was one in the infinite past (up to small PN corrections due to the radiation). 
Thus, it is not entirely surprising that there could be a nonsmooth limit when going from the quasielliptic to the quasicircular scenario, although a smooth limit would of course more satisfactory. Second, in most cases, only the center-of-inertia is physically relevant, and a constant shift between two frames is unlikely to affect any physical observable. Thus, it should be possible to  renormalize this apparent divergence by a physically irrelevant, constant shift. Instead of working in the center-of-mass frame, defined uniquely by $G_i + \Gamma_i^{s} + \Gamma_i = 0$, one would then work in a center-of-inertia frame, defined up to an arbitrary constant shift by $P_i + \Pi_i^s  + \Pi_i =0$.

\subsection{The scalar dipolar moment as 2.5PN in the CM frame}
\label{subsec:scalar_dipole_CM}

When computing the 1.5PN energy flux in the CM frame \cite{BBT22}, the scalar dipole was obtained at 2.5PN order in a generic frame. But only $\p_t^2 {\dI}{}_i^s$ was needed with 2.5PN accuracy in the CM frame, and not ${\dI}{}_i^s$, so only the 1.5PN-accurate formula for the passage to the CM was required (see App. B1 of \citetalias{BBT22} for an explanation). However, the angular momentum flux I will be computing hereafter involves $\p_t{\dI}{}_i^s$, so the 2.5PN-accurate formula for the passage to the CM is necessary.\footnote{\textit{A contrario}, in GR, the quadrupolar moment  $\dI_{ij}$ has a particular structure that avoids the appearance of this nonlocal term at 3.5PN order, see \cite{BFT24}.} The nonlocal term \eqref{eq:defGammaGammas} will thus propagate into the expression of the scalar dipole moment in the CM frame, which is thus split as follows:
\be {\dI}^s_i =  {\dI}^{s,\text{inst}}_i + \frac{\sqrt{\alpha}\zeta (\Sp + \Sm \delta)}{(1-\zeta)\phi_0} \Gamma_i^s  \,, \ee
where ${\dI}^{s,\text{inst}}_i $ is instantaneous and relegated to App.~\ref{app:scalarDipole}, and $\Gamma_i^s$ is defined in \eqref{eq:defGammaGammas}. 
%
%
However, the angular momentum flux does not depend directly on $\dI^s_i$, but rather on its time derivative $\partial_t{\dI}^s_i$, which is obtained thanks to the identity $\dd \Gamma^s_i  / \dd t = \Pi^s_i + \mathcal{F}^i_{s,\bm{G}}$, where $\Pi_i^s$ is defined in \eqref{eq:defPiPis} and $\mathcal{F}^i_{s,\bm{G}}$ is explicitly presented in \eqref{seq:expressionGi}. The first derivative of the scalar dipole is then split as follows:
\be \overset{(1)}{\dI}{\!}^s_i =  \overset{(1)}{\dI}{\!}^{s,\text{inst}}_i + \frac{\sqrt{\alpha}\zeta (\Sp + \Sm \delta)}{(1-\zeta)\phi_0}\Big(\mathcal{F}^i_{s,\bm{G}}+ \Pi_i^s \Big)\,. \ee
Finally, using the identity $\dd \Pi^s_i  / \dd t =  \mathcal{F}^i_{s,\bm{P}}$, I find that 
\be \label{def:dt2Isi_localized} \overset{(2)}{\dI}{\!}^s_i =  \overset{(2)}{\dI}{\!}^{s,\text{inst}}_i + \frac{\sqrt{\alpha}\zeta (\Sp + \Sm \delta)}{(1-\zeta)\phi_0}\Big(\overset{(1)}{\mathcal{F}}{}^i_{s,\bm{G}}+  \mathcal{F}^i_{s,\bm{P}} \Big)\,, \ee
which is purely instantaneous. Importantly, I have checked that this result is in agreement with the corresponding expression used in \citetalias{BBT22}, which was derived without the knowledge of the 2.5PN-accurate formula for the passage to the CM frame.
\section{The flux of energy and angular momentum}
\label{sec:flux}

This section is devoted to the derivation of the orbit-averaged fluxes of energy and angular momentum sourced by quasielliptic binary systems. These quantities are very important as they are required for the derivation of the secular evolution of the orbital elements. They were obtained at Newtonian order in \citetalias{T24_QK} (i.e.  at relative 1PN order beyond to the leading $-$1PN order dipolar fluxes), and I will now extend this result to 1.5PN order, which is the same order at which the energy flux generated by \emph{circular orbits} was derived in \citetalias{BBT22}. Due to the presence of hereditary tail and memory integrals, the difficulty of the computation is comparable to the 2.5PN flux in general relativity~\cite{ABIQ08tail, ABIQ08, ABIS09}.

\subsection{General expressions in terms of the source moments}
Recall that the expressions of the fluxes in terms of the radiative moments read~\cite{Thorne:1980ru, Blanchet:2013haa, COS19, BBT22, T24_QK}

\bse \begin{align}
\mathcal{F} &= \sum_{\ell=2}^{\infty} \frac{G \phi_0}{c^{2\ell+1}} \Big(\alpha_\ell\, \overset{(1)}{\mathcal{U}}_L\overset{(1)}{\mathcal{U}}_L  + \frac{\beta_\ell}{c^2} \overset{(1)}{\mathcal{V}}_L\overset{(1)}{\mathcal{V}}_L   \Big) \,,\\
\mathcal{F}^s &= \sum_{\ell=0}^{\infty} \frac{G \phi_0 (3+2\omega_0)}{c^{2\ell+1}} \gamma_\ell \overset{(1)}{\mathcal{U}}{}^s_L\overset{(1)}{\mathcal{U}}{}^s_L \label{seq:scalar_energy_flux_3} \,,\\
\mathcal{G}_i &= \epsilon_{ijk}\sum_{\ell=2}^{\infty}\frac{G \phi_0}{c^{2\ell+1}}  \Big(\alpha'_\ell \,\mathcal{U}_{jL-1}\overset{(1)}{\mathcal{U}}_{kL-1}  + \frac{\beta'_\ell}{c^2}  \mathcal{V}_{jL-1}\overset{(1)}{\mathcal{V}}_{kL-1}\Big) \,,\\
\mathcal{G}^s_i &= \epsilon_{ijk}\sum_{\ell=1}^{\infty} \frac{G \phi_0 (3+2\omega_0)}{c^{2\ell+1}}\gamma'_\ell\,\mathcal{U}^s_{jL-1} \overset{(1)}{\mathcal{U}}{}^s_{kL-1}  	  \label{seq:scalar_angular_momentum_flux_3}\,.
\end{align}\ese
where
\begin{align}
\alpha_\ell &= \frac{(\ell+1)(\ell+2)}{(\ell-1)\ell \ell!(2\ell+1)!!}\,,&\qquad
\beta_\ell &= \frac{4 \ell (\ell+2)}{(\ell-1)(\ell+1)!(2\ell+1)!!}\,,&\qquad
\gamma_\ell &= \frac{1}{\ell! (2\ell+1)!!}\,,
\end{align}
and $\alpha'_\ell = \ell\,\alpha_\ell$, $\beta'_\ell = \ell\,\beta_\ell$, and $\gamma'_\ell = \ell\,\gamma_\ell$.

Anticipating the fact that nonspinning binaries do not undergo precession, I will define $\mathcal{G} = \mathcal{G}_i \ell_i$ and  $\mathcal{G}^s = \mathcal{G}^s_i \ell_i$. Indeed, the angular momentum vector will always be along $\ell_i$, which is a constant vector of norm $1$, and always orthogonal to the orbital plane.
Replacing the radiative moments by their expressions in terms of the source moments [given by (3.28-29) of \cite{BBT22}], I find that, at 1.5PN order beyond quadrupolar radiation, the fluxes can be divided in different pieces: instantaneous, tails, memory, and the new nonlocal $\bm{\Pi}^s$ contribution. This decomposition reads:

\bse\begin{align}
\mathcal{F} &=\mathcal{F}^\mathrm{inst} + \mathcal{F}^\mathrm{tail} \,, \\*
\mathcal{F}^s &=\mathcal{F}^{s,\mathrm{inst}} + \mathcal{F}^{s,\mathrm{tail}} \,, \\*
\mathcal{G}_i &= \mathcal{G}_i^\mathrm{inst} + \mathcal{G}_i^\mathrm{tail} + \mathcal{G}_i^\mathrm{mem} \,,\\*
\mathcal{G}^s_i &= \mathcal{G}_i^{s,\mathrm{inst}} + \mathcal{G}_i^{s,\bm{\Pi}^s} + \mathcal{G}_i^{s,\mathrm{tail}}  \,.
\end{align}\ese

The memory term, present only in the tensor angular momentum flux, reads
\begin{align}\label{eq:memory_flux_source_moments}
\mathcal{G}_i^\mathrm{mem}&=  - \frac{2 G^2 \phi_0 (3+2\omega_0)}{15 c^8}\epsilon_{ik(j}\!\!\overset{(3)}{\dI}\!{}_{a)k}\!\int_0^\infty\!\!\! \dd\tau\big[\!\overset{(2)}{\dI}\!{}^s_a \overset{(2)}{\dI}\!{}^s_j\big](u-\tau) \,,
\end{align}
which is in agreement with (4.11) of \cite{JR24}. 
%
%

The tails terms  read:
 \label{eq:tail_fluxes_source_moments}
\bse\begin{align}
\mathcal{F}^\mathrm{tail} &= \frac{4 G^2 \dM}{5 c^8} \overset{(3)}{\dI}\!{}_{ij} \int_0^\infty \dd \tau \overset{(5)}{\dI}\!{}_{ij}(u-\tau) \left[\ln\left(\frac{c \tau}{2b_0}\right)+ \frac{11}{12}\right] \,,\\
\mathcal{F}^{s,\mathrm{tail}} &= \frac{4 G^2 \dM (3+2\omega_0)}{c^6}\Bigg\{\frac{1}{3} \overset{(2)}{\dI}\!{}^s_{i} \int_0^\infty \dd\tau  \overset{(4)}{\dI}\!{}^s_{i}(u-\tau) \left[\ln\left(\frac{c \tau}{2b_0}\right)+1\right] + \frac{\overset{(1)}{E}{}^s}{c^2\phi_0^2}\int_0^\infty \dd\tau \overset{(3)}{E}{}^s (u-\tau) \ln\left(\frac{c \tau}{2b_0}\right)\nn\\
&\qquad\qquad\qquad\qquad + \frac{1}{30 c^2}\overset{(3)}{\dI}\!{}^s_{ij}\int_0^\infty \dd\tau\,  \overset{(5)}{\dI}\!{}^s_{ij}(u-\tau)  \left[\ln\left(\frac{c \tau}{2b_0}\right)+\frac{3}{2}\right]   \Bigg\} \,, \\
\mathcal{G}_i^\mathrm{tail} &= \frac{4 G^2 \dM}{5 c^8} \epsilon_{ijk} \Bigg\{ \overset{(2)}{\dI}\!{}_{ja} \int_0^\infty \dd \tau \overset{(5)}{\dI}\!{}_{ka}(u-\tau) \left[\ln\left(\frac{c \tau}{2b_0}\right)+ \frac{11}{12}\right] -  \overset{(3)}{\dI}\!{}_{ja} \int_0^\infty \dd \tau \overset{(4)}{\dI}\!{}_{ka}(u-\tau) \left[\ln\left(\frac{c \tau}{2b_0}\right)+ \frac{11}{12}\right]  \Bigg\} \,, \\
\mathcal{G}_i^{s,\mathrm{tail}} &= \frac{2 G^2 \dM (3+2\omega_0)}{3 c^6}\epsilon_{ijk}\Bigg\{ \overset{(1)}{\dI}\!{}^s_{j} \int_0^\infty \dd\tau  \overset{(4)}{\dI}\!{}^s_{k}(u-\tau) \left[\ln\left(\frac{c \tau}{2b_0}\right)+1\right] - \overset{(2)}{\dI}\!{}^s_{j} \int_0^\infty \dd\tau  \overset{(3)}{\dI}\!{}^s_{k}(u-\tau) \left[\ln\left(\frac{c \tau}{2b_0}\right)+1\right] \nn\\
&\qquad\qquad\quad + \frac{1}{5 c^2}\overset{(2)}{\dI}\!{}^s_{ja}\int_0^\infty \dd\tau\,  \overset{(5)}{\dI}\!{}^s_{ka}(u-\tau)  \left[\ln\left(\frac{c \tau}{2b_0}\right)+\frac{3}{2}\right]  - \frac{1}{5 c^2}\overset{(3)}{\dI}\!{}^s_{ja}\int_0^\infty \dd\tau\,  \overset{(4)}{\dI}\!{}^s_{ka}(u-\tau)  \left[\ln\left(\frac{c \tau}{2b_0}\right)+\frac{3}{2}\right]   \Bigg\}  \,,
\end{align}\ese
 which is also in agreement with (4.10) and (5.17) of \cite{JR24}. 

The nonlocal $\bm{\Pi}^s$ contribution to the angular momentum flux arises from the nonlocal contribution in the passage to the CM, see Sec.~\ref{sec:CM}. Note that in the energy flux, both dipolar moments entering at Newtonian order bear two time derivatives, which has the effect of ``localizing'' the~$\bm{\Pi}^s$ contribution, see \eqref{def:dt2Isi_localized}. However, the angular momentum flux contains a contribution (entering at Newtonian order) from the first time derivative of the dipolar moment, which will thus feature~$\bm{\Pi}^s$. This contribution reads:
\begin{align}
\mathcal{G}_i^{s,\bm{\Pi}^s} &= \frac{G (3+2\omega_0)}{3 c^3}\frac{\sqrt{\alpha}\zeta (\Sp + \Sm \delta)}{(1-\zeta)}  \epsilon_{ijk}\, \Pi_j^s  \! \overset{(2)}{\dI}{\!}_k^{s}  
\end{align}

Finally, the instantaneous fluxes are defined as what remains, namely
\bse\begin{align}	
\mathcal{F}^\mathrm{inst} &=
 \frac{G \phi_0}{5 c^5} \overset{(3)}{\dI}{\!}_{ij}  \overset{(3)}{\dI}{\!}_{ij} 
 + \frac{G \phi_0}{c^7}\Bigg[\frac{16}{45} \overset{(3)}{\dJ}{\!}_{ij}  \overset{(3)}{\dJ}{\!}_{ij}  + \frac{1}{189} \overset{(4)}{\dI}{\!}_{ijk}  \overset{(4)}{\dI}{\!}_{ijk} \Bigg]  
 \nn\\*
 &
\qquad - \frac{G^2 \phi_0 (3+2\omega_0)}{15 c^8}\Bigg[
 4 \overset{(1)}{\dI}{\!}^s_{i}  \overset{(3)}{\dI}{\!}^s_{j}  \overset{(3)}{\dI}{\!}_{ij} 
+ 2 \, \dI^s_{i}  \overset{(4)}{\dI}{\!}^s_{j}  \overset{(3)}{\dI}{\!}_{ij}
+ \frac{m_s}{\phi_0} \overset{(3)}{\dI}{\!}_{ij} \overset{(4)}{\dI}{\!}^s_{ij} \Bigg] 
+ \mathcal{O}\left(\frac{1}{c^9}\right) \,, \\
\mathcal{F}^{s,\mathrm{inst}} &= 
\frac{G \phi_0(3+2\omega_0)}{3 c^3} \overset{(2)}{\dI}{\!}_i^{s}  \overset{(2)}{\dI}{\!}_i^{s} 
+ \frac{G \phi_0 (3+2\omega_0) }{c^5}\Bigg[ \frac{1}{30} \overset{(3)}{\dI}{\!}^s_{ij}  \overset{(3)}{\dI}{\!}^s_{ij} + \frac{\overset{(1)}{E}{\!}^s  \overset{(1)}{E}{\!}^s }{\phi_0^2}   \Bigg] 
+ \frac{G \phi_0 (3+2\omega_0)}{630c^7}  \overset{(4)}{\dI}{\!}^s_{ijk}  \overset{(4)}{\dI}{\!}^s_{ijk} \nn\\
&  + \frac{G^2 \phi_0 (3+2\omega_0)}{45  c^8}\Bigg[
- 180 \,  \frac{m_s \overset{(1)}{E}{\!}^s \overset{(2)}{E}{\!}^s}{\phi_0^3}
- 120  \frac{\overset{(2)}{E}{\!}^s}{\phi_0} \overset{(1)}{\dI}{\!}^s_a \overset{(2)}{\dI}{\!}^s_a
- 60 \frac{\overset{(1)}{E}{\!}^s}{\phi_0} \overset{(2)}{\dI}{\!}^s_a \overset{(2)}{\dI}{\!}^s_a
- 240 \overset{(2)}{\dI}{\!}^s_a \overset{(2)}{\dI}{\!}^s_a  \overset{(2)}{\dW}{}^s
- 60 \frac{\overset{(3)}{E}{\!}^s}{\phi_0} {\dI}{}^s_a  \overset{(2)}{\dI}{\!}^s_a
\nn\\&\qquad\qquad\qquad\qquad\ \,
- 12 \overset{(2)}{\dI}{\!}^s_a \overset{(2)}{\dI}{\!}^s_b \overset{(3)}{\dI}{\!}_{ab}
+ 20 \frac{\overset{(1)}{E}{\!}^s}{\phi_0}  \overset{(1)}{\dI}{\!}^s_a  \overset{(3)}{\dI}{\!}^s_a
- 120  \overset{(2)}{\dI}{\!}^s_a  \overset{(3)}{\dI}{\!}^s_a  \overset{(1)}{\dW}{}
+ 12  \overset{(2)}{\dI}{\!}^s_a  \overset{(3)}{\dI}{\!}^s_b \overset{(2)}{\dI}{\!}_{ab}
+ 12 \overset{(2)}{\dI}{\!}^s_{a} \overset{(3)}{\dI}{\!}^s_{b} \overset{(2)}{\dI}{\!}^s_{ab}
\nn\\&\qquad\qquad\qquad\qquad\ \,
- 120 \overset{(1)}{\dI}{\!}^s_{a}  \overset{(2)}{\dI}{\!}^s_{a}  \overset{(3)}{\dW}{\!}
+120 \frac{m_s}{\phi_0}  \overset{(2)}{\dI}{\!}^s_{a}  \overset{(3)}{\dY}{\!}_{a}
- 6  \overset{(1)}{\dI}{\!}^s_{a}  \overset{(2)}{\dI}{\!}^s_{b}  \overset{(4)}{\dI}{\!}_{ab}
- 3 \frac{m_s}{\phi_0} \overset{(3)}{\dI}{\!}^s_{ab} \overset{(4)}{\dI}{\!}_{ab}
+ 20 \frac{\overset{(1)}{E}{\!}^s}{\phi_0} {\dI}^s_a \overset{(4)}{\dI}{\!}^s_{a}
\nn\\&\qquad\qquad\qquad\qquad\ \,
+ 36 \overset{(2)}{\dI}{\!}^s_{a} \overset{(4)}{\dI}{\!}^s_{b}  \overset{(1)}{\dI}{\!}_{ab}
+ 24 \overset{(2)}{\dI}{\!}^s_{a} \overset{(4)}{\dI}{\!}^s_{b} \overset{(1)}{\dI}{\!}^s_{ab}
+ 30 \,\epsilon_{abi}\, \dJ_a  \overset{(2)}{\dI}{\!}^s_{b}   \overset{(4)}{\dI}{\!}^s_{i}
+ 18 \overset{(2)}{\dI}{\!}^s_{a}  \overset{(5)}{\dI}{\!}^s_{b} \, \dI_{ab} 
+ 12 \overset{(2)}{\dI}{\!}^s_{a}  \overset{(5)}{\dI}{\!}^s_{b} \, \dI^s_{ab}
\nn\\&\qquad\qquad\qquad\qquad\ \,
+ \frac{4\omega_0' \phi_0}{3+2\omega_0} \Bigg( 
   45  \frac{m_s \overset{(1)}{E}{\!}^s \overset{(2)}{E}{\!}^s}{\phi_0^3}
+ 30 \frac{\overset{(2)}{E}{\!}^s}{\phi_0} \overset{(1)}{\dI}{\!}^s_{a} \overset{(2)}{\dI}{\!}^s_{a}
+ 15  \frac{\overset{(1)}{E}{\!}^s}{\phi_0} \overset{(2)}{\dI}{\!}^s_{a} \overset{(2)}{\dI}{\!}^s_{a}
+ 15  \frac{\overset{(3)}{E}{\!}^s}{\phi_0} \, {\dI}^s_{a} \overset{(2)}{\dI}{\!}^s_{a}
-   5 \frac{\overset{(1)}{E}{\!}^s}{\phi_0}  \overset{(1)}{\dI}{\!}^s_{a} \overset{(3)}{\dI}{\!}^s_{a}
\nn\\&\qquad\qquad\qquad\qquad\qquad\qquad\qquad
 - 3  \overset{(2)}{\dI}{\!}^s_{a}  \overset{(3)}{\dI}{\!}^s_{b} \overset{(2)}{\dI}{\!}^s_{ab}
- 5 \frac{\overset{(1)}{E}{\!}^s}{\phi_0}  \dI_a^s  \overset{(4)}{\dI}{\!}^s_{a}
- 6  \overset{(2)}{\dI}{\!}^s_{a}  \overset{(4)}{\dI}{\!}^s_{b}  \overset{(1)}{\dI}{\!}^s_{ab}
- 3  \overset{(2)}{\dI}{\!}^s_{a}  \overset{(5)}{\dI}{\!}^s_{b}\, \dI^s_{ab}
\Bigg)
 \Bigg]  
 + \mathcal{O}\left(\frac{1}{c^9}\right)  \,, \\
\mathcal{G}_i^\mathrm{inst} &= \frac{2 G \phi_0}{5 c^5}\epsilon_{ijk}  \overset{(2)}{\dI}{\!}_{ja}  \overset{(3)}{\dI}{\!}_{ka} + \frac{G \phi_0}{c^7}\epsilon_{ijk}\Bigg[\frac{32}{45}  \overset{(2)}{\dJ}{\!}_{ja} \overset{(3)}{\dJ}{\!}_{ka}  + \frac{1}{63}  \overset{(3)}{\dI}{\!}_{jab}  \overset{(4)}{\dI}{\!}_{kab}  \Bigg] \nn\\
& + \frac{G^2 \phi_0 (3+2\omega) \epsilon_{ijk} }{15 c^8} \Bigg[
- \overset{(1)}{\dI}{\!}^s_{j}  \overset{(2)}{\dI}{\!}^s_{a}  \overset{(3)}{\dI}{\!}_{ak} 
+ 2 \overset{(1)}{\dI}{\!}^s_{j}  \overset{(3)}{\dI}{\!}^s_{a}  \overset{(2)}{\dI}{\!}_{ak} 
- \dI^s_{j}  \overset{(3)}{\dI}{\!}^s_{a}  \overset{(3)}{\dI}{\!}_{ak} 
+ \dI^s_{j}  \overset{(4)}{\dI}{\!}^s_{a}  \overset{(2)}{\dI}{\!}_{ak} 
-  \overset{(1)}{\dI}{\!}^s_{a}  \overset{(2)}{\dI}{\!}^s_{j}  \overset{(3)}{\dI}{\!}_{ak}
+ 2  \overset{(1)}{\dI}{\!}^s_{a}  \overset{(3)}{\dI}{\!}^s_{j}  \overset{(2)}{\dI}{\!}_{ak}
\nn\\&\qquad\qquad\qquad\qquad\quad\ \,
-  \dI^s_{a}  \overset{(3)}{\dI}{\!}^s_{j}  \overset{(3)}{\dI}{\!}_{ak}
+ \frac{m_s}{\phi_0} \overset{(3)}{\dI}{\!}^s_{ak}  \overset{(3)}{\dI}{\!}_{aj}
+ \dI^s_{a}  \overset{(4)}{\dI}{\!}^s_{j}  \overset{(2)}{\dI}{\!}_{ak}
- \frac{m_s}{\phi_0} \overset{(2)}{\dI}{\!}_{aj} \overset{(4)}{\dI}{\!}^s_{ak}
\Bigg]  + \mathcal{O}\left(\frac{1}{c^9}\right)  \,, \\
\mathcal{G}_i^{s,\mathrm{inst}} &=\frac{G \phi_0 (3+2\omega_0)}{3 c^3}\epsilon_{ijk}\Bigg( \overset{(1)}{\dI}{\!}_j^{s,\text{inst}} +\frac{\sqrt{\alpha}\zeta (\Sp + \Sm \delta)}{(1-\zeta)\phi_0} \mathcal{F}^j_{s,\bm{G}}\Bigg)  \overset{(2)}{\dI}{\!}_k^{s} \nn\\
 &    + \frac{G \phi_0 (3+2\omega_0)}{15 c^5}\epsilon_{ijk}  \overset{(2)}{\dI}{\!}^s_{ja}  \overset{(3)}{\dI}{\!}^s_{ka}  + \frac{G \phi_0 (3+2\omega_0)}{210 c^7}\epsilon_{ijk}  \overset{(3)}{\dI}{\!}^s_{jab}  \overset{(4)}{\dI}{\!}^s_{kab}   \nn\\
 & + \frac{G^2 \phi_0 (3+2\omega) }{15 c^8} \Bigg[ \epsilon_{ijk} \Bigg\{ 
 - 20 \frac{\overset{(1)}{E}{\!}^s}{\phi_0}  \overset{(1)}{\dI}{\!}^s_j \overset{(2)}{\dI}{\!}^s_k
 - 10 \frac{\overset{(2)}{E}{\!}^s}{\phi_0}  {\dI}^s_j  \overset{(2)}{\dI}{\!}^s_{k}
 - 60  \overset{(1)}{\dI}{\!}^s_{j} \overset{(2)}{\dI}{\!}^s_{k} \overset{(2)}{\dW}
 - 20 \frac{m_s}{\phi_0} \overset{(2)}{\dI}{\!}^s_{j} \overset{(2)}{\dY}{\!}_{k}
+10 \frac{\overset{(3)}{E}{\!}^s}{\phi_0} {\dI}^s_j  \overset{(1)}{\dI}{\!}^s_{k} 
- 20 \overset{(1)}{\dI}{\!}^s_{j}  \overset{(3)}{\dI}{\!}^s_{k}  \overset{(1)}{\dW}
\nn\\&\qquad\qquad\qquad\qquad\qquad
+ 20 \frac{m_s}{\phi_0} \overset{(1)}{\dI}{\!}^s_{j}  \overset{(3)}{\dY}{\!}_{k} 
+ \overset{(2)}{\dI}{\!}^s_{a} \overset{(2)}{\dI}{\!}^s_{j}   \overset{(2)}{\dI}{\!}_{ka} 
+ \overset{(1)}{\dI}{\!}^s_{a} \overset{(2)}{\dI}{\!}^s_{j}   \overset{(3)}{\dI}{\!}_{ka} 
- \frac{m_s}{\phi_0}  \overset{(3)}{\dI}{\!}_{ja}  \overset{(3)}{\dI}{\!}^s_{ka} 
-  \overset{(1)}{\dI}{\!}^s_{a} \overset{(1)}{\dI}{\!}^s_{j}  \overset{(4)}{\dI}{\!}_{ka} 
- \frac{m_s}{\phi_0} \overset{(2)}{\dI}{\!}^s_{ja}  \overset{(4)}{\dI}{\!}_{ka} 
\nn\\&\qquad\qquad\qquad\qquad\qquad
- 2  \overset{(1)}{\dI}{\!}^s_{j}  \overset{(2)}{\dI}{\!}^s_{a}  \overset{(1)}{\dI}{\!}_{ka}
+ 2   \overset{(1)}{\dI}{\!}^s_{j}  \overset{(3)}{\dI}{\!}^s_{a}  \overset{(2)}{\dI}{\!}_{ka}
- 3 \overset{(2)}{\dI}{\!}^s_{j} \overset{(3)}{\dI}{\!}^s_{a}  \overset{(1)}{\dI}{\!}_{ka}
- 2 \overset{(2)}{\dI}{\!}^s_{j} \overset{(3)}{\dI}{\!}^s_{a}  \overset{(1)}{\dI}{\!}^s_{ka}
+ 2 \overset{(1)}{\dI}{\!}^s_{j} \overset{(3)}{\dI}{\!}^s_{a}  \overset{(2)}{\dI}{\!}^s_{ka}
+6  \overset{(1)}{\dI}{\!}^s_{j} \overset{(4)}{\dI}{\!}^s_{a}  \overset{(1)}{\dI}{\!}_{ka}
\nn\\&\qquad\qquad\qquad\qquad\qquad
+ 4 \overset{(1)}{\dI}{\!}^s_{j} \overset{(4)}{\dI}{\!}^s_{a}  \overset{(1)}{\dI}{\!}^s_{ka}
- 3  \overset{(2)}{\dI}{\!}^s_{j} \overset{(4)}{\dI}{\!}^s_{a}  \dI_{ka}
-2 \overset{(2)}{\dI}{\!}^s_{j} \overset{(4)}{\dI}{\!}^s_{a}  \dI^s_{ka}
+3 \overset{(1)}{\dI}{\!}^s_{j} \overset{(5)}{\dI}{\!}^s_{a}  \dI_{ka}
+2 \overset{(1)}{\dI}{\!}^s_{j} \overset{(5)}{\dI}{\!}^s_{a}  \dI^s_{ka}
 \nn\\&\qquad\qquad\qquad\qquad\qquad
 + \frac{\omega_0'\phi_0}{3+2\omega_0}\Bigg(
20 \frac{\overset{(1)}{E}{\!}^s}{\phi_0} \overset{(1)}{\dI}{\!}^s_{j} \overset{(2)}{\dI}{\!}^s_{k} 
 + 10 \frac{\overset{(2)}{E}{\!}^s}{\phi_0}  \dI^s_{j} \overset{(2)}{\dI}{\!}^s_{k} 
 - 10 \frac{\overset{(3)}{E}{\!}^s}{\phi_0} \dI^s_{j} \overset{(1)}{\dI}{\!}^s_{k} 
 + 2 \overset{(2)}{\dI}{\!}^s_{j} \overset{(3)}{\dI}{\!}^s_{a}  \overset{(1)}{\dI}{\!}^s_{ka}
 - 2 \overset{(1)}{\dI}{\!}^s_{j} \overset{(3)}{\dI}{\!}^s_{a}  \overset{(2)}{\dI}{\!}^s_{ka} 
  \nn\\&\qquad\qquad\qquad\qquad\qquad\qquad\qquad\quad
 - 4  \overset{(1)}{\dI}{\!}^s_{j} \overset{(4)}{\dI}{\!}^s_{a}  \overset{(1)}{\dI}{\!}^s_{ka} 
 +2 \overset{(2)}{\dI}{\!}^s_{j} \overset{(4)}{\dI}{\!}^s_{a}  \dI^s_{ka} 
 - 2 \overset{(1)}{\dI}{\!}^s_{j} \overset{(5)}{\dI}{\!}^s_{a}  \dI^s_{ka} 
 \Bigg)    \Bigg\} 
  \nn\\&\qquad\qquad\qquad\qquad
   + 5 \dJ_i \overset{(2)}{\dI}{\!}^s_{a} \overset{(3)}{\dI}{\!}^s_{a}
   - 5 \dJ_a \overset{(2)}{\dI}{\!}^s_{a} \overset{(3)}{\dI}{\!}^s_{i}
   - 5 \dJ_i \overset{(1)}{\dI}{\!}^s_{a} \overset{(4)}{\dI}{\!}^s_{a}
   + 5 \dJ_a \overset{(1)}{\dI}{\!}^s_{a} \overset{(4)}{\dI}{\!}^s_{i}
  \Bigg] + \mathcal{O}\left(\frac{1}{c^9}\right) \,.
\end{align}\ese

\subsection{The instananeous fluxes}

The instantaneous fluxes are first computed at 1.5PN order in terms of the phase space variables $(\bm{y}_1, \bm{y}_2, \bm{v}_1, \bm{v}_2)$ by (i) replacing the source source multipolar moments by their expressions in the CM frame, and (ii) computing the time derivatives using the equations of motion in the CM frame. The required source moments are almost all provided in App. B1 of~\citetalias{BBT22}, except for the scalar dipole moment $\dI^s_i$, which is presented at 2PN order instead of 2.5PN (see the discussion therein). The full scalar dipole at 2.5PN order is presented for the first time in App.~\ref{app:scalarDipole}, and it features a nonlocal piece arising from the passage to the CM frame. I also provide all the source moments in machine-readable form in the Supplemental Material~\cite{SuppMaterial}. The equations of motion (both in a general frame and in the CM frame) are given at required order in \cite{MW13}, and I have  provided these in a machine-readable form in the Supplemental Material~\cite{SuppMaterial} as well.

The expressions of the energy fluxes in terms of the orbital variables in the CM frame, namely $\bm{x} = r \bm{n}$ and $\bm{v}$,  were already provided at 1.5PN order in (4.10) and (C2) of~\citetalias{BBT22}, confirming the previously known 1PN result in the tensor sector [(6.5) of \cite{Lang:2014osa}] and correcting the 1PN result in the scalar sector [(6.7-8) of \cite{Lang:2014osa}].  Here, I provide for the angular momentum fluxes at 1.5PN in terms of the orbital variables in the CM frame:
\bse\label{eq:inst_Gi_Gsi_orbital_vars}\begin{align}
\label{seq:inst_Gi_orbital_vars}
\mathcal{G}_i^{\mathrm{inst}} &= \frac{16 (1 -  \zeta)  m \nu^2}{5 \alpha c^5}\Big(\frac{\tilde{G}\alpha m}{r}\Big)^2   \epsilon_{ijk} n_{j} v_{k} \Bigg\{
 \frac{\tilde{G}\alpha m}{r} -  \frac{3}{2} (nv)^2 + v^2 \nn\\
&+ \frac{1}{c^2} \Bigg[ \Big(- \frac{745}{84} - 4 \bep -  \frac{7}{2} \gam + 4 \bem \delta + \frac{1}{42} \nu\Big)  \Big(\frac{\tilde{G}\alpha m}{r}\Big)^2  + \Big(\frac{31}{7} + 4 \bep + \gam - 4 \bem \delta + \frac{197}{84} \nu \Big)  \Big(\frac{\tilde{G}\alpha m}{r}\Big)  (nv)^2 \nn\\
&\qquad + \Big(\frac{95}{56} -  \frac{45}{7} \nu \Big) (nv)^4 +  \Big(- \frac{29}{21} - 2 \bep + 2 \bem \delta -  \frac{95}{42} \nu \Big) \Big(\frac{\tilde{G}\alpha m}{r}\Big)  v^2 \nn\\
&\qquad + \Big(- \frac{37}{14} + \frac{277}{28} \nu \Big) (nv)^2 v^2 + \Big(\frac{307}{168} + \frac{1}{2} \gam -  \frac{137}{42} \nu \Big) v^{4}\Bigg]\nn\\
& + \frac{1}{c^3}\Bigg[ \Big(\frac{\tilde{G}\alpha m}{r}\Big)^2 \Big(- \frac{1}{4} \gam -  \zeta \Sm^2 \nu\Big) (nv) + \frac{5}{8} \gam  \Big(\frac{\tilde{G}\alpha m}{r}\Big) (nv)^3 -  \frac{1}{2} \gam  \Big(\frac{\tilde{G}\alpha m}{r}\Big) (nv) v^2\Bigg]\Bigg\} + \calO\left(\frac{1}{c^9}\right) \,,\\[0.6cm]
\label{seq:inst_Gsi_orbital_vars}
\mathcal{G}_i^{s, \mathrm{inst}} &= \frac{m \nu^2}{3 c^3}\Big(\frac{\tilde{G}\alpha m}{r}\Big)^2   \epsilon_{ijk} n_{j} v_{k} \Bigg\{\widetilde{A}^{-1\text{PN}} + \frac{1}{c^2}\Bigg[\widetilde{B}_1^{\text{N}} \Big(\frac{\tilde{G}\alpha m}{r}\Big) + \widetilde{B}_2^{\text{N}} (nv)^2 + \widetilde{B}_3^{\text{N}} v^2\Bigg]+ \frac{1}{c^3} \widetilde{C}^{0.5\text{PN}}  \Big(\frac{\tilde{G}\alpha m}{r}\Big) (nv) \nn\\
& + \frac{1}{c^4} \Bigg[\widetilde{D}_1^{1\text{PN}} \Big(\frac{\tilde{G}\alpha m}{r}\Big)^2 + \widetilde{D}_2^{1\text{PN}} \Big(\frac{\tilde{G}\alpha m}{r}\Big) (nv)^2 + \widetilde{D}_3^{1\text{PN}}  \Big(\frac{\tilde{G}\alpha m}{r}\Big) v^2 + \widetilde{D}_4^{1\text{PN}} (nv)^4 +  \widetilde{D}_5^{1\text{PN}}(nv)^2 v^2 + \widetilde{D}_6^{1\text{PN}} v^4  \Bigg] \nn\\
& + \frac{1}{c^5} \Bigg[ \widetilde{E}_1^{1.5\text{PN}} \Big(\frac{\tilde{G}\alpha m}{r}\Big)^2 (nv) + \widetilde{E}_2^{1.5\text{PN}} \Big(\frac{\tilde{G}\alpha m}{r}\Big)(nv)^3 + \widetilde{E}_3^{1.5\text{PN}} \Big(\frac{\tilde{G}\alpha m}{r}\Big) (nv) v^2 \Bigg]  \Bigg\} + \calO\left(\frac{1}{c^9}\right) \,,
\end{align}\ese
where the coefficients associated to \eqref{seq:inst_Gsi_orbital_vars} are given in App.~\ref{app:inst_Gsi_orbital_vars}.
The tensor flux \eqref{seq:inst_Gi_orbital_vars}  is in full agreement with (4.7-8) of~\cite{JR24} at 1.5PN, and the scalar flux \eqref{seq:inst_Gsi_orbital_vars}  is in full agreement with (5.00-15) of~\cite{JR24} at 1PN. The 1.5PN piece of  $\mathcal{G}^{s,\mathrm{inst}}_i$ was computed for the first time in this work. All these results can also be found in machine-readable form in the Supplemental Material~\cite{SuppMaterial}.

Using the identities $(nv) = \dot{r}$,  \mbox{$v^2=\dot{r}^2 + r^2\dot{\phi}^2$}, and \mbox{$\epsilon_{ijk}n_j v_k= r \dot{\phi} \ell^i$}, I then obtain the fluxes in terms of~$(r,\phi,\dot{r},\dot{\phi})$. Specializing to the case of quasielliptic orbits, I can then use the identities $\dot{r}=\mathcal{R}(r^{-1},E, J)$  and $\dot{\phi}=\mathcal{S}(r^{-1},E, J)$, given in (3.1) of~\citetalias{T24_QK}, along with the quasi-Keplerian representation \eqref{eq:PK_equations}, to obtain the fluxes only in terms of the eccentric anomaly $u$ and the QK parameters.
When computing $\left\langle \mathcal{F}^\mathrm{inst} \right\rangle$,  $\left\langle \mathcal{G}^\mathrm{inst} \right\rangle$,  $\left\langle \mathcal{F}^{s,\mathrm{inst}} \right\rangle$ and  $\left\langle \mathcal{G}^{s,\mathrm{inst}} \right\rangle$,  all the QK parameters are first reexpressed in terms only of $x$ and $e_t$ [using (C1) of~\citetalias{T24_QK}], and one can then always use trigonometric relations to write the orbit averaged fluxes as a linear combinations of the following master integrals,
\bse\begin{align} \frac{1}{2\pi}\int_0^{2\pi} \frac{\dd u}{(1-e_t \cos u)^n} &= \frac{1}{(1-e_t^2)^{n/2}} P_{n-1}\left(\frac{1}{\sqrt{1-e_t^2}}\right)\,,  \\*
\frac{1}{2\pi}\int_0^{2\pi} \frac{\dd u\,\sin u}{(1-e_t \cos u )^n} &= 0 \,,
\end{align}\ese
where $n \ge 1$ and $P_n(x)$ is the $n$-th Legendre polynomial. Finally, I find that the instantaneous fluxes read, for the tensor sector,
\bse\label{tensor_flux_orbit_averaged_inst_final} \begin{align}
&\left\langle \mathcal{F}^\text{inst} \right\rangle = \frac{32 c^5 (1 +  \gam/2) x^5\nu^2}{5  \tilde{G} \alpha} \Bigg\{\frac{1 + \tfrac{73}{24} e_t^2 + \tfrac{37}{96} e_t^4}{(1 -  e_t^2)^{7/2}} \nn\\*
&\ \ \qquad + \frac{x}{(1 -  e_t^2)^{9/2}} \Bigg[- \tfrac{1247}{336} -  \tfrac{8}{3} \bep -  \tfrac{4}{3} \gam + \tfrac{8}{3} \bem \delta  -  \tfrac{35}{12}\nu  + e_t^2 \bigg( \tfrac{10475}{672} -  \tfrac{571}{36} \bep + \tfrac{379}{36} \gam + \tfrac{571}{36} \bem \delta -  \tfrac{1081}{36} \nu\bigg) \nn\\*
&\ \ \quad \qquad+ e_t^4 \bigg(\tfrac{10043}{384} -  \tfrac{1003}{144} \bep + \tfrac{955}{72} \gam + \tfrac{1003}{144} \bem \delta -  \tfrac{311}{12} \nu\bigg)  + e_t^6 \bigg(\tfrac{2179}{1792} + \tfrac{131}{288} \gam -  \tfrac{851}{576} \nu\bigg) \Bigg] + \mathcal{O}(x^2)\Bigg\} \,,  \\ 
&\ \ \quad\left\langle \mathcal{G}^\text{inst} \right\rangle = \frac{32 c^2 (1 + \gam/2) m \nu^2 x^{7/2} }{5}  \Bigg\{\frac{1 + \tfrac{7}{8} e_t^2}{(1 -  e_t^2)^2} \nn\\*
&\  \qquad\quad + \frac{x}{(1 -  e_t^2)^3} \Bigg[- \tfrac{1247}{336} -  \tfrac{8}{3} \bep -  \tfrac{4}{3} \gam + \tfrac{8}{3} \bem \delta -  \tfrac{35}{12} \nu  + e_t^2 \bigg(\tfrac{3019}{336} -  \tfrac{35}{6} \bep + \tfrac{21}{4} \gam + \tfrac{35}{6} \bem \delta -  \tfrac{335}{24} \nu \bigg) \nn\\*
&\ \ \quad \qquad \qquad+ e_t^4 \bigg(\tfrac{8399}{2688} -  \tfrac{1}{4} \bep + \tfrac{67}{48} \gam + \tfrac{1}{4} \bem \delta -  \tfrac{275}{96} \nu \bigg) \Bigg]  + \mathcal{O}(x^2)\Bigg\}  \,,
\end{align}\ese
and are given in the scalar scalar by
\bse\label{scalar_flux_orbit_averaged_inst_final}\begin{align}
\left\langle \mathcal{F}^{s,\mathrm{inst}} \right\rangle &= \frac{c^5 \nu^2 \zeta x^5 }{3 \tilde{G} \alpha} \Bigg(4 \Sm^2 x^{-1} \frac{1+\frac{1}{2}e_t^2}{(1-e_t^2)^{5/2}} + \frac{\frak{f}_1 + \frak{f}_2  e_t^2 + \frak{f}_3  e_t^4}{(1-e_t^2)^{7/2}}  + x \Bigg[ \frac{\frak{f}_4 + \frak{f}_5  e_t^2 + \frak{f}_6 e_t^4 + \frak{f}_7  e_t^6 }{(1-e_t^2)^{9/2}} +  \frac{\frak{f}_8 + \frak{f}_9  e_t^2 + \frak{f}_{10}  e_t^4}{(1-e_t^2)^{4}}\Bigg]+ \mathcal{O}(x^2) \Bigg) \,, \\
\left\langle \mathcal{G}^{s,\mathrm{inst}} \right\rangle &= \frac{c^2 m \nu^2 \zeta x^{7/2}}{3} \Bigg( \frac{4\Sm^2 x^{-1}}{1-e_t^2} + \frac{\frak{g}^s_1 + \frak{g}^s_2 e_t^2}{(1-e_t^2)^2} +  x \Bigg[ \frac{\frak{g}_3 + \frak{g}_4  e_t^2 + \frak{g}_5  e_t^4}{(1-e_t^2)^{3}} +  \frac{\frak{g}_6 + \frak{g}_6  e_t^2 + \frak{g}_7  e_t^4}{(1-e_t^2)^{7/2}}\Bigg]  + \mathcal{O}(x^2)  \Bigg) \,,
\end{align}\ese
where the expressions of the coefficients $(\frak{f}_n , \frak{g}_n)$ are relegated to App.~\ref{app:orbitAveragedScalarFluxes}. These fluxes are also presented in machine-readable form the Supplemental Material~\cite{SuppMaterial}.

\subsection{The $\mathbf{\Pi}^s$ flux}

As we know from Sec.~\ref{subsec:nonlocal_term_elliptic}, one can decompose $\Pi^s_i$ into an AC and a DC contribution. We propagate this decomposition to the flux, namely
$ \mathcal{G}_i^{s,\bm{\Pi}^s} =  \mathcal{G}_i^{s,\bm{\Pi}^{s}_\text{AC}} + \mathcal{G}_i^{s,\bm{\Pi}^{s}_\text{DC}}$ where 
\bse\begin{align}
\mathcal{G}_i^{s,\bm{\Pi}^{s}_\text{AC}}  &= \frac{G  (3+2\omega_0)}{3 c^3}\frac{\sqrt{\alpha}\zeta (\Sp + \Sm \delta)}{(1-\zeta)}  \epsilon_{ijk}\, \Pi_j^{s,\text{AC}}  \, \overset{(2)}{\dI}{\!}_k^{s} \,, \\
\mathcal{G}_i^{s,\bm{\Pi}^{s}_\text{DC}}  &= \frac{G (3+2\omega_0)}{3 c^3}\frac{\sqrt{\alpha}\zeta (\Sp + \Sm \delta)}{(1-\zeta)}  \epsilon_{ijk}\, \Pi_j^{s,\text{DC}}  \, \overset{(2)}{\dI}{\!}_k^{s}   \,.
\end{align}\ese
The orbit averaged DC piece of the flux vanishes, because
\begin{align}
\Big\langle\mathcal{G}_i^{s,\bm{\Pi}^s_\text{DC}}\Big\rangle &\propto  \ \epsilon_{ijk}\, \Pi_j^{s,\text{DC}} \,  \Big\langle \overset{(2)}{\dI}{\!}_k^{s} \,\Big\rangle =  0
\end{align}
where I have use the fact that $\bm{\Pi}^{s}_{\text{DC}}$ is approximately constant over an orbit and that time derivatives of multipolar moments vanish upon orbit averaging. Morover, even before orbit averaging, this contribution is finite and vanishes for $e \rightarrow 0$.

Conversely, the AC flux is best expressed by replacing the multipolar moments by their expression as a Fourier series (see Sec.~\ref{sec:Fourier}). After performing explicitly the time integral in $\Pi_j^{s,\text{AC}}$, I explicitly find 
\begin{align}
 \mathcal{G}_i^{s,\bm{\Pi}^s_\text{AC}}  &=    \frac{2 G^2 \phi_0 (3+2\omega_0)^2 \sqrt{\alpha} \zeta  (\Sp + \Sm \delta) }{9 (1-\zeta) c^8}  \nn\\*
 &\qquad \times \epsilon_{ijk}\sum_{r\in\mathbb{Z}} \sum_{p +q \neq 0} \de^{\di (p+q+r) \ell} \frac{ (\di r n)^2}{\di  (p+q)n} \Big( {}_r\widetilde{\dI}^s_{k} \Big)  \bigg[\frac{ (\di p n )(\di q n)^2}{ \phi_0}  \Big({}_p \widetilde{E}^s\Big)  \Big( {}_q\widetilde{\dI}^s_{j} \Big) + \frac{(\di p n)^2 (\di q n)^3}{5}  \Big( {}_p \widetilde{\dI}^s_{a}\Big) \Big( {}_q \widetilde{\dI}^s_{ja}\Big) \bigg] .
\end{align}
Orbit averaging the previous equation is quite simple thanks to the identity $\langle \de^{\di (p+q+r) \ell}\rangle = \delta_{p+q+r}^0 $, where $\delta^a_b$ is the Kronecker symbol. I then obtain
\begin{align}
\Big\langle\ \mathcal{G}_i^{s,\bm{\Pi}^s_\text{AC}} \Big\rangle\ &=    \frac{2 G^2 \phi_0 (3+2\omega_0)^2 \sqrt{\alpha} \zeta  (\Sp + \Sm \delta) }{9 (1-\zeta) c^8}   \nn\\*
 &\quad \times \epsilon_{ijk} \sum_{p + q \neq 0}  \bigg[\frac{n^4 p  q^2 (p+q)}{ \phi_0}  \Big({}_p \widetilde{E}^s\Big)  \Big( {}_q\widetilde{\dI}^s_{j} \Big) \Big( {}_{-p-q}\widetilde{\dI}^s_{k}\Big) - \frac{n^6 p^2 q^3 (p+q)}{5}  \Big( {}_p \widetilde{\dI}^s_{a}\Big) \Big( {}_q \widetilde{\dI}^s_{ja}\Big)  \Big( {}_{-p-q}\widetilde{\dI}^s_{k}\Big) \bigg] .
\end{align}
Since the motion is planar, only the $z$ component of the angular momentum flux is nonvanishing, and a simple analysis involving the explicit expressions of the Fourier coefficients [given by Eqs.~\eqref{eq:normalizedFourierCoefficients}] leads to the conclusion that $\big\langle\ \mathcal{G}_z^{s,\bm{\Pi}^s_\text{AC}} \big\rangle \in \di \mathbb{R}$. But of course, it is also true that $\big\langle\ \mathcal{G}_z^{s,\bm{\Pi}^s_\text{AC}} \big\rangle \in  \mathbb{R}$, which implies that 
\be\Big\langle\ \mathcal{G}_i^{s,\bm{\Pi}^s_\text{AC}} \Big\rangle\ = 0  \,.\ee

In summary, I have found that the $\bm{\Pi}^s$ contribution to the angular momentum flux entire vanishes after orbit averaging, namely 
\be\Big\langle\ \mathcal{G}_i^{s,\bm{\Pi}^s} \Big\rangle\ = 0 \,.  \ee

\subsection{The tail fluxes}

The tail fluxes, given by \eqref{eq:tail_fluxes_source_moments}, feature the difficulty of being nonlocal in time with a logarithmic kernel. These are best treated \cite{BS93, ABIQ08tail, LY16} by replacing all multipolar moments by their Fourier decomposition, as given in Sec.~\ref{sec:Fourier}. Once this replacement is performed, the nonlocal integrals can all be performed using the master formula~\cite{ABIQ08tail, LY16}
\begin{align}
&\int_0^{+\infty}\dd\tau\, \de^{\di(p + rk) n \tau} \ln\left(\frac{c\tau}{2b_0}\right) = -  \frac{1}{p n} \left( 1- \frac{rk}{p}\right)\left[\frac{\pi}{2}\mathrm{sg}(p)+\di \gamma_\mathrm{E} + \di \ln\left(\frac{2|p|n b_0}{c}\right) \right]  - \frac{\di r k }{p^2 n} + \mathcal{O}(k^2) \,,
\end{align} 
which simplifies straightforwardly at Newtonian order (i.e. in the $k \rightarrow 0$ limit). Note that the case $p=0$ never occurs, because the multipolar moments in the nonlocal integrals always bear enough time derivatives, which is equivalent to multiplying the Fourier coefficients by some power of $\di n p$. The flux can be written as a sum over the relative integers $p$, $q$, $r$ and $s$, where each term is proportional to $\de^{\di \ell  [(r+s)k + p + q]}$ [$r+s =0$ is allowed]. It is then straightforward to perform the time averaging thanks to the formula \cite{ABIQ08tail, LY16}
\begin{align}
\left\langle \de^{\di \ell (p + r k)}  \right\rangle &= \Bigg\{\begin{matrix} 1 + \di \pi r k \qquad \text{if $p=0$} \\ \frac{r}{p} k \qquad\qquad\,\  \text{if $p\neq0$} \end{matrix} \Bigg\} + \mathcal{O}(k^2) \,.
\end{align}
After some manipulating some sums, I find as expected that the flux is real-valued, and that all arbitrary constants~$b_0$ vanish from the final result, which read
\bse\label{eq:tail_fluxes_orbit_averaged_intermediate}\begin{align}
 \left\langle\mathcal{F}^\mathrm{tail}\right\rangle  &= \frac{4 \pi G^2 \dM n^7}{5 c^8} \left(\mathcal{I}_2\right)^2 \rho_2(e_t)  \,, \\
  \left\langle\mathcal{G}_i^\mathrm{tail}\right\rangle  &= -\frac{8 \di \pi G^2 \dM n^6}{5 c^8} \left(\mathcal{I}_2\right)^2   \tilde{\rho}_2(e_t)\,, \\
 \left\langle\mathcal{F}^{s,\mathrm{tail}}\right\rangle  &=  \frac{4 \pi G^2 \dM(3+2\omega_0)}{c^6} \Bigg\{\frac{n^5}{3} \left(\mathcal{I}^s_1\right)^2  \left[ \rho^s_1(e_t) + 5 k \sigma^s_1(e_t)\right] + \frac{n^7}{30 c^2} \left(\mathcal{I}^s_2\right)^2 \rho_2^s(e_t)  + \frac{n^3}{\phi_0^2 c^2} \left(\mathcal{E}^s_0\right)^2 \rho_0^s(e_t) \Bigg\} \,, \\
 \left\langle\mathcal{G}_i^{s,\mathrm{tail}}\right\rangle  &= - \frac{4 \di \pi G^2 \dM(3+2\omega_0)}{c^6}  \Bigg\{\frac{n^4}{3} \left(\mathcal{I}^s_1\right)^2 \left[\tilde{\rho}_1^s(e_t) + 4k \tilde{\sigma}_1^s(e_t)\right]  + \frac{n^6}{15 c^2} \left(\mathcal{I}^s_2\right)^2 \tilde{\rho}_2^s(e_t) \Bigg\} \,,
 \end{align}\ese
 where I have introduced in this intermediate step some \emph{unnormalized enhancement functions}, defined by
\bse \label{eq:def_unnormalized_enhancement_functions} \begin{align}
\rho_2(e_t) &=  \sum_{p=1}^\infty p^7  \left({}_p\widehat{\dI}_{ij}\right) \left({}_{p}\widehat{\dI}_{ij}\right)^{\!*}\,,&
\tilde{\rho}_2(e_t) &= \epsilon_{ijk}\sum_{p=1}^\infty p^6 \left({}_p\widehat{\dI}_{ja}\right)  \left({}_{p}\widehat{\dI}_{ka}\right)^{\!*}\,, \nn\\*
\rho_2^s(e_t) &=  \sum_{p=1}^\infty p^7  \left({}_p\widehat{\dI}^s_{ij}\right) \left({}_{p}\widehat{\dI}^s_{ij}\right)^{\!*} \,,&
\tilde{\rho}^s_2(e_t) &= \epsilon_{ijk}\sum_{p=1}^\infty p^6  \left({}_p\widehat{\dI}^s_{ja}\right)  \left({}_{p}\widehat{\dI}^s_{ka}\right)^{\!*}\,,\nn\\*
\rho_1^s(e_t) &=  \sum_{p=1}^{\infty}p^5\!\!\!\!\!\! \sum_{\substack{m\in\{-1,1\}\\s\in\{-1,1\}}} \!\!\!  \left({}^{m}_{\,\,p}\widehat{\dI}^s_{i}\right) \left({}^{m}_{\,\,p}\widehat{\dI}^s_{i}\right)^* \,, &
\tilde{\rho}_1^s(e_t) &=  \epsilon_{ijk}\sum_{p=1}^{\infty}p^4 \!\!\!\!\!\!\sum_{\substack{m\in\{-1,1\}\\s\in\{-1,1\}}} \!\!\! \left({}^{m}_{\,\,p}\widehat{\dI}^s_{j}\right) \left({}^{m}_{\,\,p}\widehat{\dI}^s_{k}\right)^*   \,,\nn\\*
\sigma_1^s(e_t) &= \sum_{p=1}^{\infty} p^4 \!\!\!  \!\!\sum_{m\in\{-1,1\}} \!\!\!\!\!\! m  \left({}^{m}_{\,\,p}\widehat{\dI}^s_{i}\right) \left({}^{m}_{\,\,p}\widehat{\dI}^s_{i}\right)^* \,,&
\tilde{\sigma}_1^s(e_t) &= \epsilon_{ijk} \sum_{p=1}^{\infty} p^3\!\!\!  \!\!\sum_{m\in\{-1,1\}} \!\!\!\!\!\! m  \left({}^{m}_{\,\,p}\widehat{\dI}^s_{j}\right) \left({}^{m}_{\,\,p}\widehat{\dI}^s_{k}\right)^* \,, \nn\\*
\rho_0^s(e_t) &= \sum_{p=1}^\infty p^3 \left({}_p\widehat{E}^s \right) \left({}_{p}\widehat{E}^s \right)^{\!*}  \,. &
\end{align}\ese
It will prove useful to decompose the dipolar unnormalized enhancement functions as 
\begin{align*}
\rho_1^s(e_t) &= \rho_1^{s,00}(e_t) + x\Big(\rho_1^{s,01}(e_t) +\nu \rho_1^{s,11}(e_t)\Big)\\*
\tilde{\rho}_1^s(e_t) &= \tilde{\rho}_1^{s,00}(e_t) + x\Big(\tilde{\rho}_1^{s,01}(e_t) +\nu \tilde{\rho}_1^{s,11}(e_t)\Big)
\end{align*}
where
\bse\begin{align}
\rho_1^{s,00}(e_t) &=  \sum_{p=1}^{\infty}p^5\!\!\!\!\!\! \sum_{\substack{m\in\{-1,1\}\\s\in\{-1,1\}}} \!\!\!  \left({}^{m}_{\,\,p}\widehat{\dI}^{s,00}_{i}\right) \left({}^{m}_{\,\,p}\widehat{\dI}^{s,00}_{i}\right)^* \\
\tilde{\rho}_1^{s,00}(e_t) &=  \epsilon_{ijk}\sum_{p=1}^{\infty}p^4 \!\!\!\!\!\!\sum_{\substack{m\in\{-1,1\}\\s\in\{-1,1\}}} \!\!\! \left({}^{m}_{\,\,p}\widehat{\dI}^{s,00}_{j}\right) \left({}^{m}_{\,\,p}\widehat{\dI}^{s,00}_{k}\right)^* \,, \\
\rho_1^{s,01}(e_t) &=  \sum_{p=1}^{\infty}p^5\!\!\!\!\!\! \sum_{\substack{m\in\{-1,1\}\\s\in\{-1,1\}}} \Bigg[ \left({}^{m}_{\,\,p}\widehat{\dI}^{s,00}_{i}\right) \left({}^{m}_{\,\,p}\widehat{\dI}^{s,01}_{i}\right)^*  +  \left({}^{m}_{\,\,p}\widehat{\dI}^{s,01}_{i}\right) \left({}^{m}_{\,\,p}\widehat{\dI}^{s,00}_{i}\right)^*  \Bigg] \,, \\
\tilde{\rho}_1^{s,01}(e_t) &=  \epsilon_{ijk}\sum_{p=1}^{\infty}p^4 \!\!\!\!\!\!\sum_{\substack{m\in\{-1,1\}\\s\in\{-1,1\}}} \Bigg[ \left({}^{m}_{\,\,p}\widehat{\dI}^{s,00}_{j}\right) \left({}^{m}_{\,\,p}\widehat{\dI}^{s,01}_{k}\right)^* + \left({}^{m}_{\,\,p}\widehat{\dI}^{s,01}_{j}\right) \left({}^{m}_{\,\,p}\widehat{\dI}^{s,00}_{k}\right)^*  \Bigg] \,, \\
\rho_1^{s,11}(e_t) &=  \sum_{p=1}^{\infty}p^5\!\!\!\!\!\! \sum_{\substack{m\in\{-1,1\}\\s\in\{-1,1\}}} \Bigg[ \left({}^{m}_{\,\,p}\widehat{\dI}^{s,00}_{i}\right) \left({}^{m}_{\,\,p}\widehat{\dI}^{s,11}_{i}\right)^*  +  \left({}^{m}_{\,\,p}\widehat{\dI}^{s,11}_{i}\right) \left({}^{m}_{\,\,p}\widehat{\dI}^{s,00}_{i}\right)^*  \Bigg]  \,,\\
\tilde{\rho}_1^{s,11}(e_t) &=  \epsilon_{ijk}\sum_{p=1}^{\infty}p^4 \!\!\!\!\!\!\sum_{\substack{m\in\{-1,1\}\\s\in\{-1,1\}}} \Bigg[ \left({}^{m}_{\,\,p}\widehat{\dI}^{s,00}_{j}\right) \left({}^{m}_{\,\,p}\widehat{\dI}^{s,11}_{k}\right)^*  + \left({}^{m}_{\,\,p}\widehat{\dI}^{s,11}_{j}\right) \left({}^{m}_{\,\,p}\widehat{\dI}^{s,00}_{k}\right)^*  \Bigg]
\end{align}\ese
The limits of these unnormalized enhancement functions when $e_t \rightarrow 0$ are straightforward to compute, and read:
\begin{align}\label{eq:decomposition_rho1s_tilderho1s}
\rho_2(0) &= 32 \,, &  \rho_2^s(0) &= 32 \,,& \rho_1^{s,00}(0) &= \frac{1}{2} &  \rho_1^{s,01}(0) &= - \frac{9}{5}\mathcal{A} \,, & \rho_1^{s,11}(0) &=  \frac{14}{15} \,, & \sigma_1^{s}(0) &= \frac{1}{2} \,,& \rho_0^s(0) &=0 \,, \nn\\*
\tilde\rho_2(0) &= 16\di \,, &  \tilde\rho_2^s(0) &= 16\di \,,& \tilde\rho_1^{s,00}(0) &=  \frac{\di}{2} &  \tilde\rho_1^{s,01}(0) &=  - \frac{9\di}{5} \mathcal{A} \,, & \tilde\rho_1^{s,11}(0) &= \frac{14\di}{15}\,, & \tilde\sigma_1^{s}(0) &= \frac{\di}{2}  \,,
\end{align}
where I have introduced the recurring parameter\begin{align}\label{eq:A_denominator}
 \mathcal{A} &\equiv 1 +\frac{5}{27}\gam +  \frac{10}{27}\bep - \frac{10}{9} \frac{\bep}{ \gam} - \frac{10}{9} \frac{\bem \Sp}{\gam \Sm} + \delta \left[- \frac{10}{27}\bem + \frac{10}{9} \frac{\bem}{\gam} - \frac{4}{9} \frac{\Sp}{\Sm}+ \frac{10}{9} \frac{\bep \Sp}{\gam \Sm}\right] \,.
 \end{align}
It will prove useful to note that $\rho_0^s(e_t) \sim e_t/4$ as $e_t \rightarrow 0$. I then express Eqs.~\eqref{eq:tail_fluxes_orbit_averaged_intermediate} only in terms of the pair of variable $x$ and $e_t$. For this, I use the 1PN QK parametrization \cite{T24_QK} to expand the ADM mass, the mean motion and the periastron advance:
\bse\begin{align}
\dM &= m\left(1-\frac{x \nu}{2} + \calO(x^2)\right) \,,\\*
n &= \frac{c^3 x^{3/2}}{\tilde{G}\alpha m}\left(1 -\frac{x  \mathcal{B}}{1-e_t^2} + \calO(x^2)\right) \,, \\*
k &= \frac{x  \mathcal{B}}{1-e_t^2} + \calO(x^2)\,,
\end{align}\ese	
and where the reader is reminded that $\mathcal{B}$ was defined in Eq.~\eqref{eq:calB}.

The expression for the fluxes thus obtained can finally be written similarly to the fluxes for circular orbits, in which some extra \emph{normalized} enhancement functions of the eccentricity have been inserted. In the tensor sector, they are given by
\bse \label{tensor_flux_orbit_averaged_tail_final}\begin{align}
\left\langle \mathcal{F}^\mathrm{tail}\right\rangle &= \frac{32 c^5 x^5 \nu^2 (1+ \gam/2)}{5 \tilde{G} \alpha}\times 4\pi (1+ \gam/2) x^{3/2} \varphi_2(e_t) + \calO(x^7) \,,\\
\left\langle \mathcal{G}^\mathrm{tail}\right\rangle &= \frac{32 c^2 x^{7/2} m \nu^2 (1+ \gam/2)}{5}\times 4\pi (1+ \gam/2) x^{3/2} \tilde{\varphi}_2(e_t) + \calO(x^7)\,,
\end{align}\ese
where the normalized enhancement functions read
\begin{align}\label{eq:def_normalized_enhancement_functions_tensor}
\varphi_2(e_t) = \frac{1}{32}\rho_2(e_t) \qquad\qquad\text{and}\qquad\qquad\qquad\tilde{\varphi}_2(e_t) =  -\frac{\di}{16}\tilde{\rho}_2(e_t)\,,
\end{align}
where the reader is reminded  that the unnormalized enhancement functions are given by \eqref{eq:def_unnormalized_enhancement_functions}.

Similarly, the scalar sector is given by
\bse\label{scalar_flux_orbit_averaged_tail_final}\begin{align}
\left\langle \mathcal{F}^{s,\mathrm{tail}}\right\rangle &= \frac{ c^5 x^5 \nu^2 \zeta}{3\tilde{G} \alpha}\times 4\pi (1+ \gam/2) \sqrt{x}  \Bigg\{  2\Sm^2 \varphi^s_1(e_t) \nn\\*
& + x\Bigg[ \frac{8}{5} \varphi_2^s(e_t) \bigg( 4 \Sm^2 - 4 \Sm\Sp \delta  -   \gam \zeta^{-1}  - 8 \Sm^2 \nu  \bigg) +\frac{41}{15}  \Sm^2 \nu \theta^s_1(e_t)  \nn\\*
&\qquad +  \frac{4}{15}\alpha^s_1(e_t) \bigg(-27 \Sm^2 - 10 \bep \Sm^2 + 30 \bep \Sm^2 \gam^{-1} - 5 \gam \Sm^2 + 30 \bem \Sm \Sp \gam^{-1}\nn \\*
& \qquad\qquad\qquad\qquad + \delta \left[10 \bem \Sm^2 -30 \bem \Sm^2 \gam^{-1} + 12 \Sm \Sp - 30 \bep \Sm \Sp \gam^{-1} \right]  \bigg) \nn\\*
&\qquad+ \frac{1}{3} e_t^2 \varphi^s_{0}(e_t) \bigg(48 \bep \zeta^{-1} - 72 \bep^2 \zeta^{-1} \gam^{-1} - 8  \gam \zeta^{-1} + 17 \Sm^2 + 144 \bem^2 \Sm^2 \gam^{-2} \nn\\*
& \qquad\qquad\qquad\qquad+ 144  \bep^2 \Sm^2 \gam^{-2}- 96 \bep \Sm^2 \gam^{-1} + 288 \bem \bep \Sm\Sp \gam^{-2} - 96 \bem \Sm \Sp \gam^{-1} \nn\\*
& \qquad\qquad \qquad\qquad+ \delta\left[24 \bem  \Sm^2 \gam^{-1} - 8 \Sm \Sp+ 24 \bep \Sm \Sp \gam^{-1}\right] - 4 \Sm^{2} \nu \bigg) \Bigg]\Bigg\} + \calO(x^7)  \,,\\
\left\langle \mathcal{G}^{s,\mathrm{tail}}\right\rangle &= \frac{c^2 x^{7/2} m \nu^2 \zeta }{3}\times 4\pi (1+ \gam/2) \sqrt{x} \Bigg\{  2\Sm^2 \tilde\varphi^s_1(e_t) \nn\\*
& + x\Bigg[ \frac{8}{5} \tilde\varphi_2^s(e_t)\bigg( 4 \Sm^2 - 4 \Sm\Sp \delta  -   \gam \zeta^{-1}  - 8 \Sm^2 \nu  \bigg)  +\frac{41}{15}  \Sm^2 \nu \tilde\theta^s_1(e_t)  \nn\\*
&\qquad +  \frac{4}{15}\tilde\alpha^s_1(e_t) \bigg(-27 \Sm^2 - 10 \bep \Sm^2 + 30 \bep \Sm^2 \gam^{-1} - 5 \gam \Sm^2 + 30 \bem \Sm \Sp \gamma^{-1}\nn \\*
& \qquad\qquad\qquad\qquad +\delta \left[10 \bem \Sm^2 -30 \bem \Sm^2 \gam^{-1} + 12 \Sm \Sp - 30 \bep \Sm \Sp \gam^{-1} \right]  \bigg)  \Bigg]\Bigg\} + \calO(x^7) \,,
\end{align}\ese
where the enhancement functions read 
\begin{align}\label{eq:def_normalized_enhancement_functions_scalar}
&\varphi^s_0(e_t) = \frac{4\, \rho^s_0(e_t)}{e_t^2}\,, &\qquad&\nn\\*
&\varphi^s_2(e_t) = \frac{1}{32} \, \rho^s_2(e_t) \,, \quad& \tilde{\varphi}^s_2(e_t) &= - \frac{\di}{16} \tilde{\rho}^s_2(e_t)\,, \nn\\*
&\varphi_1^s(e_t) = 2\, \rho_1^{s,00}(e_t)\,,\qquad& \tilde{\varphi}^s_1(e_t) &= -2\di  \tilde{\rho}_1^{s,00}(e_t)\,, \nn\\*
&\alpha_1^s(e_t) = - \frac{5}{9\mathcal{A}}\Big[\rho_1^{s,01}(e_t)  + \frac{5 \mathcal{B}}{1-e_t^2}\left(\sigma_1^s(e_t) - \rho_1^{s,00}(e_t) \right)\Big] \,,  \qquad& \tilde{\alpha}^s_1(e_t) &=  \frac{5\di }{9\mathcal{A}}\Big[\tilde{\rho}_1^{s,01}(e_t)  + \frac{4 \mathcal{B}}{1-e_t^2}\left(\tilde{\sigma}_1^s(e_t) - \tilde{\rho}_1^{s,00}(e_t) \right)\Big] \,,\nn\\*
&\theta^s_1(e_t) = - \frac{30}{41}\left(\rho_1^{s,00}(e_t) - 2  \rho_1^{s,11}(e_t)\right)\,, \qquad&\tilde{\theta}^s_1(e_t) & =  \frac{30\di}{41}\left(\tilde{\rho}_1^{s,00}(e_t) - 2 \tilde{\rho}_1^{s,11}(e_t)\right)\,, 
\end{align}
and where the reader is reminded  that the unnormalized enhancement functions are given by \eqref{eq:def_unnormalized_enhancement_functions}.
The normalized enhancement functions are defined such that they tend to $1$ when $e_t \rightarrow 0$. I chose to normalize it such that  $\varphi_0^s(e_t) \sim e_t^2$ when $e_t \rightarrow 0$).
Thus, taking the $e_t  \rightarrow 0$ limit and setting all enhancement functions to $1$ in \eqref{tensor_flux_orbit_averaged_tail_final} and \eqref{scalar_flux_orbit_averaged_tail_final} 
 immediately yields the expressions for the tail sector of the circular energy flux. For the energy fluxes, it can be verified that the circular limit for the tail sector  agrees with the odd PN-order terms in (5.3) of \citetalias{BBT22}. Moreover, it is immediate to verify that in the circular limit, $\big\langle \mathcal{F}_\text{circ}^{\mathrm{tail}} \big\rangle= \omega \,\big\langle \mathcal{G}_\text{circ}^{\mathrm{tail}} \big\rangle $ and $\big\langle \mathcal{F}_\text{circ}^{s,\mathrm{tail}}\, \big\rangle = \omega\, \big\langle \mathcal{G}_\text{circ}^{s,\mathrm{tail}}\big\rangle $. 
 
Of particular interest will be the small-eccentricity expansions of these enhancement functions: 
\bse\label{eq:expansion_normalized_enhancement_functions}\begin{align}
\varphi_2(e_t) = &= 1 + \frac{2335}{192} e_t^2 + \frac{42955}{768}e_t^4 + \frac{6204647}{36864}e_t^6 + \frac{352891481}{884736}e_t^8 + \frac{286907786543}{353894400} e_t^{10} + \mathcal{O}(e_t^{12}) \,,\\
\tilde{\varphi}_2(e_t) &=  1 + \frac{209}{32} e_t^2 + \frac{2415}{128}e_t^4 + \frac{730751}{18432}e_t^6 + \frac{10355719}{147456}e_t^8 + \frac{6594861233}{58982400} e_t^{10} + \mathcal{O}(e_t^{12}) \,,\\
\varphi^s_2(e_t) &= 1 + \frac{2335}{192} e_t^2 + \frac{42955}{768}e_t^4 + \frac{6204647}{36864}e_t^6 + \frac{352891481}{884736}e_t^8 + \frac{286907786543}{353894400} e_t^{10} + \mathcal{O}(e_t^{12}) \,,\\
\tilde{\varphi}^s_2(e_t) &=  1 + \frac{209}{32} e_t^2 + \frac{2415}{128}e_t^4 + \frac{730751}{18432}e_t^6 + \frac{10355719}{147456}e_t^8 + \frac{6594861233}{58982400} e_t^{10} + \mathcal{O}(e_t^{12})\,,\\
\varphi^s_1(e_t) &=   1+ 7 e_t^2 + \frac{717}{32}e_t^4+ \frac{7435}{144}e_t^6 + \frac{7305575}{73728}e_t^8 + \frac{103947697}{614400} e_t^{10} + \mathcal{O}(e_t^{12}) \,, \\
\tilde{\varphi}^s_1(e_t) &=  1+ 3 e_t^2 + \frac{179}{32}e_t^4+ \frac{1249}{144}e_t^6 + \frac{99715}{8192}e_t^8 + \frac{29573111}{1843200} e_t^{10} + \mathcal{O}(e_t^{12}) \,,\\
\alpha^s_1(e_t) &=  1+ \frac{1}{\mathcal{A}^2} \sum_{n=2}^{10} \mathcal{C}_n e_t^n + \mathcal{O}(e_t^2) \,,\\
\tilde{\alpha}^s_1(e_t) &=  1+ \frac{1}{\mathcal{A}^2} \sum_{n=2}^{10} \widetilde{\mathcal{C}}_n e_t^n + \mathcal{O}(e_t^2)  \,, \\
\theta^s_1(e_t) &= 1+ \frac{317}{41} e_t^2 + \frac{36837}{1312}e_t^4 + \frac{214765}{2952} e_t^6 + \frac{469593535}{3022848}e_t^8 + \frac{7363736527}{25190400}e_t^{10}+ \mathcal{O}(e_t^{12})\,,\\
\tilde{\theta}^s_1(e_t) &=  1+ \frac{153}{41} e_t^2 + \frac{11179}{1312}e_t^4 + \frac{46327}{2952} e_t^6 + \frac{25733108}{1007616}e_t^8 + \frac{2895685801}{75571200} e_t^{10}+ \mathcal{O}(e_t^{12}) \,,\\
\varphi^s_0(e_t) &= 1 + \frac{31}{4} e_t^2 + \frac{2771}{96}e_t^4 + \frac{177151}{2304}e_t^6 + \frac{62037619}{368640}e_t^{8} + \mathcal{O}(e_t^{10}) \,,
\end{align}\ese
where $\mathcal{A}$ is given in \eqref{eq:A_denominator}, and the coefficients $\mathcal{C}_n$ and $\widetilde{\mathcal{C}}_n$ are given explicitly in App.~\ref{app:smallEccentricityExpansion} [they depend in a complicated manner on the scalar-tensor parameters and are therefore very lengthy]. 

Although these expansions are formally only valid for small eccentricities, they can be resummed by factoring out $(1-e_t^2)^{-N/2}$, where $N$ is an integer to be determined for each enhancement function. This determination can either be done heuristically, by comparison to the numerically-evaluated function, or using asymptotic analysis in the lines of IV.C. of Ref.~\cite{Forseth:2015oua} (see also \cite{LY16} for an alternative asymptotic analysis). Despite their simplicity, these resummations can dramatically improve the accuracy of the approximation for high eccentricities.

\subsection{The memory flux}

Finally, I tackle the case of the memory integral, given by \eqref{eq:memory_flux_source_moments}. Such an integral appears in GR for the first time at 2.5PN, and was first studied in Ref.~\cite{ABIS09}. The latter work found that the magnitude of the memory effect was infinite in the limit where the initial eccentricity of the system is $1$, i.e., in the limit where the binary was formed in the infinite past [see their Eq.~(A11)]. To mitigate this problem, the authors then invoke the astrophysical argument that in practice, the binary must have formed by capture with some initial eccentricity. Still, this suggests that the waveform is dominated by the remote past, and the zero-eccentricity limit is not smooth. That method was recently adapted to ST theories in Ref.~\cite{JR24}, whose Eq.~(4.19) also diverges\footnote{
In (4.19) of~\cite{JR24}, replace $a$ by its expression in terms of $e$ given by  \eqref{seq:relation_a_e}, and perform the change of variables $t \rightarrow e$ using \eqref{seq:dedt}. The integral then reads (for $\kappa \equiv {(\tilde{G}\alpha m)^2 \zeta\mathcal{S}_-^2 \nu}/{c^3}$):
\begin{align*}
&\int_{t_1}^{t_0} \dd t \frac{e^2(e^2+6)}{a^4(e^2-1)^4} =  -\frac{1}{2 \kappa c_0} \int_{e_1}^{e_0} \dd e \frac{6+e^2}{(1-e^2)^{3/2}e^{1/3}} = -\frac{1}{2 \kappa c_0} \left[e^{2/3}\left\{\frac{7}{\sqrt{1-e^2}} + 2 F\left(\frac{1}{3}, \frac{1}{2}; \frac{4}{3}; e^2\right)\right\}\right]^{e_0}_{e_1}   \sim \frac{7}{2 c_0 \kappa} \frac{1}{\sqrt{1-e_1^2}}\quad\text{as}\  e_1\rightarrow 1 \,.
\end{align*}
}
as the initial eccentricity goes to~$1$.  

When reexamining this problem, I found that these aforementioned divergences are cured by a careful split between AC and DC pieces. I will now describe in detail the full treatment of the memory integral in the case of ST  theory, and the complete treatment of the GR case is relegated to Appendix~\ref{app:memory_in_GR}.

For a nonprecessing orbit, the direction of the angular momentum flux is constant and points in the $z$-direction. It is then straightforward to explicitly express its norm \mbox{$\mathcal{G} = |\vec{\mathcal{G}}|= \mathcal{G}_z $} in terms of the $(x,y,z)$ components, namely

\begin{align}
\mathcal{G}^\mathrm{mem}(u) &= \frac{G^2 \phi_0 (3+2\omega_0)}{15 c^{8}} \Bigg\{\left(\dI_{xx}^{(3)}(u) -\dI_{yy}^{(3)}(u)\right) \int_0^{\infty} \dd\tau\,  f(u-\tau) - \dI_{xy}^{(3)}(u)   \int_0^{\infty} \dd\tau\, g(u-\tau) \Bigg\}\,,
\end{align}
where I have used the fact that $\dI_{xz} = \dI_{yz}=0$ and defined
\be
f =  \overset{(2)}{\dI} {\!}_{x}^{s} \overset{(2)}{\dI} {\!}_{y}^{s} \qquad\qquad\text{and}\qquad\qquad
g =  \overset{(2)}{\dI} {\!}_{x}^{s}  \overset{(2)}{\dI} {\!}_{x}^{s} -  \overset{(2)}{\dI} {\!}_{y}^{s}  \overset{(2)}{\dI} {\!}_{y}^{s} \,.
\ee

In order to distinguish the AC and DC contributions, I decompose these functions into orbit-averaged and zero-average contributions, namely \mbox{$f =  \left\langle f \right\rangle\!+ \tilde{f}$} and \mbox{$g =  \left\langle g \right\rangle\! + \tilde{g}$}. Consequently, the functions $\tilde{f}$ and $\tilde{g}$ contain purely oscillatory terms which average out to zero over a single orbit. Thus, it is now possible to define precisely the AC and DC contributions: 
\bse \label{eq:flux_memory_AC_DC} \begin{align}
\label{seq:flux_memory_DC} \mathcal{G}_\mathrm{DC}^\mathrm{mem}(u) &= \frac{G^2 \phi_0 (3+2\omega_0)}{15 c^{8}} \Bigg\{\left(\dI_{xx}^{(3)}(u) -\dI_{yy}^{(3)}(u)\right) \int_0^{\infty} \dd\tau\, \left\langle f \right\rangle\!(u-\tau)\  - \dI_{xy}^{(3)}(u)   \int_0^{\infty} \dd\tau\, \left\langle g \right\rangle\! (u-\tau) \Bigg\} \,,\\
 \label{seq:flux_memory_AC} \mathcal{G}_\mathrm{AC}^\mathrm{mem}(u) &= \frac{G^2 \phi_0 (3+2\omega_0)}{15 c^{8}} \Bigg\{\left(\dI_{xx}^{(3)}(u) -\dI_{yy}^{(3)}(u)\right) \int_0^{\infty} \dd\tau\,  \tilde{f}(u-\tau)  - \dI_{xy}^{(3)}(u)   \int_0^{\infty} \dd\tau\, \tilde{g}(u-\tau) \Bigg\}\,.
\end{align} \ese

\subsubsection{The AC part}
First, I will focus on the AC terms. Replacing the components of the quadrupole moments by their Fourier series following the notations of Sec.~\ref{sec:Fourier}, I find that 
\bse\begin{align}
f &= n^4  \sum_{p \in \mathbb{Z}} \sum_{q \in \mathbb{Z}} \de^{\di(p+q)\ell} \,p^2 q^2 \ {}_p\widetilde{\dI}_{x}^s\ {}_p\widetilde{\dI}_{y}^s \,, \\
g &= n^4   \sum_{p \in \mathbb{Z}} \sum_{q \in \mathbb{Z}} \de^{\di(p+q)\ell} \, p^2 q^2 \left( {}_p\widetilde{\dI}_{x}^s \ {}_p\widetilde{\dI}_{x}^s - {}_p\widetilde{\dI}_{y}^s \ {}_p\widetilde{\dI}_{y}^s\right) \,.
\end{align}\ese

By definition of the orbital average, the DC component exactly corresponds to the terms for which $p+q=0$, so I can now simply reexpress the AC term as 
\bse\begin{align}
\tilde{f} &=  n^4  \sum_{p + q \neq0} \de^{\di(p+q)\ell} \,p^2 q^2 \ {}_p\widetilde{\dI}_{x}^s\ {}_p\widetilde{\dI}_{x}^s \,,\\
\tilde{g} &= n^4    \sum_{p + q \neq 0} \de^{\di(p+q)\ell} \, p^2 q^2 \left( {}_p\widetilde{\dI}_{x}^s \ {}_p\widetilde{\dI}_{x}^s - {}_p\widetilde{\dI}_{y}^s \ {}_p\widetilde{\dI}_{y}^s\right)\,.
\end{align}\ese

Replacing all moments by their expressions as Fourier series \eqref{eq:genericNewtonianFourierSeries}, replacing the normalizations of the Fourier coefficients by their expressions \eqref{eq:FourierNormalizationFactor}, performing the time integrals explicitly,  and substituting at leading order $n = c^3 x^{3/2}/(\tilde{G}\alpha m)$, I readily obtain 
\begin{align}\label{eq:GmemAC_exp}
\mathcal{G}_\mathrm{AC}^\mathrm{mem} &= - \frac{4}{15} \Sm^2 \zeta (1+\gam/2) m \nu^3 c^2 x^5 \sum_{r\in\mathbb{Z}} \sum_{p + q \neq 0}e^{\di(r+p+q)\ell}\,\frac{p^2q ^2 r^3}{p+q}\, \mathcal{S}_{r,p,q} \,,
\end{align}
where
\begin{align}
\mathcal{S}_{r,p,q} &=  \Big({}_r\widehat{\dI}_{xx} - {}_r\widehat{\dI}_{yy} \Big)  {}_p\widehat{\dI}_{x}^s \ {}_q\widehat{\dI}_{y}^s - {}_r\widehat{\dI}_{xy} \Big({}_p\widehat{\dI}_{x}^s \ {}_q\widehat{\dI}_{x}^s - {}_p\widehat{\dI}_{y}^s\ {}_q\widehat{\dI}_{y}^s \Big)\,.
\end{align}
Here, I expressed $\mathcal{S}_{r,p,q}$ in terms of the normalized Fourier coefficients, given by \eqref{eq:normalizedFourierCoefficients}.  Thanks to the latter expressions, it is immediate to see that $\mathcal{S}_{r,p,q} \in \di \mathbb{R}$. Since $\mathcal{G}_\mathrm{AC}^\mathrm{mem} \in \mathbb{R}$ by construction, taking the real part of \eqref{eq:GmemAC_exp} yields the explicitly real-valued expression:
\begin{align}\label{eq:GmemAC_sin}
\mathcal{G}_\mathrm{AC}^\mathrm{mem} &=  - \frac{4}{15} \Sm^2 \zeta (1+\gam/2) m \nu^3 c^2 x^5 \sum_{r\in\mathbb{Z}} \sum_{p + q \neq 0}\sin\Big((r+p+q)\ell\Big)\,\frac{p^2q ^2 r^3}{p+q}\, \di\mathcal{S}_{r,p,q} \,.
\end{align}
In order to compute the orbit-average of $\mathcal{G}_\mathrm{AC}^\mathrm{mem}$, it is in fact more practical to start from the expression \eqref{eq:GmemAC_exp}, and to find 
\begin{align}
\left\langle \mathcal{G}_\mathrm{AC}^\mathrm{mem} \right\rangle  &=  \frac{4}{15} \Sm^2 \zeta (1+\gam/2) m \nu^3 c^2 x^5  \sum_{p + q \neq 0} p^2 q^2 (p+q)^2\, \mathcal{S}_{-p-q,p,q}\,.
\end{align}
Since $\mathcal{S}_{-p-q,p,q}$ is purely imaginary, it follows that $\left\langle \mathcal{G}_\mathrm{AC}^\mathrm{mem} \right\rangle \in \di \mathbb{R}$, but of course it is also true by construction that $\left\langle \mathcal{G}_\mathrm{AC}^\mathrm{mem} \right\rangle \in \mathbb{R}$. All this implies that, necessarily,
\begin{align}
\left\langle \mathcal{G}_\mathrm{AC}^\mathrm{mem} \right\rangle = 0 \,.
\end{align} 

\subsubsection{The DC part}
I now turn to the DC part.  Thanks to the explicit expression of the quadrupole moment and upon careful computation of the orbital average, I obtain $\langle f \rangle= 0$ and 
\begin{align}
\langle g \rangle &= \frac{\tilde{G}^2 \alpha^3 \zeta^2 \Sm^2 m^4 \nu^2}{(1-\zeta)^2 \phi_0^2}\times \frac{e^2}{a^4 (1-e^2)^{5/2}} \,.
\end{align}
Thanks to the Eqs.~\eqref{eq:PetersMathews_ST}, I can reexpress the nonvanishing memory integral as an integral over eccentricity. In the far past, the eccentricity asymptotes towards one, so the memory integral reads
\begin{align}\label{eq:integral_orbit_averaged_g}
\int_0^{\infty} \!\!\dd\tau \ \langle g \rangle(u-\tau)&=  \frac{c^3 \alpha \zeta  m^2 \nu}{2 (1-\zeta)^2 \phi_0^2  c_0}\int_{e}^1 \frac{\dd \tilde{e}}{\tilde{e}^{1/3}}= \frac{3 c^3 \alpha \zeta  m^2 \nu}{4 (1-\zeta)^2 \phi_0^2  c_0} (1-e^{2/3})\,.
\end{align}
Finally, plugging \eqref{eq:integral_orbit_averaged_g} into \eqref{seq:flux_memory_DC} leads to 
\begin{align}\label{eq:G_DC_mem_final}
\mathcal{G}_\mathrm{DC}^\mathrm{mem}&=   - \frac{(\tilde{G} \alpha m)^2 \nu  \phi_0 (1+\gam/2) }{20 c^5}  \frac{e^{4/3}}{a(1+e^{2/3}+e^{4/3})}\, \dI_{xy}^{(3)} \,,
\end{align} 
where $c_0$ has been replaced by its expression \eqref{seq:relation_a_e} in terms of $a$ and $e$. Thus, one can now see that $\mathcal{G}_\mathrm{DC}^\mathrm{mem} \rightarrow 0 $ in the circularized limit $e \rightarrow 0$. Remarkably, this conclusion is perfectly consistent with the result one would have  obtained if one had instead computed the DC memory integral using an eternally quasicircular model \cite{BFIJ02}.

Finally, since the orbital parameters are (approximately) constant over one orbit, and since $\langle \dI_{ij}^{(3)} \rangle = 0$, I finally find that 
\be
\left\langle \mathcal{G}_\mathrm{DC}^\mathrm{mem} \right\rangle = 0 \,.
\ee

\subsection{Recapitulation}

Having reviewed the different contributions to the orbit-averaged energy and angular momentum flux, let me summarize the results:
\bse\begin{align}
\left\langle \mathcal{F} \right\rangle &= \left\langle \mathcal{F}^{\mathrm{inst}}\right\rangle + \left\langle \mathcal{F}^{\mathrm{tail}}\right\rangle \,, \\
\left\langle \mathcal{G} \right\rangle &=  \left\langle \mathcal{G}^{\mathrm{inst}}\right\rangle + \left\langle \mathcal{G}^{\mathrm{tail}}\right\rangle \,,\\
\left\langle \mathcal{F}^s \right\rangle &= \left\langle \mathcal{F}^{s,\mathrm{inst}}\right\rangle + \left\langle \mathcal{F}^{s,\mathrm{tail}}\right\rangle \,, \\
\left\langle \mathcal{G}^s \right\rangle &=  \left\langle \mathcal{G}^{s,\mathrm{inst}}\right\rangle + \left\langle \mathcal{G}^{s,\mathrm{tail}}\right\rangle \,.
\end{align}\ese
The \emph{tensorial instantaneous} fluxes  $\left\langle \mathcal{F}^{\mathrm{inst}}\right\rangle$ and $\left\langle \mathcal{G}^{\mathrm{inst}}\right\rangle$ are given by \eqref{tensor_flux_orbit_averaged_inst_final}, whereas the \emph{scalar instantaneous} fluxes $\left\langle \mathcal{F}^{s,\mathrm{inst}}\right\rangle$ and $\left\langle \mathcal{G}^{s,\mathrm{inst}}\right\rangle$ are given  \eqref{scalar_flux_orbit_averaged_inst_final}. 
The \emph{tensorial tail} fluxes $\left\langle \mathcal{F}^{s,\mathrm{tail}}\right\rangle $ and $ \left\langle \mathcal{G}^{s,\mathrm{tail}}\right\rangle$  are instead given by \eqref{tensor_flux_orbit_averaged_tail_final},  whereas the \emph{scalar tail} fluxes $\left\langle \mathcal{F}^{s,\mathrm{tail}}\right\rangle $ and $ \left\langle \mathcal{G}^{s,\mathrm{tail}}\right\rangle$ are given by \eqref{scalar_flux_orbit_averaged_tail_final}. Note that the tail fluxes are given in terms of some \emph{enhancement} functions of the eccentricity, defined exactly in \eqref{eq:def_normalized_enhancement_functions_tensor} and \eqref{eq:def_normalized_enhancement_functions_scalar}, and as a small eccentricity expansion in \eqref{eq:expansion_normalized_enhancement_functions}. 
All other nonlocal terms (namely, $\mathcal{G}_i^{s,\bm{\Pi}^s}$ and $\mathcal{G}_\text{mem}$) vanish upon orbit averaging.  Note that \emph{after orbit averaging} and at this order, there is a clean separation between the instantaneous terms, which enter at integer PN order, and the hereditary tail terms, which appear at half-integer PN orders. This is analogous to the observation made in general relativity \cite{ABIQ08}, and one can expect this split to be broken at the level of the tails-of-tails.

\section{Evolution of the orbital elements}
\label{sec:orbital_elements}
Thanks to both the QK parametrization of Section~\ref{sec:PK} and the fluxes obtained in Section~\ref{sec:flux}, it is now possible to obtain from flux-balance arguments the secular evolution of the QK parameters under the effect of radiation reaction at 1.5PN order, i.e. at relative 2.5PN beyond the leading radiation reaction.

Consider an orbital element, which is denoted generically $\xi$. Its orbit-averaged evolution can be expressed as 
\be \label{eq:secular_evolution_orbital_parameters} \left\langle \frac{\dd \xi}{\dd t}\right\rangle = \frac{\p \xi}{\p E} \left\langle \frac{\dd E}{\dd t}\right\rangle  + \frac{\p \xi}{\p J} \left\langle  \frac{\dd J}{\dd t}\right\rangle  \,. \ee
where $J=|\mathbf{J}|$.
Since the expressions of the orbital elements in terms of $E$ and $J$ were computed in Sec.~\ref{sec:PK} and the energy and angular momentum fluxes were computed in Sec.~\ref{sec:flux}, we now have all the ingredients to compute the expressions of the orbital elements, assuming that the following balance equations hold:
\be
 \label{eq:orbit_average_balance_equations}
   \left\langle \frac{\dd E}{\dd t}\right\rangle  = -\left\langle \mathcal{F}\right\rangle-\left\langle \mathcal{F}^s\right\rangle    \quad \mathrm{and}\quad \left\langle  \frac{\dd J}{\dd t}\right\rangle = - \left\langle \mathcal{G}\right\rangle-\left\langle \mathcal{G}^s\right\rangle \,.
\ee

Only the evolution equations for the pair of orbital elements $(x,e_t)$ is presented --- once these are solved (numerically), the other orbital elements can be straightforwardly deduced through the QK parametrization of~\citetalias{T24_QK}. Note that the 2PN-accurate QK parametrization must also be used to express $(E,J)$ in terms of $(x,e_t)$. The evolution equations read

\bse\label{eq:evolution_x_et}\begin{align}
\label{seq:evolution_x} \left\langle \frac{\dd x}{\dd t} \right\rangle  &= \frac{2c^3 \zeta \nu x^4}{3 \tilde{G}\alpha m}\Bigg\{ \frac{4 \Sm^2\left(1+ \tfrac{1}{2}e_t^2\right)}{(1-e_t^2)^{5/2}}
+ \frac{x}{15(1-e_t^2)^{7/2}}\Big( \frak{X}_1 + e_t^2 \frak{X}_2 + e_t^4 \frak{X}_3 \Big) 
+ 8 \pi\left(1 + \tfrac{1}{2} \gam \right) \Sm^2  \varphi^s_1(e_t) \,x^{3/2} \nn\\*
&\qquad\qquad + x^2 \Bigg( \frac{\frak{X}_4 + e_t^2 \frak{X}_5 + e_t^4 \frak{X}_6+  e_t^6 \frak{X}_7 }{(1-e_t^2)^{9/2}} +  \frac{\frak{X}_8 + e_t^2 \frak{X}_9 + e_t^4 \frak{X}_{10} }{(1-e_t^2)^{4}}\Bigg)
 \nn\\*
&\qquad\qquad + 4\pi \left(1 + \tfrac{1}{2} \gam \right)  x^{5/2} \Bigg( \mathcal{X}_{11} \varphi_2(e_t) + \mathcal{X}_{12} \varphi^s_2(e_t) + \mathcal{X}_{13} \alpha^s_1(e_t) + \mathcal{X}_{14} \theta^s_1(e_t)    \nn\\*
&\qquad\qquad\qquad\qquad\qquad\qquad  + \Big(\mathcal{X}_{15} + e_t^2 \mathcal{X}_{16} \Big)  \frac{\varphi^s_1(e_t)}{1-e_t^2}  + \mathcal{X}_{17} \frac{\tilde{\varphi}^s_1(e_t)}{(1-e_t^2)^{3/2}}   + \mathcal{X}_{18} e_t^2 \varphi^s_0(e_t)   \Bigg) + \mathcal{O}(x^3) \Bigg\} \,, \\
\label{seq:evolution_et}  \left\langle \frac{\dd e_t}{\dd t} \right\rangle  &= -\frac{c^3 \zeta \nu x^3 e_t}{\tilde{G}\alpha m}\Bigg\{ \frac{2 \Sm^2}{(1-e_t^2)^{3/2}} 
+ \frac{x}{15(1-e_t^2)^{5/2}}\Big( \frak{E}_1 + e_t^2 \frak{E}_2\Big)
 \nn\\*
&\qquad\qquad+ \frac{8 \pi}{3}\left(1 + \tfrac{1}{2} \gam \right) \Sm^2  \frac{1-e_t^2}{e_t^2} \,\Big(\varphi^s_1(e_t) - \frac{\tilde{\varphi}^s_1}{\sqrt{1-e_t^2}}\Big)x^{3/2}+ x^2 \Bigg( \frac{\frak{E}_3 + e_t^2 \frak{E}_4 + e_t^4 \frak{E}_5 }{(1-e_t^2)^{7/2}} +  \frac{\frak{E}_6 + e_t^2 \frak{E}_7  }{(1-e_t^2)^{3}}\Bigg)
 \nn\\*
&\qquad\qquad + 4\pi \left(1 + \tfrac{1}{2} \gam \right)  x^{5/2} \Bigg[ 
   \frak{E}_{8} \frac{1-e_t^2}{e_t^2}\Big(\varphi_2(e_t) - \frac{\tilde{\varphi}_2}{\sqrt{1-e_t^2}}\Big) 
+ \frak{E}_{9} \frac{1-e_t^2}{e_t^2}\Big(\varphi^s_2(e_t) - \frac{\tilde{\varphi}^s_2}{\sqrt{1-e_t^2}}\Big) \nn\\*
&\qquad\qquad\qquad\qquad\qquad\qquad + \frak{E}_{10} \frac{1-e_t^2}{e_t^2}\Big(\alpha^s_1(e_t) - \frac{\tilde{\alpha}^s_2}{\sqrt{1-e_t^2}}\Big) 
+\frak{E}_{11} \frac{1-e_t^2}{e_t^2}\Big(\theta^s_1(e_t) - \frac{\tilde{\theta}^s_1}{\sqrt{1-e_t^2}}\Big) \nn\\*
&\qquad\qquad\qquad\qquad\qquad\qquad  +  \frac{ \frak{E}_{12}}{e_t^2}\Big(\varphi^s_1(e_t) - \frac{\tilde{\varphi}^s_1}{\sqrt{1-e_t^2}}\Big) + \frak{E}_{13} \varphi_1^s(e_t) + \frac{\frak{E}_{14}}{\sqrt{1-e_t^2}} \tilde{\varphi}_1^s(e_t) \nn \\*
& \qquad\qquad\qquad\qquad\qquad\qquad+ \frak{E}_{15} (1-e_t^2) \varphi_0^s(e_t) \Bigg] + \mathcal{O}(x^3)\Bigg\} \,,
\end{align}\ese
where the expressions of the coefficients $\frak{X}_n$ and  $\frak{E}_n$ are relegated to App.~\ref{app:orbitalEvolution}.

In Eq.~\eqref{seq:evolution_et}, at first sight, there seems to be some divergent terms when $e_t \rightarrow 0$ for the half-integer PN orders. However, this is fortunately not the case, because the enhancement functions, denoted here generically by $\varphi(e_t)$ and $\tilde{\varphi}(e_t)$, appear as the combination\footnote{The global prefactor $e_t$ has been included.} \mbox{$e_t^{-1}[\varphi(e_t) - (1-e_t^2)^{-1/2} \tilde{\varphi}(e_t) ]$}, which behaves as~$\mathcal{O}(e_t)$ in the $e_t \rightarrow 0$ limit. This means not only that the limit for quasicircular orbits is well-defined, but also that eccentricity evolution is quenched when going to the quasicircular orbit limit, which means that quasicircular orbits are indeed stable. 

\section{Conclusion}
\label{sec:conclusion}

In this work on ST theories, I obtained the orbit-averaged fluxes of energy and angular momentum for quasielliptic orbits at 1.5PN beyond quadrupolar radiation, i.e. relative 2.5PN beyond the leading $-1$PN dipolar radiation. This extends the results of \citetalias{T24_QK} where these results were obtained at Newtonian order, as well as those of \citetalias{BBT22}, where these results were obtained in the limiting case of circular orbits at 1.5PN order. From these results, I was able to derive the 1.5PN secular evolution of the orbital elements  $x$ and $e_t$ (the evolution of the other orbital elements are then straightforward to obtain thanks to the results of  \citetalias{T24_QK} ). The secular evolution of the orbital parameters under radiation reaction  is the eccentric-orbit analog of the frequency chirp in the case of circular orbits, as is hence the main observable in the GW signal. These results are part of a general program aiming at obtaining a full GW template bank for alternative theories of gravity. In the future, these results should be included into EOB models~\cite{Julie:2017pkb, Julie:2017ucp, Julie:2022qux, Jain:2022nxs, Jain:2023fvt} and hybridized with NR models~\cite{Ma:2023sok,Corman:2024vlk} in order to obtain full inspiral-merger-ringdown waveform templates. Although these theories are already heavily constrained by binary pulsar experiments, these waveform models could help constrain these theories even further, in particular with third-generation gravitational wave detectors. 

Finally, I would like to stress that studying such alternative theories of gravity has a theoretical interest in its own right, and helps with our understanding of GR. This work was the occasion to revisit the classical GR derivation of the memory contribution to the angular momentum flux \cite{ABIS09}. It was found that the spurious divergence in the initial eccentricity is cured by a more careful split between DC and AC contributions. The simplicity of the Peters formula in ST tensor theory~\eqref{eq:PetersMathews_ST}, with respect to its GR counterpart \eqref{eq:PetersMathews_GR}, means that using it to study hereditary effects provides a useful toy model to better understand GR phenomenology --- compare the simple ST expression \eqref{eq:G_DC_mem_final} with the GR expression \eqref{eq:G_DC_mem_final_GR}, which is expressed in terms of hypergeometric functions. This is particularly true for the study of the hereditary terms arising from the passage to the center of mass, see Sec.~\ref{sec:CM}. These terms were recently discovered in GR~\cite{BFT24} and evaluated for a quasicircular inspiral model. In this work, they appeared in the ST context, and were studied for the first time using an quasielliptic inspiral model. In the term associated to the translation between the matter content and the center of mass frame, I found that taking the vanishing eccentricity limit  did not smoothly recover the quasicircular model, see Sec.~\ref{subsec:nonlocal_term_elliptic}. This prompted me to revisit this problem in the GR scenario, for which the expressions are much more complicated and require evaluating integrals involving hypergeometric functions, and I found a completely analogous phenomenology \cite{TrestiniUnpublished24}, which will prompt further investigations. This work was also the occasion to introduce some technology which could prove useful in GR calculations in the future, such as the special function $\Theta^{p}_{q}(e_t)$ involved in the 1PN Fourier coefficients of the scalar dipole moment.

Future developments could include: computing the `post-adiabatic' short period contributions to the orbital phasing, in the lines of \cite{Damour:2004bz};  completing the full waveform amplitude modes consistently at 1.5PN order; extending these results to the quasihyperbolic case; etc.  Another avenue is to extend these templates to more complicated theories, such as Einstein-scalar-Gauss-Bonnet.


\subsection*{Note}
While this paper was in preparation, I became aware that a similar computation was being performed by Jain and Rettegno \cite{JR24}. Before the publication of~\cite{JR24}, we compared our results for the generic expression of the \emph{instantaneous} part of the angular momentum flux at (i) 1PN in the scalar sector [Eqs.  (5.10-15) of~\cite{JR24} and the 1PN truncation of Eq.~\eqref{seq:inst_Gsi_orbital_vars} of this paper]; and (ii) 1.5PN in the tensor sector  [\mbox{Eqs. (4.7-8) of~\cite{JR24}} and Eq.~\eqref{seq:inst_Gi_orbital_vars} of this paper], and found perfect agreement.
\acknowledgments
The author thanks Tamanna Jain and Piero Rettegno for cross-checking parts of the general expressions of the instantaneous fluxes, Eve Dones for her help in the derivation of \eqref{eq:FsP_FsG_radiative_moments}, and Luc Blanchet for discussions about the memory contribution and the nonlocal terms appearing the passage to the center-of-mass frame. The author also thanks the Institut d'Astrophysique de Paris for its hospitality. Most computations were performed using the xAct library~\cite{xtensor} for \emph{Mathematica}~\cite{Mathematica}. The author acknowledges support from the Czech Academy of Sciences under the Grant No. LQ100102101 and from the ERC Consolidator/UKRI Frontier
Research Grant GWModels (selected by the ERC and funded by UKRI [grant number
EP/Y008251/1]).

\appendix
\section{Revisiting the memory contribution to the angular momentum flux in general relativity}\label{app:memory_in_GR}

The present work in scalar-tensor theory has been the occasion to revisit the problem of the angular momentum flux in general relativity, and correct some inconsistencies in the literature. Thus, in this section of the Appendix, I shall assume that the underlying theory of gravity is general relativity. The first study of the memory contribution to the angular momentum flux was done by Arun~\emph{et~al.}~\cite{ABIS09}. In that work, it is claimed that the AC contribution vanishes upon orbit averaging, but that the DC contribution is nonvanishing, even after orbit averaging. Strikingly, it is found in their Eq.~(A12) that the orbit-averaged DC contribution depends strongly on some initial eccentricity $e_1$ at which the binary was formed, and \emph{diverges} as~$\sim (1-e_1^2)^{-1}$ when $e_1 \rightarrow 1$ (see their Eq.~(A11). Despite the argument that a compact binary cannot have existed eternally, their conclusions seem unphysical from the viewpoint of the formal two-body problem. Later, it was found by Loutrel and Yunes~\cite{LY16} that the orbit-averaged DC contribution should instead vanish upon orbit-averaging (see the end of their Sec. 6.2), using the argument that the DC contribution is proportional to $\dI_{ij}^{(3)}$, which trivially vanishes after orbit averaging. However, they do not provide a satisfying explanation for the discrepancy (they invoke the choice of the initial phase, but this choice is arbitrary and should not affect the physical result). Moreover, when studying the DC contribution \emph{before} orbit averaging, Loutrel and Yunes use a discrete model, valid for large eccentricities: the orbital parameters $(a,e)$ are modeled as constant over an orbit, and updated discretely at every periastron passage, where most of the gravitational waves are emitted [see their Eq. (208)]. However, they do not address the convergence of their model when summing over the infinite past.

In the following, I will prove the following statements: (i) the AC and DC memory contributions vanish upon orbit averaging, and (ii) the DC contribution is nonvanishing but finite before orbit averaging. Moreover, the finite DC contribution vanishes in the limit of a circularized orbit ($e \rightarrow 0$), regardless of whether one uses a quasi-Keplerian or quasi-circular model for the past inspiral. The reason for the discrepancy is that Arun~\emph{et~al.}~\cite{ABIS09} decomposed their integrand as follows [see their Eq. (A4)]:
\be\mathcal{G}^\mathrm{mem} = \sum_n \int_{t_1}^{t_0} \dd t f_n(a,e) e^{\di n \phi} \,,\ee
and identified the DC contribution with the $n=0$ component [see their Eq. (A5)]. However, this implicitly assumes that the orbit-averaging procedure is $\frac{1}{2\pi}\int_0^{2\pi} \dd \phi $, when instead it should be\footnote{I have used the expression of $\dot{\phi}$ given by  (A3c) of~\cite{ABIS09}, and used the fact that the orbital (radial) period reads $P = \frac{2\pi a^{3/2}}{\sqrt{G m}}$ }
\be\label{eq:orbit_average_corrected}\frac{1}{P} \int_0^{P} \dd t = \frac{1}{P} \int_0^{2\pi} \frac{\dd \phi}{\dot{\phi}} = \frac{(1-e^2)^{3/2}}{2\pi} \int_{0}^{2\pi} \frac{\dd \phi }{(1+e\cos\phi)^2}\,.\ee
In the following, I revisit the computation of Arun~\emph{et~al.}~\cite{ABIS09} by carefully implementing of the orbit averaging procedure following \eqref{eq:orbit_average_corrected}.
At 3.5PN order, recall that the memory contribution to the angular momentum flux reads \cite{ABIS09}
\begin{align}
\mathcal{G}_i^\mathrm{mem} = \frac{4 G^2}{35 c^{10}} \epsilon_{ijk} \dI_{ja}^{(3)} \int_0^{\infty} \dd \tau\, \left[ \dI_{kb}^{(3)} \dI_{ab}^{(3)}\right](t_\text{ret}-\tau) \,.
\end{align}
For a nonprecessing orbit, the direction of the angular momentum flux is constant and points in the $z$-direction. It is then straightforward to explicitly express its norm \mbox{$\mathcal{G} = |\vec{\mathcal{G}}|= \mathcal{G}_z $} in terms of the $(x,y,z)$ components:
\begin{align}\label{eq:G_mem_in_xyz_components_GR}
\mathcal{G}^\mathrm{mem} &= \frac{4 G^2}{35 c^{10}}\Bigg\{\left(\dI_{xx}^{(3)} -\dI_{yy}^{(3)}\right) \int_0^{\infty} \dd\tau\,  f(t_\text{ret}-\tau) - \dI_{xy}^{(3)}   \int_0^{\infty} \dd\tau\, g(t_\text{ret}-\tau) \Bigg\} \,,
\end{align}
where I used the fact that $\dI_{xz} = \dI_{yz}=0$ and where I define 
\begin{subequations}\begin{align}
f &=  \dI_{xy}^{(3)}\left( \dI_{xx}^{(3)} +  \dI_{yy}^{(3)}  \right) \,,\\
g &= \left( \dI_{xx}^{(3)}\right)^2 - \left(\dI_{yy}^{(3)}\right)^2 \,.
\end{align}\end{subequations}
In order to distinguish the AC and DC contributions, I decompose these functions into \mbox{$f =  \left\langle f \right\rangle\!+ \tilde{f}$} and \mbox{$g =  \left\langle g \right\rangle\! + \tilde{g}$}, where 
I have defined the orbit average as $\left\langle h \right\rangle =\frac{1}{P}\int_0^{P}\dd  t \,h(t) =  \frac{1}{2\pi} \int_0^{2\pi}\dd \ell \,h(\ell)$. Here, I have introduced the mean anomaly \mbox{$\ell = n(t-t_0) = \frac{2\pi}{P}(t-t_0)$}, which will prove useful for the analysis. Consequently, the functions $\tilde{f}$ and $\tilde{g}$ contain purely oscillatory terms which average out to zero over a single orbit. Thus, I can now define precisely the AC and DC contributions: 
\begin{subequations}\begin{align}
\mathcal{G}_\mathrm{DC}^\mathrm{mem} &= \frac{4 G^2}{35 c^{10}}\Bigg\{\left(\dI_{xx}^{(3)} -\dI_{yy}^{(3)}\right) \int_0^{\infty} \dd\tau\,  \left\langle f \right\rangle\!(t_\text{ret}-\tau)  - \dI_{xy}^{(3)}(u)   \int_0^{\infty} \dd\tau\,  \left\langle g \right\rangle\!(t_\text{ret}-\tau) \Bigg\} \,,\\
\mathcal{G}_\mathrm{AC}^\mathrm{mem} &= \frac{4 G^2}{35 c^{10}}\Bigg\{\left(\dI_{xx}^{(3)} -\dI_{yy}^{(3)}\right) \int_0^{\infty} \dd\tau\,  \tilde{f}(t_\text{ret}-\tau) - \dI_{xy}^{(3)}(u)   \int_0^{\infty} \dd\tau\, \tilde{g}(t_\text{ret}-\tau) \Bigg\} \,.
\end{align} \end{subequations}

\subsection{The AC part}

First, let me focus on the AC terms. I replace the components of the quadrupole moments by their Fourier series, which I denote as $\dI_{ij} = \sum_{p\in\mathbb{Z}} \de^{\di p \ell} \ {}_p \widetilde{\dI}_{ij}$. I find that 
\begin{subequations}\begin{align}
f &= - n^6  \sum_{p \in \mathbb{Z}} \sum_{q \in \mathbb{Z}} \de^{\di(p+q)\ell} \,p^3 q^3 \ {}_p\widetilde{\dI}_{xy}\left( {}_q\widetilde{\dI}_{xx} + {}_q\widetilde{\dI}_{yy} \right) \\
g &= - n^6   \sum_{p \in \mathbb{Z}} \sum_{q \in \mathbb{Z}} \de^{\di(p+q)\ell} \, p^3 q^3 \left( {}_p\widetilde{\dI}_{xx}\  {}_q\widetilde{\dI}_{xx}-  {}_p\widetilde{\dI}_{yy} \ {}_q\widetilde{\dI}_{yy} \right)
\end{align}\end{subequations}

By definition of the orbital average, the DC component is exactly defined by the terms for which $p+q=0$, so I can now simply reexpress the AC term as 
\begin{subequations}\begin{align}
\tilde{f} &= - n^6  \sum_{p + q \neq 0 }  \de^{\di(p+q)\ell} \,p^3 q^3 \ {}_p\widetilde{\dI}_{xy}\left( {}_q\widetilde{\dI}_{xx} + {}_q\widetilde{\dI}_{yy} \right) \\
\tilde{g} &= - n^6 \sum_{p + q \neq 0 } \de^{\di(p+q)\ell} \, p^3 q^3 \left( {}_p\widetilde{\dI}_{xx}\  {}_q\widetilde{\dI}_{xx}-  {}_p\widetilde{\dI}_{yy} \ {}_q\widetilde{\dI}_{yy} \right)
\end{align}\end{subequations}

Using \mbox{$\int_0^{\infty} \dd\tau \, \de^{\di k n (t-t_0-\tau)} =  \de^{\di k n (t-t_0)}/(i k n)$} for \mbox{$k\neq 0$}, replacing all moments by their Fourier expansions, and substituting at leading order $n = c^3 x^{3/2}/(Gm)$ I readily obtain 
\begin{align}\label{eq:GR_Gmem_AC}
\mathcal{G}_\mathrm{AC}^\mathrm{mem} = \frac{4m \nu^3 c^2 x^6}{35}\sum_{r\in\mathbb{Z}} \sum_{p + q \neq 0}e^{\di(p+q+r)\ell}\,\frac{p^3q^3 r^3}{p+q}\, \mathcal{S}_{r,p,q} \,,
\end{align}
where
\begin{align}
\mathcal{S}_{r,p,q} &= \left({}_r\widehat{\dI}_{xx} - {}_r\widehat{\dI}_{yy} \right) {}_p\widehat{\dI}_{xy} \left({}_q\widehat{\dI}_{xx} + {}_q\widehat{\dI}_{yy} \right) - {}_r\widehat{\dI}_{xy} \left({}_p\widehat{\dI}_{xx}\ {}_q\widehat{\dI}_{xx}  - {}_p\widehat{\dI}_{yy}\ {}_q\widehat{\dI}_{yy} \right)  \,.
\end{align}

Here, I expressed $\mathcal{S}_{r,p,q}$ in terms of the normalized Fourier coefficients ${}_p\widehat{\dI}_{ij} \equiv (m\nu a^2)^{-1} {}_p\widetilde{\dI}_{ij}$, which are explicitly given by (A3) of \cite{ABIQ08tail}. Thanks to the latter expressions, it is immediate to see that $\mathcal{S}_{r,p,q} \in \di \mathbb{R}$. Since $\mathcal{G}_\mathrm{AC}^\mathrm{mem} \in \mathbb{R}$ by construction, taking the real part of \eqref{eq:GR_Gmem_AC} yields the explicitly real-valued expression: 
\begin{align}\label{eq:GR_Gmem_AC_real}
\mathcal{G}_\mathrm{AC}^\mathrm{mem} = \frac{4m \nu^3 c^2 x^6}{35}\sum_{r\in\mathbb{Z}} \sum_{p + q \neq 0}\sin\Big((p+q+r)\ell\Big) \,\frac{p^3q^3 r^3}{p+q}\, \di\mathcal{S}_{r,p,q} \,,
\end{align}
In order to compute the orbit-average of $\mathcal{G}_\mathrm{AC}^\mathrm{mem}$, it is in fact more practical to start directly from the expression \eqref{eq:GR_Gmem_AC}, and to find
\begin{align}
\left\langle \mathcal{G}_\mathrm{AC}^\mathrm{mem} \right\rangle  = -\frac{4m \nu^3 c^2 x^6}{35} \sum_{p + q \neq 0} p^3q^3 (p+q)^2\, \mathcal{S}_{-p-q,p,q} \,.
\end{align}
Noting that on the one hand, $\left\langle \mathcal{G}_\mathrm{AC}^\mathrm{mem} \right\rangle$ is necessarily a real number, but on the other hand, it also purely imaginary (because $\mathcal{S}_{-p-q,p,q} \in \di \mathbb{R}$), I finally conclude that 
\begin{align}
\left\langle \mathcal{G}_\mathrm{AC}^\mathrm{mem} \right\rangle = 0 \,.
\end{align} 

\subsection{The DC part}

\begin{figure}
\includegraphics[width=0.45\textwidth]{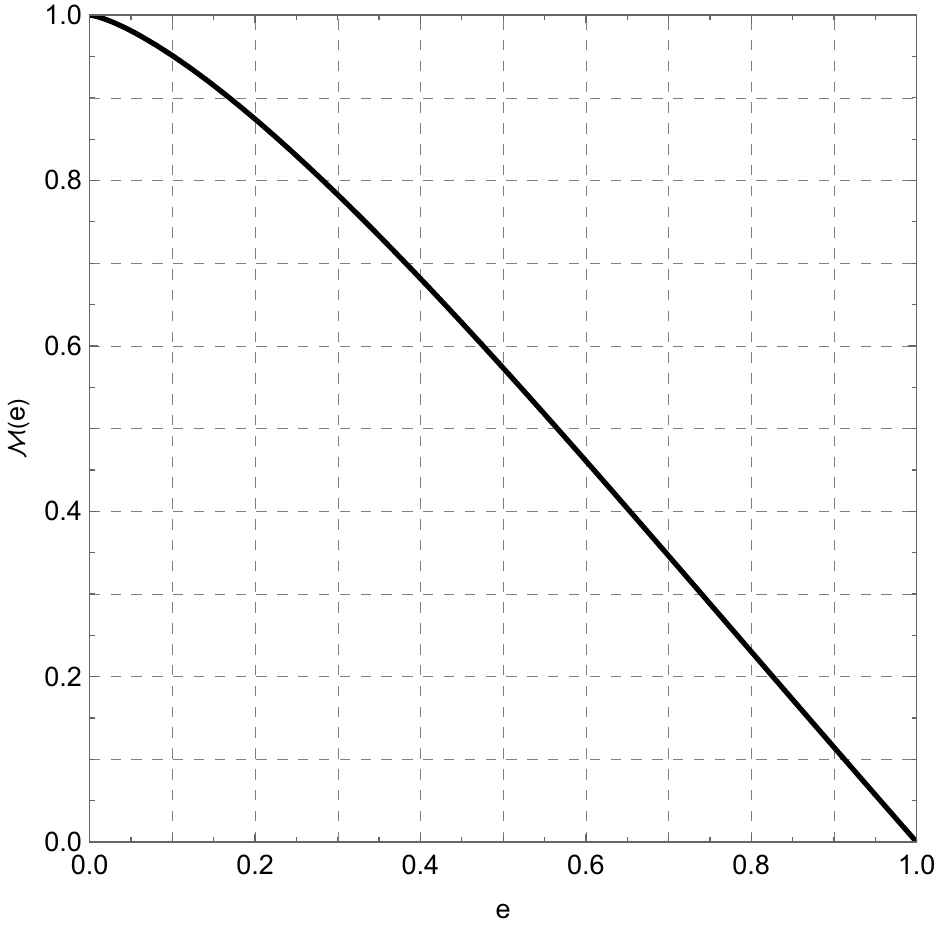}\caption{Plot of $\mathcal{M}(e)$}\label{fig:M(e)}
\end{figure}

I now turn to the DC part.  Thanks to the explicit expression of the quadrupole moment and upon careful computation of the orbital average, I obtain $\langle f \rangle= 0$ and 
\begin{align}
\langle g \rangle &= -\frac{G^3 m^5 \nu^2 e^2(13+2 e^2)}{3 a^5(1-e^2)^{7/2}}  \,.
\end{align}
Thanks to the standard relations derived by Peters \cite{Peters64}
\begin{subequations}\label{eq:PetersMathews_GR}
\begin{align}
\left\langle \frac{\dd e}{\dd t} \right\rangle &= - \frac{304 G^3 m^3 \nu e}{15 c^5 a^4 (1-e^2)^{5/2}} \left(1+ \frac{121}{304} e^2\right) \,, \label{eq:PetersMathews_GR_de_dt} \\
a(e) &= \frac{c_0 e^{12/19}}{1-e^2}\left(1 + \frac{121}{304} e^2\right)^\frac{870}{2299} \,,\label{eq:PetersMathews_GR_a_of_e}
\end{align}
\end{subequations}
I can reexpress the nonvanishing memory integral as an integral over eccentricity. In the far past, the eccentricity asymptotes towards one, so the memory integral reads

\begin{align}\label{eq:integral_of_averaged_g_GR}
\int_0^{\infty} \!\!\dd\tau \ \langle g \rangle(u-\tau)&= - \frac{5 c^5 m^2 \nu}{304 c_0}\int_{e}^1 \dd \tilde{e} \,  \frac{\tilde{e} ^{7/19}(13+2\tilde{e}^2)}{\left(1+ \frac{121}{304}\tilde{e}^2\right)^\frac{3169}{2299}} =  - \frac{5 c^5 m^2 \nu}{304 c_0} \mathcal{K} \,\mathcal{M}(e) \,,
\end{align}
where I have introduced the constant
\begin{align}
\mathcal{K} &= \int_{0}^1 \dd \tilde{e} \,  \frac{\tilde{e} ^{7/19}(13+2\tilde{e}^2)}{\left(1+ \frac{121}{304}\tilde{e}^2\right)^\frac{3169}{2299}} = \frac{19}{2}\, F\!\left(\frac{13}{19}, \frac{3169}{2299}; \frac{32}{19}; - \frac{121}{304}\right)+ \frac{19}{32}\,  F\!\left(\frac{3169}{2299}, \frac{32}{19} ; \frac{51}{19} ; - \frac{121}{304}\right) \,,
\end{align}
as well as the following function of eccentricity:
\begin{align}
\mathcal{M}(e) &= 1-\frac{1}{\mathcal{K}} \int_{e}^1 \dd \tilde{e} \,  \frac{\tilde{e} ^{7/19}(13+2\tilde{e}^2)}{\left(1+ \frac{121}{304}\tilde{e}^2\right)^{3169/2299}}\nn\\*
&=1- \frac{e^{26/19}}{\mathcal{K}}\Bigg\{ \frac{19}{2}\, F\!\left(\frac{13}{19}, \frac{3169}{2299}; \frac{32}{19}; - \frac{121}{304}e^2\right)  + \frac{19}{32} e^2 \,  F\!\left(\frac{3169}{2299}, \frac{32}{19} ; \frac{51}{19} ; - \frac{121}{304} e^2\right)\Bigg\} \,.
\end{align}
In these expressions, $F(a,b;c;d)$ is the ordinary (or Gaussian) hypergeometric function, and I have chosen to normalize $\mathcal{M}(e)$ such that $\mathcal{M}(0) = 1$ and $\mathcal{M}(1) = 0$ (see Fig.~\ref{fig:M(e)}). The constant $\mathcal{K}$ takes the numerical value $\mathcal{K}  \approx 8.29969$, and it is possible to approximate $\mathcal{M}(e)$ with the following expansion (which leads to absolute errors smaller than~$0.0015$ over the whole range of $e\in[0,1]$)
\begin{equation}
\mathcal{M}(e)\approx 1 -   e^{26/19} \Big( 1.145 - 0.184\, e^{2} + 0.0511\, e^{4} - 0.0163\, e^{6} + 0.00548\, e^{8}\Big) \,.
\end{equation}
Plugging \eqref{eq:integral_of_averaged_g_GR} into \eqref{eq:G_mem_in_xyz_components_GR} leads to the final expression
\begin{align}\label{eq:G_DC_mem_final_GR}
\mathcal{G}_\mathrm{DC}^\mathrm{mem} &=\!  \frac{G^2  m^2 \nu \mathcal{K} }{532 c^{5}} \frac{e^{12/29}}{a(1-e^2)}\!\left(\!1+ \frac{121}{304} e^2\!\right) ^\frac{870}{2299}\!\!\!\!\mathcal{M}(e) \, \dI_{xy}^{(3)} \,,
\end{align} 
where $c_0$ has been replaced by its expression \eqref{eq:PetersMathews_GR_a_of_e} in terms of $a$ and $e$. Thus, one can now see that $\mathcal{G}_\mathrm{DC}^\mathrm{mem} \rightarrow 0 $ in the circularized limit $e \rightarrow 0$. Remarkably, this conclusion is perfectly consistent with the result one would have  obtained if one had instead computed the DC memory integral using an eternally quasicircular model, in which case the angular momentum flux is proportional to the energy flux and not does feature any memory-like nonlocal term.

Finally, since the orbital parameters are (approximately) constant over one orbit, and since $\langle \dI_{ij}^{(3)} \rangle = 0$, I find that 
\be
\left\langle \mathcal{G}_\mathrm{DC}^\mathrm{mem} \right\rangle = 0 \,.
\ee

Since the memory contribution vanishes under orbit-averaging, it does not lead to any \emph{secular} contributions to the evolution of the orbital parameters. However, \emph{before orbit averaging}, the memory contribution is nonvanishing and finite, so it can lead to some \emph{post-adiabatic, oscillatory} contributions to the evolution of the orbital parameters, along the lines of \cite{Damour:2004bz}. I will not attempt to compute these contributions here, but let me highlight that the results obtained above prove that these contributions should be finite even for an eternally isolated binary system.  
Finally, let me mention that a complete treatment of this memory integral in the case of \emph{unbound} quasihyperbolic orbits can be found in Ref.~\cite{Bini:2021qvf}.

\section{Details concerning the passage to the center of mass}
\label{app:passageCM}

\bse\begin{align}
p_1 &= \frac{ (\tilde{G} \alpha m)^3  m \nu}{15 c^5} \Bigg[ 20 \bem \zeta \Sm^2 - 50 \zeta \Sm \Sp - 20 \bep \zeta \Sm \Sp - 20 \zeta \gam \Sm \Sp \nn\\
& \qquad\qquad\qquad+ \delta \Big(-24 - 10 \gam - 50 \zeta \Sm^2 - 20 \bep \zeta \Sm^2 - 20 \zeta \gam \Sm^2 + 20 \bem \zeta \Sm \Sp \Big)    \nn\\
& \qquad\qquad\qquad +  \nu \Big(-80 \bem \zeta \Sm^2 + 160 \bem \zeta \Sm^2 \gamInv - 26 \zeta \Sm \Sp + 160 \bep \zeta \Sm \Sp\gamInv \Big) +   6 \zeta \Sm^2  \nu \delta \,\Bigg] \\
p_2 &=  \frac{(\tilde{G} \alpha m)^2  m \nu}{15 c^5}  \Bigg[ 30 \bem - 15 \zeta \Sm \Sp    +  \delta \Big(-30 \bep + 15 \gam - 15 \zeta \Sm^2 \Big)  + 45 \zeta \Sm \Sp \nu + 15 \zeta \Sm^2  \nu \delta  \Bigg] \\
p_3 &= \frac{(\tilde{G} \alpha m)^2  m \nu}{15 c^5}  \Bigg[-10 \bem + 20 \zeta \Sm \Sp + 10 \zeta \gam \Sm \Sp + \delta \Big(12 + 10 \bep + 20 \zeta \Sm^2 + 10 \zeta \gam \Sm^2\Big)  \Bigg] \\
p_4 &=  \frac{ (\tilde{G} \alpha m)^2  m \nu }{15 c^5} \Bigg[-20 \bem - 30 \zeta \Sm \Sp - 20 \zeta \gam \Sm \Sp    + \delta \Big(20 \bep - 10 \gam - 30 \zeta \Sm^2 - 20 \zeta \gam \Sm^2\Big)   \nn\\
&\qquad\qquad\qquad\  -10 \zeta \Sm \Sp \nu  - 10 \zeta \Sm^2   \nu  \delta \Bigg]\\
q_1 &= \frac{ (\tilde{G} \alpha m)^2  m \nu }{30 c^5} \Bigg[-20 \bem + 5 \zeta \Sm \Sp   + (20 \bep - 10 \gam + 5 \zeta \Sm^2) \delta  -8 \zeta \Sm \Sp  \nu - 12 \zeta \Sm^2  \nu \delta\Bigg] \\
q_2 &= \frac{  (\tilde{G} \alpha m)^2  m \nu }{30 c^5}  \Bigg[ 20 \bem + 5 \zeta \Sm \Sp    + \delta \Big(48 - 20 \bep + 30 \gam + 5 \zeta \Sm^2\Big)    - 16 \zeta \Sm^2  \nu \delta \nn\\
&\qquad\qquad\quad + \nu \Big(- 320 \bem \zeta \Sm^2\gamInv + 76 \zeta \Sm \Sp -  320 \bep \zeta \Sm \Sp \gamInv \Big)    \Bigg]\\
q_3 &= \frac{  (\tilde{G} \alpha m) m \nu }{30 c^5} \Bigg[-5 \zeta \Sm \Sp    + \delta \Big(-24 - 10 \gam - 5 \zeta \Sm^2\Big)     + 16 \zeta \Sm \Sp \nu + 4 \zeta \Sm^2  \nu \delta\Bigg]
\end{align}\ese

\section{Coefficients for the small eccentricity expansion of $\alpha^s_1(e_t)$ and $\tilde{\alpha}^s_1(e_t)$}

\label{app:smallEccentricityExpansion}

The coefficients associated to $\alpha_1^s(e_t)$ read
\bse\begin{align}
\mathcal{C}_2 &= \frac{5611}{324} + \frac{350}{729} \bem^2 -  \frac{685}{486} \bep + \frac{350}{729} \bep^2 + \frac{2900 \bem^2}{81 \gam^2} + \frac{2900 \bep^2}{81 \gam^2} -  \frac{200 \bem^2}{27 \gam} -  \frac{7615 \bep}{162 \gam} -  \frac{200 \bep^2}{27 \gam} + \frac{2285}{324} \gam + \frac{1375}{729} \bep \gam \nn\\
& + \frac{200}{243} \gam^2 + \frac{1235 \bep}{162 \zeta \Sm^2} -  \frac{725 \bem^2}{81 \zeta \gam \Sm^2} -  \frac{725 \bep^2}{81 \zeta \gam \Sm^2} -  \frac{131 \gam}{81 \zeta \Sm^2} -  \frac{1130 \bem \Sp}{243 \Sm} + \frac{5800 \bem \bep \Sp}{81 \gam^2 \Sm} -  \frac{7615 \bem \Sp}{162 \gam \Sm} -  \frac{400 \bem \bep \Sp}{27 \gam \Sm} \nn\\
& +\delta \bigg(\frac{685}{486} \bem -  \frac{700}{729} \bem \bep -  \frac{5800 \bem \bep}{81 \gam^2} + \frac{7615 \bem}{162 \gam} + \frac{400 \bem \bep}{27 \gam} -  \frac{1375}{729} \bem \gam -  \frac{1235 \bem}{162 \zeta \Sm^2} + \frac{1450 \bem \bep}{81 \zeta \gam \Sm^2} -  \frac{731 \Sp}{54 \Sm} \nn\\
&\qquad + \frac{1130 \bep \Sp}{243 \Sm} -  \frac{2900 \bem^2 \Sp}{81 \gam^2 \Sm} -  \frac{2900 \bep^2 \Sp}{81 \gam^2 \Sm} + \frac{200 \bem^2 \Sp}{27 \gam \Sm} + \frac{7615 \bep \Sp}{162 \gam \Sm} + \frac{200 \bep^2 \Sp}{27 \gam \Sm} -  \frac{1615 \gam \Sp}{486 \Sm}\bigg)  \nn\\
& + \nu \bigg(- \frac{1048}{81} -  \frac{1400}{729} \bem^2 -  \frac{5800 \bem^2}{81 \gam^2} -  \frac{5800 \bep^2}{81 \gam^2} + \frac{800 \bem^2}{27 \gam} + \frac{4940 \bep}{81 \gam} -  \frac{2470 \bep}{81 \zeta \Sm^2} + \frac{2900 \bep^2}{81 \zeta \gam \Sm^2} + \frac{524 \gam}{81 \zeta \Sm^2} \nn\\
&\qquad -  \frac{1060 \bem \Sp}{81 \Sm} -  \frac{11600 \bem \bep \Sp}{81 \gam^2 \Sm} + \frac{4940 \bem \Sp}{81 \gam \Sm} + \frac{800 \bem \bep \Sp}{27 \gam \Sm}\bigg)  \,, \\
\mathcal{C}_4 &= \frac{89797}{1296} + \frac{5975}{2916} \bem^2 -  \frac{4925}{972} \bep + \frac{5975}{2916} \bep^2 + \frac{56875 \bem^2}{324 \gam^2} + \frac{56875 \bep^2}{324 \gam^2} -  \frac{68825 \bem^2}{1944 \gam} -  \frac{277255 \bep}{1296 \gam} -  \frac{68825 \bep^2}{1944 \gam} \nn\\
& + \frac{184855}{7776} \gam + \frac{24125}{3888} \bep \gam + \frac{60425}{23328} \gam^2 + \frac{16465 \bep}{432 \zeta \Sm^2} -  \frac{56875 \bem^2}{1296 \zeta \gam \Sm^2} -  \frac{56875 \bep^2}{1296 \zeta \gam \Sm^2} -  \frac{5329 \gam}{648 \zeta \Sm^2} -  \frac{3025 \bem \Sp}{216 \Sm} \nn\\
&+ \frac{56875 \bem \bep \Sp}{162 \gam^2 \Sm} -  \frac{277255 \bem \Sp}{1296 \gam \Sm} -  \frac{68825 \bem \bep \Sp}{972 \gam \Sm} \nn\\
&+ \delta \bigg(\frac{4925}{972} \bem -  \frac{5975}{1458} \bem \bep -  \frac{56875 \bem \bep}{162 \gam^2} + \frac{277255 \bem}{1296 \gam}+ \frac{68825 \bem \bep}{972 \gam} -  \frac{24125}{3888} \bem \gam -  \frac{16465 \bem}{432 \zeta \Sm^2} + \frac{56875 \bem \bep}{648 \zeta \gam \Sm^2} \nn\\
&\qquad -  \frac{78397 \Sp}{1296 \Sm} + \frac{3025 \bep \Sp}{216 \Sm} -  \frac{56875 \bem^2 \Sp}{324 \gam^2 \Sm} -  \frac{56875 \bep^2 \Sp}{324 \gam^2 \Sm}+ \frac{68825 \bem^2 \Sp}{1944 \gam \Sm} + \frac{277255 \bep \Sp}{1296 \gam \Sm} + \frac{68825 \bep^2 \Sp}{1944 \gam \Sm} \nn\\
&\qquad -  \frac{50815 \gam \Sp}{3888 \Sm}\bigg)  \nn\\
& + \nu\bigg(- \frac{5329}{81} -  \frac{5975}{729} \bem^2 -  \frac{56875 \bem^2}{162 \gam^2} -  \frac{56875 \bep^2}{162 \gam^2} + \frac{68825 \bem^2}{486 \gam} + \frac{16465 \bep}{54 \gam} -  \frac{16465 \bep}{108 \zeta \Sm^2} + \frac{56875 \bep^2}{324 \zeta \gam \Sm^2} + \frac{5329 \gam}{162 \zeta \Sm^2} \nn\\
&\qquad -  \frac{10475 \bem \Sp}{162 \Sm} -  \frac{56875 \bem \bep \Sp}{81 \gam^2 \Sm} + \frac{16465 \bem \Sp}{54 \gam \Sm}+ \frac{68825 \bem \bep \Sp}{486 \gam \Sm}\bigg)  \,,\\
\mathcal{C}_6 &= \frac{16954805}{93312} + \frac{757075}{104976} \bem^2 -  \frac{787255}{139968} \bep + \frac{757075}{104976} \bep^2 + \frac{3181825 \bem^2}{5832 \gam^2} + \frac{3181825 \bep^2}{5832 \gam^2} -  \frac{984725 \bem^2}{8748 \gam} -  \frac{29334965 \bep}{46656 \gam}\nn\\
& -  \frac{984725 \bep^2}{8748 \gam} + \frac{1521485}{31104} \gam + \frac{2693225}{209952} \bep \gam + \frac{968075}{209952} \gam^2 + \frac{5578585 \bep}{46656 \zeta \Sm^2} -  \frac{3181825 \bem^2}{23328 \zeta \gam \Sm^2} -  \frac{3181825 \bep^2}{23328 \zeta \gam \Sm^2} -  \frac{606625 \gam}{23328 \zeta \Sm^2} \nn\\
&-  \frac{739595 \bem \Sp}{34992 \Sm} + \frac{3181825 \bem \bep \Sp}{2916 \gam^2 \Sm} -  \frac{29334965 \bem \Sp}{46656 \gam \Sm} -  \frac{984725 \bem \bep \Sp}{4374 \gam \Sm} \nn\\
& + \delta\bigg(\frac{787255}{139968} \bem -  \frac{757075}{52488} \bem \bep -  \frac{3181825 \bem \bep}{2916 \gam^2} + \frac{29334965 \bem}{46656 \gam} + \frac{984725 \bem \bep}{4374 \gam} -  \frac{2693225}{209952} \bem \gam -  \frac{5578585 \bem}{46656 \zeta \Sm^2} \nn\\
&\qquad+ \frac{3181825 \bem \bep}{11664 \zeta \gam \Sm^2} -  \frac{905435 \Sp}{5184 \Sm} + \frac{739595 \bep \Sp}{34992 \Sm} -  \frac{3181825 \bem^2 \Sp}{5832 \gam^2 \Sm} -  \frac{3181825 \bep^2 \Sp}{5832 \gam^2 \Sm} + \frac{984725 \bem^2 \Sp}{8748 \gam \Sm} + \frac{29334965 \bep \Sp}{46656 \gam \Sm} \nn\\
&\qquad+ \frac{984725 \bep^2 \Sp}{8748 \gam \Sm} -  \frac{4582045 \gam \Sp}{139968 \Sm}\bigg)  \nn\\
&+ \nu \bigg(- \frac{606625}{2916} -  \frac{757075}{26244} \bem^2 -  \frac{3181825 \bem^2}{2916 \gam^2} -  \frac{3181825 \bep^2}{2916 \gam^2} + \frac{984725 \bem^2}{2187 \gam} + \frac{5578585 \bep}{5832 \gam} -  \frac{5578585 \bep}{11664 \zeta \Sm^2}\nn\\
& \qquad+ \frac{3181825 \bep^2}{5832 \zeta \gam \Sm^2} + \frac{606625 \gam}{5832 \zeta \Sm^2} -  \frac{3638785 \bem \Sp}{17496 \Sm} -  \frac{3181825 \bem \bep \Sp}{1458 \gam^2 \Sm} + \frac{5578585 \bem \Sp}{5832 \gam \Sm} + \frac{984725 \bem \bep \Sp}{2187 \gam \Sm}\bigg)   \,,\\
\mathcal{C}_8 &=  \frac{2280254495}{5971968} + \frac{133124075}{6718464} \bem^2 + \frac{67506595}{8957952} \bep + \frac{133124075}{6718464} \bep^2 + \frac{984057275 \bem^2}{746496 \gam^2} + \frac{984057275 \bep^2}{746496 \gam^2} -  \frac{138922825 \bem^2}{497664 \gam} \nn\\
&  - \! \frac{2175881095 \bep}{1492992 \gam} - \! \frac{138922825 \bep^2}{497664 \gam} + \frac{24773035}{331776} \gam + \frac{539614825}{26873856} \bep \gam + \frac{30374075}{5971968} \gam^2 + \frac{433693805 \bep}{1492992 \zeta \Sm^2} -  \frac{984057275 \bem^2}{2985984 \zeta \gam \Sm^2} \nn\\
& - \! \frac{984057275 \bep^2}{2985984 \zeta \gam \Sm^2} - \! \frac{23688235 \gam}{373248 \zeta \Sm^2} -\!  \frac{48448015 \bem \Sp}{4478976 \Sm} + \frac{984057275 \bem \bep \Sp}{373248 \gam^2 \Sm} -  \frac{2175881095 \bem \Sp}{1492992 \gam \Sm} -  \frac{138922825 \bem \bep \Sp}{248832 \gam \Sm}  \nn\\
&+ \delta \bigg(\! \!- \frac{67506595}{8957952} \bem -  \! \frac{133124075}{3359232} \bem \bep -  \frac{984057275 \bem \bep}{373248 \gam^2} + \frac{2175881095 \bem}{1492992 \gam} + \frac{138922825 \bem \bep}{248832 \gam} -  \frac{539614825}{26873856} \bem \gam  \nn\\
&\qquad-  \frac{433693805 \bem}{1492992 \zeta \Sm^2} + \frac{984057275 \bem \bep}{1492992 \zeta \gam \Sm^2} -  \frac{66169445 \Sp}{165888 \Sm} + \frac{48448015 \bep \Sp}{4478976 \Sm} -  \frac{984057275 \bem^2 \Sp}{746496 \gam^2 \Sm} -  \frac{984057275 \bep^2 \Sp}{746496 \gam^2 \Sm}  \nn\\
&\qquad+ \frac{138922825 \bem^2 \Sp}{497664 \gam \Sm} + \frac{2175881095 \bep \Sp}{1492992 \gam \Sm} + \frac{138922825 \bep^2 \Sp}{497664 \gam \Sm} -  \frac{291555685 \gam \Sp}{4478976 \Sm}\bigg)  \nn\\
&+ \nu \bigg(- \frac{23688235}{46656} -  \frac{133124075}{1679616} \bem^2 -  \frac{984057275 \bem^2}{373248 \gam^2} -  \frac{984057275 \bep^2}{373248 \gam^2} + \frac{138922825 \bem^2}{124416 \gam} + \frac{433693805 \bep}{186624 \gam}  \nn\\
&\qquad -  \frac{433693805 \bep}{373248 \zeta \Sm^2} + \frac{984057275 \bep^2}{746496 \zeta \gam \Sm^2} + \frac{23688235 \gam}{93312 \zeta \Sm^2} -  \frac{24177665 \bem \Sp}{46656 \Sm} -  \frac{984057275 \bem \bep \Sp}{186624 \gam^2 \Sm} + \frac{433693805 \bem \Sp}{186624 \gam \Sm}  \nn\\
&\qquad+ \frac{138922825 \bem \bep \Sp}{124416 \gam \Sm}\bigg)  \,,\\
\mathcal{C}_{10} &= \frac{417420389873}{597196800} + \frac{1216325399}{26873856} \bem^2 + \frac{9152412233}{179159040} \bep + \frac{1216325399}{26873856} \bep^2  + \frac{4052305187 \bem^2}{1492992 \gam^2} + \frac{4052305187 \bep^2}{1492992 \gam^2}\nn\\
& -  \frac{2634315293 \bem^2}{4478976 \gam} -  \frac{173559800129 \bep}{59719680 \gam} -  \frac{2634315293 \bep^2}{4478976 \gam} + \frac{30980827673}{358318080} \gam + \frac{147842587}{5971968} \bep \gam + \frac{28564471}{26873856} \gam^2 \nn\\
& + \! \frac{11949020167 \bep}{19906560 \zeta \Sm^2} -  \! \frac{4052305187 \bem^2}{5971968 \zeta \gam \Sm^2} - \! \frac{4052305187 \bep^2}{5971968 \zeta \gam \Sm^2} -  \frac{19637839753 \gam}{149299200 \zeta \Sm^2} + \frac{611720999 \bem \Sp}{14929920 \Sm} + \frac{4052305187 \bem \bep \Sp}{746496 \gam^2 \Sm} \nn\\
& -  \frac{173559800129 \bem \Sp}{59719680 \gam \Sm} -  \frac{2634315293 \bem \bep \Sp}{2239488 \gam \Sm} \nn\\
& +\delta \bigg(- \frac{9152412233}{179159040} \bem -  \frac{1216325399}{13436928} \bem \bep -  \frac{4052305187 \bem \bep}{746496 \gam^2} + \frac{173559800129 \bem}{59719680 \gam} + \frac{2634315293 \bem \bep}{2239488 \gam}\nn\\
&\qquad  -  \frac{147842587}{5971968} \bem \gam -  \frac{11949020167 \bem}{19906560 \zeta \Sm^2} + \frac{4052305187 \bem \bep}{2985984 \zeta \gam \Sm^2} -  \frac{234588929299 \Sp}{298598400 \Sm} -  \frac{611720999 \bep \Sp}{14929920 \Sm} \nn\\
& \qquad -  \frac{4052305187 \bem^2 \Sp}{1492992 \gam^2 \Sm}  -  \frac{4052305187 \bep^2 \Sp}{1492992 \gam^2 \Sm} + \frac{2634315293 \bem^2 \Sp}{4478976 \gam \Sm} + \frac{173559800129 \bep \Sp}{59719680 \gam \Sm} + \frac{2634315293 \bep^2 \Sp}{4478976 \gam \Sm} \nn\\
&\qquad  -  \frac{20094871289 \gam \Sp}{179159040 \Sm}\bigg)  \nn\\
& + \nu \bigg(- \frac{19637839753}{18662400} -  \frac{1216325399}{6718464} \bem^2 -  \frac{4052305187 \bem^2}{746496 \gam^2} -  \frac{4052305187 \bep^2}{746496 \gam^2} + \frac{2634315293 \bem^2}{1119744 \gam} + \frac{11949020167 \bep}{2488320 \gam} \nn\\
&\qquad-  \frac{11949020167 \bep}{4976640 \zeta \Sm^2} + \frac{4052305187 \bep^2}{1492992 \zeta \gam \Sm^2} + \frac{19637839753 \gam}{37324800 \zeta \Sm^2} -  \frac{2722571261 \bem \Sp}{2488320 \Sm} -  \frac{4052305187 \bem \bep \Sp}{373248 \gam^2 \Sm} \nn\\
&\qquad + \frac{11949020167 \bem \Sp}{2488320 \gam \Sm} + \frac{2634315293 \bem \bep \Sp}{1119744 \gam \Sm}\bigg)  \,.\\
\end{align}\ese
The coefficients associated to $\tilde{\alpha}_1^s(e_t)$ read
\bse \begin{align}
\tilde{\mathcal{C}}_2 &=  \frac{2299}{324} + \frac{250}{729} \bem^2 + \frac{5}{162} \bep + \frac{250}{729} \bep^2 + \frac{1300 \bem^2}{81 \gam^2} + \frac{1300 \bep^2}{81 \gam^2} -  \frac{100 \bem^2}{27 \gam} -  \frac{3295 \bep}{162 \gam} -  \frac{100 \bep^2}{27 \gam} + \frac{2455}{972} \gam + \frac{175}{243} \bep \gam \nn\\
& + \frac{200}{729} \gam^2 + \frac{185 \bep}{54 \zeta \Sm^2} -  \frac{325 \bem^2}{81 \zeta \gam \Sm^2} -  \frac{325 \bep^2}{81 \zeta \gam \Sm^2} -  \frac{59 \gam}{81 \zeta \Sm^2} -  \frac{110 \bem \Sp}{81 \Sm} + \frac{2600 \bem \bep \Sp}{81 \gam^2 \Sm} -  \frac{3295 \bem \Sp}{162 \gam \Sm} -  \frac{200 \bem \bep \Sp}{27 \gam \Sm} \nn\\
&+ \delta\bigg(- \frac{5}{162} \bem -  \frac{500}{729} \bem \bep -  \frac{2600 \bem \bep}{81 \gam^2} + \frac{3295 \bem}{162 \gam} + \frac{200 \bem \bep}{27 \gam} -  \frac{175}{243} \bem \gam -  \frac{185 \bem}{54 \zeta \Sm^2} + \frac{650 \bem \bep}{81 \zeta \gam \Sm^2} -  \frac{937 \Sp}{162 \Sm} \nn\\
& \qquad+ \frac{110 \bep \Sp}{81 \Sm} -  \frac{1300 \bem^2 \Sp}{81 \gam^2 \Sm} -  \frac{1300 \bep^2 \Sp}{81 \gam^2 \Sm} + \frac{100 \bem^2 \Sp}{27 \gam \Sm} + \frac{3295 \bep \Sp}{162 \gam \Sm} + \frac{100 \bep^2 \Sp}{27 \gam \Sm} -  \frac{205 \gam \Sp}{162 \Sm}\bigg)  \nn\\
& + \nu \bigg(- \frac{472}{81} -  \frac{1000}{729} \bem^2 -  \frac{2600 \bem^2}{81 \gam^2} -  \frac{2600 \bep^2}{81 \gam^2} + \frac{400 \bem^2}{27 \gam} + \frac{740 \bep}{27 \gam} -  \frac{370 \bep}{27 \zeta \Sm^2} + \frac{1300 \bep^2}{81 \zeta \gam \Sm^2} + \frac{236 \gam}{81 \zeta \Sm^2} -  \frac{1580 \bem \Sp}{243 \Sm}\nn\\
&\qquad -  \frac{5200 \bem \bep \Sp}{81 \gam^2 \Sm} + \frac{740 \bem \Sp}{27 \gam \Sm} + \frac{400 \bem \bep \Sp}{27 \gam \Sm}\bigg)  \,,\\
\tilde{\mathcal{C}}_4 &= \frac{4867}{288} + \frac{10625}{11664} \bem^2 + \frac{415}{864} \bep + \frac{10625}{11664} \bep^2 + \frac{15025 \bem^2}{324 \gam^2} + \frac{15025 \bep^2}{324 \gam^2} -  \frac{40675 \bem^2}{3888 \gam} -  \frac{11875 \bep}{216 \gam} -  \frac{40675 \bep^2}{3888 \gam} + \frac{2335}{486} \gam \nn\\
&  + \frac{32075}{23328} \bep \gam + \frac{3575}{7776} \gam^2 + \frac{13075 \bep}{1296 \zeta \Sm^2} -  \frac{15025 \bem^2}{1296 \zeta \gam \Sm^2} -  \frac{15025 \bep^2}{1296 \zeta \gam \Sm^2} -  \frac{157 \gam}{72 \zeta \Sm^2} -  \frac{3685 \bem \Sp}{1944 \Sm} + \frac{15025 \bem \bep \Sp}{162 \gam^2 \Sm} \nn\\
&-  \frac{11875 \bem \Sp}{216 \gam \Sm} -  \frac{40675 \bem \bep \Sp}{1944 \gam \Sm} \nn\\
& + \delta\bigg(- \frac{415}{864} \bem -  \frac{10625}{5832} \bem \bep -  \frac{15025 \bem \bep}{162 \gam^2} + \frac{11875 \bem}{216 \gam} + \frac{40675 \bem \bep}{1944 \gam} -  \frac{32075}{23328} \bem \gam -  \frac{13075 \bem}{1296 \zeta \Sm^2} + \frac{15025 \bem \bep}{648 \zeta \gam \Sm^2} \nn\\
&\qquad-  \frac{19939 \Sp}{1296 \Sm} + \frac{3685 \bep \Sp}{1944 \Sm} -  \frac{15025 \bem^2 \Sp}{324 \gam^2 \Sm} -  \frac{15025 \bep^2 \Sp}{324 \gam^2 \Sm} + \frac{40675 \bem^2 \Sp}{3888 \gam \Sm} + \frac{11875 \bep \Sp}{216 \gam \Sm} + \frac{40675 \bep^2 \Sp}{3888 \gam \Sm} \nn\\
&\qquad-  \frac{3785 \gam \Sp}{1296 \Sm}\bigg)  \nn\\
&+ \nu\bigg(- \frac{157}{9} -  \frac{10625}{2916} \bem^2 -  \frac{15025 \bem^2}{162 \gam^2} -  \frac{15025 \bep^2}{162 \gam^2} + \frac{40675 \bem^2}{972 \gam} + \frac{13075 \bep}{162 \gam} -  \frac{13075 \bep}{324 \zeta \Sm^2} + \frac{15025 \bep^2}{324 \zeta \gam \Sm^2} + \frac{157 \gam}{18 \zeta \Sm^2} \nn\\
&\qquad -  \frac{4595 \bem \Sp}{243 \Sm} -  \frac{15025 \bem \bep \Sp}{81 \gam^2 \Sm} + \frac{13075 \bem \Sp}{162 \gam \Sm} + \frac{40675 \bem \bep \Sp}{972 \gam \Sm}\bigg)  \,,\\
\tilde{\mathcal{C}}_6 &= \frac{2822579}{93312} + \frac{13450}{6561} \bem^2 + \frac{365885}{139968} \bep + \frac{13450}{6561} \bep^2 + \frac{568675 \bem^2}{5832 \gam^2} + \frac{568675 \bep^2}{5832 \gam^2}  -  \frac{783875 \bem^2}{34992 \gam} -  \frac{5129315 \bep}{46656 \gam}\nn\\
& -  \frac{783875 \bep^2}{34992 \gam} + \frac{191795}{31104} \gam + \frac{196675}{104976} \bep \gam + \frac{89075}{209952} \gam^2 + \frac{998635 \bep}{46656 \zeta \Sm^2} -  \frac{568675 \bem^2}{23328 \zeta \gam \Sm^2} -  \frac{568675 \bep^2}{23328 \zeta \gam \Sm^2} -  \frac{108739 \gam}{23328 \zeta \Sm^2} \nn\\
& -  \frac{15485 \bem \Sp}{34992 \Sm} + \frac{568675 \bem \bep \Sp}{2916 \gam^2 \Sm} -  \frac{5129315 \bem \Sp}{46656 \gam \Sm} -  \frac{783875 \bem \bep \Sp}{17496 \gam \Sm} \nn\\
&+ \delta\bigg(- \frac{365885}{139968} \bem -  \frac{26900}{6561} \bem \bep -  \frac{568675 \bem \bep}{2916 \gam^2} + \frac{5129315 \bem}{46656 \gam} + \frac{783875 \bem \bep}{17496 \gam} -  \frac{196675}{104976} \bem \gam -  \frac{998635 \bem}{46656 \zeta \Sm^2} \nn\\
&\qquad + \frac{568675 \bem \bep}{11664 \zeta \gam \Sm^2} -  \frac{156953 \Sp}{5184 \Sm} + \frac{15485 \bep \Sp}{34992 \Sm} -  \frac{568675 \bem^2 \Sp}{5832 \gam^2 \Sm} -  \frac{568675 \bep^2 \Sp}{5832 \gam^2 \Sm} \nn\\
&\qquad+ \frac{783875 \bem^2 \Sp}{34992 \gam \Sm} + \frac{5129315 \bep \Sp}{46656 \gam \Sm} + \frac{783875 \bep^2 \Sp}{34992 \gam \Sm} -  \frac{686215 \gam \Sp}{139968 \Sm}\bigg)  \nn\\
&+ \nu\bigg(- \frac{108739}{2916} -  \frac{53800}{6561} \bem^2 -  \frac{568675 \bem^2}{2916 \gam^2} -  \frac{568675 \bep^2}{2916 \gam^2} + \frac{783875 \bem^2}{8748 \gam} + \frac{998635 \bep}{5832 \gam} -  \frac{998635 \bep}{11664 \zeta \Sm^2} + \frac{568675 \bep^2}{5832 \zeta \gam \Sm^2} \nn\\
&\qquad + \frac{108739 \gam}{5832 \zeta \Sm^2} -  \frac{715855 \bem \Sp}{17496 \Sm} -  \frac{568675 \bem \bep \Sp}{1458 \gam^2 \Sm} + \frac{998635 \bem \Sp}{5832 \gam \Sm} + \frac{783875 \bem \bep \Sp}{8748 \gam \Sm}\bigg)  \,,\\
\tilde{\mathcal{C}}_8 &=  \frac{279720785}{5971968} + \frac{52915775}{13436928} \bem^2 + \frac{10793545}{1492992} \bep + \frac{52915775}{13436928} \bep^2 + \frac{129014575 \bem^2}{746496 \gam^2} + \frac{129014575 \bep^2}{746496 \gam^2} -  \frac{30321725 \bem^2}{746496 \gam}\nn\\
& -  \frac{140005055 \bep}{746496 \gam} -  \frac{30321725 \bep^2}{746496 \gam} + \frac{25701095}{4478976} \gam + \frac{3024625}{1492992} \bep \gam + \frac{1527475}{53747712} \gam^2 + \frac{2108695 \bep}{55296 \zeta \Sm^2} -  \frac{129014575 \bem^2}{2985984 \zeta \gam \Sm^2} \nn\\
& -  \frac{129014575 \bep^2}{2985984 \zeta \gam \Sm^2} -  \frac{3113185 \gam}{373248 \zeta \Sm^2} + \frac{224185 \bem \Sp}{55296 \Sm} + \frac{129014575 \bem \bep \Sp}{373248 \gam^2 \Sm} -  \frac{140005055 \bem \Sp}{746496 \gam \Sm} -  \frac{30321725 \bem \bep \Sp}{373248 \gam \Sm} \nn\\
&+ \delta\bigg(- \frac{10793545}{1492992} \bem -  \frac{52915775}{6718464} \bem \bep -  \frac{129014575 \bem \bep}{373248 \gam^2} + \frac{140005055 \bem}{746496 \gam} + \frac{30321725 \bem \bep}{373248 \gam} -  \frac{3024625}{1492992} \bem \gam \nn\\
&\qquad  -  \frac{2108695 \bem}{55296 \zeta \Sm^2} + \frac{129014575 \bem \bep}{1492992 \zeta \gam \Sm^2} -  \frac{76048315 \Sp}{1492992 \Sm} -  \frac{224185 \bep \Sp}{55296 \Sm} -  \frac{129014575 \bem^2 \Sp}{746496 \gam^2 \Sm} -  \frac{129014575 \bep^2 \Sp}{746496 \gam^2 \Sm} \nn\\
&\qquad + \frac{30321725 \bem^2 \Sp}{746496 \gam \Sm} + \frac{140005055 \bep \Sp}{746496 \gam \Sm} + \frac{30321725 \bep^2 \Sp}{746496 \gam \Sm} -  \frac{10479115 \gam \Sp}{1492992 \Sm}\bigg)   \nn\\
&+ \nu\bigg(- \frac{3113185}{46656} -  \frac{52915775}{3359232} \bem^2 -  \frac{129014575 \bem^2}{373248 \gam^2} -  \frac{129014575 \bep^2}{373248 \gam^2} + \frac{30321725 \bem^2}{186624 \gam} + \frac{2108695 \bep}{6912 \gam} -  \frac{2108695 \bep}{13824 \zeta \Sm^2} \nn\\
&\qquad +\! \frac{129014575 \bep^2}{746496 \zeta \gam \Sm^2} +\! \frac{3113185 \gam}{93312 \zeta \Sm^2} - \! \frac{41715005 \bem \Sp}{559872 \Sm} - \! \frac{129014575 \bem \bep \Sp}{186624 \gam^2 \Sm} +\! \frac{2108695 \bem \Sp}{6912 \gam \Sm} + \! \frac{30321725 \bem \bep \Sp}{186624 \gam \Sm}\!\bigg)  \,,\\
\tilde{\mathcal{C}}_{10} &=  \frac{13218658111}{199065600} +\! \frac{180939719}{26873856} \bem^2 + \! \frac{897726671}{59719680} \bep + \! \frac{180939719}{26873856} \bep^2 + \! \frac{410983607 \bem^2}{1492992 \gam^2} + \frac{410983607 \bep^2}{1492992 \gam^2} -  \frac{295961663 \bem^2}{4478976 \gam} \nn\\
& -  \frac{5769772463 \bep}{19906560 \gam} -  \frac{295961663 \bep^2}{4478976 \gam} + \frac{987177893}{358318080} \gam + \frac{89549603}{53747712} \bep \gam -  \frac{7615843}{8957952} \gam^2 + \frac{3639706241 \bep}{59719680 \zeta \Sm^2} -  \frac{410983607 \bem^2}{5971968 \zeta \gam \Sm^2} \nn\\
&-\!  \frac{410983607 \bep^2}{5971968 \zeta \gam \Sm^2}\! -  \! \frac{665257271 \gam}{49766400 \zeta \Sm^2} \!+ \! \frac{560781617 \bem \Sp}{44789760 \Sm} \!+ \! \frac{410983607 \bem \bep \Sp}{746496 \gam^2 \Sm} \!- \! \frac{5769772463 \bem \Sp}{19906560 \gam \Sm} - \! \frac{295961663 \bem \bep \Sp}{2239488 \gam \Sm} \nn\\
&+\delta \bigg(\!\!- \frac{897726671}{59719680} \bem -  \frac{180939719}{13436928} \bem \bep -  \frac{410983607 \bem \bep}{746496 \gam^2} + \frac{5769772463 \bem}{19906560 \gam} + \frac{295961663 \bem \bep}{2239488 \gam} -  \frac{89549603}{53747712} \bem \gam \nn\\
&\qquad \!-\!  \frac{3639706241 \bem}{59719680 \zeta \Sm^2}\! + \! \frac{410983607 \bem \bep}{2985984 \zeta \gam \Sm^2} \!- \! \frac{23226346279 \Sp}{298598400 \Sm}\! -  \! \frac{560781617 \bep \Sp}{44789760 \Sm}\! - \! \frac{410983607 \bem^2 \Sp}{1492992 \gam^2 \Sm} \!-\!  \frac{410983607 \bep^2 \Sp}{1492992 \gam^2 \Sm} \nn\\
&\qquad+ \frac{295961663 \bem^2 \Sp}{4478976 \gam \Sm} + \frac{5769772463 \bep \Sp}{19906560 \gam \Sm} + \frac{295961663 \bep^2 \Sp}{4478976 \gam \Sm} -  \frac{543403783 \gam \Sp}{59719680 \Sm}\bigg)  \nn\\
&+ \nu \bigg(\!\!- \frac{665257271}{6220800} -  \frac{180939719}{6718464} \bem^2 -  \frac{410983607 \bem^2}{746496 \gam^2} -  \frac{410983607 \bep^2}{746496 \gam^2} + \frac{295961663 \bem^2}{1119744 \gam} + \frac{3639706241 \bep}{7464960 \gam} \nn\\
& \qquad-  \frac{3639706241 \bep}{14929920 \zeta \Sm^2} + \frac{410983607 \bep^2}{1492992 \zeta \gam \Sm^2} + \frac{665257271 \gam}{12441600 \zeta \Sm^2} -  \frac{2719530689 \bem \Sp}{22394880 \Sm} -  \frac{410983607 \bem \bep \Sp}{373248 \gam^2 \Sm} \nn\\
& \qquad + \frac{3639706241 \bem \Sp}{7464960 \gam \Sm} + \frac{295961663 \bem \bep \Sp}{1119744 \gam \Sm}\bigg)   \,.
\end{align}\ese

\section{Expression of the orbit-averaged instantaneous scalar fluxes} 
\label{app:orbitAveragedScalarFluxes}

\bse\begin{align}
\frak{f}_1 =& - \frac{8 \gam}{5 \zeta} - 8 \Sm^2 -  \frac{16}{3} \bep \Sm^2 + \frac{16 \bep \Sm^2}{\gam} -  \frac{8}{3} \gam \Sm^2 + \frac{16 \bem \Sm \Sp}{\gam} \nn\\
& + \bigg[\frac{16}{3} \bem \Sm^2 -  \frac{16 \bem \Sm^2}{\gam} -  \frac{16 \bep \Sm \Sp}{\gam}\bigg] \delta -  \frac{16}{3} \Sm^2 \nu  \,,\\
\frak{f}_2 =& \frac{32 \bep}{\zeta} -  \frac{48 \bep^2}{\zeta \gam} -  \frac{51 \gam}{5 \zeta} + 72 \Sm^2 -  \frac{56}{3} \bep \Sm^2 + \frac{96 \bem^2 \Sm^2}{\gam^2} + \frac{96 \bep^2 \Sm^2}{\gam^2} -  \frac{16 \bep \Sm^2}{\gam} + 36 \gam \Sm^2  \nn\\
& + \frac{192 \bem \bep \Sm \Sp}{\gam^2} -  \frac{16 \bem \Sm \Sp}{\gam} + \bigg[\frac{56}{3} \bem \Sm^2 -  \frac{32 \bem \Sm^2}{\gam} -  \frac{32 \bep \Sm \Sp}{\gam}\bigg] \delta -  \frac{124}{3} \Sm^2 \nu  \,,\\
\frak{f}_3 =& \frac{8 \bep}{\zeta} -  \frac{12 \bep^2}{\zeta \gam} -  \frac{39 \gam}{20 \zeta} + 13 \Sm^2 + \frac{24 \bem^2 \Sm^2}{\gam^2} + \frac{24 \bep^2 \Sm^2}{\gam^2} -  \frac{10 \bep \Sm^2}{\gam} + \frac{14}{3} \gam \Sm^2 + \frac{48 \bem \bep \Sm \Sp}{\gam^2} -  \frac{10 \bem \Sm \Sp}{\gam}  \nn\\
&  + \bigg[ - \frac{2 \bem \Sm^2}{\gam} -  \frac{2 \bep \Sm \Sp}{\gam} \bigg] \delta -  \frac{25}{3} \Sm^2 \nu \,,\\
\frak{f}_4 =& - \frac{32 \bep}{5 \zeta} -  \frac{8 \bem^2}{\zeta \gam} -  \frac{8 \bep^2}{\zeta \gam} + \frac{97 \gam}{14 \zeta} + \frac{64 \bep \gam}{15 \zeta} + \frac{211 \gam^2}{120 \zeta} -  \frac{3 \gam^3}{16 \zeta} - 68 \Sm^2 -  \frac{16}{3} \bem^2 \Sm^2 - 8 \bep \Sm^2 -  \frac{16}{3} \bep^2 \Sm^2 + \frac{32 \bem^2 \Sm^2}{\gam^2}  \nn\\
&  + \frac{32 \bep^2 \Sm^2}{\gam^2} -  \frac{32 \bem^2 \Sm^2}{3 \gam} -  \frac{16 \bep \Sm^2}{\gam} -  \frac{32 \bep^2 \Sm^2}{3 \gam} -  \frac{325}{6} \gam \Sm^2 -  \frac{45}{4} \gam^2 \Sm^2 -  \frac{16}{3} \chip \Sm^2 + \frac{16 \chip \Sm^2}{\gam} + \frac{21}{2} \zeta \Sm^4  \nn\\
& + \frac{21}{4} \zeta \gam \Sm^4 -  \frac{16}{3} \bem \Sm \Sp + \frac{64 \bem \bep \Sm \Sp}{\gam^2} -  \frac{16 \bem \Sm \Sp}{\gam} -  \frac{64 \bem \bep \Sm \Sp}{3 \gam} + \frac{16 \chim \Sm \Sp}{\gam} + \frac{3}{2} \zeta \Sp^4 + \frac{3}{4} \zeta \gam \Sp^4  \nn\\
& + \bigg[\frac{32 \bem}{5 \zeta} + \frac{16 \bem \bep}{\zeta \gam} -  \frac{64 \bem \gam}{15 \zeta} + 8 \bem \Sm^2 + \frac{32}{3} \bem \bep \Sm^2 -  \frac{64 \bem \bep \Sm^2}{\gam^2} + \frac{16 \bem \Sm^2}{\gam} + \frac{64 \bem \bep \Sm^2}{3 \gam}  \nn\\
&  + \frac{16}{3} \chim \Sm^2 -  \frac{16 \chim \Sm^2}{\gam} + 22 \Sm \Sp + \frac{16}{3} \bep \Sm \Sp -  \frac{32 \bem^2 \Sm \Sp}{\gam^2} -  \frac{32 \bep^2 \Sm \Sp}{\gam^2} + \frac{32 \bem^2 \Sm \Sp}{3 \gam} + \frac{16 \bep \Sm \Sp}{\gam}  \nn\\
& + \frac{32 \bep^2 \Sm \Sp}{3 \gam} + 15 \gam \Sm \Sp + \frac{3}{2} \gam^2 \Sm \Sp -  \frac{16 \chip \Sm \Sp}{\gam} + 6 \zeta \Sm^3 \Sp + 3 \zeta \gam \Sm^3 \Sp + 6 \zeta \Sm \Sp^3 + 3 \zeta \gam \Sm \Sp^3\bigg] \delta  \nn\\
& + \bigg[\frac{32 \bep^2}{\zeta \gam} + \frac{14 \gam}{3 \zeta} + \frac{\gam^2}{3 \zeta} + \frac{\gam^3}{6 \zeta} + \frac{142}{3} \Sm^2 + \frac{64}{3} \bem^2 \Sm^2 + \frac{140}{3} \bep \Sm^2 -  \frac{192 \bem^2 \Sm^2}{\gam^2} + \frac{64 \bep^2 \Sm^2}{\gam^2} + \frac{320 \bem^2 \Sm^2}{3 \gam}  \nn\\
& -  \frac{224 \bep \Sm^2}{3 \gam}  -  \frac{64 \bep^2 \Sm^2}{\gam} + \frac{28}{3} \gam \Sm^2 -  \frac{2}{3} \gam^2 \Sm^2 + \frac{32}{3} \chip \Sm^2 -  \frac{32 \chip \Sm^2}{\gam} -  \frac{28}{3} \zeta \Sm^4 -  \frac{14}{3} \zeta \gam \Sm^4 -  \frac{128 \bem \bep \Sm \Sp}{\gam^2} \nn\\
&  -  \frac{224 \bem \Sm \Sp}{3 \gam}  + \frac{128 \bem \bep \Sm \Sp}{3 \gam} -  \frac{32 \chim \Sm \Sp}{\gam} -  \frac{4}{3} \zeta \Sp^4 -  \frac{2}{3} \zeta \gam \Sp^4  \bigg] \nu \nn\\
& + \bigg[- \frac{44}{3} \bem \Sm^2 + \frac{80 \bem \Sm^2}{3 \gam} + \frac{80 \bep \Sm \Sp}{3 \gam}\bigg] \delta\nu + \frac{8}{3} \Sm^2 \nu^2 \,,\\
\frak{f}_5 =& \frac{4 \bep}{\zeta} -  \frac{64 \bep^2}{3 \zeta} + \frac{384 \bem^2 \bep}{\zeta \gam^2} -  \frac{384 \bep^3}{\zeta \gam^2} -  \frac{204 \bem^2}{\zeta \gam} + \frac{164 \bep^2}{\zeta \gam} -  \frac{160 \bep^3}{\zeta \gam} -  \frac{5641 \gam}{140 \zeta} + \frac{1394 \bep \gam}{15 \zeta} -  \frac{3803 \gam^2}{120 \zeta} -  \frac{5 \gam^3}{16 \zeta}  \nn\\
& + \frac{32 \chip}{\zeta}   -  \frac{96 \bep \chip}{\zeta \gam} - 29 \Sm^2 - 16 \bem^2 \Sm^2 - 96 \bep \Sm^2 - 16 \bep^2 \Sm^2 -  \frac{768 \bem^2 \bep \Sm^2}{\gam^3} + \frac{768 \bep^3 \Sm^2}{\gam^3} + \frac{272 \bem^2 \Sm^2}{\gam^2}  \nn\\
& + \frac{320 \bem^2 \bep \Sm^2}{\gam^2}   -  \frac{112 \bep^2 \Sm^2}{\gam^2} + \frac{320 \bep^3 \Sm^2}{\gam^2} -  \frac{256 \bem^2 \Sm^2}{3 \gam} + \frac{264 \bep \Sm^2}{\gam} -  \frac{160 \bep^2 \Sm^2}{3 \gam} + \frac{183}{2} \gam \Sm^2 -  \frac{392}{3} \bep \gam \Sm^2  \nn\\
& + \frac{631}{12} \gam^2 \Sm^2   + \frac{192 \bem \chim \Sm^2}{\gam^2} -  \frac{160}{3} \chip \Sm^2 + \frac{192 \bep \chip \Sm^2}{\gam^2} + \frac{16 \chip \Sm^2}{\gam} + \frac{35}{2} \zeta \Sm^4 + \frac{35}{4} \zeta \gam \Sm^4 + 128 \bem \Sm \Sp   \nn\\
&  -  \frac{768 \bem^3 \Sm \Sp}{\gam^3} + \frac{768 \bem \bep^2 \Sm \Sp}{\gam^3} + \frac{160 \bem \bep \Sm \Sp}{\gam^2} + \frac{640 \bem \bep^2 \Sm \Sp}{\gam^2} + \frac{264 \bem \Sm \Sp}{\gam} -  \frac{416 \bem \bep \Sm \Sp}{3 \gam}  \nn\\
& + \frac{192 \bep \chim \Sm \Sp}{\gam^2}  + \frac{16 \chim \Sm \Sp}{\gam} + \frac{192 \bem \chip \Sm \Sp}{\gam^2} + \frac{5}{2} \zeta \Sp^4 + \frac{5}{4} \zeta \gam \Sp^4 + \bigg[\frac{92 \bem}{5 \zeta} -  \frac{32 \bem \bep}{3 \zeta} + \frac{136 \bem \bep}{\zeta \gam}  \nn\\
& + \frac{160 \bem \bep^2}{\zeta \gam} -  \frac{198 \bem \gam}{5 \zeta}   -  \frac{32 \chim}{\zeta} + \frac{96 \bep \chim}{\zeta \gam} - 32 \bem \Sm^2 + 32 \bem \bep \Sm^2 -  \frac{768 \bem^3 \Sm^2}{\gam^3} + \frac{768 \bem \bep^2 \Sm^2}{\gam^3} -  \frac{320 \bem^3 \Sm^2}{\gam^2}  \nn\\
& -  \frac{544 \bem \bep \Sm^2}{\gam^2}   -  \frac{320 \bem \bep^2 \Sm^2}{\gam^2} -  \frac{360 \bem \Sm^2}{\gam} + \frac{800 \bem \bep \Sm^2}{3 \gam} + \frac{392}{3} \bem \gam \Sm^2 + \frac{160}{3} \chim \Sm^2 -  \frac{192 \bep \chim \Sm^2}{\gam^2}  \nn\\
& -  \frac{16 \chim \Sm^2}{\gam} -  \frac{192 \bem \chip \Sm^2}{\gam^2} + 73 \Sm \Sp - 256 \bep \Sm \Sp -  \frac{768 \bem^2 \bep \Sm \Sp}{\gam^3} + \frac{768 \bep^3 \Sm \Sp}{\gam^3} -  \frac{80 \bem^2 \Sm \Sp}{\gam^2}  \nn\\
& -  \frac{640 \bem^2 \bep \Sm \Sp}{\gam^2} -  \frac{464 \bep^2 \Sm \Sp}{\gam^2} + \frac{352 \bem^2 \Sm \Sp}{3 \gam} -  \frac{360 \bep \Sm \Sp}{\gam} + \frac{448 \bep^2 \Sm \Sp}{3 \gam} + 55 \gam \Sm \Sp + \frac{5}{2} \gam^2 \Sm \Sp  \nn\\
& -  \frac{192 \bem \chim \Sm \Sp}{\gam^2} -  \frac{192 \bep \chip \Sm \Sp}{\gam^2} -  \frac{16 \chip \Sm \Sp}{\gam} + 10 \zeta \Sm^3 \Sp + 5 \zeta \gam \Sm^3 \Sp + 10 \zeta \Sm \Sp^3 + 5 \zeta \gam \Sm \Sp^3\bigg] \delta  \nn\\
& + \bigg[- \frac{448 \bep}{3 \zeta} + \frac{352 \bep^2}{\zeta \gam} + \frac{1094 \gam}{15 \zeta} + \frac{10 \gam^2}{3 \zeta} + \frac{5 \gam^3}{3 \zeta} -  \frac{473}{3} \Sm^2 + 64 \bem^2 \Sm^2 + \frac{1228}{3} \bep \Sm^2 -  \frac{1216 \bem^2 \Sm^2}{\gam^2} -  \frac{192 \bep^2 \Sm^2}{\gam^2}  \nn\\
& + \frac{3136 \bem^2 \Sm^2}{3 \gam} -  \frac{872 \bep \Sm^2}{3 \gam} -  \frac{640 \bep^2 \Sm^2}{\gam} - 172 \gam \Sm^2 -  \frac{20}{3} \gam^2 \Sm^2 + \frac{320}{3} \chip \Sm^2 -  \frac{128 \chip \Sm^2}{\gam} -  \frac{280}{3} \zeta \Sm^4 -  \frac{140}{3} \zeta \gam \Sm^4  \nn\\
&  -  \frac{1408 \bem \bep \Sm \Sp}{\gam^2} -  \frac{872 \bem \Sm \Sp}{3 \gam} + \frac{1216 \bem \bep \Sm \Sp}{3 \gam} -  \frac{128 \chim \Sm \Sp}{\gam} -  \frac{40}{3} \zeta \Sp^4 -  \frac{20}{3} \zeta \gam \Sp^4 \bigg] \nu \nn\\
& + \bigg[- \frac{484}{3} \bem \Sm^2 + \frac{968 \bem \Sm^2}{3 \gam} + \frac{968 \bep \Sm \Sp}{3 \gam}\bigg] \delta  \nu + \frac{364}{3} \Sm^2 \nu^2 \,,\\
\frak{f}_6 =& \frac{1286 \bep}{5 \zeta} -  \frac{832 \bep^2}{3 \zeta} + \frac{288 \bem^2 \bep}{\zeta \gam^2} -  \frac{288 \bep^3}{\zeta \gam^2} -  \frac{165 \bem^2}{\zeta \gam} -  \frac{273 \bep^2}{\zeta \gam} -  \frac{40 \bep^3}{\zeta \gam} -  \frac{7269 \gam}{80 \zeta} + \frac{5921 \bep \gam}{30 \zeta} -  \frac{9835 \gam^2}{192 \zeta} -  \frac{43 \gam^3}{384 \zeta}  \nn\\
& + \frac{24 \chip}{\zeta} -  \frac{72 \bep \chip}{\zeta \gam} + 489 \Sm^2 + 6 \bem^2 \Sm^2 - 337 \bep \Sm^2 + 6 \bep^2 \Sm^2 -  \frac{576 \bem^2 \bep \Sm^2}{\gam^3} + \frac{576 \bep^3 \Sm^2}{\gam^3} + \frac{1020 \bem^2 \Sm^2}{\gam^2}  \nn\\
& + \frac{80 \bem^2 \bep \Sm^2}{\gam^2} + \frac{732 \bep^2 \Sm^2}{\gam^2} + \frac{80 \bep^3 \Sm^2}{\gam^2} + \frac{1196 \bem^2 \Sm^2}{3 \gam} -  \frac{206 \bep \Sm^2}{\gam} + \frac{1508 \bep^2 \Sm^2}{3 \gam} + \frac{7835}{16} \gam \Sm^2 -  \frac{250}{3} \bep \gam \Sm^2  \nn\\
& + \frac{3801}{32} \gam^2 \Sm^2 + \frac{144 \bem \chim \Sm^2}{\gam^2} - 6 \chip \Sm^2 + \frac{144 \bep \chip \Sm^2}{\gam^2} -  \frac{18 \chip \Sm^2}{\gam} + \frac{301}{48} \zeta \Sm^4 + \frac{301}{96} \zeta \gam \Sm^4 -  \frac{370}{3} \bem \Sm \Sp  \nn\\
& -  \frac{576 \bem^3 \Sm \Sp}{\gam^3} + \frac{576 \bem \bep^2 \Sm \Sp}{\gam^3} + \frac{1752 \bem \bep \Sm \Sp}{\gam^2} + \frac{160 \bem \bep^2 \Sm \Sp}{\gam^2} -  \frac{206 \bem \Sm \Sp}{\gam} + \frac{2704 \bem \bep \Sm \Sp}{3 \gam}  \nn\\
& + \frac{144 \bep \chim \Sm \Sp}{\gam^2} -  \frac{18 \chim \Sm \Sp}{\gam} + \frac{144 \bem \chip \Sm \Sp}{\gam^2} + \frac{43}{48} \zeta \Sp^4 + \frac{43}{96} \zeta \gam \Sp^4 + \bigg[\frac{136 \bem \bep}{3 \zeta} + \frac{90 \bem \bep}{\zeta \gam} + \frac{40 \bem \bep^2}{\zeta \gam}  \nn\\
&   -  \frac{307 \bem \gam}{10 \zeta} -  \frac{24 \chim}{\zeta} + \frac{72 \bep \chim}{\zeta \gam} + 69 \bem \Sm^2 - 12 \bem \bep \Sm^2 -  \frac{576 \bem^3 \Sm^2}{\gam^3} + \frac{576 \bem \bep^2 \Sm^2}{\gam^3} -  \frac{80 \bem^3 \Sm^2}{\gam^2}  \nn\\
&  -  \frac{360 \bem \bep \Sm^2}{\gam^2} -  \frac{80 \bem \bep^2 \Sm^2}{\gam^2} -  \frac{214 \bem \Sm^2}{\gam} + \frac{80 \bem \bep \Sm^2}{3 \gam} + \frac{250}{3} \bem \gam \Sm^2 + 6 \chim \Sm^2 -  \frac{144 \bep \chim \Sm^2}{\gam^2} + \frac{18 \chim \Sm^2}{\gam}  \nn\\
&  -  \frac{144 \bem \chip \Sm^2}{\gam^2} -  \frac{9}{4} \Sm \Sp -  \frac{434}{3} \bep \Sm \Sp -  \frac{576 \bem^2 \bep \Sm \Sp}{\gam^3} + \frac{576 \bep^3 \Sm \Sp}{\gam^3} -  \frac{36 \bem^2 \Sm \Sp}{\gam^2} -  \frac{160 \bem^2 \bep \Sm \Sp}{\gam^2}  \nn\\
&  -  \frac{324 \bep^2 \Sm \Sp}{\gam^2} -  \frac{116 \bem^2 \Sm \Sp}{3 \gam} -  \frac{214 \bep \Sm \Sp}{\gam} + \frac{196 \bep^2 \Sm \Sp}{3 \gam} + \frac{223}{24} \gam \Sm \Sp + \frac{43}{48} \gam^2 \Sm \Sp  \nn\\
&  -  \frac{144 \bem \chim \Sm \Sp}{\gam^2} -  \frac{144 \bep \chip \Sm \Sp}{\gam^2} + \frac{18 \chip \Sm \Sp}{\gam} + \frac{43}{12} \zeta \Sm^3 \Sp + \frac{43}{24} \zeta \gam \Sm^3 \Sp + \frac{43}{12} \zeta \Sm \Sp^3 + \frac{43}{24} \zeta \gam \Sm \Sp^3\bigg] \delta  \nn\\
&  + \bigg[- \frac{296 \bep}{\zeta} + \frac{480 \bep^2}{\zeta \gam} + \frac{454 \gam}{5 \zeta} + \frac{3 \gam^2}{8 \zeta} + \frac{3 \gam^3}{16 \zeta} -  \frac{1101}{2} \Sm^2 - 24 \bem^2 \Sm^2 + \frac{973}{6} \bep \Sm^2 -  \frac{1104 \bem^2 \Sm^2}{\gam^2} -  \frac{816 \bep^2 \Sm^2}{\gam^2}   \nn\\
& + \frac{568 \bem^2 \Sm^2}{3 \gam} + \frac{152 \bep \Sm^2}{\gam} -  \frac{72 \bep^2 \Sm^2}{\gam} -  \frac{1607}{6} \gam \Sm^2 -  \frac{3}{4} \gam^2 \Sm^2 + 12 \chip \Sm^2 -  \frac{36 \chip \Sm^2}{\gam} -  \frac{21}{2} \zeta \Sm^4 -  \frac{21}{4} \zeta \gam \Sm^4   \nn\\
& -  \frac{1920 \bem \bep \Sm \Sp}{\gam^2} + \frac{152 \bem \Sm \Sp}{\gam} + \frac{352 \bem \bep \Sm \Sp}{3 \gam} -  \frac{36 \chim \Sm \Sp}{\gam} -  \frac{3}{2} \zeta \Sp^4 -  \frac{3}{4} \zeta \gam \Sp^4\bigg] \nu  \nn\\
& + \bigg[- \frac{793}{6} \bem \Sm^2 + \frac{238 \bem \Sm^2}{\gam} + \frac{238 \bep \Sm \Sp}{\gam}\bigg] \delta  \nu + \frac{397}{2} \Sm^2 \nu^2 \,,\\
\frak{f}_7 =& \frac{137 \bep}{10 \zeta} -  \frac{8 \bep^2}{\zeta} -  \frac{9 \bem^2}{2 \zeta \gam} -  \frac{57 \bep^2}{2 \zeta \gam} -  \frac{771 \gam}{224 \zeta} + \frac{10 \bep \gam}{3 \zeta} -  \frac{19 \gam^2}{20 \zeta} + \frac{231}{8} \Sm^2 -  \frac{8}{3} \bep \Sm^2 + \frac{66 \bem^2 \Sm^2}{\gam^2} + \frac{66 \bep^2 \Sm^2}{\gam^2} + \frac{16 \bem^2 \Sm^2}{\gam}  \nn\\
&  -  \frac{19 \bep \Sm^2}{\gam} + \frac{16 \bep^2 \Sm^2}{\gam} + \frac{67}{3} \gam \Sm^2 + \frac{14}{3} \gam^2 \Sm^2 -  \frac{8}{3} \bem \Sm \Sp + \frac{132 \bem \bep \Sm \Sp}{\gam^2} -  \frac{19 \bem \Sm \Sp}{\gam} + \frac{32 \bem \bep \Sm \Sp}{\gam}   \nn\\
& + \bigg[- \frac{3 \bem}{10 \zeta} + \frac{3 \bem \bep}{\zeta \gam} + \frac{2}{3} \bem \Sm^2 -  \frac{12 \bem \bep \Sm^2}{\gam^2} + \frac{\bem \Sm^2}{\gam} -  \frac{21}{8} \Sm \Sp + \frac{2}{3} \bep \Sm \Sp -  \frac{6 \bem^2 \Sm \Sp}{\gam^2} -  \frac{6 \bep^2 \Sm \Sp}{\gam^2}  \nn\\
&  + \frac{\bep \Sm \Sp}{\gam} -  \frac{5}{4} \gam \Sm \Sp\bigg] \delta + \bigg[- \frac{92 \bep}{3 \zeta} + \frac{46 \bep^2}{\zeta \gam} + \frac{299 \gam}{40 \zeta} -  \frac{455}{12} \Sm^2 -  \frac{92 \bem^2 \Sm^2}{\gam^2} -  \frac{92 \bep^2 \Sm^2}{\gam^2} + \frac{115 \bep \Sm^2}{3 \gam}  \nn\\
	& -  \frac{37}{3} \gam \Sm^2  -  \frac{184 \bem \bep \Sm \Sp}{\gam^2} + \frac{115 \bem \Sm \Sp}{3 \gam} \bigg]\nu + \bigg[\frac{23 \bem \Sm^2}{3 \gam} + \frac{23 \bep \Sm \Sp}{3 \gam}\bigg] \delta \nu + \frac{75}{4} \Sm^2 \nu^2 \,,\\
\frak{f}_8 =& \ \Sm^2 \Bigg\{50 + \frac{155}{3} \gam + \frac{40}{3} \gam^2 -  \frac{20}{3} \zeta \Sm^2 -  \frac{10}{3} \zeta \gam \Sm^2 \nn\\
& + \bigg[- \frac{20}{3} \zeta \Sm \Sp -  \frac{10}{3} \zeta \gam \Sm \Sp\bigg] \delta + \bigg[-20 + \frac{20}{3} \bep -  \frac{40}{3} \gam\bigg]\nu -  \frac{20}{3} \bem \delta \nu\Bigg\} \,,\\
\frak{f}_9 =&\  \Sm^2  \Bigg\{155 + \frac{961}{6} \gam + \frac{124}{3} \gam^2 -  \frac{62}{3} \zeta \Sm^2 -  \frac{31}{3} \zeta \gam \Sm^2 \nn\\
& + \bigg[- \frac{62}{3} \zeta \Sm \Sp -  \frac{31}{3} \zeta \gam \Sm \Sp\bigg] \delta + \bigg[-62 + \frac{62}{3} \bep -  \frac{124}{3} \gam\bigg]\nu -  \frac{62}{3} \bem \delta \nu \Bigg\} \,,\\
\frak{f}_{10} =& \ \Sm^2  \Bigg\{20 + \frac{62}{3} \gam + \frac{16}{3} \gam^2 -  \frac{8}{3} \zeta \Sm^2 -  \frac{4}{3} \zeta \gam \Sm^2 \nn\\
& + \bigg[- \frac{8}{3} \zeta \Sm \Sp -  \frac{4}{3} \zeta \gam \Sm \Sp \bigg]\delta + \bigg[-8 + \frac{8}{3} \bep -  \frac{16}{3} \gam \bigg]\nu -  \frac{8}{3} \bem \delta \nu \Bigg\}\,.
\end{align}\ese

\bse\begin{align}
 \frak{g}_1 =& - \frac{8 \gam}{5 \zeta} - 8 \Sm^2 -  \frac{16}{3} \bep \Sm^2 + \frac{16 \bep \Sm^2}{\gam} -  \frac{8}{3} \gam \Sm^2 + \frac{16 \bem \Sm \Sp}{\gam} \nn\\
& + \bigg[\frac{16}{3} \bem \Sm^2 -  \frac{16 \bem \Sm^2}{\gam} -  \frac{16 \bep \Sm \Sp}{\gam}\bigg] \delta -  \frac{16}{3} \Sm^2 \nu  \,,\\
 \frak{g}_2 =& - \frac{7 \gam}{5 \zeta} + 26 \Sm^2 - 4 \bep \Sm^2 + \frac{8 \bep \Sm^2}{\gam} + \frac{40}{3} \gam \Sm^2 + \frac{8 \bem \Sm \Sp}{\gam} \nn\\
& + \bigg[4 \bem \Sm^2 -  \frac{8 \bem \Sm^2}{\gam} -  \frac{8 \bep \Sm \Sp}{\gam}\bigg] \delta -  \frac{44}{3} \Sm^2 \nu \,, \\
 \frak{g}_3 =& - \frac{32 \bep}{5 \zeta} -  \frac{8 \bem^2}{\zeta \gam} -  \frac{8 \bep^2}{\zeta \gam} + \frac{97 \gam}{14 \zeta} + \frac{64 \bep \gam}{15 \zeta} + \frac{113 \gam^2}{60 \zeta} -  \frac{\gam^3}{8 \zeta} - 38 \Sm^2 -  \frac{16}{3} \bem^2 \Sm^2 - 8 \bep \Sm^2 -  \frac{16}{3} \bep^2 \Sm^2 + \frac{32 \bem^2 \Sm^2}{\gam^2} \nn\\
& + \frac{32 \bep^2 \Sm^2}{\gam^2} -  \frac{32 \bem^2 \Sm^2}{3 \gam} -  \frac{16 \bep \Sm^2}{\gam} -  \frac{32 \bep^2 \Sm^2}{3 \gam} -  \frac{71}{3} \gam \Sm^2 -  \frac{7}{2} \gam^2 \Sm^2 -  \frac{16}{3} \chip \Sm^2 + \frac{16 \chip \Sm^2}{\gam} + 7 \zeta \Sm^4 + \frac{7}{2} \zeta \gam \Sm^4 \nn\\
& -  \frac{16}{3} \bem \Sm \Sp + \frac{64 \bem \bep \Sm \Sp}{\gam^2} -  \frac{16 \bem \Sm \Sp}{\gam} -  \frac{64 \bem \bep \Sm \Sp}{3 \gam} + \frac{16 \chim \Sm \Sp}{\gam} + \zeta \Sp^4 + \frac{1}{2} \zeta \gam \Sp^4 \nn\\
& + \bigg[\frac{32 \bem}{5 \zeta} + \frac{16 \bem \bep}{\zeta \gam} -  \frac{64 \bem \gam}{15 \zeta} + 8 \bem \Sm^2 + \frac{32}{3} \bem \bep \Sm^2 -  \frac{64 \bem \bep \Sm^2}{\gam^2} + \frac{16 \bem \Sm^2}{\gam} + \frac{64 \bem \bep \Sm^2}{3 \gam} + \frac{16}{3} \chim \Sm^2 \nn\\
& -  \frac{16 \chim \Sm^2}{\gam} + 22 \Sm \Sp + \frac{16}{3} \bep \Sm \Sp -  \frac{32 \bem^2 \Sm \Sp}{\gam^2} -  \frac{32 \bep^2 \Sm \Sp}{\gam^2} + \frac{32 \bem^2 \Sm \Sp}{3 \gam} + \frac{16 \bep \Sm \Sp}{\gam} + \frac{32 \bep^2 \Sm \Sp}{3 \gam} \nn\\
& + 14 \gam \Sm \Sp + \gam^2 \Sm \Sp -  \frac{16 \chip \Sm \Sp}{\gam} + 4 \zeta \Sm^3 \Sp + 2 \zeta \gam \Sm^3 \Sp + 4 \zeta \Sm \Sp^3 + 2 \zeta \gam \Sm \Sp^3\bigg] \delta \nn\\
& + \bigg[\frac{32 \bep^2}{\zeta \gam} + \frac{14 \gam}{3 \zeta} + \frac{\gam^2}{3 \zeta} + \frac{\gam^3}{6 \zeta} + \frac{106}{3} \Sm^2 + \frac{64}{3} \bem^2 \Sm^2 + \frac{152}{3} \bep \Sm^2 -  \frac{192 \bem^2 \Sm^2}{\gam^2} + \frac{64 \bep^2 \Sm^2}{\gam^2} + \frac{320 \bem^2 \Sm^2}{3 \gam} -  \frac{224 \bep \Sm^2}{3 \gam} \nn\\
& -  \frac{64 \bep^2 \Sm^2}{\gam} + \frac{4}{3} \gam \Sm^2 -  \frac{2}{3} \gam^2 \Sm^2 + \frac{32}{3} \chip \Sm^2 -  \frac{32 \chip \Sm^2}{\gam} -  \frac{28}{3} \zeta \Sm^4 -  \frac{14}{3} \zeta \gam \Sm^4 -  \frac{128 \bem \bep \Sm \Sp}{\gam^2} -  \frac{224 \bem \Sm \Sp}{3 \gam} \nn\\
& + \frac{128 \bem \bep \Sm \Sp}{3 \gam} -  \frac{32 \chim \Sm \Sp}{\gam} -  \frac{4}{3} \zeta \Sp^4 -  \frac{2}{3} \zeta \gam \Sp^4 \bigg]\nu+ \bigg[- \frac{56}{3} \bem \Sm^2 + \frac{80 \bem \Sm^2}{3 \gam} + \frac{80 \bep \Sm \Sp}{3 \gam}  \bigg] \delta\nu + \frac{8}{3} \Sm^2 \nu^2  \,,\\
 \frak{g}_4 =& - \frac{44 \bep}{5 \zeta} -  \frac{24 \bem^2}{\zeta \gam} -  \frac{24 \bep^2}{\zeta \gam} -  \frac{173 \gam}{14 \zeta} + \frac{28 \bep \gam}{3 \zeta} -  \frac{1013 \gam^2}{120 \zeta} -  \frac{\gam^3}{48 \zeta} - 68 \Sm^2 -  \frac{16}{3} \bem^2 \Sm^2 + \frac{152}{3} \bep \Sm^2 -  \frac{16}{3} \bep^2 \Sm^2 \nn\\
& + \frac{96 \bem^2 \Sm^2}{\gam^2} + \frac{96 \bep^2 \Sm^2}{\gam^2} -  \frac{112 \bem^2 \Sm^2}{3 \gam} + \frac{184 \bep \Sm^2}{\gam} -  \frac{112 \bep^2 \Sm^2}{3 \gam} -  \frac{63}{2} \gam \Sm^2 -  \frac{104}{3} \bep \gam \Sm^2 + \frac{17}{12} \gam^2 \Sm^2 - 16 \chip \Sm^2 \nn\\
& + \frac{24 \chip \Sm^2}{\gam} + \frac{7}{6} \zeta \Sm^4 + \frac{7}{12} \zeta \gam \Sm^4 + 104 \bem \Sm \Sp + \frac{192 \bem \bep \Sm \Sp}{\gam^2} + \frac{184 \bem \Sm \Sp}{\gam} -  \frac{224 \bem \bep \Sm \Sp}{3 \gam} \nn\\
& + \frac{24 \chim \Sm \Sp}{\gam} + \frac{1}{6} \zeta \Sp^4 + \frac{1}{12} \zeta \gam \Sp^4 + \bigg[\frac{44 \bem}{5 \zeta} + \frac{48 \bem \bep}{\zeta \gam} -  \frac{28 \bem \gam}{3 \zeta} -  \frac{152}{3} \bem \Sm^2 + \frac{32}{3} \bem \bep \Sm^2 -  \frac{192 \bem \bep \Sm^2}{\gam^2} \nn\\
& -  \frac{184 \bem \Sm^2}{\gam} + \frac{224 \bem \bep \Sm^2}{3 \gam} + \frac{104}{3} \bem \gam \Sm^2 + 16 \chim \Sm^2 -  \frac{24 \chim \Sm^2}{\gam} + 34 \Sm \Sp - 104 \bep \Sm \Sp -  \frac{96 \bem^2 \Sm \Sp}{\gam^2} \nn\\
& -  \frac{96 \bep^2 \Sm \Sp}{\gam^2} + \frac{112 \bem^2 \Sm \Sp}{3 \gam} -  \frac{184 \bep \Sm \Sp}{\gam} + \frac{112 \bep^2 \Sm \Sp}{3 \gam} + \frac{61}{3} \gam \Sm \Sp + \frac{1}{6} \gam^2 \Sm \Sp -  \frac{24 \chip \Sm \Sp}{\gam} \nn\\
& + \frac{2}{3} \zeta \Sm^3 \Sp + \frac{1}{3} \zeta \gam \Sm^3 \Sp + \frac{2}{3} \zeta \Sm \Sp^3 + \frac{1}{3} \zeta \gam \Sm \Sp^3\bigg] \delta + \bigg[- \frac{32 \bep}{\zeta} + \frac{96 \bep^2}{\zeta \gam} + \frac{83 \gam}{3 \zeta} + \frac{\gam^2}{\zeta} + \frac{\gam^3}{2 \zeta} - 76 \Sm^2 + \frac{64}{3} \bem^2 \Sm^2 \nn\\
& + 144 \bep \Sm^2 -  \frac{384 \bem^2 \Sm^2}{\gam^2} + \frac{1024 \bem^2 \Sm^2}{3 \gam} -  \frac{536 \bep \Sm^2}{3 \gam} -  \frac{192 \bep^2 \Sm^2}{\gam} -  \frac{212}{3} \gam \Sm^2 - 2 \gam^2 \Sm^2 + 32 \chip \Sm^2 -  \frac{48 \chip \Sm^2}{\gam} \nn\\
& - 28 \zeta \Sm^4 - 14 \zeta \gam \Sm^4 -  \frac{384 \bem \bep \Sm \Sp}{\gam^2} -  \frac{536 \bem \Sm \Sp}{3 \gam} + \frac{448 \bem \bep \Sm \Sp}{3 \gam} -  \frac{48 \chim \Sm \Sp}{\gam} - 4 \zeta \Sp^4 - 2 \zeta \gam \Sp^4  \bigg] \nu\nn\\
& +  \bigg[-68 \bem \Sm^2 + \frac{464 \bem \Sm^2}{3 \gam} + \frac{464 \bep \Sm \Sp}{3 \gam}\bigg] \delta  \nu + 64 \Sm^2 \nu^2 \,, \\
 \frak{g}_5 =& \frac{\bep}{5 \zeta} -  \frac{3 \bem^2}{\zeta \gam} -  \frac{3 \bep^2}{\zeta \gam} -  \frac{533 \gam}{112 \zeta} + \frac{2 \bep \gam}{5 \zeta} -  \frac{521 \gam^2}{240 \zeta} + \frac{\gam^3}{32 \zeta} + \frac{181}{4} \Sm^2 + \frac{4}{3} \bep \Sm^2 + \frac{12 \bem^2 \Sm^2}{\gam^2} + \frac{12 \bep^2 \Sm^2}{\gam^2} + \frac{20 \bep \Sm^2}{\gam} \nn\\
& + \frac{467}{12} \gam \Sm^2 -  \frac{8}{3} \bep \gam \Sm^2 + \frac{63}{8} \gam^2 \Sm^2 -  \frac{7}{4} \zeta \Sm^4 -  \frac{7}{8} \zeta \gam \Sm^4 + \frac{28}{3} \bem \Sm \Sp + \frac{24 \bem \bep \Sm \Sp}{\gam^2} + \frac{20 \bem \Sm \Sp}{\gam} -  \frac{1}{4} \zeta \Sp^4 \nn\\
& -  \frac{1}{8} \zeta \gam \Sp^4 + \bigg[- \frac{\bem}{5 \zeta} + \frac{6 \bem \bep}{\zeta \gam} -  \frac{2 \bem \gam}{5 \zeta} -  \frac{4}{3} \bem \Sm^2 -  \frac{24 \bem \bep \Sm^2}{\gam^2} -  \frac{20 \bem \Sm^2}{\gam} + \frac{8}{3} \bem \gam \Sm^2 + \frac{1}{4} \Sm \Sp -  \frac{28}{3} \bep \Sm \Sp\nn\\
&  -  \frac{12 \bem^2 \Sm \Sp}{\gam^2} -  \frac{12 \bep^2 \Sm \Sp}{\gam^2} -  \frac{20 \bep \Sm \Sp}{\gam} -  \frac{1}{4} \gam^2 \Sm \Sp -  \zeta \Sm^3 \Sp -  \frac{1}{2} \zeta \gam \Sm^3 \Sp -  \zeta \Sm \Sp^3 -  \frac{1}{2} \zeta \gam \Sm \Sp^3\bigg] \delta \nn\\
& + \bigg[- \frac{8 \bep}{\zeta} + \frac{12 \bep^2}{\zeta \gam} + \frac{71 \gam}{12 \zeta} -  \frac{395}{6} \Sm^2 + \frac{34}{3} \bep \Sm^2 -  \frac{24 \bem^2 \Sm^2}{\gam^2} -  \frac{24 \bep^2 \Sm^2}{\gam^2} -  \frac{38 \bep \Sm^2}{3 \gam} -  \frac{98}{3} \gam \Sm^2 -  \frac{48 \bem \bep \Sm \Sp}{\gam^2}\nn\\
&  -  \frac{38 \bem \Sm \Sp}{3 \gam}\bigg]\nu + \bigg[- \frac{34}{3} \bem \Sm^2 + \frac{74 \bem \Sm^2}{3 \gam} + \frac{74 \bep \Sm \Sp}{3 \gam}  \bigg] \delta\nu + \frac{179}{6} \Sm^2 \nu^2 \,,\\
 \frak{g}_6 =& \Sm^2 \Bigg\{20 + \frac{62}{3} \gam + \frac{16}{3} \gam^2 -  \frac{8}{3} \zeta \Sm^2 -  \frac{4}{3} \zeta \gam \Sm^2 \nn\\
& + \bigg[ - \frac{8}{3} \zeta \Sm \Sp -  \frac{4}{3} \zeta \gam \Sm \Sp\Bigg] \delta + \Bigg[-8 + \frac{8}{3} \bep -  \frac{16}{3} \gam \bigg]\nu-  \frac{8}{3} \bem \delta \nu\Bigg\} \,,\\
 \frak{g}_7 =& \Sm^2 \Bigg\{20 + \frac{62}{3} \gam + \frac{16}{3} \gam^2 -  \frac{8}{3} \zeta \Sm^2 -  \frac{4}{3} \zeta \gam \Sm^2 \nn\\
& + \bigg[ - \frac{8}{3} \zeta \Sm \Sp -  \frac{4}{3} \zeta \gam \Sm \Sp \Bigg] \delta + \Bigg[-8 + \frac{8}{3} \bep -  \frac{16}{3} \gam \bigg] \nu -  \frac{8}{3} \bem \delta  \nu\Bigg\} \,,\\
 \frak{g}_8 &= \Sm^2 \Bigg\{-40 -  \frac{124}{3} \gam -  \frac{32}{3} \gam^2 + \frac{16}{3} \zeta \Sm^2 + \frac{8}{3} \zeta \gam \Sm^2 \nn\\
& + \bigg[\frac{16}{3} \zeta \Sm \Sp + \frac{8}{3} \zeta \gam \Sm \Sp\bigg] \delta + \bigg[16 -  \frac{16}{3} \bep + \frac{32}{3} \gam\bigg]\nu + \frac{16}{3} \bem \delta \nu\Bigg\} \,.
\end{align}\ese

\section{Coefficients entering the expression of the evolution of the orbital parameters}

\label{app:orbitalEvolution}

The coefficients associated to the frequency evolution equation \eqref{seq:evolution_x} read
\bse\begin{align}
\frak{X}_1 &= 288 \bar{\zeta}^{-1} + 120 \bar{\zeta}^{-1} \gam - 30 \Sm^2 - 160 \bep \Sm^2 + 40 \gam \Sm^2 + 240 \bep \bar{\gamma}^{-1} \Sm^2 + 240 \bem \bar{\gamma}^{-1} \Sm \Sp \nn\\
&\quad + \delta \bigg(160 \bem \Sm^2 - 240 \bem \bar{\gamma}^{-1} \Sm^2 - 240 \bep \bar{\gamma}^{-1} \Sm \Sp \bigg)  - 70 \Sm^2 \nu\,,\\
\frak{X}_2 &= 876 \bar{\zeta}^{-1} + 480 \bep \bar{\zeta}^{-1} + 285 \bar{\zeta}^{-1} \gam - 720 \bep^2 \bar{\zeta}^{-1} \bar{\gamma}^{-1} + 1095 \Sm^2 - 260 \bep \Sm^2 + 540 \gam \Sm^2 - 240 \bep \bar{\gamma}^{-1} \Sm^2 \nn\\
&\quad + 1440 \bem^2 \bar{\gamma}^{-2} \Sm^2 + 1440 \bep^2 \bar{\gamma}^{-2} \Sm^2 - 240 \bem \bar{\gamma}^{-1} \Sm \Sp + 2880 \bem \bep \bar{\gamma}^{-2} \Sm \Sp \nn\\
&\quad +\delta \bigg(260 \bem \Sm^2 - 480 \bem \bar{\gamma}^{-1} \Sm^2 - 480 \bep \bar{\gamma}^{-1} \Sm \Sp\bigg)  - 625 \Sm^2 \nu \,,\\
\frak{X}_3 &= 111 \bar{\zeta}^{-1} + 120 \bep \bar{\zeta}^{-1} + \frac{105}{4} \bar{\zeta}^{-1} \gam - 180 \bep^2 \bar{\zeta}^{-1} \bar{\gamma}^{-1} + 270 \Sm^2 + 110 \gam \Sm^2 - 150 \bep \bar{\gamma}^{-1} \Sm^2 + 360 \bem^2 \bar{\gamma}^{-2} \Sm^2 \nn\\
&\quad + 360 \bep^2 \bar{\gamma}^{-2} \Sm^2 - 150 \bem \bar{\gamma}^{-1} \Sm \Sp + 720 \bem \bep \bar{\gamma}^{-2} \Sm \Sp + \delta \Big(-30 \bem \bar{\gamma}^{-1} \Sm^2 - 30 \bep \bar{\gamma}^{-1} \Sm \Sp\Big)  - 130 \Sm^2 \nu \,,\\
\frak{X}_4 &= - \frac{1486}{35} \bar{\zeta}^{-1} -  \frac{416}{5} \bep \bar{\zeta}^{-1} -  \frac{167}{10} \bar{\zeta}^{-1} \gam - 32 \bep \bar{\zeta}^{-1} \gam + \frac{1}{8} \bar{\zeta}^{-1} \gam^2 + \frac{1}{16} \bar{\zeta}^{-1} \gam^3 - 8 \bem^2 \bar{\zeta}^{-1} \bar{\gamma}^{-1} - 8 \bep^2 \bar{\zeta}^{-1} \bar{\gamma}^{-1} + \frac{59}{2} \Sm^2 \nn\\
&\quad -  \frac{64}{9} \bem^2 \Sm^2 - 12 \bep \Sm^2 -  \frac{64}{9} \bep^2 \Sm^2 + \frac{391}{6} \gam \Sm^2 -  \frac{304}{9} \bep \gam \Sm^2 + \frac{983}{36} \gam^2 \Sm^2 - 32 \bem^2 \bar{\gamma}^{-1} \Sm^2 + 8 \bep \bar{\gamma}^{-1} \Sm^2 \nn\\
&\quad - 32 \bep^2 \bar{\gamma}^{-1} \Sm^2 + 32 \bem^2 \bar{\gamma}^{-2} \Sm^2 + 32 \bep^2 \bar{\gamma}^{-2} \Sm^2 -  \frac{64}{3} \chip \Sm^2 + 16 \bar{\gamma}^{-1} \chip \Sm^2 -  \frac{7}{2} \zeta \Sm^4 -  \frac{7}{4} \zeta \gam \Sm^4 + 16 \bem \Sm \Sp \nn\\
&\quad+ 8 \bem \bar{\gamma}^{-1} \Sm \Sp - 64 \bem \bep \bar{\gamma}^{-1} \Sm \Sp + 64 \bem \bep \bar{\gamma}^{-2} \Sm \Sp + 16 \bar{\gamma}^{-1} \chim \Sm \Sp -  \frac{1}{2} \zeta \Sp^4 -  \frac{1}{4} \zeta \gam \Sp^4 \nn\\
&\quad + \delta \bigg(\frac{416}{5} \bem \bar{\zeta}^{-1} + 32 \bem \bar{\zeta}^{-1} \gam + 16 \bem \bep \bar{\zeta}^{-1} \bar{\gamma}^{-1} + 12 \bem \Sm^2 + \frac{128}{9} \bem \bep \Sm^2 + \frac{304}{9} \bem \gam \Sm^2 - 8 \bem \bar{\gamma}^{-1} \Sm^2 \nn\\
&\quad\qquad + 64 \bem \bep \bar{\gamma}^{-1} \Sm^2 - 64 \bem \bep \bar{\gamma}^{-2} \Sm^2 + \frac{64}{3} \chim \Sm^2 - 16 \bar{\gamma}^{-1} \chim \Sm^2 + 22 \Sm \Sp - 16 \bep \Sm \Sp + 11 \gam \Sm \Sp \nn\\
&\quad\qquad -  \frac{1}{2} \gam^2 \Sm \Sp + 32 \bem^2 \bar{\gamma}^{-1} \Sm \Sp - 8 \bep \bar{\gamma}^{-1} \Sm \Sp + 32 \bep^2 \bar{\gamma}^{-1} \Sm \Sp - 32 \bem^2 \bar{\gamma}^{-2} \Sm \Sp - 32 \bep^2 \bar{\gamma}^{-2} \Sm \Sp \nn\\
&\quad\qquad - 16 \bar{\gamma}^{-1} \chip \Sm \Sp - 2 \zeta \Sm^3 \Sp -  \zeta \gam \Sm^3 \Sp - 2 \zeta \Sm \Sp^3 -  \zeta \gam \Sm \Sp^3\bigg)  \nn\\
&\quad  + \nu \bigg(- \frac{264}{5} \bar{\zeta}^{-1} - 22 \bar{\zeta}^{-1} \gam + \frac{4}{3} \bar{\zeta}^{-1} \gam^2 + \frac{2}{3} \bar{\zeta}^{-1} \gam^3 + 32 \bep^2 \bar{\zeta}^{-1} \bar{\gamma}^{-1} -  \frac{25}{2} \Sm^2 + \frac{256}{9} \bem^2 \Sm^2 + \frac{1216}{9} \bep \Sm^2 -  \frac{472}{9} \gam \Sm^2 \nn\\
&\quad\qquad -  \frac{8}{3} \gam^2 \Sm^2 + 384 \bem^2 \bar{\gamma}^{-1} \Sm^2 - 72 \bep \bar{\gamma}^{-1} \Sm^2 - 256 \bep^2 \bar{\gamma}^{-1} \Sm^2 - 192 \bem^2 \bar{\gamma}^{-2} \Sm^2 + 64 \bep^2 \bar{\gamma}^{-2} \Sm^2 + \frac{128}{3} \chip \Sm^2\nn\\
&\quad\qquad  - 32 \bar{\gamma}^{-1} \chip \Sm^2 -  \frac{112}{3} \zeta \Sm^4 -  \frac{56}{3} \zeta \gam \Sm^4 - 72 \bem \bar{\gamma}^{-1} \Sm \Sp + 128 \bem \bep \bar{\gamma}^{-1} \Sm \Sp - 128 \bem \bep \bar{\gamma}^{-2} \Sm \Sp \nn\\
&\quad\qquad - 32 \bar{\gamma}^{-1} \chim \Sm \Sp -  \frac{16}{3} \zeta \Sp^4 -  \frac{8}{3} \zeta \gam \Sp^4\bigg)   + \delta \nu \bigg(- \frac{280}{9} \bem \Sm^2 + 24 \bem \bar{\gamma}^{-1} \Sm^2 + 24 \bep \bar{\gamma}^{-1} \Sm \Sp\bigg)  + \frac{43}{18} \Sm^2 \nu^2 \,,\\
\frak{X}_5 &= \frac{2193}{7} \bar{\zeta}^{-1} -  \frac{5308}{15} \bep \bar{\zeta}^{-1} -  \frac{128}{3} \bep^2 \bar{\zeta}^{-1} + \frac{20543}{60} \bar{\zeta}^{-1} \gam -  \frac{214}{3} \bep \bar{\zeta}^{-1} \gam + \frac{3769}{48} \bar{\zeta}^{-1} \gam^2 -  \frac{13}{32} \bar{\zeta}^{-1} \gam^3 - 204 \bem^2 \bar{\zeta}^{-1} \bar{\gamma}^{-1} \nn\\
&\quad  + 188 \bep^2 \bar{\zeta}^{-1} \bar{\gamma}^{-1} - 128 \bep^3 \bar{\zeta}^{-1} \bar{\gamma}^{-1} + 384 \bem^2 \bep \bar{\zeta}^{-1} \bar{\gamma}^{-2} - 384 \bep^3 \bar{\zeta}^{-1} \bar{\gamma}^{-2} + 32 \bar{\zeta}^{-1} \chip - 96 \bep \bar{\zeta}^{-1} \bar{\gamma}^{-1} \chip -  \frac{49}{4} \Sm^2  \nn\\
&\quad -  \frac{8}{9} \bem^2 \Sm^2 -  \frac{484}{3} \bep \Sm^2 -  \frac{8}{9} \bep^2 \Sm^2 + \frac{1621}{12} \gam \Sm^2 -  \frac{488}{3} \bep \gam \Sm^2 + \frac{4885}{72} \gam^2 \Sm^2 -  \frac{304}{3} \bem^2 \bar{\gamma}^{-1} \Sm^2 + 296 \bep \bar{\gamma}^{-1} \Sm^2  \nn\\
&\quad -  \frac{112}{3} \bep^2 \bar{\gamma}^{-1} \Sm^2 + 224 \bem^2 \bar{\gamma}^{-2} \Sm^2 + 256 \bem^2 \bep \bar{\gamma}^{-2} \Sm^2 - 160 \bep^2 \bar{\gamma}^{-2} \Sm^2 + 256 \bep^3 \bar{\gamma}^{-2} \Sm^2 - 768 \bem^2 \bep \bar{\gamma}^{-3} \Sm^2  \nn\\
&\quad + 768 \bep^3 \bar{\gamma}^{-3} \Sm^2 + 192 \bem \bar{\gamma}^{-2} \chim \Sm^2 -  \frac{136}{3} \chip \Sm^2 + 16 \bar{\gamma}^{-1} \chip \Sm^2 + 192 \bep \bar{\gamma}^{-2} \chip \Sm^2 + \frac{91}{4} \zeta \Sm^4 + \frac{91}{8} \zeta \gam \Sm^4  \nn\\
&\quad + \frac{416}{3} \bem \Sm \Sp + 296 \bem \bar{\gamma}^{-1} \Sm \Sp -  \frac{416}{3} \bem \bep \bar{\gamma}^{-1} \Sm \Sp + 64 \bem \bep \bar{\gamma}^{-2} \Sm \Sp + 512 \bem \bep^2 \bar{\gamma}^{-2} \Sm \Sp  \nn\\
&\quad - 768 \bem^3 \bar{\gamma}^{-3} \Sm \Sp + 768 \bem \bep^2 \bar{\gamma}^{-3} \Sm \Sp + 16 \bar{\gamma}^{-1} \chim \Sm \Sp + 192 \bep \bar{\gamma}^{-2} \chim \Sm \Sp  \nn\\
&\quad+ 192 \bem \bar{\gamma}^{-2} \chip \Sm \Sp + \frac{13}{4} \zeta \Sp^4 + \frac{13}{8} \zeta \gam \Sp^4 \nn\\
&\quad +  \delta \bigg(\frac{5404}{15} \bem \bar{\zeta}^{-1} + \frac{32}{3} \bem \bep \bar{\zeta}^{-1} + \frac{374}{3} \bem \bar{\zeta}^{-1} \gam + 136 \bem \bep \bar{\zeta}^{-1} \bar{\gamma}^{-1} + 128 \bem \bep^2 \bar{\zeta}^{-1} \bar{\gamma}^{-1} - 32 \bar{\zeta}^{-1} \chim  \nn\\
&\quad\qquad+ 96 \bep \bar{\zeta}^{-1} \bar{\gamma}^{-1} \chim + \frac{100}{3} \bem \Sm^2 + \frac{16}{9} \bem \bep \Sm^2 + \frac{488}{3} \bem \gam \Sm^2 - 368 \bem \bar{\gamma}^{-1} \Sm^2 + \frac{800}{3} \bem \bep \bar{\gamma}^{-1} \Sm^2 \nn\\
&\quad\qquad- 256 \bem^3 \bar{\gamma}^{-2} \Sm^2 - 544 \bem \bep \bar{\gamma}^{-2} \Sm^2 - 256 \bem \bep^2 \bar{\gamma}^{-2} \Sm^2 - 768 \bem^3 \bar{\gamma}^{-3} \Sm^2 + 768 \bem \bep^2 \bar{\gamma}^{-3} \Sm^2 + \frac{136}{3} \chim \Sm^2 \nn\\
&\quad\qquad - 16 \bar{\gamma}^{-1} \chim \Sm^2 - 192 \bep \bar{\gamma}^{-2} \chim \Sm^2 - 192 \bem \bar{\gamma}^{-2} \chip \Sm^2 + 73 \Sm \Sp -  \frac{800}{3} \bep \Sm \Sp + \frac{113}{2} \gam \Sm \Sp \nn\\
&\quad\qquad + \frac{13}{4} \gam^2 \Sm \Sp + \frac{304}{3} \bem^2 \bar{\gamma}^{-1} \Sm \Sp - 368 \bep \bar{\gamma}^{-1} \Sm \Sp + \frac{496}{3} \bep^2 \bar{\gamma}^{-1} \Sm \Sp - 80 \bem^2 \bar{\gamma}^{-2} \Sm \Sp  \nn\\
&\quad\qquad- 512 \bem^2 \bep \bar{\gamma}^{-2} \Sm \Sp - 464 \bep^2 \bar{\gamma}^{-2} \Sm \Sp - 768 \bem^2 \bep \bar{\gamma}^{-3} \Sm \Sp + 768 \bep^3 \bar{\gamma}^{-3} \Sm \Sp - 192 \bem \bar{\gamma}^{-2} \chim \Sm \Sp\nn\\
&\quad\qquad - 16 \bar{\gamma}^{-1} \chip \Sm \Sp - 192 \bep \bar{\gamma}^{-2} \chip \Sm \Sp + 13 \zeta \Sm^3 \Sp + \frac{13}{2} \zeta \gam \Sm^3 \Sp + 13 \zeta \Sm \Sp^3 + \frac{13}{2} \zeta \gam \Sm \Sp^3\bigg)  \nn\\
&\quad + \nu \bigg(-570 \bar{\zeta}^{-1} - 144 \bep \bar{\zeta}^{-1} -  \frac{427}{2} \bar{\zeta}^{-1} \gam + \frac{17}{6} \bar{\zeta}^{-1} \gam^2 + \frac{17}{12} \bar{\zeta}^{-1} \gam^3 + 344 \bep^2 \bar{\zeta}^{-1} \bar{\gamma}^{-1} -  \frac{2095}{12} \Sm^2 + \frac{32}{9} \bem^2 \Sm^2 \nn\\
&\quad\qquad + \frac{3718}{9} \bep \Sm^2 -  \frac{1682}{9} \gam \Sm^2 -  \frac{17}{3} \gam^2 \Sm^2 + \frac{3040}{3} \bem^2 \bar{\gamma}^{-1} \Sm^2 - 296 \bep \bar{\gamma}^{-1} \Sm^2 - 544 \bep^2 \bar{\gamma}^{-1} \Sm^2 - 1200 \bem^2 \bar{\gamma}^{-2} \Sm^2 \nn\\
&\quad\qquad - 176 \bep^2 \bar{\gamma}^{-2} \Sm^2 + \frac{272}{3} \chip \Sm^2 - 128 \bar{\gamma}^{-1} \chip \Sm^2 -  \frac{238}{3} \zeta \Sm^4 -  \frac{119}{3} \zeta \gam \Sm^4 - 296 \bem \bar{\gamma}^{-1} \Sm \Sp \nn\\
&\quad\qquad+ \frac{1408}{3} \bem \bep \bar{\gamma}^{-1} \Sm \Sp - 1376 \bem \bep \bar{\gamma}^{-2} \Sm \Sp - 128 \bar{\gamma}^{-1} \chim \Sm \Sp -  \frac{34}{3} \zeta \Sp^4 -  \frac{17}{3} \zeta \gam \Sp^4 \bigg)  \nn\\
&\quad+  \delta \nu \bigg(- \frac{1810}{9} \bem \Sm^2 + 320 \bem \bar{\gamma}^{-1} \Sm^2 + 320 \bep \bar{\gamma}^{-1} \Sm \Sp \bigg) + \frac{1373}{12} \Sm^2 \nu^2 \,,\\
\frak{X}_6 &= \frac{12217}{20} \bar{\zeta}^{-1} + \frac{3086}{15} \bep \bar{\zeta}^{-1} -  \frac{1040}{3} \bep^2 \bar{\zeta}^{-1} + \frac{120713}{240} \bar{\zeta}^{-1} \gam + \frac{530}{3} \bep \bar{\zeta}^{-1} \gam + \frac{5895}{64} \bar{\zeta}^{-1} \gam^2 -  \frac{43}{384} \bar{\zeta}^{-1} \gam^3 \nn\\
&\quad - 165 \bem^2 \bar{\zeta}^{-1} \bar{\gamma}^{-1} - 387 \bep^2 \bar{\zeta}^{-1} \bar{\gamma}^{-1} - 32 \bep^3 \bar{\zeta}^{-1} \bar{\gamma}^{-1} + 288 \bem^2 \bep \bar{\zeta}^{-1} \bar{\gamma}^{-2} - 288 \bep^3 \bar{\zeta}^{-1} \bar{\gamma}^{-2} + 24 \bar{\zeta}^{-1} \chip\nn\\
&\quad - 72 \bep \bar{\zeta}^{-1} \bar{\gamma}^{-1} \chip + \frac{1173}{2} \Sm^2 + \frac{10}{3} \bem^2 \Sm^2 - 391 \bep \Sm^2 + \frac{10}{3} \bep^2 \Sm^2 + \frac{28097}{48} \gam \Sm^2 -  \frac{842}{9} \bep \gam \Sm^2 + \frac{40865}{288} \gam^2 \Sm^2 \nn\\
&\quad + \frac{1592}{3} \bem^2 \bar{\gamma}^{-1} \Sm^2 - 257 \bep \bar{\gamma}^{-1} \Sm^2 + \frac{1928}{3} \bep^2 \bar{\gamma}^{-1} \Sm^2 + 1248 \bem^2 \bar{\gamma}^{-2} \Sm^2 + 64 \bem^2 \bep \bar{\gamma}^{-2} \Sm^2 + 960 \bep^2 \bar{\gamma}^{-2} \Sm^2  \nn\\
&\quad + 64 \bep^3 \bar{\gamma}^{-2} \Sm^2 - 576 \bem^2 \bep \bar{\gamma}^{-3} \Sm^2 + 576 \bep^3 \bar{\gamma}^{-3} \Sm^2 + 144 \bem \bar{\gamma}^{-2} \chim \Sm^2 - 6 \chip \Sm^2 - 18 \bar{\gamma}^{-1} \chip \Sm^2  \nn\\
&\quad + 144 \bep \bar{\gamma}^{-2} \chip \Sm^2 + \frac{301}{48} \zeta \Sm^4 + \frac{301}{96} \zeta \gam \Sm^4 -  \frac{466}{3} \bem \Sm \Sp - 257 \bem \bar{\gamma}^{-1} \Sm \Sp + \frac{3520}{3} \bem \bep \bar{\gamma}^{-1} \Sm \Sp  \nn\\
&\quad + 2208 \bem \bep \bar{\gamma}^{-2} \Sm \Sp + 128 \bem \bep^2 \bar{\gamma}^{-2} \Sm \Sp - 576 \bem^3 \bar{\gamma}^{-3} \Sm \Sp + 576 \bem \bep^2 \bar{\gamma}^{-3} \Sm \Sp - 18 \bar{\gamma}^{-1} \chim \Sm \Sp  \nn\\
&\quad + 144 \bep \bar{\gamma}^{-2} \chim \Sm \Sp + 144 \bem \bar{\gamma}^{-2} \chip \Sm \Sp + \frac{43}{48} \zeta \Sp^4 + \frac{43}{96} \zeta \gam \Sp^4  \nn\\
&\quad+ \delta \bigg(\frac{1912}{15} \bem \bar{\zeta}^{-1} + \frac{152}{3} \bem \bep \bar{\zeta}^{-1} + \frac{98}{3} \bem \bar{\zeta}^{-1} \gam + 90 \bem \bep \bar{\zeta}^{-1} \bar{\gamma}^{-1} + 32 \bem \bep^2 \bar{\zeta}^{-1} \bar{\gamma}^{-1} - 24 \bar{\zeta}^{-1} \chim  \nn\\
&\quad\qquad + 72 \bep \bar{\zeta}^{-1} \bar{\gamma}^{-1} \chim + 59 \bem \Sm^2 -  \frac{20}{3} \bem \bep \Sm^2 + \frac{842}{9} \bem \gam \Sm^2 - 277 \bem \bar{\gamma}^{-1} \Sm^2 + \frac{32}{3} \bem \bep \bar{\gamma}^{-1} \Sm^2 - 64 \bem^3 \bar{\gamma}^{-2} \Sm^2  \nn\\
&\quad\qquad- 360 \bem \bep \bar{\gamma}^{-2} \Sm^2 - 64 \bem \bep^2 \bar{\gamma}^{-2} \Sm^2 - 576 \bem^3 \bar{\gamma}^{-3} \Sm^2 + 576 \bem \bep^2 \bar{\gamma}^{-3} \Sm^2 + 6 \chim \Sm^2 + 18 \bar{\gamma}^{-1} \chim \Sm^2 \nn\\
&\quad\qquad - 144 \bep \bar{\gamma}^{-2} \chim \Sm^2 - 144 \bem \bar{\gamma}^{-2} \chip \Sm^2 -  \frac{9}{4} \Sm \Sp -  \frac{530}{3} \bep \Sm \Sp + \frac{223}{24} \gam \Sm \Sp + \frac{43}{48} \gam^2 \Sm \Sp \nn\\
&\quad\qquad -  \frac{152}{3} \bem^2 \bar{\gamma}^{-1} \Sm \Sp - 277 \bep \bar{\gamma}^{-1} \Sm \Sp + \frac{184}{3} \bep^2 \bar{\gamma}^{-1} \Sm \Sp - 36 \bem^2 \bar{\gamma}^{-2} \Sm \Sp - 128 \bem^2 \bep \bar{\gamma}^{-2} \Sm \Sp \nn\\
&\quad\qquad - 324 \bep^2 \bar{\gamma}^{-2} \Sm \Sp - 576 \bem^2 \bep \bar{\gamma}^{-3} \Sm \Sp + 576 \bep^3 \bar{\gamma}^{-3} \Sm \Sp - 144 \bem \bar{\gamma}^{-2} \chim \Sm \Sp + 18 \bar{\gamma}^{-1} \chip \Sm \Sp \nn\\
&\quad\qquad - 144 \bep \bar{\gamma}^{-2} \chip \Sm \Sp + \frac{43}{12} \zeta \Sm^3 \Sp + \frac{43}{24} \zeta \gam \Sm^3 \Sp + \frac{43}{12} \zeta \Sm \Sp^3 + \frac{43}{24} \zeta \gam \Sm \Sp^3 \bigg)  \nn\\
&\quad + \nu \bigg(- \frac{5061}{10} \bar{\zeta}^{-1} - 300 \bep \bar{\zeta}^{-1} -  \frac{1287}{8} \bar{\zeta}^{-1} \gam + \frac{3}{8} \bar{\zeta}^{-1} \gam^2 + \frac{3}{16} \bar{\zeta}^{-1} \gam^3 + 486 \bep^2 \bar{\zeta}^{-1} \bar{\gamma}^{-1} -  \frac{3701}{6} \Sm^2 -  \frac{40}{3} \bem^2 \Sm^2 \nn\\
&\quad\qquad + \frac{2819}{18} \bep \Sm^2 -  \frac{5395}{18} \gam \Sm^2 -  \frac{3}{4} \gam^2 \Sm^2 + \frac{520}{3} \bem^2 \bar{\gamma}^{-1} \Sm^2 + 153 \bep \bar{\gamma}^{-1} \Sm^2 - 72 \bep^2 \bar{\gamma}^{-1} \Sm^2 - 1116 \bem^2 \bar{\gamma}^{-2} \Sm^2 \nn\\
&\quad\qquad - 828 \bep^2 \bar{\gamma}^{-2} \Sm^2 + 12 \chip \Sm^2 - 36 \bar{\gamma}^{-1} \chip \Sm^2 -  \frac{21}{2} \zeta \Sm^4 -  \frac{21}{4} \zeta \gam \Sm^4 + 153 \bem \bar{\gamma}^{-1} \Sm \Sp + \frac{304}{3} \bem \bep \bar{\gamma}^{-1} \Sm \Sp \nn\\
&\quad\qquad- 1944 \bem \bep \bar{\gamma}^{-2} \Sm \Sp - 36 \bar{\gamma}^{-1} \chim \Sm \Sp -  \frac{3}{2} \zeta \Sp^4 -  \frac{3}{4} \zeta \gam \Sp^4\bigg)  \nn\\
&\quad + \delta \nu \bigg(- \frac{2279}{18} \bem \Sm^2 + 243 \bem \bar{\gamma}^{-1} \Sm^2 + 243 \bep \bar{\gamma}^{-1} \Sm \Sp \bigg)  + 204 \Sm^2 \nu^2  \,,\\
\frak{X}_7 &= \frac{11717}{280} \bar{\zeta}^{-1} + \frac{337}{10} \bep \bar{\zeta}^{-1} - 24 \bep^2 \bar{\zeta}^{-1} + \frac{4993}{160} \bar{\zeta}^{-1} \gam + 14 \bep \bar{\zeta}^{-1} \gam + \frac{23}{4} \bar{\zeta}^{-1} \gam^2 -  \frac{9}{2} \bem^2 \bar{\zeta}^{-1} \bar{\gamma}^{-1} -  \frac{117}{2} \bep^2 \bar{\zeta}^{-1} \bar{\gamma}^{-1} \nn\\
&\quad + \frac{561}{8} \Sm^2 - 16 \bep \Sm^2 + \frac{176}{3} \gam \Sm^2 + \frac{112}{9} \gam^2 \Sm^2 + 48 \bem^2 \bar{\gamma}^{-1} \Sm^2 - 44 \bep \bar{\gamma}^{-1} \Sm^2 + 48 \bep^2 \bar{\gamma}^{-1} \Sm^2  \nn\\
&\quad  + 126 \bem^2 \bar{\gamma}^{-2} \Sm^2 + 126 \bep^2 \bar{\gamma}^{-2} \Sm^2 - 16 \bem \Sm \Sp - 44 \bem \bar{\gamma}^{-1} \Sm \Sp + 96 \bem \bep \bar{\gamma}^{-1} \Sm \Sp + 252 \bem \bep \bar{\gamma}^{-2} \Sm \Sp  \nn\\
&\quad  + \delta \bigg(- \frac{3}{10} \bem \bar{\zeta}^{-1} + 3 \bem \bep \bar{\zeta}^{-1} \bar{\gamma}^{-1} - 2 \bem \Sm^2 - 4 \bem \bar{\gamma}^{-1} \Sm^2 - 12 \bem \bep \bar{\gamma}^{-2} \Sm^2 -  \frac{21}{8} \Sm \Sp - 2 \bep \Sm \Sp \nn\\
&\quad\qquad  -  \frac{5}{4} \gam \Sm \Sp - 4 \bep \bar{\gamma}^{-1} \Sm \Sp - 6 \bem^2 \bar{\gamma}^{-2} \Sm \Sp - 6 \bep^2 \bar{\gamma}^{-2} \Sm \Sp \bigg)  \nn\\
&\quad + \nu \bigg(- \frac{148}{5} \bar{\zeta}^{-1} - 32 \bep \bar{\zeta}^{-1} - 7 \bar{\zeta}^{-1} \gam + 48 \bep^2 \bar{\zeta}^{-1} \bar{\gamma}^{-1} -  \frac{353}{6} \Sm^2 -  \frac{208}{9} \gam \Sm^2 + 40 \bep \bar{\gamma}^{-1} \Sm^2 - 96 \bem^2 \bar{\gamma}^{-2} \Sm^2  \nn\\
&\quad\qquad- 96 \bep^2 \bar{\gamma}^{-2} \Sm^2 + 40 \bem \bar{\gamma}^{-1} \Sm \Sp - 192 \bem \bep \bar{\gamma}^{-2} \Sm \Sp \bigg)     + \delta \nu \bigg(8 \bem \bar{\gamma}^{-1} \Sm^2 + 8 \bep \bar{\gamma}^{-1} \Sm \Sp \bigg)  + \frac{184}{9} \Sm^2 \nu^2 \,,\\
\frak{X}_8 &= -10 \Sm^2 \Bigg[1 + \frac{31}{30} \gam + \frac{4}{15} \gam^2 -  \frac{2}{15} \zeta \Sm^2 -  \frac{1}{15} \zeta \gam \Sm^2 +  \delta \bigg(- \frac{2}{15} \zeta \Sm \Sp -  \frac{1}{15} \zeta \gam \Sm \Sp \bigg) \nn\\
&\qquad\qquad + \nu \bigg(- \frac{2}{5} + \frac{2}{15} \bep -  \frac{4}{15} \gam \bigg)  -  \frac{2}{15} \bem \delta \nu\Bigg] \,,\\
\frak{X}_9 &= 200 \Sm^2 \Bigg[1 + \frac{31}{30} \gam + \frac{4}{15} \gam^2 -  \frac{2}{15} \zeta \Sm^2 -  \frac{1}{15} \zeta \gam \Sm^2 + \delta \bigg(- \frac{2}{15} \zeta \Sm \Sp -  \frac{1}{15} \zeta \gam \Sm \Sp \bigg)  \nn\\
&\qquad\qquad + \nu \bigg(- \frac{2}{5} + \frac{2}{15} \bep -  \frac{4}{15} \gam \bigg)   -  \frac{2}{15} \bem \delta \nu\Bigg] \,,\\
\frak{X}_{10} &= 35 \Sm^2 \Bigg[1 + \frac{31}{30} \gam + \frac{4}{15} \gam^2 -  \frac{2}{15} \zeta \Sm^2 -  \frac{1}{15} \zeta \gam \Sm^2 + \delta \bigg(- \frac{2}{15} \zeta \Sm \Sp -  \frac{1}{15} \zeta \gam \Sm \Sp \bigg)  \nn\\
&\qquad\qquad + \nu \bigg(- \frac{2}{5} + \frac{2}{15} \bep -  \frac{4}{15} \gam \bigg)  -  \frac{2}{15} \bem \delta \nu \Bigg] \,,\\
\frak{X}_{11} &=  \frac{96}{5} \bar{\zeta}^{-1}\bigg(1 + \frac{1}{2} \gam\bigg) \,,\\
\frak{X}_{12} &= - \frac{8}{5} \bar{\zeta}^{-1} \gam + \frac{32}{5} \Sm^2 -  \frac{32}{5} \Sm \Sp \delta -  \frac{64}{5} \Sm^2 \nu \,,\\
\frak{X}_{13} &= - \frac{36}{5} \Sm^2 -  \frac{8}{3} \bep \Sm^2 -  \frac{4}{3} \gam \Sm^2 + 8 \bep \bar{\gamma}^{-1} \Sm^2 + 8 \bem \bar{\gamma}^{-1} \Sm \Sp \nn\\
& \quad+ \delta \bigg(\frac{8}{3} \bem \Sm^2 - 8 \bem \bar{\gamma}^{-1} \Sm^2 + \frac{16}{5} \Sm \Sp - 8 \bep \bar{\gamma}^{-1} \Sm \Sp\bigg)   \,,\\
\frak{X}_{14} &= \frac{41}{15} \Sm^2 \nu \,,\\
\frak{X}_{15} &= - \Sm^2 \Bigg[1 + \frac{4}{3} \bep -  \frac{4}{3} \bem \delta -  \frac{1}{3} \nu \Bigg] \,,\\
\frak{X}_{16} &= 5 \Sm^2 \Bigg[1 + \frac{8}{15} \gam -  \frac{1}{15} \nu \Bigg] \,,\\
\frak{X}_{17} &= 4 \Sm^2 \Bigg[ 1 -  \frac{1}{3} \bep + \frac{2}{3} \gam + \frac{1}{3} \bem \delta \Bigg] \,,\\
\frak{X}_{18} &= 16 \bep \bar{\zeta}^{-1} -  \frac{8}{3} \bar{\zeta}^{-1} \gam - 24 \bep^2 \bar{\zeta}^{-1} \bar{\gamma}^{-1} + \frac{17}{3} \Sm^2 - 32 \bep \bar{\gamma}^{-1} \Sm^2 + 48 \bem^2 \bar{\gamma}^{-2} \Sm^2 + 48 \bep^2 \bar{\gamma}^{-2} \Sm^2 - 32 \bem \bar{\gamma}^{-1} \Sm \Sp \nn\\
&\quad + 96 \bem \bep \bar{\gamma}^{-2} \Sm \Sp + \delta \bigg(8 \bem \bar{\gamma}^{-1} \Sm^2 -  \frac{8}{3} \Sm \Sp + 8 \bep \bar{\gamma}^{-1} \Sm \Sp \bigg)  -  \frac{4}{3} \Sm^2 \nu \,.
\end{align}\ese
The coefficients associated to the eccentricity evolution equation \eqref{seq:evolution_et} read
\bse\begin{align}
\frak{E}_1 &= 304 \bar{\zeta}^{-1} + 160 \bep \bar{\zeta}^{-1} + 100 \bar{\zeta}^{-1} \gam - 240 \bep^2 \bar{\zeta}^{-1} \bar{\gamma}^{-1} + 165 \Sm^2 - 120 \bep \Sm^2 + 100 \gam \Sm^2 \nn\\
&\quad  - 40 \bep \bar{\gamma}^{-1} \Sm^2 + 480 \bem^2 \bar{\gamma}^{-2} \Sm^2 + 480 \bep^2 \bar{\gamma}^{-2} \Sm^2 - 40 \bem \bar{\gamma}^{-1} \Sm \Sp + 960 \bem \bep \bar{\gamma}^{-2} \Sm \Sp \nn\\
&\quad +\delta  \bigg(120 \bem \Sm^2 - 200 \bem \bar{\gamma}^{-1} \Sm^2 - 200 \bep \bar{\gamma}^{-1} \Sm \Sp \bigg)  - 125 \Sm^2 \nu \,,\\
\frak{E}_2 &= 121 \bar{\zeta}^{-1} + 40 \bep \bar{\zeta}^{-1} + \frac{175}{4} \bar{\zeta}^{-1} \gam - 60 \bep^2 \bar{\zeta}^{-1} \bar{\gamma}^{-1} + 280 \Sm^2 - 20 \bep \Sm^2 + 130 \gam \Sm^2 - 10 \bep \bar{\gamma}^{-1} \Sm^2 \nn\\
&\quad + 120 \bem^2 \bar{\gamma}^{-2} \Sm^2 + 120 \bep^2 \bar{\gamma}^{-2} \Sm^2 - 10 \bem \bar{\gamma}^{-1} \Sm \Sp + 240 \bem \bep \bar{\gamma}^{-2} \Sm \Sp \nn\\
&\quad + \delta\bigg(20 \bem \Sm^2 - 50 \bem \bar{\gamma}^{-1} \Sm^2 - 50 \bep \bar{\gamma}^{-1} \Sm \Sp\bigg)  - 150 \Sm^2 \nu \,,\\
\frak{E}_3 &= - \frac{939}{35} \bar{\zeta}^{-1} -  \frac{5848}{45} \bep \bar{\zeta}^{-1} + \frac{64}{9} \bep^2 \bar{\zeta}^{-1} + \frac{109}{180} \bar{\zeta}^{-1} \gam - 30 \bep \bar{\zeta}^{-1} \gam + \frac{289}{144} \bar{\zeta}^{-1} \gam^2 -  \frac{31}{288} \bar{\zeta}^{-1} \gam^3 -  \frac{188}{3} \bem^2 \bar{\zeta}^{-1} \bar{\gamma}^{-1} \nn\\
&\quad + 116 \bep^2 \bar{\zeta}^{-1} \bar{\gamma}^{-1} -  \frac{128}{3} \bep^3 \bar{\zeta}^{-1} \bar{\gamma}^{-1} + 128 \bem^2 \bep \bar{\zeta}^{-1} \bar{\gamma}^{-2} - 128 \bep^3 \bar{\zeta}^{-1} \bar{\gamma}^{-2} + \frac{32}{3} \bar{\zeta}^{-1} \chip - 32 \bep \bar{\zeta}^{-1} \bar{\gamma}^{-1} \chip -  \frac{295}{12} \Sm^2  \nn\\
&\quad  -  \frac{8}{3} \bem^2 \Sm^2 -  \frac{302}{9} \bep \Sm^2 -  \frac{8}{3} \bep^2 \Sm^2 + \frac{69}{4} \gam \Sm^2 -  \frac{320}{9} \bep \gam \Sm^2 + \frac{1055}{72} \gam^2 \Sm^2 -  \frac{640}{9} \bem^2 \bar{\gamma}^{-1} \Sm^2 + 52 \bep \bar{\gamma}^{-1} \Sm^2  \nn\\
&\quad  -  \frac{448}{9} \bep^2 \bar{\gamma}^{-1} \Sm^2 -  \frac{128}{3} \bem^2 \bar{\gamma}^{-2} \Sm^2 + \frac{256}{3} \bem^2 \bep \bar{\gamma}^{-2} \Sm^2 -  \frac{512}{3} \bep^2 \bar{\gamma}^{-2} \Sm^2 + \frac{256}{3} \bep^3 \bar{\gamma}^{-2} \Sm^2 - 256 \bem^2 \bep \bar{\gamma}^{-3} \Sm^2  \nn\\
&\quad  + 256 \bep^3 \bar{\gamma}^{-3} \Sm^2 + 64 \bem \bar{\gamma}^{-2} \chim \Sm^2 -  \frac{152}{9} \chip \Sm^2 + \frac{8}{3} \bar{\gamma}^{-1} \chip \Sm^2 + 64 \bep \bar{\gamma}^{-2} \chip \Sm^2 + \frac{217}{36} \zeta \Sm^4 + \frac{217}{72} \zeta \gam \Sm^4  \nn\\
&\quad + \frac{184}{9} \bem \Sm \Sp + 52 \bem \bar{\gamma}^{-1} \Sm \Sp -  \frac{1088}{9} \bem \bep \bar{\gamma}^{-1} \Sm \Sp -  \frac{640}{3} \bem \bep \bar{\gamma}^{-2} \Sm \Sp + \frac{512}{3} \bem \bep^2 \bar{\gamma}^{-2} \Sm \Sp  \nn\\
&\quad - 256 \bem^3 \bar{\gamma}^{-3} \Sm \Sp + 256 \bem \bep^2 \bar{\gamma}^{-3} \Sm \Sp + \frac{8}{3} \bar{\gamma}^{-1} \chim \Sm \Sp + 64 \bep \bar{\gamma}^{-2} \chim \Sm \Sp  \nn\\
&\quad + 64 \bem \bar{\gamma}^{-2} \chip \Sm \Sp + \frac{31}{36} \zeta \Sp^4 + \frac{31}{72} \zeta \gam \Sp^4   \nn\\
&\quad  + \delta\bigg(\frac{4504}{45} \bem \bar{\zeta}^{-1} + \frac{32}{9} \bem \bep \bar{\zeta}^{-1} + \frac{302}{9} \bem \bar{\zeta}^{-1} \gam + \frac{104}{3} \bem \bep \bar{\zeta}^{-1} \bar{\gamma}^{-1} + \frac{128}{3} \bem \bep^2 \bar{\zeta}^{-1} \bar{\gamma}^{-1} -  \frac{32}{3} \bar{\zeta}^{-1} \chim  \nn\\
&\quad\qquad + 32 \bep \bar{\zeta}^{-1} \bar{\gamma}^{-1} \chim + \frac{110}{9} \bem \Sm^2 + \frac{16}{3} \bem \bep \Sm^2 + \frac{320}{9} \bem \gam \Sm^2 - 28 \bem \bar{\gamma}^{-1} \Sm^2 + \frac{704}{9} \bem \bep \bar{\gamma}^{-1} \Sm^2  \nn\\
&\quad\qquad -  \frac{256}{3} \bem^3 \bar{\gamma}^{-2} \Sm^2 -  \frac{416}{3} \bem \bep \bar{\gamma}^{-2} \Sm^2 -  \frac{256}{3} \bem \bep^2 \bar{\gamma}^{-2} \Sm^2 - 256 \bem^3 \bar{\gamma}^{-3} \Sm^2 + 256 \bem \bep^2 \bar{\gamma}^{-3} \Sm^2 + \frac{152}{9} \chim \Sm^2  \nn\\
&\quad\qquad -  \frac{8}{3} \bar{\gamma}^{-1} \chim \Sm^2 - 64 \bep \bar{\gamma}^{-2} \chim \Sm^2 - 64 \bem \bar{\gamma}^{-2} \chip \Sm^2 + \frac{61}{3} \Sm \Sp -  \frac{376}{9} \bep \Sm \Sp + \frac{283}{18} \gam \Sm \Sp + \frac{31}{36} \gam^2 \Sm \Sp  \nn\\
&\quad\qquad + \frac{256}{9} \bem^2 \bar{\gamma}^{-1} \Sm \Sp - 28 \bep \bar{\gamma}^{-1} \Sm \Sp + \frac{448}{9} \bep^2 \bar{\gamma}^{-1} \Sm \Sp -  \frac{16}{3} \bem^2 \bar{\gamma}^{-2} \Sm \Sp -  \frac{512}{3} \bem^2 \bep \bar{\gamma}^{-2} \Sm \Sp  \nn\\
&\quad\qquad -  \frac{400}{3} \bep^2 \bar{\gamma}^{-2} \Sm \Sp - 256 \bem^2 \bep \bar{\gamma}^{-3} \Sm \Sp + 256 \bep^3 \bar{\gamma}^{-3} \Sm \Sp - 64 \bem \bar{\gamma}^{-2} \chim \Sm \Sp -  \frac{8}{3} \bar{\gamma}^{-1} \chip \Sm \Sp  \nn\\
&\quad\qquad - 64 \bep \bar{\gamma}^{-2} \chip \Sm \Sp + \frac{31}{9} \zeta \Sm^3 \Sp + \frac{31}{18} \zeta \gam \Sm^3 \Sp + \frac{31}{9} \zeta \Sm \Sp^3 + \frac{31}{18} \zeta \gam \Sm \Sp^3\bigg)    \nn\\
&\quad  + \nu\bigg(- \frac{4084}{45} \bar{\zeta}^{-1} -  \frac{176}{9} \bep \bar{\zeta}^{-1} -  \frac{311}{9} \bar{\zeta}^{-1} \gam + \frac{19}{18} \bar{\zeta}^{-1} \gam^2 + \frac{19}{36} \bar{\zeta}^{-1} \gam^3 + \frac{200}{3} \bep^2 \bar{\zeta}^{-1} \bar{\gamma}^{-1} + \frac{157}{12} \Sm^2 + \frac{32}{3} \bem^2 \Sm^2  \nn\\
&\quad\qquad + \frac{1036}{9} \bep \Sm^2 -  \frac{82}{3} \gam \Sm^2 -  \frac{19}{9} \gam^2 \Sm^2 + \frac{3040}{9} \bem^2 \bar{\gamma}^{-1} \Sm^2 -  \frac{652}{9} \bep \bar{\gamma}^{-1} \Sm^2 -  \frac{608}{3} \bep^2 \bar{\gamma}^{-1} \Sm^2 -  \frac{848}{3} \bem^2 \bar{\gamma}^{-2} \Sm^2  \nn\\
&\quad\qquad + 16 \bep^2 \bar{\gamma}^{-2} \Sm^2 + \frac{304}{9} \chip \Sm^2 -  \frac{112}{3} \bar{\gamma}^{-1} \chip \Sm^2 -  \frac{266}{9} \zeta \Sm^4 -  \frac{133}{9} \zeta \gam \Sm^4 -  \frac{652}{9} \bem \bar{\gamma}^{-1} \Sm \Sp  \nn\\
&\quad\qquad + \frac{1216}{9} \bem \bep \bar{\gamma}^{-1} \Sm \Sp -  \frac{800}{3} \bem \bep \bar{\gamma}^{-2} \Sm \Sp -  \frac{112}{3} \bar{\gamma}^{-1} \chim \Sm \Sp -  \frac{38}{9} \zeta \Sp^4 -  \frac{19}{9} \zeta \gam \Sp^4\bigg)  \nn\\
&\quad + \delta \nu  \bigg(- \frac{316}{9} \bem \Sm^2 + \frac{412}{9} \bem \bar{\gamma}^{-1} \Sm^2 + \frac{412}{9} \bep \bar{\gamma}^{-1} \Sm \Sp\bigg) + \frac{223}{36} \Sm^2 \nu^2 \,,\\
\frak{E}_4 &= \frac{29917}{105} \bar{\zeta}^{-1} + \frac{2087}{45} \bep \bar{\zeta}^{-1} -  \frac{1184}{9} \bep^2 \bar{\zeta}^{-1} + \frac{44279}{180} \bar{\zeta}^{-1} \gam + \!\frac{464}{9} \bep \bar{\zeta}^{-1} \gam + \!\frac{28129}{576} \bar{\zeta}^{-1} \gam^2 - \! \frac{7}{128} \bar{\zeta}^{-1} \gam^3  - 62 \bem^2 \bar{\zeta}^{-1} \bar{\gamma}^{-1}  \nn\\
&\quad - 172 \bep^2 \bar{\zeta}^{-1} \bar{\gamma}^{-1} -  \frac{32}{3} \bep^3 \bar{\zeta}^{-1} \bar{\gamma}^{-1} + 96 \bem^2 \bep \bar{\zeta}^{-1} \bar{\gamma}^{-2} - 96 \bep^3 \bar{\zeta}^{-1} \bar{\gamma}^{-2} + 8 \bar{\zeta}^{-1} \chip - 24 \bep \bar{\zeta}^{-1} \bar{\gamma}^{-1} \chip + \frac{717}{4} \Sm^2  \nn\\
&\quad + \frac{10}{9} \bem^2 \Sm^2 -  \frac{451}{3} \bep \Sm^2  + \frac{10}{9} \bep^2 \Sm^2 + \frac{29357}{144} \gam \Sm^2 -  \frac{178}{3} \bep \gam \Sm^2 + \frac{16031}{288} \gam^2 \Sm^2 + \frac{1736}{9} \bem^2 \bar{\gamma}^{-1} \Sm^2   - 37 \bep \bar{\gamma}^{-1} \Sm^2  \nn\\
&\quad + \frac{2072}{9} \bep^2 \bar{\gamma}^{-1} \Sm^2 + 516 \bem^2 \bar{\gamma}^{-2} \Sm^2 + \frac{64}{3} \bem^2 \bep \bar{\gamma}^{-2} \Sm^2 + 420 \bep^2 \bar{\gamma}^{-2} \Sm^2 + \frac{64}{3} \bep^3 \bar{\gamma}^{-2} \Sm^2  - 192 \bem^2 \bep \bar{\gamma}^{-3} \Sm^2  \nn\\
&\quad +\! 192 \bep^3 \bar{\gamma}^{-3} \Sm^2 + \! 48 \bem \bar{\gamma}^{-2} \chim \Sm^2 - \!  \frac{22}{3} \chip \Sm^2 + 2 \bar{\gamma}^{-1} \chip \Sm^2 + \! 48 \bep \bar{\gamma}^{-2} \chip \Sm^2 + \! \frac{49}{16} \zeta \Sm^4 +\! \frac{49}{32} \zeta \gam \Sm^4 -\!  \frac{190}{9} \bem \Sm \Sp  \nn\\
&\quad - 37 \bem \bar{\gamma}^{-1} \Sm \Sp + \frac{3808}{9} \bem \bep \bar{\gamma}^{-1} \Sm \Sp + 936 \bem \bep \bar{\gamma}^{-2} \Sm \Sp + \frac{128}{3} \bem \bep^2 \bar{\gamma}^{-2} \Sm \Sp   - 192 \bem^3 \bar{\gamma}^{-3} \Sm \Sp  \nn\\
&\quad + 192 \bem \bep^2 \bar{\gamma}^{-3} \Sm \Sp + 2 \bar{\gamma}^{-1} \chim \Sm \Sp + 48 \bep \bar{\gamma}^{-2} \chim \Sm \Sp + 48 \bem \bar{\gamma}^{-2} \chip \Sm \Sp + \frac{7}{16} \zeta \Sp^4 + \frac{7}{32} \zeta \gam \Sp^4   \nn\\
&\quad  + \delta \bigg(\frac{3991}{45} \bem \bar{\zeta}^{-1} + \frac{152}{9} \bem \bep \bar{\zeta}^{-1} + \frac{260}{9} \bem \bar{\zeta}^{-1} \gam + 44 \bem \bep \bar{\zeta}^{-1} \bar{\gamma}^{-1} + \frac{32}{3} \bem \bep^2 \bar{\zeta}^{-1} \bar{\gamma}^{-1} - 8 \bar{\zeta}^{-1} \chim   \nn\\
&\quad\qquad+ 24 \bep \bar{\zeta}^{-1} \bar{\gamma}^{-1} \chim  + \frac{71}{3} \bem \Sm^2 -  \frac{20}{9} \bem \bep \Sm^2 + \frac{178}{3} \bem \gam \Sm^2 - 177 \bem \bar{\gamma}^{-1} \Sm^2 + \frac{320}{9} \bem \bep \bar{\gamma}^{-1} \Sm^2   \nn\\
&\quad\qquad -  \frac{64}{3} \bem^3 \bar{\gamma}^{-2} \Sm^2 - 176 \bem \bep \bar{\gamma}^{-2} \Sm^2 -  \frac{64}{3} \bem \bep^2 \bar{\gamma}^{-2} \Sm^2 - 192 \bem^3 \bar{\gamma}^{-3} \Sm^2 + 192 \bem \bep^2 \bar{\gamma}^{-3} \Sm^2 + \frac{22}{3} \chim \Sm^2   \nn\\
&\quad\qquad - 2 \bar{\gamma}^{-1} \chim \Sm^2 - 48 \bep \bar{\gamma}^{-2} \chim \Sm^2 - 48 \bem \bar{\gamma}^{-2} \chip \Sm^2 + \frac{21}{2} \Sm \Sp -  \frac{950}{9} \bep \Sm \Sp + \frac{79}{8} \gam \Sm \Sp + \frac{7}{16} \gam^2 \Sm \Sp   \nn\\
&\quad\qquad -  \frac{8}{9} \bem^2 \bar{\gamma}^{-1} \Sm \Sp - 177 \bep \bar{\gamma}^{-1} \Sm \Sp + \frac{328}{9} \bep^2 \bar{\gamma}^{-1} \Sm \Sp - 40 \bem^2 \bar{\gamma}^{-2} \Sm \Sp -  \frac{128}{3} \bem^2 \bep \bar{\gamma}^{-2} \Sm \Sp   \nn\\
&\quad\qquad - 136 \bep^2 \bar{\gamma}^{-2} \Sm \Sp - 192 \bem^2 \bep \bar{\gamma}^{-3} \Sm \Sp + 192 \bep^3 \bar{\gamma}^{-3} \Sm \Sp - 48 \bem \bar{\gamma}^{-2} \chim \Sm \Sp - 2 \bar{\gamma}^{-1} \chip \Sm \Sp   \nn\\
&\quad\qquad - 48 \bep \bar{\gamma}^{-2} \chip \Sm \Sp + \frac{7}{4} \zeta \Sm^3 \Sp + \frac{7}{8} \zeta \gam \Sm^3 \Sp + \frac{7}{4} \zeta \Sm \Sp^3 + \frac{7}{8} \zeta \gam \Sm \Sp^3 \bigg)    \nn\\
&\quad + \nu \bigg(- \frac{7753}{30} \bar{\zeta}^{-1} -  \frac{364}{3} \bep \bar{\zeta}^{-1} -  \frac{2099}{24} \bar{\zeta}^{-1} \gam + \frac{11}{24} \bar{\zeta}^{-1} \gam^2 + \frac{11}{48} \bar{\zeta}^{-1} \gam^3 + 210 \bep^2 \bar{\zeta}^{-1} \bar{\gamma}^{-1} -  \frac{4165}{18} \Sm^2 -  \frac{40}{9} \bem^2 \Sm^2  \nn\\
&\quad\qquad  + \frac{2069}{18} \bep \Sm^2 -  \frac{2321}{18} \gam \Sm^2 -  \frac{11}{12} \gam^2 \Sm^2 + \frac{1672}{9} \bem^2 \bar{\gamma}^{-1} \Sm^2 + \frac{5}{3} \bep \bar{\gamma}^{-1} \Sm^2 - 88 \bep^2 \bar{\gamma}^{-1} \Sm^2 - 532 \bem^2 \bar{\gamma}^{-2} \Sm^2   \nn\\
&\quad\qquad - 308 \bep^2 \bar{\gamma}^{-2} \Sm^2 + \frac{44}{3} \chip \Sm^2 - 28 \bar{\gamma}^{-1} \chip \Sm^2 -  \frac{77}{6} \zeta \Sm^4 -  \frac{77}{12} \zeta \gam \Sm^4 + \frac{5}{3} \bem \bar{\gamma}^{-1} \Sm \Sp + \frac{880}{9} \bem \bep \bar{\gamma}^{-1} \Sm \Sp   \nn\\
&\quad\qquad - 840 \bem \bep \bar{\gamma}^{-2} \Sm \Sp - 28 \bar{\gamma}^{-1} \chim \Sm \Sp -  \frac{11}{6} \zeta \Sp^4 -  \frac{11}{12} \zeta \gam \Sp^4\bigg)    \nn\\
&\quad +  \delta \nu\bigg(- \frac{1433}{18} \bem \Sm^2 + \frac{415}{3} \bem \bar{\gamma}^{-1} \Sm^2 + \frac{415}{3} \bep \bar{\gamma}^{-1} \Sm \Sp\bigg) + \frac{728}{9} \Sm^2 \nu^2 \,,\\
\frak{E}_5 &= \frac{13929}{280} \bar{\zeta}^{-1} + \frac{177}{10} \bep \bar{\zeta}^{-1} -  \frac{40}{3} \bep^2 \bar{\zeta}^{-1} + \frac{57629}{1440} \bar{\zeta}^{-1} \gam + \frac{68}{9} \bep \bar{\zeta}^{-1} \gam + \frac{1151}{144} \bar{\zeta}^{-1} \gam^2 + \frac{1}{96} \bar{\zeta}^{-1} \gam^3 -  \frac{5}{2} \bem^2 \bar{\zeta}^{-1} \bar{\gamma}^{-1}   \nn\\
&\quad   -  \frac{65}{2} \bep^2 \bar{\zeta}^{-1} \bar{\gamma}^{-1} + \frac{1679}{24} \Sm^2 -  \frac{86}{9} \bep \Sm^2 + \frac{2171}{36} \gam \Sm^2 -  \frac{16}{9} \bep \gam \Sm^2 + \frac{925}{72} \gam^2 \Sm^2 + \frac{80}{3} \bem^2 \bar{\gamma}^{-1} \Sm^2 - 14 \bep \bar{\gamma}^{-1} \Sm^2   \nn\\
&\quad + \frac{80}{3} \bep^2 \bar{\gamma}^{-1} \Sm^2 + 70 \bem^2 \bar{\gamma}^{-2} \Sm^2 + 70 \bep^2 \bar{\gamma}^{-2} \Sm^2 -  \frac{7}{12} \zeta \Sm^4 -  \frac{7}{24} \zeta \gam \Sm^4 -  \frac{44}{9} \bem \Sm \Sp - 14 \bem \bar{\gamma}^{-1} \Sm \Sp    \nn\\
&\quad + \frac{160}{3} \bem \bep \bar{\gamma}^{-1} \Sm \Sp + 140 \bem \bep \bar{\gamma}^{-2} \Sm \Sp -  \frac{1}{12} \zeta \Sp^4 -  \frac{1}{24} \zeta \gam \Sp^4   \nn\\
&\quad  + \delta \bigg(\frac{43}{30} \bem \bar{\zeta}^{-1} + \frac{2}{3} \bem \bar{\zeta}^{-1} \gam + 3 \bem \bep \bar{\zeta}^{-1} \bar{\gamma}^{-1} -  \frac{16}{9} \bem \Sm^2 + \frac{16}{9} \bem \gam \Sm^2 - 14 \bem \bar{\gamma}^{-1} \Sm^2 - 12 \bem \bep \bar{\gamma}^{-2} \Sm^2  \nn\\
&\quad\qquad  -  \frac{19}{24} \Sm \Sp -  \frac{58}{9} \bep \Sm \Sp -  \frac{5}{12} \gam \Sm \Sp -  \frac{1}{12} \gam^2 \Sm \Sp - 14 \bep \bar{\gamma}^{-1} \Sm \Sp - 6 \bem^2 \bar{\gamma}^{-2} \Sm \Sp - 6 \bep^2 \bar{\gamma}^{-2} \Sm \Sp  \nn\\
&\quad\qquad  -  \frac{1}{3} \zeta \Sm^3 \Sp -  \frac{1}{6} \zeta \gam \Sm^3 \Sp -  \frac{1}{3} \zeta \Sm \Sp^3 -  \frac{1}{6} \zeta \gam \Sm \Sp^3 \bigg)    \nn\\
&\quad  +\nu \bigg(- \frac{1664}{45} \bar{\zeta}^{-1} -  \frac{160}{9} \bep \bar{\zeta}^{-1} -  \frac{112}{9} \bar{\zeta}^{-1} \gam + \frac{80}{3} \bep^2 \bar{\zeta}^{-1} \bar{\gamma}^{-1} -  \frac{1247}{18} \Sm^2 + \frac{44}{9} \bep \Sm^2 -  \frac{280}{9} \gam \Sm^2 + \frac{112}{9} \bep \bar{\gamma}^{-1} \Sm^2  \nn\\
&\quad\qquad  -  \frac{160}{3} \bem^2 \bar{\gamma}^{-2} \Sm^2 -  \frac{160}{3} \bep^2 \bar{\gamma}^{-2} \Sm^2 + \frac{112}{9} \bem \bar{\gamma}^{-1} \Sm \Sp -  \frac{320}{3} \bem \bep \bar{\gamma}^{-2} \Sm \Sp \bigg)    \nn\\
&\quad  +\delta \nu  \bigg(- \frac{44}{9} \bem \Sm^2 + \frac{128}{9} \bem \bar{\gamma}^{-1} \Sm^2 + \frac{128}{9} \bep \bar{\gamma}^{-1} \Sm \Sp \bigg)  + \frac{80}{3} \Sm^2 \nu^2 \,,\\
\frak{E}_6 &= 55 \Sm^2 \Bigg[ 1 + \frac{31}{30} \gam + \frac{4}{15} \gam^2 -  \frac{2}{15} \zeta \Sm^2 -  \frac{1}{15} \zeta \gam \Sm^2 + \delta\bigg(- \frac{2}{15} \zeta \Sm \Sp -  \frac{1}{15} \zeta \gam \Sm \Sp\bigg)  \nn\\
&\qquad\qquad + \nu \bigg(- \frac{2}{5} + \frac{2}{15} \bep -  \frac{4}{15} \gam \bigg)  -  \frac{2}{15} \bem \delta \nu \Bigg]  \,,\\
\frak{E}_7 &= 35 \Sm^2 \Bigg[ 1 + \frac{31}{30} \gam + \frac{4}{15} \gam^2 -  \frac{2}{15} \zeta \Sm^2 -  \frac{1}{15} \zeta \gam \Sm^2 +  \delta\bigg(- \frac{2}{15} \zeta \Sm \Sp -  \frac{1}{15} \zeta \gam \Sm \Sp \bigg)  \nn\\
&\qquad\qquad + \nu \bigg(- \frac{2}{5} + \frac{2}{15} \bep -  \frac{4}{15} \gam\bigg)  -  \frac{2}{15} \bem \delta \nu \Bigg] \,,\\
\frak{E}_8 &= \frac{32}{5} \bar{\zeta}^{-1} \bigg( 1 + \frac{1}{2} \gam \bigg) \,,\\
\frak{E}_9 &= - \frac{8}{15} \bar{\zeta}^{-1} \gam + \frac{32}{15} \Sm^2 -  \frac{32}{15} \Sm \Sp \delta -  \frac{64}{15} \Sm^2 \nu \,,\\
\frak{E}_{10} &= - \frac{12}{5} \Sm^2 -  \frac{8}{9} \bep \Sm^2 -  \frac{4}{9} \gam \Sm^2 + \frac{8}{3} \bep \bar{\gamma}^{-1} \Sm^2 + \frac{8}{3} \bem \bar{\gamma}^{-1} \Sm \Sp \nn\\
&\quad + \delta \bigg(\frac{8}{9} \bem \Sm^2 -  \frac{8}{3} \bem \bar{\gamma}^{-1} \Sm^2 + \frac{16}{15} \Sm \Sp -  \frac{8}{3} \bep \bar{\gamma}^{-1} \Sm \Sp \bigg)   \,,\\
\frak{E}_{11} &= \frac{41}{45} \Sm^2 \nu\,,\\
\frak{E}_{12} &= - \frac{7}{3} \Sm^2 \Bigg[1 + \frac{4}{21} \bep + \frac{8}{21} \gam -  \frac{4}{21} \bem \delta -  \frac{11}{21} \nu \Bigg]\,,\\
\frak{E}_{13} &= \frac{11}{3} \Sm^2 \Bigg[1 + \frac{16}{33} \gam -  \frac{1}{3} \nu \Bigg] \,,\\
\frak{E}_{14} &= - \Sm^2 \Bigg[1 + \frac{4}{9} \gam -  \frac{5}{9} \nu \Bigg] \,,\\
\frak{E}_{15} &= \frac{16}{3} \bep \bar{\zeta}^{-1} -  \frac{8}{9} \bar{\zeta}^{-1} \gam - 8 \bep^2 \bar{\zeta}^{-1} \bar{\gamma}^{-1} + \frac{17}{9} \Sm^2 -  \frac{32}{3} \bep \bar{\gamma}^{-1} \Sm^2 + 16 \bem^2 \bar{\gamma}^{-2} \Sm^2 + 16 \bep^2 \bar{\gamma}^{-2} \Sm^2 -  \frac{32}{3} \bem \bar{\gamma}^{-1} \Sm \Sp \nn\\
&\quad  + 32 \bem \bep \bar{\gamma}^{-2} \Sm \Sp + \delta\bigg(\frac{8}{3} \bem \bar{\gamma}^{-1} \Sm^2 -  \frac{8}{9} \Sm \Sp + \frac{8}{3} \bep \bar{\gamma}^{-1} \Sm \Sp \bigg)  -  \frac{4}{9} \Sm^2 \nu \,.
\end{align}\ese
\section{Expression of the scalar dipole at 2.5PN order}
\label{app:scalarDipole}

Recall that the scalar dipole is decomposed into an instantaneous and a hereditary contribution:
\be \dI_i^{s}  = \dI_i^{s,\text{inst}}   + \dI_i^{s,\bm{\Gamma}} \,.\ee
The hereditary contribution appears at 2.5PN order and is given by
\be \dI_i^{s,\bm{\Gamma}} =  \frac{\sqrt{\alpha}\zeta (\Sp + \Sm \delta)}{(1-\zeta)\phi_0} \Gamma_i^s \,, \ee
while the instantaneous contribution is given with 2.5PN accuracy by
\begin{align}
\dI_i^{s,\text{inst}} &= - \frac{\alpha^{1/2} \zeta m \nu r}{(1 -  \zeta) \phi_0} \Biggl(2 \Sm n^{i} \nn\\
& + \frac{1}{5 c^2} \Bigg\{\left(\frac{\tilde{G}\alpha m}{r}\right)n^{i} \Bigg[-9 \Sm + 20 \bep \bar{\gamma}^{-1} \Sm + 20 \bem \bar{\gamma}^{-1} \Sp + 13 \Sm \nu +\delta  \Big(-20 \bem \bar{\gamma}^{-1} \Sm + 4 \Sp - 20 \bep \bar{\gamma}^{-1} \Sp \Big)  \Bigg] \nn\\
&\qquad\qquad + n^{i} v^2   \Big[\Sm + 4 \Sp \delta - 7 \Sm \nu \Big] +v^{i}  (nv)  \Big[2 \Sm - 2 \Sp \delta - 4 \Sm \nu \Big]\Bigg\} \nn\\
& + \frac{1}{3 c^3}  \left(\frac{\tilde{G}\alpha m}{r}\right)  v^{i} \Bigg\{ - \gam \Sm + 4 \zeta \Sm^3 + 4 \zeta \Sm^2 \Sp \delta - 8 \zeta \Sm^3 \nu \Bigg\} \nn\\
& + \frac{1}{140 c^4} \Bigg\{  \left(\frac{\tilde{G}\alpha m}{r}\right)^2 n^{i} \Bigg[ -38 \Sm + 224 \bep \Sm - 56 \gam \Sm - 672 \bep \bar{\gamma}^{-1} \Sm + 560 \bar{\gamma}^{-1} \chip \Sm + 224 \bem \Sp - 672 \bem \bar{\gamma}^{-1} \Sp \nn\\
& \qquad + 560 \bar{\gamma}^{-1} \chim \Sp + \delta \Big(-224 \bem \Sm + 672 \bem \bar{\gamma}^{-1} \Sm - 560 \bar{\gamma}^{-1} \chim \Sm - 32 \Sp - 224 \bep \Sp + 56 \gam \Sp + 672 \bep \bar{\gamma}^{-1} \Sp  \nn\\
& \qquad - 560 \bar{\gamma}^{-1} \chip \Sp \Big)  + \nu \Big(-1234 \Sm + 112 \bep \Sm - 728 \gam \Sm - 224 \bep \bar{\gamma}^{-1} \Sm - 4480 \bem^2 \bar{\gamma}^{-2} \Sm + 4480 \bep^2 \bar{\gamma}^{-2} \Sm \nn\\
& \qquad  - 1120 \bar{\gamma}^{-1} \chip \Sm - 896 \bem \Sp - 224 \bem \bar{\gamma}^{-1} \Sp - 1120 \bar{\gamma}^{-1} \chim \Sp  \Big)  \nn\\
& \qquad + \nu \delta \Big(448 \bem \Sm - 672 \bem \bar{\gamma}^{-1} \Sm + 120 \Sp - 672 \bep \bar{\gamma}^{-1} \Sp \Big)   + 470 \Sm \nu^2\Bigg] \nn\\
&\quad + n^{i}  v^{4} \Bigg[ 33 \Sm + 72 \Sp \delta  -279 \Sm\nu  - 312 \Sp \delta \nu  + 549 \Sm \nu^2\Bigg]\nn\\
&\quad +  n^{i}\left(\frac{\tilde{G}\alpha m}{r}\right) (nv)^2 \Bigg[ 105 \Sm + 56 \gam \Sm - 56 \bep \bar{\gamma}^{-1} \Sm - 56 \bem \bar{\gamma}^{-1} \Sp  - 245 \Sm \nu^2 \nn\\
& \qquad  + \delta\Big(56 \bem \bar{\gamma}^{-1} \Sm + 140 \Sp + 84 \gam \Sp + 56 \bep \bar{\gamma}^{-1} \Sp \Big)  + \nu \Big(-49 \Sm - 112 \gam \Sm - 112 \bep \bar{\gamma}^{-1} \Sm - 112 \bem \bar{\gamma}^{-1} \Sp  \Big)  \nn\\
& \qquad   + \nu \delta \Big(-336 \bem \bar{\gamma}^{-1} \Sm + 224 \Sp - 336 \bep \bar{\gamma}^{-1} \Sp \Big)  \Bigg]  \nn\\
&\quad  + n^{i}\left(\frac{\tilde{G}\alpha m}{r}\right)  v^2\Bigg[65 \Sm + 28 \gam \Sm - 224 \bep \bar{\gamma}^{-1} \Sm - 224 \bem \bar{\gamma}^{-1} \Sp - 675 \Sm \nu^2  \nn\\
& \qquad   + \delta \Big(224 \bem \bar{\gamma}^{-1} \Sm + 320 \Sp + 252 \gam \Sp + 224 \bep \bar{\gamma}^{-1} \Sp \Big)  + \nu\delta  \Big(336 \bem \bar{\gamma}^{-1} \Sm + 228 \Sp + 336 \bep \bar{\gamma}^{-1} \Sp \Big)  \nn\\
& \qquad    + \nu \Big(-1303 \Sm - 756 \gam \Sm + 672 \bep \bar{\gamma}^{-1} \Sm + 672 \bem \bar{\gamma}^{-1} \Sp  \Big)  \Bigg] \nn\\
&\quad +   v^{i} \left(\frac{\tilde{G}\alpha m}{r}\right) (nv) \Bigg[ 310 \Sm + 168 \gam \Sm + 112 \bep \bar{\gamma}^{-1} \Sm + 112 \bem \bar{\gamma}^{-1} \Sp - 514 \Sm \nu^2 \nn\\
&\qquad + \delta \Big(-112 \bem \bar{\gamma}^{-1} \Sm - 240 \Sp - 168 \gam \Sp - 112 \bep \bar{\gamma}^{-1} \Sp \Big)   + \nu \delta (112 \bem \bar{\gamma}^{-1} \Sm - 192 \Sp + 112 \bep \bar{\gamma}^{-1} \Sp)    \nn\\
&\qquad  + \nu \Big(1182 \Sm + 504 \gam \Sm - 336 \bep \bar{\gamma}^{-1} \Sm - 336 \bem \bar{\gamma}^{-1} \Sp \Big) \Bigg]  \nn\\
& \quad + v^{i} (nv) v^2  \Bigg[ 20 \Sm - 20 \Sp \delta + -164 \Sm \nu + 124 \Sp \delta + 376 \Sm \nu^2\Bigg]  \Bigg\}\nn\\
& + \frac{1}{60 c^5 (1 -  \zeta)}\Bigg\{ \left(\frac{\tilde{G}\alpha m}{r}\right)^2 n^{i} (nv) \Bigg[-40 \bep \Sm + 40 \bep \zeta \Sm + 30 \gam \Sm - 30 \zeta \gam \Sm - 40 \zeta \Sm^3 + 40 \zeta^2 \Sm^3 + 40 \bem \Sp \nn\\
&\qquad - 40 \bem \zeta \Sp +\delta \Big(40 \bem \Sm - 40 \bem \zeta \Sm - 40 \bep \Sp + 40 \bep \zeta \Sp + 15 \gam \Sp - 15 \zeta \gam \Sp - 30 \zeta \Sm^2 \Sp + 30 \zeta^2 \Sm^2 \Sp \nn\\
&\qquad  - 10 \zeta \Sp^3 + 10 \zeta^2 \Sp^3\Big)  + \nu\Big(-256 \Sm + 1120 \bep \Sm + 256 \zeta \Sm - 1120 \bep \zeta \Sm - 546 \gam \Sm + 546 \zeta \gam \Sm \nn\\
& \qquad - 2240 \bep \bar{\gamma}^{-1} \Sm  + 2240 \bep \zeta \bar{\gamma}^{-1} \Sm + 3840 \bem^2 \bar{\gamma}^{-2} \Sm + 7680 \bep^2 \bar{\gamma}^{-2} \Sm - 3840 \bem^2 \zeta \bar{\gamma}^{-2} \Sm - 7680 \bep^2 \zeta \bar{\gamma}^{-2} \Sm \nn\\
& \qquad+ 816 \zeta \Sm^3 - 816 \zeta^2 \Sm^3 - 1920 \bep \zeta \bar{\gamma}^{-1} \Sm^3 + 1920 \bep \zeta^2 \bar{\gamma}^{-1} \Sm^3 + 3840 \bep \zeta \bar{\gamma}^{-2} \Sm^3 - 3840 \bep \zeta^2 \bar{\gamma}^{-2} \Sm^3 \nn\\
& \qquad- 15360 \bem^2 \zeta \bar{\gamma}^{-3} \Sm^3 - 15360 \bep^2 \zeta \bar{\gamma}^{-3} \Sm^3 + 15360 \bem^2 \zeta^2 \bar{\gamma}^{-3} \Sm^3 + 15360 \bep^2 \zeta^2 \bar{\gamma}^{-3} \Sm^3 - 800 \bem \bar{\gamma}^{-1} \Sp \nn\\
& \qquad + 1280 \bem \zeta \bar{\gamma}^{-1} \Sp + 3840 \bem \bep \bar{\gamma}^{-2} \Sp - 3840 \bem \bep \zeta \bar{\gamma}^{-2} \Sp - 1920 \bem \zeta \bar{\gamma}^{-1} \Sm^2 \Sp + 1920 \bem \zeta^2 \bar{\gamma}^{-1} \Sm^2 \Sp \nn\\
& \qquad+ 2880 \bem \zeta \bar{\gamma}^{-2} \Sm^2 \Sp - 3840 \bem \zeta^2 \bar{\gamma}^{-2} \Sm^2 \Sp - 30720 \bem \bep \zeta \bar{\gamma}^{-3} \Sm^2 \Sp + 30720 \bem \bep \zeta^2 \bar{\gamma}^{-3} \Sm^2 \Sp \nn\\
& \qquad + 960 \bem \zeta \bar{\gamma}^{-2} \Sp^3  \Big) + \nu  \delta \Big(18 \gam \Sp - 18 \zeta \gam \Sp + 108 \zeta \Sm^2 \Sp - 108 \zeta^2 \Sm^2 \Sp + 36 \zeta \Sp^3 - 36 \zeta^2 \Sp^3 \Big)   - 448  \zeta(1  - \zeta)\Sm^3 \nu^2 \nn\\
& \qquad + \alpha^{1/2} \nu \Bigg(480 \bem + 240 \bem \gam + 480 \bem \zeta \Sm^2 + 960 \bem \zeta \bar{\gamma}^{-1} \Sm^2 - 592 \zeta \Sm \Sp + 960 \bep \zeta \Sm \Sp - 296 \zeta \gam \Sm \Sp \nn\\
& \qquad\qquad\quad + 1920 \bep \zeta \bar{\gamma}^{-1} \Sm \Sp + 480 \bem \zeta \Sp^2 + 960 \bem \zeta \bar{\gamma}^{-1} \Sp^2 - 960 \bem \Sm^2 \lambda_1 - 1920 \bem \bar{\gamma}^{-1} \Sm^2 \lambda_1 + 592 \Sm \Sp \lambda_1 \nn\\
& \qquad\qquad\quad - 960 \bep \Sm \Sp \lambda_1 + 296 \gam \Sm \Sp \lambda_1 - 1920 \bep \bar{\gamma}^{-1} \Sm \Sp \lambda_1  + \delta \Big(28 \gam + 14 \gam^2 + 56 \zeta \Sm^2 + 28 \zeta \gam \Sm^2 + 56 \zeta \Sp^2 \nn\\
& \qquad\qquad\quad+ 28 \zeta \gam \Sp^2 - 112 \Sm^2 \lambda_1 - 56 \gam \Sm^2 \lambda_1\Big)\Bigg)\Bigg]  \nn\\
&\quad +   v^{i}\left(\frac{\tilde{G}\alpha m}{r}\right)^2 \Bigg[-96 \Sm + 40 \bep \Sm + 96 \zeta \Sm - 40 \bep \zeta \Sm - 50 \gam \Sm + 50 \zeta \gam \Sm - 40 \zeta \Sm^3 + 40 \zeta^2 \Sm^3 - 40 \bem \Sp \nn\\
& \qquad + 40 \bem \zeta \Sp + \delta \Big(-40 \bem \Sm + 40 \bem \zeta \Sm - 96 \Sp + 40 \bep \Sp + 96 \zeta \Sp - 40 \bep \zeta \Sp - 60 \gam \Sp + 65 \zeta \gam \Sp \nn\\
& \qquad  - 40 \zeta \Sm^2 \Sp + 30 \zeta^2 \Sm^2 \Sp + 10 \zeta^2 \Sp^3 \Big)  + \nu\Big(768 \Sm - 640 \bep \Sm - 768 \zeta \Sm + 640 \bep \zeta \Sm + 522 \gam \Sm - 522 \zeta \gam \Sm \nn\\
& \qquad + 32 \zeta \Sm^3 - 32 \zeta^2 \Sm^3 + 960 \bep \zeta \bar{\gamma}^{-1} \Sm^3 - 960 \bep \zeta^2 \bar{\gamma}^{-1} \Sm^3 + 120 \bem \Sp - 120 \bem \zeta \Sp + 720 \bem \zeta \bar{\gamma}^{-1} \Sm^2 \Sp  \nn\\
& \qquad  - 720 \bem \zeta^2 \bar{\gamma}^{-1} \Sm^2 \Sp + 240 \bem \zeta \bar{\gamma}^{-1} \Sp^3 - 240 \bem \zeta^2 \bar{\gamma}^{-1} \Sp^3 \Big)  + \delta \nu \Big(960 \bem \zeta \bar{\gamma}^{-1} \Sm^3 - 960 \bem \zeta^2 \bar{\gamma}^{-1} \Sm^3 + 120 \bep \Sp   \nn\\
& \qquad  - 120 \bep \zeta \Sp + 4 \zeta \gam \Sp - 32 \zeta \Sm^2 \Sp + 24 \zeta^2 \Sm^2 \Sp + 720 \bep \zeta \bar{\gamma}^{-1} \Sm^2 \Sp - 720 \bep \zeta^2 \bar{\gamma}^{-1} \Sm^2 \Sp + 8 \zeta^2 \Sp^3  \nn\\
& \qquad + 240 \bep \zeta \bar{\gamma}^{-1} \Sp^3 - 240 \bep \zeta^2 \bar{\gamma}^{-1} \Sp^3 \Big) - 544 \zeta (  1 -  \zeta )\Sm^3 \nu^2 \nn\\
&\qquad + \alpha^{1/2} \nu \Bigg(16 \zeta \Sm \Sp + 8 \zeta \gam \Sm \Sp - 16 \Sm \Sp \lambda_1 - 8 \gam \Sm \Sp \lambda_1  \nn\\
& \qquad\qquad\quad  + \delta \Big(-4 \gam - 2 \gam^2 - 8 \zeta \Sm^2 - 4 \zeta \gam \Sm^2 - 8 \zeta \Sp^2 - 4 \zeta \gam \Sp^2 + 16 \Sm^2 \lambda_1 + 8 \gam \Sm^2 \lambda_1 \Big)\Bigg)\Bigg] \nn\\
& + v^{i} \left(\frac{\tilde{G}\alpha m}{r}\right)v^2 \Bigg[48 \Sm - 48 \zeta \Sm + 10 \gam \Sm - 10 \zeta \gam \Sm + 40 \zeta \Sm^3 - 40 \zeta^2 \Sm^3  \nn\\
&  \qquad\qquad\qquad\quad +\delta  \Big(48 \Sp - 48 \zeta \Sp + 20 \gam \Sp - 20 \zeta \gam \Sp + 40 \zeta \Sm^2 \Sp - 40 \zeta^2 \Sm^2 \Sp \Big) - 160 \zeta \Sm^2 \Sp (  1 -  \zeta  )  \nu \delta   \nn\\
&  \qquad\qquad\qquad\quad + \nu \Big(-192 \Sm + 192 \zeta \Sm - 36 \gam \Sm + 36 \zeta \gam \Sm - 240 \zeta \Sm^3 + 240 \zeta^2 \Sm^3   \Big)  + 288 \zeta \Sm^3 (1  -  \zeta ) \nu^2\Bigg] \Bigg\} \Biggr) \nn\\
& + \mathcal{O}\left(\frac{1}{c^6}\right)
\end{align}
This result agrees with \cite{TLB22} at 2PN order.

\section{Expression of the instantaneous angular momentum fluxes in terms of the phase space variables}
\label{app:inst_Gsi_orbital_vars}
 
Recall Eq.~\eqref{seq:inst_Gsi_orbital_vars}:
\begin{align*}
\mathcal{G}_i^{s, \mathrm{inst}} &= \frac{m \nu^2}{3 c^3}\Big(\frac{\tilde{G}\alpha m}{r}\Big)^2   \epsilon_{ijk} n_{j} v_{k} \Bigg\{\widetilde{A}^{-1\text{PN}} + \frac{1}{c^2}\Bigg[\widetilde{B}_1^{\text{N}} \Big(\frac{\tilde{G}\alpha m}{r}\Big) + \widetilde{B}_2^{\text{N}} (nv)^2 + \widetilde{B}_3^{\text{N}} v^2\Bigg]+ \frac{1}{c^3} \widetilde{C}^{0.5\text{PN}}  \Big(\frac{\tilde{G}\alpha m}{r}\Big) (nv) \nn\\
& + \frac{1}{c^4} \Bigg[\widetilde{D}_1^{1\text{PN}} \Big(\frac{\tilde{G}\alpha m}{r}\Big)^2 + \widetilde{D}_2^{1\text{PN}} \Big(\frac{\tilde{G}\alpha m}{r}\Big) (nv)^2 + \widetilde{D}_3^{1\text{PN}}  \Big(\frac{\tilde{G}\alpha m}{r}\Big) v^2 + \widetilde{D}_4^{1\text{PN}} (nv)^4 +  \widetilde{D}_5^{1\text{PN}}(nv)^2 v^2 + \widetilde{D}_6^{1\text{PN}} v^4  \Bigg] \nn\\
& + \frac{1}{c^5} \Bigg[ \widetilde{E}_1^{1.5\text{PN}} \Big(\frac{\tilde{G}\alpha m}{r}\Big)^2 (nv) + \widetilde{E}_2^{1.5\text{PN}} \Big(\frac{\tilde{G}\alpha m}{r}\Big)(nv)^3 + \widetilde{E}_3^{1.5\text{PN}} \Big(\frac{\tilde{G}\alpha m}{r}\Big) (nv) v^2 \Bigg]  \Bigg\} + \calO\left(\frac{1}{c^9}\right) \,.
\end{align*}
 The coefficients are given by:
\bse\begin{align}
\widetilde{A}^{-1\text{PN}} &= 4 \zeta \Sm^2 \\
\widetilde{B}_1^{\text{N}}  &= - \frac{4}{5} \gam -  \frac{98}{5} \zeta \Sm^2 - 8 \bep \zeta \Sm^2 + \
\frac{8 \bep \zeta \Sm^2}{\gam} - 8 \zeta \gam \Sm^2 + \frac{8 \bem \
\zeta \Sm \Sp}{\gam} \nn\\
&\quad  + \delta\bigg[8 \bem \zeta \Sm^2 -  \frac{8 \bem \zeta \
\Sm^2}{\gam} -  \frac{12}{5} \zeta \Sm \Sp -  \frac{8 \bep \zeta \Sm \
\Sp}{\gam}\bigg]  + \frac{6}{5} \zeta \Sm^2 \nu  \\
\widetilde{B}_2^{\text{N}}  &= \frac{6}{5} \gam + \frac{6}{5} \zeta \Sm^2 -  \frac{24 \bep \zeta \Sm^2}{\gam} -  \frac{24 \bem \zeta \Sm \Sp}{\gam} \nn\\
&\quad + \delta \bigg[\frac{24 \bem \zeta \Sm^2}{\gam} -  \frac{36}{5} \zeta \Sm \Sp + \frac{24 \bep \zeta \Sm \Sp}{\gam}\bigg]  + \frac{48}{5} \zeta \Sm^2 \nu \\
\widetilde{B}_3^{\text{N}}  &= - \frac{4}{5} \gam + \frac{38}{5} \zeta \Sm^2 + \frac{8 \bep \zeta \Sm^2}{\gam} + 4 \zeta \gam \Sm^2 + \frac{8 \bem \zeta \Sm \Sp}{\gam}\nn\\
&\quad + \delta \bigg[- \frac{8 \bem \zeta \Sm^2}{\gam} + \frac{12}{5} \zeta \Sm \Sp -  \frac{8 \bep \zeta \Sm \Sp}{\gam}\bigg]  -  \frac{26}{5} \zeta \Sm^2 \nu \\
\widetilde{C}^{0.5\text{PN}} &= \frac{1}{4} \gam^2 + \zeta \gam \Sm^2 - 7 \zeta^2 \Sm^4 -  \zeta^2 \Sp^4 + \delta \bigg[-2 \zeta \gam \Sm \Sp - 4 \zeta^2 \Sm^3 \Sp - 4 \zeta^2 \Sm \Sp^3\bigg]   \\
\widetilde{D}_1^{1\text{PN}} &= -4 \bep + \frac{251}{35} \gam + \frac{16}{5} \bep \gam + \frac{14}{5} \gam^2 -  \frac{1}{8} \gam^3 + \frac{1147}{35} \zeta \Sm^2 + \frac{116}{5} \bep \zeta \Sm^2 -  \frac{16 \bem^2 \zeta \Sm^2}{\gam} -  \frac{104 \bep \zeta \Sm^2}{5 \gam} \nn\\
&\quad -  \frac{16 \bep^2 \zeta \Sm^2}{\gam}  + \frac{166}{5} \zeta \gam \Sm^2 + 16 \bep \zeta \gam \Sm^2 + \frac{17}{2} \zeta \gam^2 \Sm^2 - 8 \zeta \chip \Sm^2 + 8 \zeta^2 \Sm^4 + \frac{7}{2} \zeta^2 \gam \Sm^4 -  \frac{144}{5} \bem \zeta \Sm \Sp \nn\\
&\quad -  \frac{104 \bem \zeta \Sm \Sp}{5 \gam} -  \frac{32 \bem \bep \zeta \Sm \Sp}{\gam} + \frac{1}{2} \zeta^2 \gam \Sp^4 \nn\\
& + \delta \bigg[4 \bem -  \frac{16}{5} \bem \gam -  \frac{116}{5} \bem \zeta \Sm^2 + \frac{104 \bem \zeta \Sm^2}{5 \gam} + \frac{32 \bem \bep \zeta \Sm^2}{\gam} - 16 \bem \zeta \gam \Sm^2 + 8 \zeta \chim \Sm^2 + \frac{848}{35} \zeta \Sm \Sp \nn\\
&\qquad + \frac{144}{5} \bep \zeta \Sm \Sp + \frac{16 \bem^2 \zeta \Sm \Sp}{\gam} + \frac{104 \bep \zeta \Sm \Sp}{5 \gam} + \frac{16 \bep^2 \zeta \Sm \Sp}{\gam} + \frac{44}{5} \zeta \gam \Sm \Sp + \zeta \gam^2 \Sm \Sp  \nn\\
&\qquad  + 8 \zeta^2 \Sm^3 \Sp + 2 \zeta^2 \gam \Sm^3 \Sp + 2 \zeta^2 \gam \Sm \Sp^3\bigg]  \nn\\
&+\nu  \bigg[ \frac{56}{5} \bep -  \frac{18}{35} \gam + \frac{1}{4} \gam^3 + \frac{1292}{35} \zeta \Sm^2 + \frac{168}{5} \bep \zeta \Sm^2 + \frac{160 \bem^2 \zeta \Sm^2}{\gam} -  \frac{208 \bep \zeta \Sm^2}{5 \gam} -  \frac{96 \bep^2 \zeta \Sm^2}{\gam} + \frac{48}{5} \zeta \gam \Sm^2 \nn\\
&\qquad -  \zeta \gam^2 \Sm^2 + 16 \zeta \chip \Sm^2 - 16 \zeta^2 \Sm^4 - 7 \zeta^2 \gam \Sm^4 + \frac{256}{5} \bem \zeta \Sm \Sp -  \frac{208 \bem \zeta \Sm \Sp}{5 \gam} + \frac{64 \bem \bep \zeta \Sm \Sp}{\gam} -  \zeta^2 \gam \Sp^4 \bigg]  \nn\\
& + \nu \delta  \bigg[- \frac{8}{5} \bem \zeta \Sm^2 + \frac{16 \bem \zeta \Sm^2}{5 \gam} -  \frac{156}{35} \zeta \Sm \Sp + \frac{16 \bep \zeta \Sm \Sp}{5 \gam}\bigg]  -  \frac{191}{35} \zeta \Sm^2 \nu^2 \\
\widetilde{D}_2^{1\text{PN}}  &= \frac{36}{5} \bep + \frac{8 \bem^2}{\gam} + \frac{8 \bep^2}{\gam} -  \frac{184}{35} \gam -  \frac{16}{5} \bep \gam -  \frac{3}{10} \gam^2 + \frac{1}{4} \gam^3 -  \frac{299}{70} \zeta \Sm^2 -  \frac{32 \bem^2 \zeta \Sm^2}{\gam^2} -  \frac{32 \bep^2 \zeta \Sm^2}{\gam^2}  \nn\\
&\quad + \frac{476 \bep \zeta \Sm^2}{5 \gam} -  \frac{22}{5} \zeta \gam \Sm^2 -  \zeta \gam^2 \Sm^2 -  \frac{64 \zeta \chip \Sm^2}{\gam} - 14 \zeta^2 \Sm^4 - 7 \zeta^2 \gam \Sm^4 - 32 \bem \zeta \Sm \Sp \nn\\
&\quad -  \frac{64 \bem \bep \zeta \Sm \Sp}{\gam^2} + \frac{476 \bem \zeta \Sm \Sp}{5 \gam} -  \frac{64 \zeta \chim \Sm \Sp}{\gam} - 2 \zeta^2 \Sp^4 -  \zeta^2 \gam \Sp^4  \nn\\
&\quad + \delta \bigg[ - \frac{36}{5} \bem -  \frac{16 \bem \bep}{\gam} + \frac{16}{5} \bem \gam + \frac{64 \bem \bep \zeta \Sm^2}{\gam^2} -  \frac{476 \bem \zeta \Sm^2}{5 \gam} + \frac{64 \zeta \chim \Sm^2}{\gam} -  \frac{498}{35} \zeta \Sm \Sp  \nn\\
&\qquad + 32 \bep \zeta \Sm \Sp+ \frac{32 \bem^2 \zeta \Sm \Sp}{\gam^2} + \frac{32 \bep^2 \zeta \Sm \Sp}{\gam^2} -  \frac{476 \bep \zeta \Sm \Sp}{5 \gam} -  \frac{138}{5} \zeta \gam \Sm \Sp - 2 \zeta \gam^2 \Sm \Sp \nn\\
&\qquad + \frac{64 \zeta \chip \Sm \Sp}{\gam}  - 8 \zeta^2 \Sm^3 \Sp  - 4 \zeta^2 \gam \Sm^3 \Sp - 8 \zeta^2 \Sm \Sp^3 - 4 \zeta^2 \gam \Sm \Sp^3 \bigg]   \nn\\
&\quad + \nu \bigg[ -8 \bep -  \frac{32 \bep^2}{\gam} + \frac{89}{35} \gam -  \frac{1821}{70} \zeta \Sm^2 - 48 \bep \zeta \Sm^2 + \frac{576 \bem^2 \zeta \Sm^2}{\gam^2} -  \frac{448 \bep^2 \zeta \Sm^2}{\gam^2} + \frac{92 \bep \zeta \Sm^2}{5 \gam} \nn\\
&\qquad + \frac{44}{5} \zeta \gam \Sm^2  + \frac{128 \zeta \chip \Sm^2}{\gam} + 128 \bem \zeta \Sm \Sp + \frac{128 \bem \bep \zeta \Sm \Sp}{\gam^2} + \frac{92 \bem \zeta \Sm \Sp}{5 \gam} + \frac{128 \zeta \chim \Sm \Sp}{\gam} \bigg]  \nn\\
&\quad+\nu  \delta \bigg[16 \bem \zeta \Sm^2 + \frac{436 \bem \zeta \Sm^2}{5 \gam} -  \frac{1608}{35} \zeta \Sm \Sp + \frac{436 \bep \zeta \Sm \Sp}{5 \gam} \bigg]   + \frac{1341}{70} \zeta \Sm^2 \nu^2 \\
\widetilde{D}_3^{1\text{PN}}  &= - \frac{4}{5} \bep -  \frac{8 \bem^2}{\gam} -  \frac{8 \bep^2}{\gam} + \frac{10}{7} \gam + \frac{8}{5} \bep \gam -  \frac{887}{70} \zeta \Sm^2 + 8 \bep \zeta \Sm^2 + \frac{32 \bem^2 \zeta \Sm^2}{\gam^2} + \frac{32 \bep^2 \zeta \Sm^2}{\gam^2} -  \frac{12 \bep \zeta \Sm^2}{\gam} \nn\\
&\quad -  \frac{32}{5} \zeta \gam \Sm^2 + \frac{16 \zeta \chip \Sm^2}{\gam} + 16 \bem \zeta \Sm \Sp + \frac{64 \bem \bep \zeta \Sm \Sp}{\gam^2} -  \frac{12 \bem \zeta \Sm \Sp}{\gam} + \frac{16 \zeta \chim \Sm \Sp}{\gam} \nn\\
&\quad + \delta \bigg[\frac{4}{5} \bem + \frac{16 \bem \bep}{\gam} -  \frac{8}{5} \bem \gam - 8 \bem \zeta \Sm^2 -  \frac{64 \bem \bep \zeta \Sm^2}{\gam^2} + \frac{12 \bem \zeta \Sm^2}{\gam} -  \frac{16 \zeta \chim \Sm^2}{\gam} + \frac{286}{35} \zeta \Sm \Sp  \nn\\
&\qquad - 16 \bep \zeta \Sm \Sp -  \frac{32 \bem^2 \zeta \Sm \Sp}{\gam^2} -  \frac{32 \bep^2 \zeta \Sm \Sp}{\gam^2} + \frac{12 \bep \zeta \Sm \Sp}{\gam} + \frac{42}{5} \zeta \gam \Sm \Sp -  \frac{16 \zeta \chip \Sm \Sp}{\gam}\bigg]  \nn\\
&\quad + \nu \bigg[- \frac{64}{5} \bep + \frac{32 \bep^2}{\gam} + \frac{114}{35} \gam + \frac{1133}{70} \zeta \Sm^2 + 16 \bep \zeta \Sm^2 -  \frac{192 \bem^2 \zeta \Sm^2}{\gam^2} + \frac{64 \bep^2 \zeta \Sm^2}{\gam^2} -  \frac{12 \bep \zeta \Sm^2}{\gam}  \nn\\
&\qquad + \frac{4}{5} \zeta \gam \Sm^2 -  \frac{32 \zeta \chip \Sm^2}{\gam}  - 32 \bem \zeta \Sm \Sp -  \frac{128 \bem \bep \zeta \Sm \Sp}{\gam^2} -  \frac{12 \bem \zeta \Sm \Sp}{\gam} -  \frac{32 \zeta \chim \Sm \Sp}{\gam} \bigg]  \nn\\
&\quad + \nu  \delta\bigg[-8 \bem \zeta \Sm^2 -  \frac{12 \bem \zeta \Sm^2}{\gam} + \frac{408}{35} \zeta \Sm \Sp -  \frac{12 \bep \zeta \Sm \Sp}{\gam}\bigg]  -  \frac{171}{70} \zeta \Sm^2 \nu^2 \\
\widetilde{D}_4^{1\text{PN}}  &= -12 \bep + \frac{33}{14} \gam -  \frac{447}{14} \zeta \Sm^2 + \frac{60 \bep \zeta \Sm^2}{\gam} - 12 \zeta \gam \Sm^2 + \frac{60 \bem \zeta \Sm \Sp}{\gam} \nn\\
&\quad +  \delta \bigg[ 12 \bem -  \frac{60 \bem \zeta \Sm^2}{\gam} -  \frac{144}{7} \zeta \Sm \Sp -  \frac{60 \bep \zeta \Sm \Sp}{\gam} - 18 \zeta \gam \Sm \Sp\bigg]   \nn\\
&\quad + \nu \bigg[ 24 \bep -  \frac{36}{7} \gam + \frac{390}{7} \zeta \Sm^2 -  \frac{120 \bep \zeta \Sm^2}{\gam} + 24 \zeta \gam \Sm^2 -  \frac{120 \bem \zeta \Sm \Sp}{\gam}\bigg]  \nn\\
&\quad +  \nu\delta  \bigg[\frac{120 \bem \zeta \Sm^2}{\gam} -  \frac{258}{7} \zeta \Sm \Sp + \frac{120 \bep \zeta \Sm \Sp}{\gam}\bigg]  + \frac{36}{7} \zeta \Sm^2 \nu^2 \\
\widetilde{D}_5^{1\text{PN}}  &= 12 \bep -  \frac{12}{7} \gam + \frac{1017}{35} \zeta \Sm^2 -  \frac{312 \bep \zeta \Sm^2}{5 \gam} + \frac{54}{5} \zeta \gam \Sm^2 -  \frac{312 \bem \zeta \Sm \Sp}{5 \gam} \nn\\
&\quad + \delta \bigg[ -12 \bem + \frac{312 \bem \zeta \Sm^2}{5 \gam} + \frac{768}{35} \zeta \Sm \Sp + \frac{312 \bep \zeta \Sm \Sp}{5 \gam} + \frac{96}{5} \zeta \gam \Sm \Sp\bigg]  \nn\\
&\quad+ \nu \bigg[ - \frac{96}{5} \bep + \frac{39}{35} \gam -  \frac{561}{35} \zeta \Sm^2 + \frac{636 \bep \zeta \Sm^2}{5 \gam} -  \frac{48}{5} \zeta \gam \Sm^2 + \frac{636 \bem \zeta \Sm \Sp}{5 \gam}\bigg]  \nn\\
&\quad+  \nu \delta  \bigg[ - \frac{852 \bem \zeta \Sm^2}{5 \gam} + \frac{1782}{35} \zeta \Sm \Sp -  \frac{852 \bep \zeta \Sm \Sp}{5 \gam}\bigg] -  \frac{906}{35} \zeta \Sm^2 \nu^2  \\
\widetilde{D}_6^{1\text{PN}}  &= - \frac{8}{5} \bep -  \frac{61}{70} \gam -  \frac{2}{5} \gam^2 + \frac{47}{10} \zeta \Sm^2 + \frac{44 \bep \zeta \Sm^2}{5 \gam} + \frac{14}{5} \zeta \gam \Sm^2 + \frac{44 \bem \zeta \Sm \Sp}{5 \gam}  \nn\\
&\quad +\delta  \bigg[\frac{8}{5} \bem -  \frac{44 \bem \zeta \Sm^2}{5 \gam} -  \frac{16}{5} \zeta \Sm \Sp -  \frac{44 \bep \zeta \Sm \Sp}{5 \gam} -  \frac{14}{5} \zeta \gam \Sm \Sp\bigg]   \nn\\
&\quad + \nu \bigg[\frac{8}{5} \bep + \frac{58}{35} \gam -  \frac{273}{10} \zeta \Sm^2 -  \frac{92 \bep \zeta \Sm^2}{5 \gam} -  \frac{58}{5} \zeta \gam \Sm^2 -  \frac{92 \bem \zeta \Sm \Sp}{5 \gam}\bigg]    \nn\\
&\quad + \nu \delta  \bigg[ \frac{164 \bem \zeta \Sm^2}{5 \gam} -  \frac{48}{5} \zeta \Sm \Sp + \frac{164 \bep \zeta \Sm \Sp}{5 \gam}\bigg]  + \frac{131}{10} \zeta \Sm^2 \nu^2 \\
\widetilde{E}_1^{1.5\text{PN}} &= - \frac{12}{5} \gam -  \frac{16}{5} \gam^2 -  \frac{2}{3} \bep \gam^2 -  \frac{2}{3} \gam^3 -  \frac{48}{5} \zeta \gam \Sm^2 -  \frac{8}{3} \bep \zeta \gam \Sm^2 -  \frac{8}{3} \zeta \gam^2 \Sm^2 + 56 \zeta^2 \Sm^4 + \frac{56}{3} \bep \zeta^2 \Sm^4  \nn\\
&\quad + \frac{56}{3} \zeta^2 \gam \Sm^4  -  \frac{16}{3} \bem \zeta \gam \Sm \Sp -  \frac{32}{3} \bem \zeta^2 \Sm^3 \Sp -  \frac{32}{3} \bem \zeta^2 \Sm \Sp^3 + 8 \zeta^2 \Sp^4 + \frac{8}{3} \bep \zeta^2 \Sp^4 + \frac{8}{3} \zeta^2 \gam \Sp^4  \nn\\
&\quad +  \delta \bigg[\frac{2}{3} \bem \gam^2 + \frac{8}{3} \bem \zeta \gam \Sm^2 -  \frac{56}{3} \bem \zeta^2 \Sm^4 + \frac{72}{5} \zeta \gam \Sm \Sp + \frac{16}{3} \bep \zeta \gam \Sm \Sp + \frac{16}{3} \zeta \gam^2 \Sm \Sp + 32 \zeta^2 \Sm^3 \Sp  \nn\\
&\qquad  + \frac{32}{3} \bep \zeta^2 \Sm^3 \Sp + \frac{32}{3} \zeta^2 \gam \Sm^3 \Sp + 32 \zeta^2 \Sm \Sp^3 + \frac{32}{3} \bep \zeta^2 \Sm \Sp^3 + \frac{32}{3} \zeta^2 \gam \Sm \Sp^3 -  \frac{8}{3} \bem \zeta^2 \Sp^4\bigg]  \nn\\
&\quad + \nu \bigg[ 24 \bep - 8 \bep^2 -  \frac{96 \bem^2}{\gam} -  \frac{96 \bep^2}{\gam} + \frac{227}{120} \gam^2 -  \gam^3 -  \frac{928}{15} \zeta \Sm^2 -  \frac{80}{3} \bep \zeta \Sm^2 -  \frac{384 \bem^2 \zeta \Sm^2}{\gam^2}  \nn\\
&\qquad -  \frac{1152 \bep^2 \zeta \Sm^2}{\gam^2}   + \frac{352 \bep \zeta \Sm^2}{\gam} -  \frac{96 \bep^2 \zeta \Sm^2}{\gam} + \frac{89}{6} \zeta \gam \Sm^2 + 4 \zeta \gam^2 \Sm^2 -  \frac{1589}{30} \zeta^2 \Sm^4 + \frac{2688 \bem^2 \zeta^2 \Sm^4}{\gam^3} \nn\\
&\qquad  + \frac{2688 \bep^2 \zeta^2 \Sm^4}{\gam^3}   + \frac{256 \bem^2 \zeta^2 \Sm^4}{\gam^2} -  \frac{672 \bep \zeta^2 \Sm^4}{\gam^2} + \frac{224 \bep^2 \zeta^2 \Sm^4}{\gam^2} + \frac{32 \bep \zeta^2 \Sm^4}{\gam} + 28 \zeta^2 \gam \Sm^4  \nn\\
&\qquad   -  \frac{768 \bem \bep \zeta \Sm \Sp}{\gam^2} + \frac{256 \bem \zeta \Sm \Sp}{\gam} + \frac{64}{3} \bem \zeta \gam \Sm \Sp + \frac{128}{3} \bem \zeta^2 \Sm^3 \Sp + \frac{6144 \bem \bep \zeta^2 \Sm^3 \Sp}{\gam^3}   \nn\\
&\qquad -  \frac{768 \bem \zeta^2 \Sm^3 \Sp}{\gam^2}   + \frac{512 \bem \bep \zeta^2 \Sm^3 \Sp}{\gam^2} + \frac{32 \bem \zeta^2 \Sm^3 \Sp}{\gam} + \frac{128}{3} \bem \zeta^2 \Sm \Sp^3 -  \frac{227}{30} \zeta^2 \Sp^4  \nn\\
&\qquad + \frac{384 \bem^2 \zeta^2 \Sp^4}{\gam^3} + \frac{384 \bep^2 \zeta^2 \Sp^4}{\gam^3}   -  \frac{96 \bep \zeta^2 \Sp^4}{\gam^2} + \frac{32 \bep^2 \zeta^2 \Sp^4}{\gam^2} + 4 \zeta^2 \gam \Sp^4\bigg]    \nn\\
&\quad + \nu \delta    \bigg[\bem \gam - 4 \bem \zeta \Sm^2 -  \frac{28 \bem \zeta^2 \Sm^4}{\gam} - 8 \bep \zeta \Sm \Sp + \frac{7}{15} \zeta \gam \Sm \Sp + \frac{14}{15} \zeta^2 \Sm^3 \Sp  \nn\\
&\qquad -  \frac{16 \bep \zeta^2 \Sm^3 \Sp}{\gam} + \frac{14}{15} \zeta^2 \Sm \Sp^3 -  \frac{16 \bep \zeta^2 \Sm \Sp^3}{\gam} -  \frac{4 \bem \zeta^2 \Sp^4}{\gam}\bigg ]  \nn\\
&\quad + \nu^2 \bigg[- \frac{17}{20} \gam^2 + \frac{17}{5} \zeta \gam \Sm^2 + \frac{119}{5} \zeta^2 \Sm^4 + \frac{17}{5} \zeta^2 \Sp^4\bigg]  \\
\widetilde{E}_2^{1.5\text{PN}} &= 6 \gam + \frac{19}{8} \gam^2 + 20 \bep \zeta \Sm^2 -  \frac{25}{2} \zeta \gam \Sm^2 + \frac{35}{2} \zeta^2 \Sm^4 - 20 \bem \zeta \Sm \Sp + \frac{5}{2} \zeta^2 \Sp^4 \nn\\
&\quad + \delta \bigg[ -20 \bem \zeta \Sm^2 + 20 \bep \zeta \Sm \Sp - 5 \zeta \gam \Sm \Sp + 10 \zeta^2 \Sm^3 \Sp + 10 \zeta^2 \Sm \Sp^3 \bigg]   \nn\\
&\quad + \nu \bigg[ 60 \bep -  \frac{240 \bem^2}{\gam} -  \frac{240 \bep^2}{\gam} + \frac{19}{2} \gam^2 + 20 \zeta \Sm^2 - 480 \bep \zeta \Sm^2 -  \frac{960 \bem^2 \zeta \Sm^2}{\gam^2} -  \frac{2880 \bep^2 \zeta \Sm^2}{\gam^2} + \frac{880 \bep \zeta \Sm^2}{\gam}  \nn\\
&\qquad + 116 \zeta \gam \Sm^2 + 60 \zeta \phi_0 \Sm^2 + 30 \zeta \gam \phi_0 \Sm^2 - 266 \zeta^2 \Sm^4 + \frac{6720 \bem^2 \zeta^2 \Sm^4}{\gam^3} + \frac{6720 \bep^2 \zeta^2 \Sm^4}{\gam^3} -  \frac{1680 \bep \zeta^2 \Sm^4}{\gam^2}  \nn\\
&\qquad+ \frac{960 \bep \zeta^2 \Sm^4}{\gam} -  \frac{1920 \bem \bep \zeta \Sm \Sp}{\gam^2} + \frac{640 \bem \zeta \Sm \Sp}{\gam} + \frac{15360 \bem \bep \zeta^2 \Sm^3 \Sp}{\gam^3} -  \frac{1920 \bem \zeta^2 \Sm^3 \Sp}{\gam^2}  \nn\\
&\qquad + \frac{960 \bem \zeta^2 \Sm^3 \Sp}{\gam} - 38 \zeta^2 \Sp^4 + \frac{960 \bem^2 \zeta^2 \Sp^4}{\gam^3} + \frac{960 \bep^2 \zeta^2 \Sp^4}{\gam^3} -  \frac{240 \bep \zeta^2 \Sp^4}{\gam^2} \bigg]  \nn\\
&\quad + \nu \delta   \bigg[ -4 \zeta \gam \Sm \Sp - 8 \zeta^2 \Sm^3 \Sp - 8 \zeta^2 \Sm \Sp^3 \bigg]   +  \nu^2 \bigg[- \gam^2 + 4 \zeta \gam \Sm^2 + 28 \zeta^2 \Sm^4 + 4 \zeta^2 \Sp^4 \bigg] \\
\widetilde{E}_3^{1.5\text{PN}} &= - \frac{24}{5} \gam - \! \frac{61}{40} \gam^2 + \! \frac{1}{4} \gam^3 - \! 12 \bep \zeta \Sm^2 + \! \frac{103}{10} \zeta \gam \Sm^2 + \! \zeta \gam^2 \Sm^2 - \!  \frac{49}{2} \zeta^2 \Sm^4  - \!7 \zeta^2 \gam \Sm^4 + \! 12 \bem \zeta \Sm \Sp - \!  \frac{7}{2} \zeta^2 \Sp^4 - \! \zeta^2 \gam \Sp^4  \nn\\
&\quad + \delta \bigg[ 12 \bem \zeta \Sm^2 - \! 12 \bep \zeta \Sm \Sp - \!  \frac{1}{5} \zeta \gam \Sm \Sp - \! 2 \zeta \gam^2 \Sm \Sp - \! 14 \zeta^2 \Sm^3 \Sp - \! 4 \zeta^2 \gam \Sm^3 \Sp -  \! 14 \zeta^2 \Sm \Sp^3 - \! 4 \zeta^2 \gam \Sm \Sp^3\bigg]   \nn\\
&\quad + \nu \bigg[  -36 \bep + \frac{144 \bem^2}{\gam} + \frac{144 \bep^2}{\gam} -  \frac{59}{10} \gam^2 -  \frac{12}{5} \zeta \Sm^2 + 288 \bep \zeta \Sm^2 + \frac{576 \bem^2 \zeta \Sm^2}{\gam^2} + \frac{1728 \bep^2 \zeta \Sm^2}{\gam^2}\nn\\
&\qquad  -  \frac{528 \bep \zeta \Sm^2}{\gam}   -  \frac{324}{5} \zeta \gam \Sm^2 - 36 \zeta \phi_0 \Sm^2 - 18 \zeta \gam \phi_0 \Sm^2 + \frac{826}{5} \zeta^2 \Sm^4 -  \frac{4032 \bem^2 \zeta^2 \Sm^4}{\gam^3} -  \frac{4032 \bep^2 \zeta^2 \Sm^4}{\gam^3}  \nn\\
&\qquad  + \frac{1008 \bep \zeta^2 \Sm^4}{\gam^2}  -  \frac{576 \bep \zeta^2 \Sm^4}{\gam} + \frac{1152 \bem \bep \zeta \Sm \Sp}{\gam^2} -  \frac{384 \bem \zeta \Sm \Sp}{\gam} -  \frac{9216 \bem \bep \zeta^2 \Sm^3 \Sp}{\gam^3}  \nn\\
&\qquad  + \frac{1152 \bem \zeta^2 \Sm^3 \Sp}{\gam^2} -  \frac{576 \bem \zeta^2 \Sm^3 \Sp}{\gam} + \frac{118}{5} \zeta^2 \Sp^4 -  \frac{576 \bem^2 \zeta^2 \Sp^4}{\gam^3} -  \frac{576 \bep^2 \zeta^2 \Sp^4}{\gam^3} + \frac{144 \bep \zeta^2 \Sp^4}{\gam^2} \bigg]   \nn\\
&\quad +   \nu \delta  \bigg[ \frac{14}{5} \zeta \gam \Sm \Sp + \frac{28}{5} \zeta^2 \Sm^3 \Sp + \frac{28}{5} \zeta^2 \Sm \Sp^3 \bigg]  + \nu^2 \bigg[ \gam^2 - 4 \zeta \gam \Sm^2 - 28 \zeta^2 \Sm^4 - 4 \zeta^2 \Sp^4 \bigg]
\end{align}\ese
 
\newpage
\bibliography{references.bib}

\end{document}